\documentclass[fleqn,usenatbib,useAMS]{mnras}

\usepackage{graphicx}	
\usepackage{amsmath}	
\usepackage{amssymb}	
\usepackage{multicol}
\usepackage{bm}		
\usepackage{pdflscape}	
\usepackage{tabularx}
\usepackage{subcaption}
\usepackage{xfrac}

\usepackage[T1]{fontenc}
\usepackage{ae,aecompl}

\usepackage{newtxtext,newtxmath}

\title[Membership by machine learning methods]{Detection of Open Cluster Members Inside and Beyond Tidal Radius by Machine Learning Methods Based on Gaia DR3}

\author[M. Noormohammadi]{
M. Noormohammadi,$^{1,2}$\thanks{E-mail: mnoormohammadi@aut.ac.ir}
M. Khakian Ghomi,$^{1}$\thanks{E-mail: khakian@aut.ac.ir}
and A. Javadi,$^{2}$\thanks{E-mail: atefeh@ipm.ir}
\\
$^{1}$Physics and Energy Engineering Department, Amirkabir University, Tehran, IRAN\\
$^{2}$School of Astronomy, Institute for Research in Fundamental Sciences (IPM), P.O.Box 1956836613, Tehran, IRAN\\
}

\date{Accepted XXX. Received YYY; in original form ZZZ}

\pubyear{2022}

\begin{document}
\label{firstpage}
\pagerange{\pageref{firstpage}--\pageref{lastpage}}
\maketitle
\color{black}
\begin{abstract}
\noindent
In our previous work, we introduced a method that combines two unsupervised algorithms: DBSCAN and GMM. We applied this method to 12 open clusters based on Gaia EDR3 data, demonstrating its effectiveness in identifying reliable cluster members within the tidal radius. However, for studying cluster morphology, we need a method capable of detecting members both inside and outside the tidal radius. By incorporating a supervised algorithm into our approach, we successfully identified members beyond the tidal radius. In our current work, we initially applied DBSCAN and GMM to identify reliable members of cluster stars. Subsequently, we trained the Random Forest algorithm using DBSCAN and GMM-selected data. Leveraging the random forest, we can identify cluster members outside the tidal radius and observe cluster morphology across a wide field of view. Our method was then applied to 15 open clusters based on Gaia DR3, which exhibit a wide range of metallicity, distances, members, and ages. Additionally, we calculated the tidal radius for each of the 15 clusters using the King profile and detected stars both inside and outside this radius. Finally, we investigated mass segregation and luminosity distribution within the clusters. Overall, our approach significantly improved the estimation of the tidal radius and detection of mass segregation compared to previous work. We found that in Collinder 463, low-mass stars do not segregate in comparison to high-mass and middle-mass stars. Additionally, we detected a peak of luminosity in the clusters, some of which were located far from the center, beyond the tidal radius.
\end{abstract}

\begin{keywords}
methods: data analysis-methods: statistical-open clusters and associations: general-stars: kinematics and dynamics
\end{keywords}

\section{INTRODUCTION} \label{section.intro}
According to accepted theories, stars are born within a single molecular cloud as a cluster. As a result, cluster members share the same physical parameters and chemical elements. Additionally, there exists an interaction between the Galaxy and clusters that affects cluster formation and morphology. To gain a comprehensive understanding of a cluster, including aspects such as the initial and present mass function, cluster morphology, planet formation theories, tidal tails, and interactions between galaxies and clusters, we must identify not only members within the tidal radius but also those outside it, such as cluster escape members~(\cite{Freeman}, \cite{friel1995}, \cite{hooger}, \cite{deBruijne}).\\
Several theories have been proposed to describe the birth and formation of stars within clusters, such as the hierarchical theory~(\cite{hierarchical}) or the centered formation theory~(\cite{lada2023}). By studying the morphology of clusters in a wide field of view, we can determine which theory is more accurate than the others~(\cite{extended-hu}).
Meanwhile, reliable cluster membership allows for the determination of the mass distribution of stars and the fraction of binary star systems within the cluster. This information can then be compared with simulation methods, such as N-body simulations~(\cite{mass-extended}, \cite{maiz-mass}, \cite{Olivares-dance}, \cite{mass-ob}).\\
Extended stellar coronae and tidal tails play an important role in the study of cluster formation, evolution, and interactions between galaxies and clusters. To achieve this, we need to study clusters across a wide field of view, covering distances of up to hundreds arcmins~(\cite{tidal-baharaja}, \cite{extended-Meingast}, \cite{extended-Tarricq}, \cite{extended-Jerabkova}, \cite{extended-tang}, \cite{extended-ye}, \cite{extended-hu}, \cite{extended-Lodieu}, \cite{extended-ngc752}).
The first and most crucial step in the study of star clusters is to identify reliable members. To achieve this goal, we require accurate and comprehensive data, along with methods that can work with this data and yield robust results. Membership determination within a star cluster occurs through two primary approaches: astrometric and photometric parameters~(\cite{Janez}, \cite{Kraus}, \cite{Gonz}, \cite{Krone-Martins}). Because stars within clusters originate from a common interstellar cloud, they share the same astronomical characteristics such as position, parallax, and proper motion. Additionally, these stars show a clear main sequence and, in the case of an old cluster, red giant branches.\\
In the current century, one of the popular and powerful methods that can identify relevant patterns within large datasets is machine learning. To achieve high accuracy, machine learning algorithms require data with high precision. The Gaia data release contains information about billions of stars in our galaxy, with high-accuracy astrometry and photometry parameters. Many studies have been conducted using machine learning methods based on the Gaia data release to identify members of star clusters, some of which are mentioned here:
\cite{cg2018} used Kmeans and UPMASK, \cite{gao-Pleiades}, \cite{gao-ngc6405}, \cite{gao-m67}, \cite{gaoclusters} used GMM and Random Forest, \cite{wilkinson-dbscan}, \cite{Morphology} used DBSCAN,\cite{knn-gmm} used KNN-GMM, \cite{noormohammadi} used DBSCAN and GMM. All these works filtered data in some way.\\
In our previous work~(\cite{noormohammadi}), we identified reliable cluster members using a combination of two unsupervised machine learning algorithms: DBSCAN and GMM. The process involved three steps. First, the data were filtered based on astrometric and photometric conditions. Next, DBSCAN identified reliable candidates using proper motion and parallax information. Finally, GMM detected reliable members from the candidates based on their position, parallax, and proper motion. We compared our method with other machine learning methods based on Gaia DR2, because those methods were applied to Gaia DR2~(\cite{cg2018}, \cite{gao-m67}, \cite{knn-gmm}). We showed that our method detected cluster members better than other methods in the cluster-dense region. Some of the members detected by DBSCAN indicated a low probability of membership by GMM because they lay outside of the cluster-dense region. Additionally, some of these outer members lie within the range of proper motion, parallax, and CMD of GMM’s high probability detection members. To identify these members, some of whom could be considered escape members, we introduce a method that combines three algorithms: DBSCAN, GMM, and Random Forest. This method can find members within a large field of view of a cluster and detect not only the cluster members but also the cluster escape members, thus presenting a better view of the cluster morphology.\\
In this work, we applied our method to 15 open clusters: nine of them were in previous work (under Gaia EDR3), and six of them are new. In Section~\ref{section.Data}, the data conditions for 15 open clusters are explained. In Section~\ref{section.Method}, the method was explained with a focus on a new step. The results in each step are shown in Section~\ref{result.Method}. In Section~\ref{descotion.Method}, we discussed our results by determining the tidal radius, studying mass segregation, and analyzing cluster luminosity. Finally, in Section~\ref{con.Data}, we summarized our work.
\section{DATA} \label{section.Data}
In 2013, Gaia was launched to provide comprehensive information about stars in the Milky Way. The first release of Gaia data (Gaia DR1) contained around 1.14 billion data sources, with more than 2 million having full astrometric parameters. The second release of Gaia data (Gaia DR2) included around 1.62 billion stars, with more than 1.33 billion having full astrometry parameters. In 2020, Gaia published the latest edition of data, which encompassed around 1.8 billion stars, with more than 1.46 billion having full astrometric parameters~\url{https://www.cosmos.esa.int/web/gaia/dr3}. The accuracy of astrometric and photometric parameters in Gaia Data Release 3 is shown in Table~\ref{uncertainties.tab}. As shown in Table~\ref{uncertainties.tab}, stars brighter than 20 magnitudes have uncertainties below 0.5 for astrometric parameters. However, by increasing the magnitude from 20 mag to 21 mag, uncertainties grow to higher than 1.00 magnitude.\\ 
The last version of the Gaia data release (GDR3) is used in this work. For high accuracy, all stars analyzed were brighter than 20 magnitudes and met the condition of completeness in position parameters (RA, DEC), proper motion (pmRA, pmDEC), parallax, G magnitude, and Bp-Rp color index.\\
Data from 15 open clusters were obtained in the Gaia Data Release\,3~(\cite{gdr3}). These clusters include NGC\,2099, M\,67, M\,41, M\,48, M\,38, M\,47, Alissi\,01, Melotte\,18, King\,06, NGC\,2343, NGC\,188, Collinder\,463, M\,34, M\,35, and NGC\,752. These clusters exhibit a variety of properties in terms of age, metallicities, and number of members, which allows for a proper evaluation of the method. For this analysis, stars within a radius of 300 arcminutes for NGC\,752 and 150 arcminutes for the other clusters, with positive parallax and magnitude brighter than 20 mag, were selected. These distances provide a wide field of view from the center of the clusters, complete information about clusters, and the ability to analyze the method in the best way.\\
Among the 15 open clusters, 9 already existed in the previous study based on Gaia EDR3~(\cite{noormohammadi}), and 6 of them are new in this study. Collinder\,463 is a poor open cluster and has an age and distance of about 270 Myr~(\cite{collinder649-bossini}), $880\pm60$ pc(\cite{collinder469-semionov}) respectively.~\cite{collinder469-semionov} studied members of Collinder\,469 and the halo based on Gaia DR2. NGC\,188 is the oldest and richest cluster that has variable stars, an X-ray binary system, and an age of around 7 Gyr~(\cite{ngc188-Gondoin2005}, \cite{ngc188-zhang2002}). M\,47 is comparable to Pleiades and has some active X-ray sources and an age of about 100 Myr~(\cite{m47-narbera}, \cite{m47-Prisinzano}). NGC\,2443 is an intermediate-age open cluster that has lithium-rich stars and giant planets and an age of about 750 Myr~(\cite{ngc2423-carlberg2016}, \cite{ngc2243-lovis2007}). Melotte\,72 is a compressed small cluster and has an age of about $1$ Gyr and a distance of 3175 pc and it is dynamically relaxed ~(\cite{melotte72-2021}, \cite{melotte72-hendi}).\\
A radius of 150 arcmin for all these clusters contains member stars and a high fraction of escape members, making it a suitable value for the search radius. 
\begin{table}
\centering
\caption{Data uncertainties in Gaia DR3}
\subcaption*{Astrometric data uncertainties}
\begin{tabular}{ccccc}
    \hline
    \hline
    Name & $G<15$ & $G=17$ & $G=20$ & $G=21$ \\
    \hline
    Position& $0.01-0.02$\,mas & $0.05$\,mas & $0.4$\,mas & $1.0$\,mas  \\
    Proper Motion & $0.02-0.03$\,mas\,yr$^{-1}$ & $0.07$\,mas\,yr & $0.5$\,mas\,yr& $1.4$\,mas\,yr\\
    Parallax & $0.02-0.03$\,mas & $0.07$\,mas & $0.5$\,mas & $1.3$\,mas\\
    \hline
  \end{tabular}
\bigskip
\subcaption*{Photometric data uncertainties}
\begin{tabular}{cccc}
    \hline
    \hline
    Name & $G<13$ & $G=17$ & $G=20$ \\
    \hline
    $G$ band & $0.3$\,mmag &$1$\,mmag &$6$\,mmag \\
    $G_{BP}$ band & $0.9$\,mmag &$12$\,mmag &$108$\,mmag\\
    $G_{RP}$ band & $0.6$\,mmag &$6$\,mmag &$52$\,mmag\\
    \hline
  \end{tabular}
  \label{uncertainties.tab}
\end{table}
\section{METHOD} \label{section.Method}
In this work, three machine-learning algorithms are used to identify star cluster members and stars that are outside the tidal radius. In the previous work~(\cite{noormohammadi}), a machine learning method was presented to identify reliable members of 12 open clusters based on the Gaia EDR3. In this work, we developed our method by adding one supervised algorithm in order to detect members beyond the cluster dense region. This new method is formed with three steps, each of them has been described in the flow:
\subsection{DBSCAN} 
DBSCAN is an unsupervised algorithm that can identify different clusters in one sample source. This algorithm has two essential parameters (input parameters) for detecting data: MinPts and Eps. The algorithm considers a circle with a radius based on Eps centered on each data point and calculates the data inside the circle. If the number of data points inside the circle is higher than MinPts, this centered data is considered a core point. Otherwise, if the data point belongs to the circle at the center of one core point, it is considered a border point; if not, it is considered noise. Before applying the algorithm to clusters, all data were normalized using the scale function from the scikit-learn library~\url{https://scikit-learn.org/stable/modules/generated/sklearn.preprocessing.scale.html}.
In the first step, DBSCAN selected candidate members in the region around the star cluster using three parameters (proper motion in RA and DEC, and Parallax). Selection of star candidates by DBSCAN causes an increased rate of cluster members compared to field stars (signal to noise). As DBSCAN has two free parameters (MinPts, Eps), we have the freedom to adjust the signal-to-noise ratio. Data detected using DBSCAN were analyzed in terms of proper motion and the CMD (Color-Magnitude Diagram) for each cluster. Having observed some indications of proper motion and the CMD of the cluster, the detection data are sent to the next step. In this work, DBSCAN could perform well with large data sizes in three dimensions.\\
\subsection{GMM (Gaussian mixture models)} 
The output of DBSCAN is used as input for GMM (Gaussian Mixture Model), which prepares the data source based on the conditions of the GMM algorithm. The GMM algorithm can detect data that have the same Gaussian distribution if the data satisfy three conditions: 1) using accurate data, 2) the rate of signal to noise must be significant, and 3) the structure of clusters among field stars must be indicated. Because of these conditions, some of the work eliminates huge volumes of data by filtering based on conditions such as astrometric parameters. However, in this work, we achieve this by using DBSCAN in the first stage.\\ 
In the next stage, the GMM algorithm was applied to 5 parameters: position in RA and DEC, proper motion in RA, DEC, and Parallax. Before applying the algorithm to clusters, all data were normalized using the scale function from the scikit-learn library~\url{https://scikit-learn.org/stable/modules/generated/sklearn.preprocessing.scale.html}. At the final stage, we analyzed the members that were detected by GMM based on proper motion and the CMD. If the selected data were without contamination (such as field stars), we returned to the DBSCAN step, increased the value of MinPts and Eps, and then applied GMM again. We must be cautious, as continuing this process may still result in contaminated data. The threshold represents the appropriate value for MinPts and Eps in DBSCAN, detecting the maximum number of reliable cluster members and optimally eliminating field stars.
Since the GMM algorithm was applied to position parameters (RA, DEC), some of the outer members were eliminated automatically. Some of this eliminated data lie in the range of proper motion and parallax of cluster members and are also consistent with the CMD of cluster members. \\
\subsection{Random Forest} 
\begin{figure}
\centering
\captionsetup[subfigure]{labelformat=empty}
\begin{subfigure}{0.25\textwidth}
        \centering
           \includegraphics[width=\textwidth]{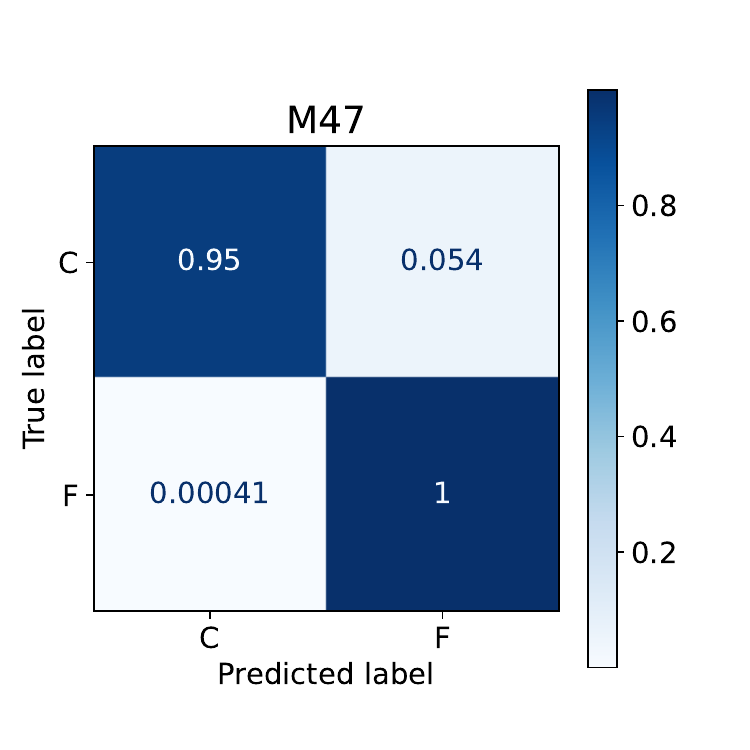}

        \end{subfigure}
        \begin{subfigure}{0.25\textwidth}

                \centering
                \includegraphics[width=\textwidth]{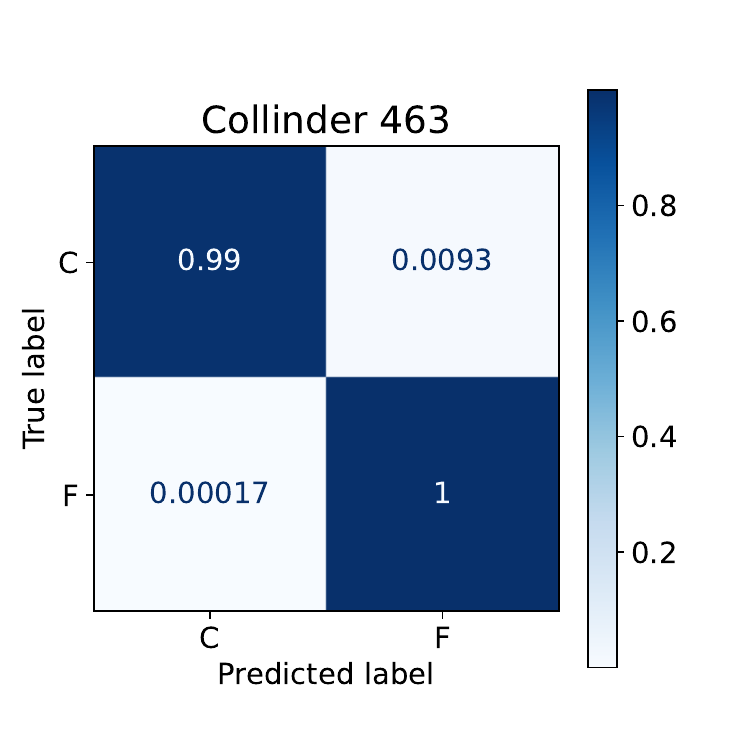}

        \end{subfigure}
        \begin{subfigure}{0.25\textwidth}
                \centering
           \includegraphics[width=\textwidth]{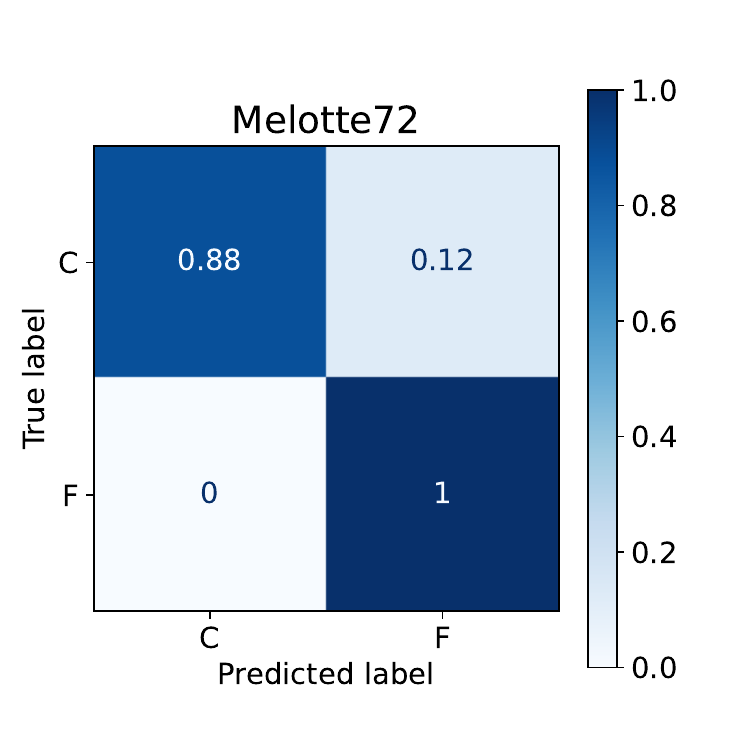}

        \end{subfigure}
        \begin{subfigure}{0.25\textwidth}
                \centering
                \includegraphics[width=\textwidth]{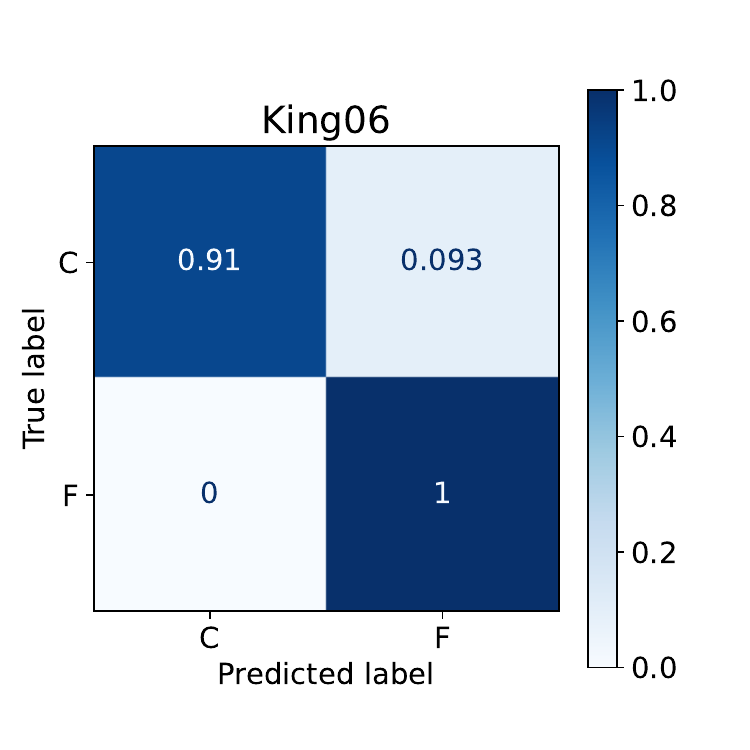}

        \end{subfigure}
        \begin{subfigure}{0.25\textwidth}
                \centering
                \includegraphics[width=\textwidth]{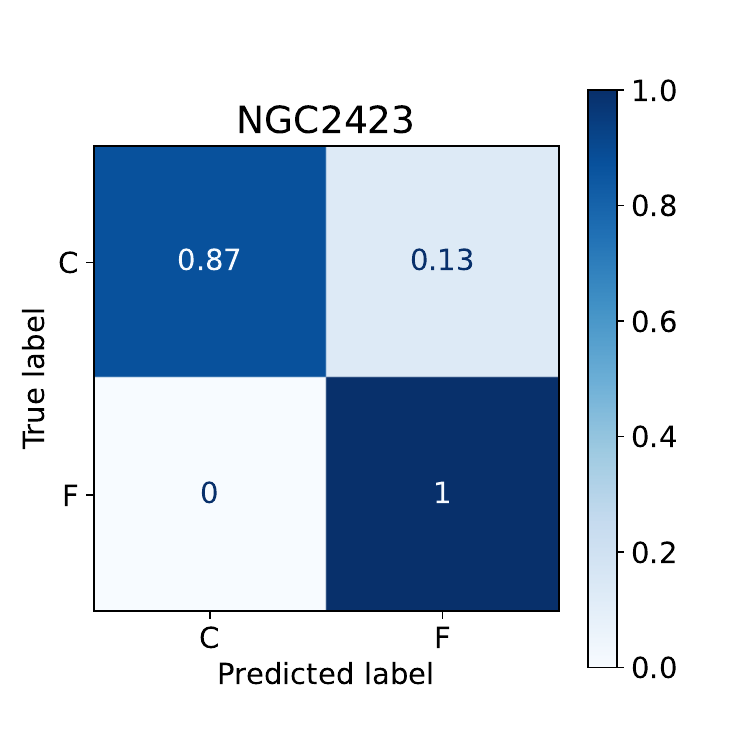}

        \end{subfigure}
        \begin{subfigure}{0.25\textwidth}
                \centering

                \includegraphics[width=\textwidth]{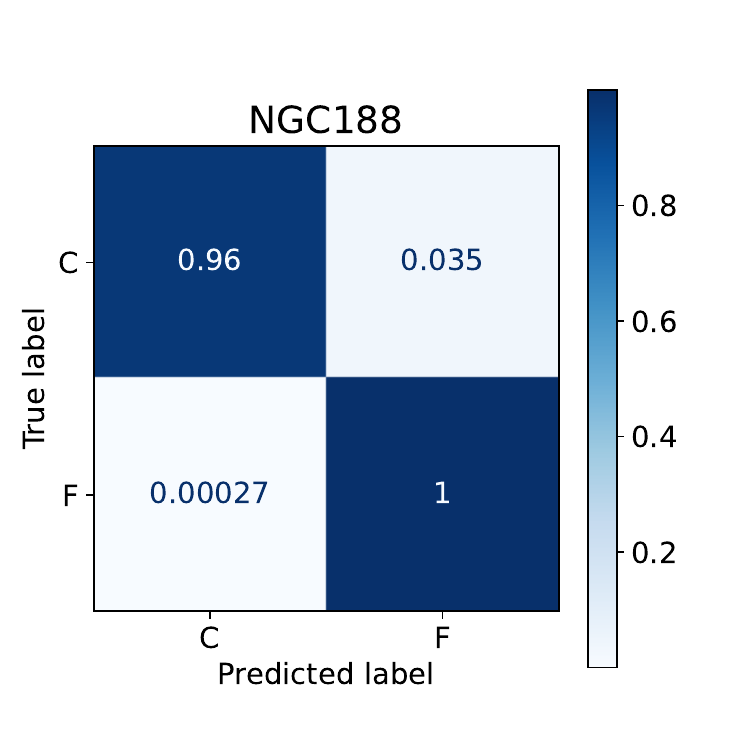}

        \end{subfigure}
        \begin{subfigure}{0.25\textwidth}
        \centering
           \includegraphics[width=\textwidth]{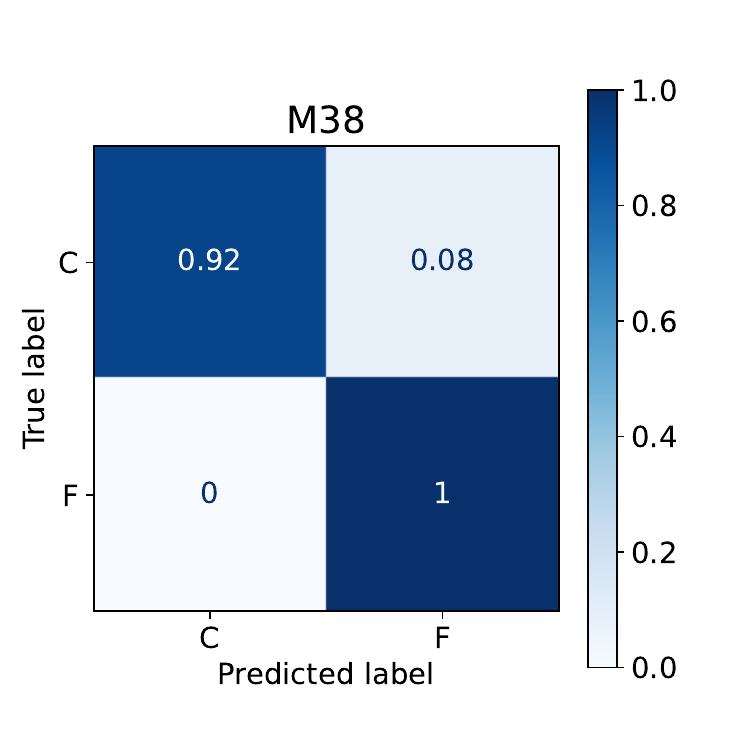}

        \end{subfigure}
        \begin{subfigure}{0.25\textwidth}
        \centering
           \includegraphics[width=\textwidth]{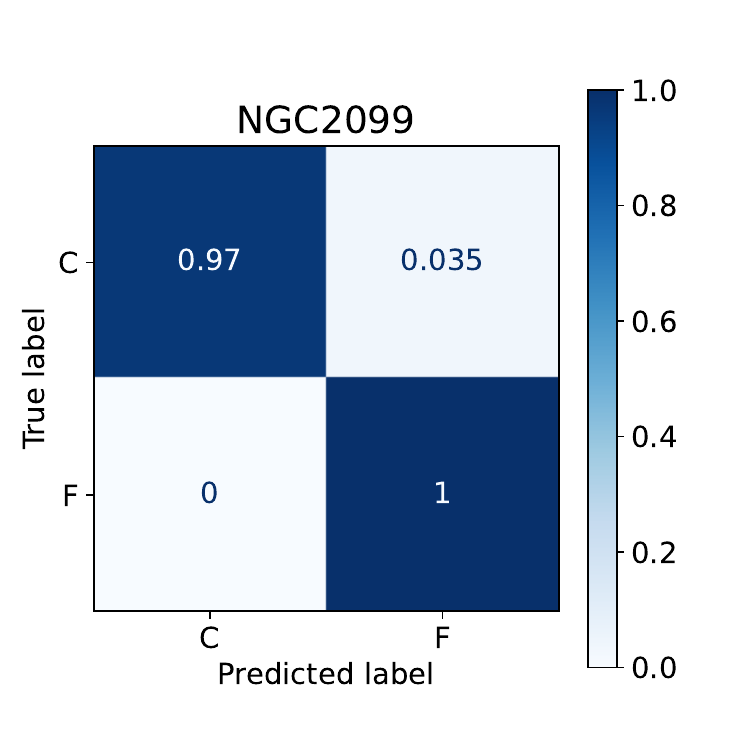}

        \end{subfigure}
        \begin{subfigure}{0.25\textwidth}
        \centering
           \includegraphics[width=\textwidth]{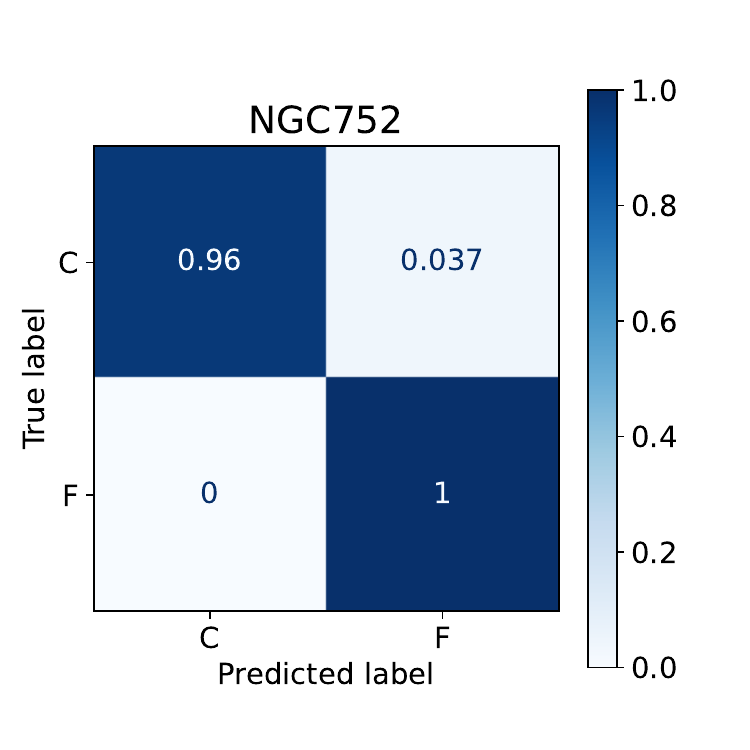}

        \end{subfigure}
        \begin{subfigure}{0.25\textwidth}
        \centering
           \includegraphics[width=\textwidth]{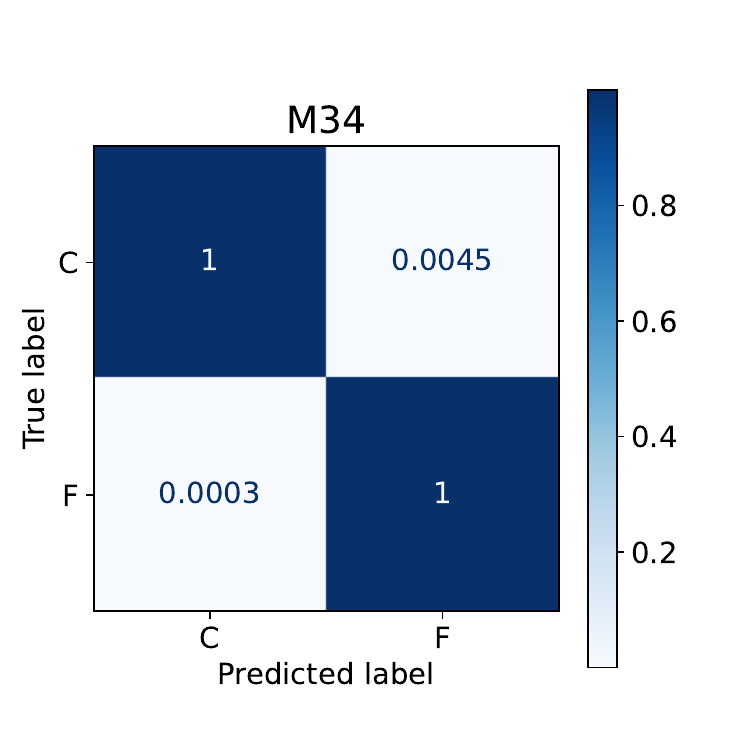}

        \end{subfigure}
        \begin{subfigure}{0.25\textwidth}
        \centering
           \includegraphics[width=\textwidth]{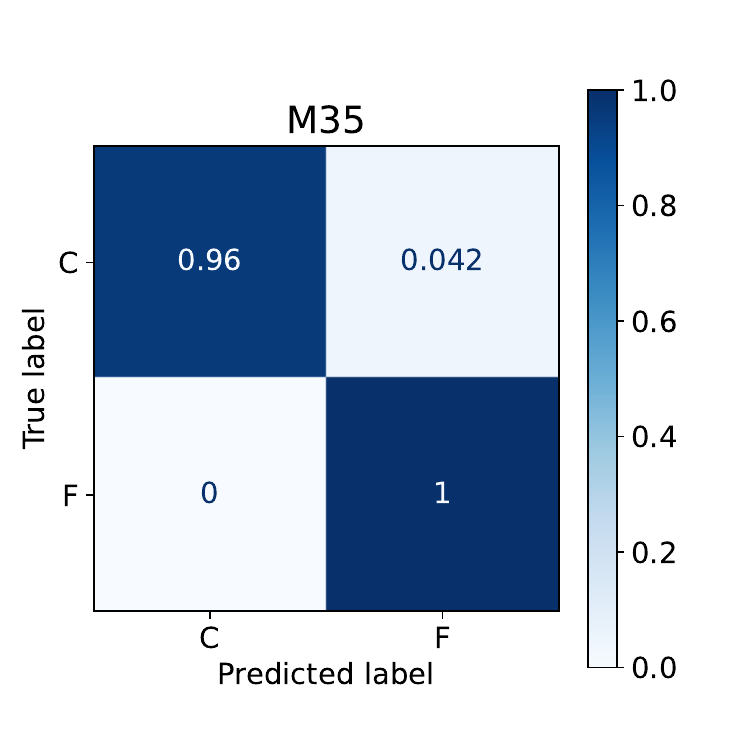}

        \end{subfigure}
        \begin{subfigure}{0.25\textwidth}
        \centering
           \includegraphics[width=\textwidth]{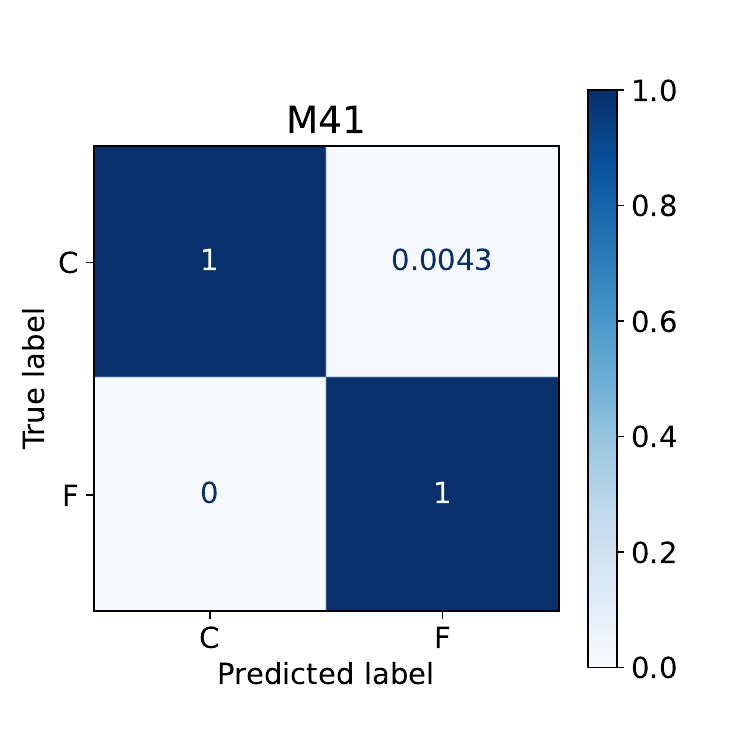}

        \end{subfigure}
        \begin{subfigure}{0.25\textwidth}
        \centering
           \includegraphics[width=\textwidth]{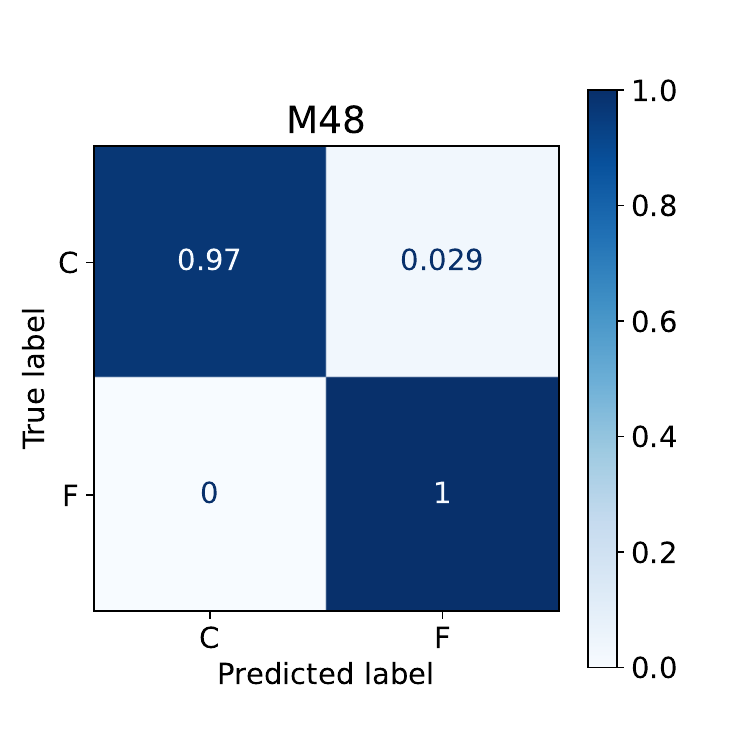}

        \end{subfigure}
        \begin{subfigure}{0.25\textwidth}
        \centering
           \includegraphics[width=\textwidth]{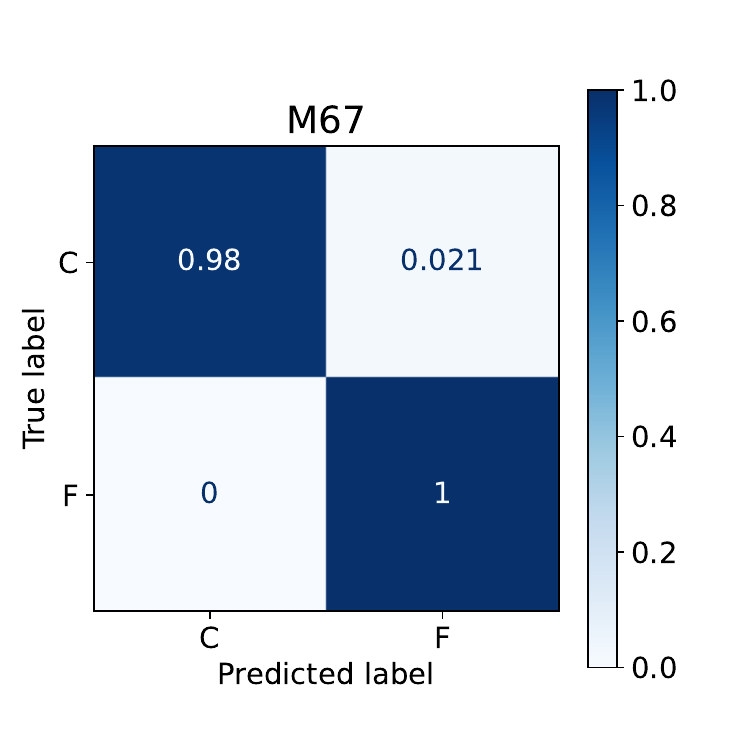}

        \end{subfigure}
        \begin{subfigure}{0.25\textwidth}
        \centering
           \includegraphics[width=\textwidth]{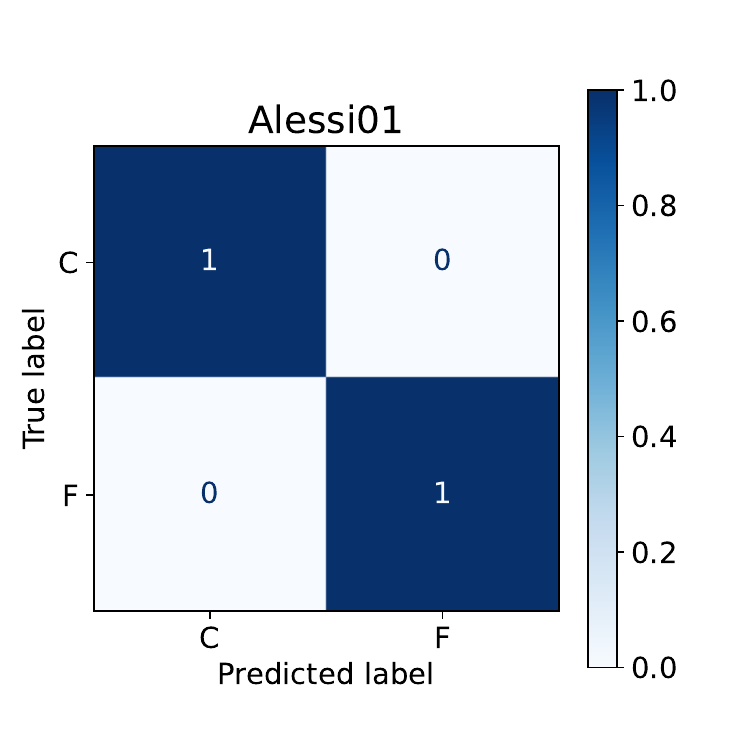}

        \end{subfigure}

  \caption{Confusion matrix. C is for cluster members and F is for field stars.}
  \label{cmatrix.fig}
\end{figure}

In this work, after reliable cluster members were found by DBSCAN and GMM, the Random Forest algorithm was used for detecting outer members that lay in the range of proper motion and parallax of cluster members and matched their CMD. Random Forest can analyze astrometric and photometric parameters and does not need to normalize data. At this stage, we can identify data points that may correspond to escaping members within the cluster. These data points typically reside in the outer layer of the cluster. Additionally, this step provides us with the optimal field of view for observing the cluster. This field of view reveals the morphology of the cluster both inside and outside the tidal radius. In the next step, the data was divided into three samples:\\
1. Data that were not detected by DBSCAN were considered as field stars. To obtain suitable data for training the Random Forest algorithm, we filtered field stars based on the range of parallax values among cluster members. This range was determined based on detected members by GMM with a probability higher than 0.8. Selection of the range of parallax is higher than the maximum parallax value among cluster members and lower than the minimum parallax value, except for Alessi01, which has few members. The details of the parallax range and the number of field stars used for training data are shown in the Table~\ref{results.tab}\\
2. The stars detected by DBSCAN but with a probability lower than 0.8 attributed by the GMM to them were considered as suspicious stars.\\
3. The stars that were detected by the GMM algorithm with a probability higher than 0.8 were considered as cluster members.\\
In step three, the Random Forest algorithm was trained using field stars and cluster members. We performed a train-test split using the train$_-$test split method from the sklearn.model$_-$selection library(\url{https://scikit-learn.org/stable/modules/generated/sklearn.model_selection.train_test_split.html}), with a 30 percent split (10 percent for Alessi01). Additionally, to investigate the best value for the random forest parameters, we calculated the F1$_-$score, which is shown in Table~\ref{results.tab}. We also analyzed the confusion matrix for each cluster, as depicted in Fig~\ref{cmatrix.fig}. The hyperparameters were chosen based on achieving high accuracy for the F1 score and the confusion matrix. After that, it was applied to suspicious stars based on five parameters: three astrometric (proper motion in RA, proper motion in DEC, and Parallax) and two photometric (G magnitude and Bp-Rp color index). The members detected by Random Forest should be evaluated in comparison with cluster members that have a probability higher than 0.8 based on proper motion, parallax, and CMD. If stars detected by Random Forest lay within the range of proper motion and parallax and on the CMD of high-probability cluster members (higher than 0.8), that were detected by GMM in five dimensions (RA, DEC, pmRA, pmDEC, Parallax), they were considered as members outside the tidal radius.\\
For all cluster we applied (n estimators=100, max depth=20, criterion=gini, random state=0) except King\,06 (n estimators=50, max depth=10, criterion=gini, random state=0) and for NGC\,2423 (n estimators=50, max depth=20, criterion=gini, random state=0) and for Melotte\,72 (n estimators=60, max depth=20, criterion=gini, random state=0). We analyzed detection data by Random Forest based on probability and selected proper data based on the Color magnitude diagram. Finally, we selected cluster member stars with a probability higher than 0.5 for Alessi\,01, King\,06, NGC\,752, M\,38, M\,41, M\,47, and M\,67 higher than 0.6 for NGC\,2423, 

\onecolumn

\begin{figure}
\centering
\captionsetup[subfigure]{labelformat=empty}
\begin{subfigure}{0.44\textwidth}
        \centering
           \includegraphics[width=\textwidth]{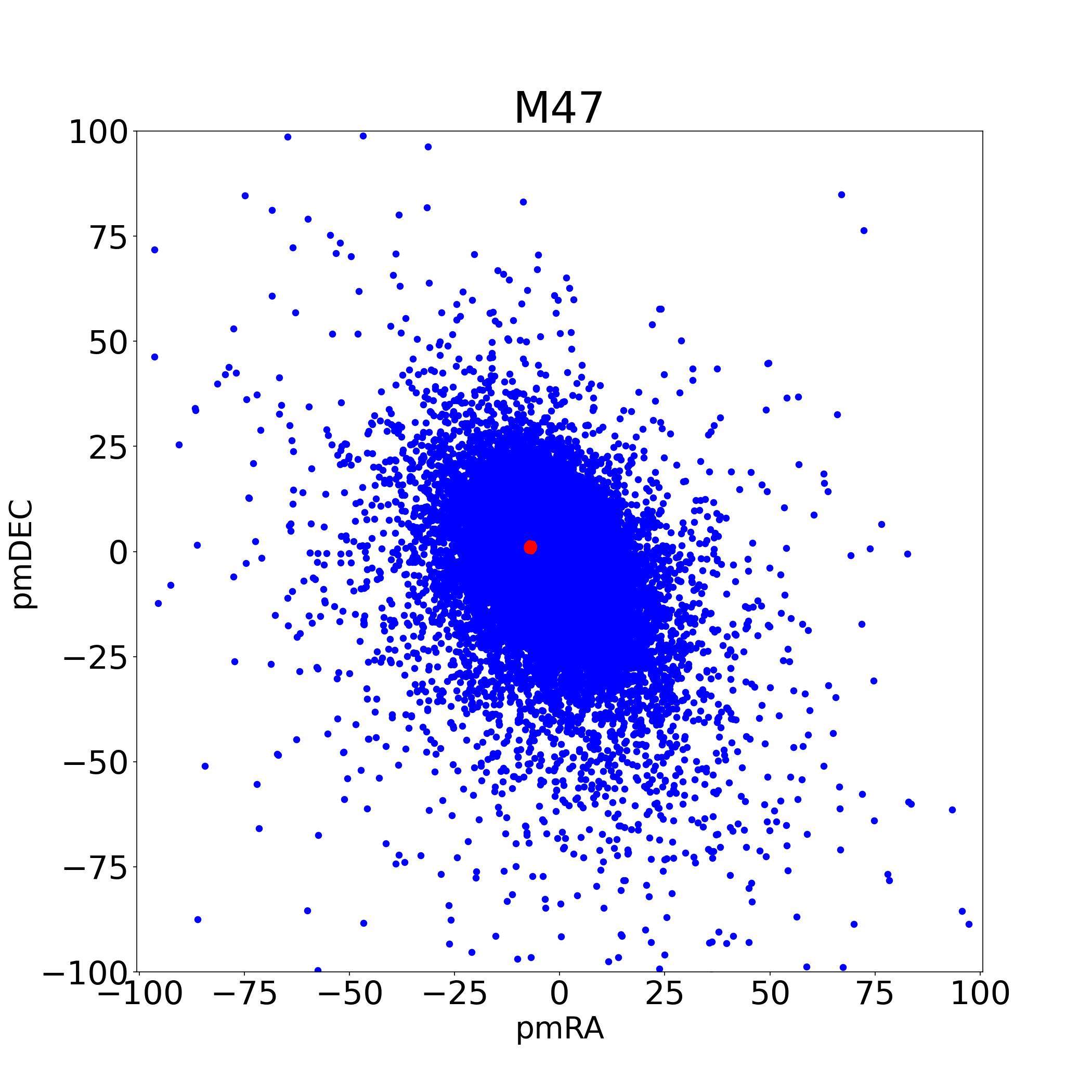}

        \end{subfigure}
        \begin{subfigure}{0.44\textwidth}

                \centering
                \includegraphics[width=\textwidth]{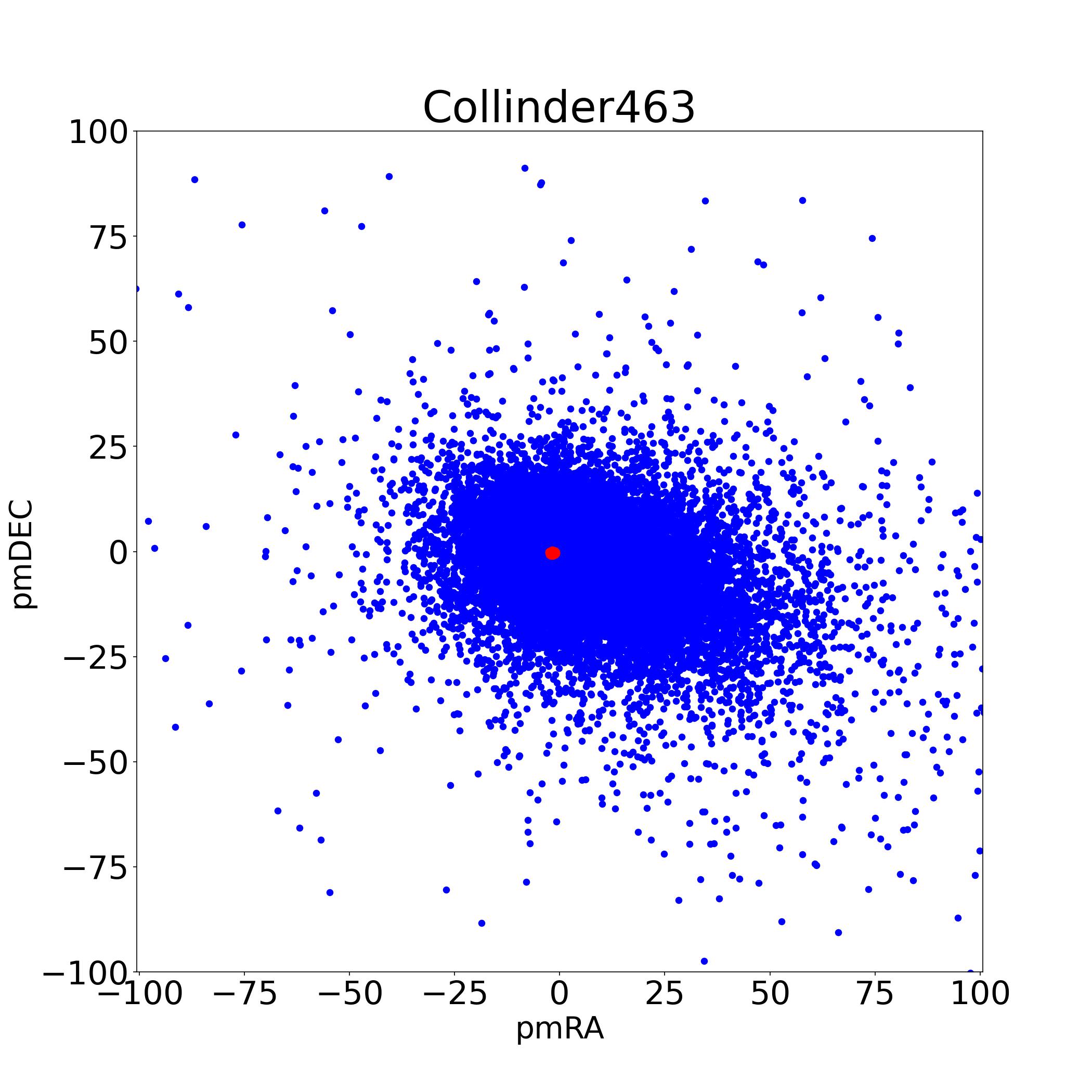}

        \end{subfigure}
        \begin{subfigure}{0.44\textwidth}
                \centering
           \includegraphics[width=\textwidth]{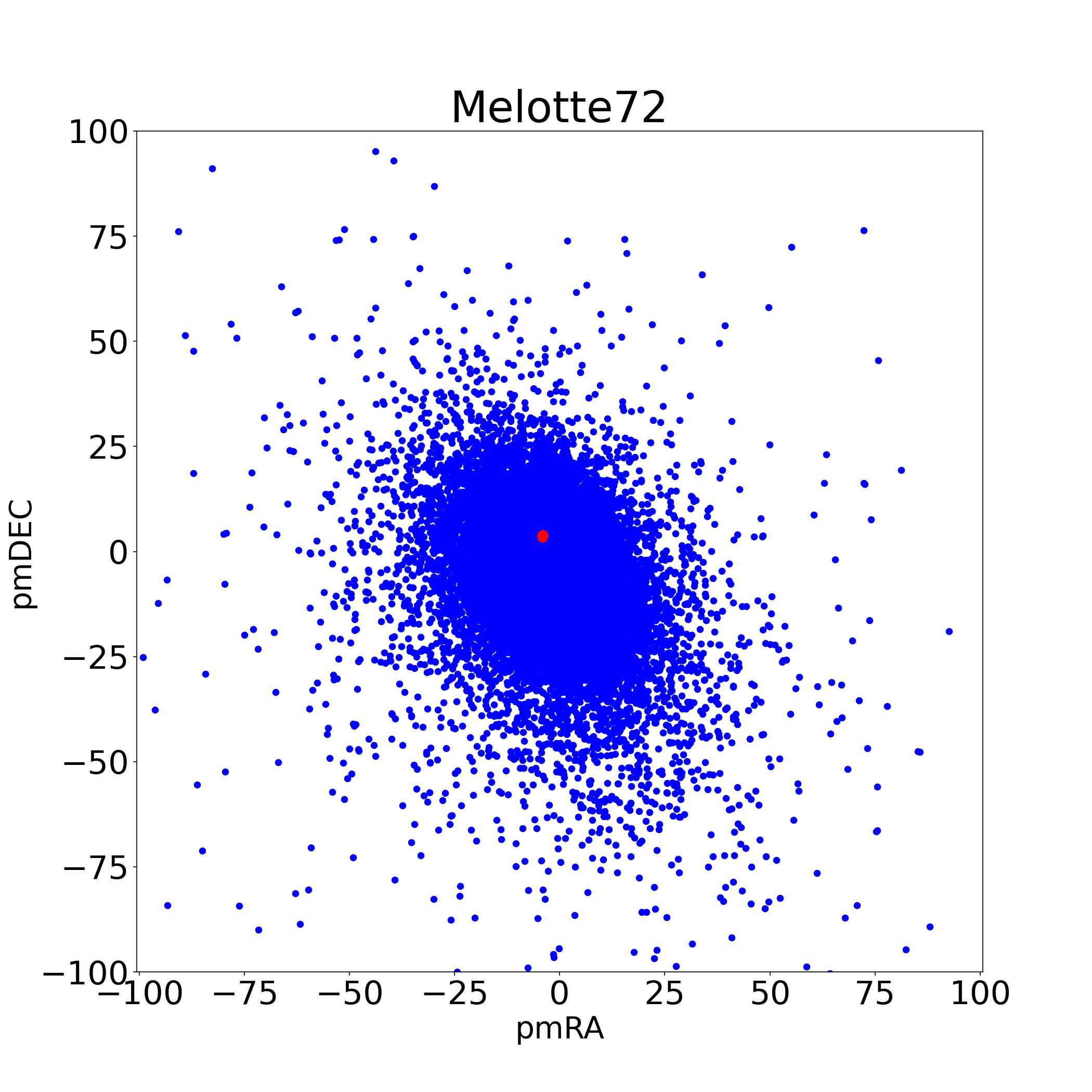}

        \end{subfigure}
        \begin{subfigure}{0.44\textwidth}
                \centering

                \includegraphics[width=\textwidth]{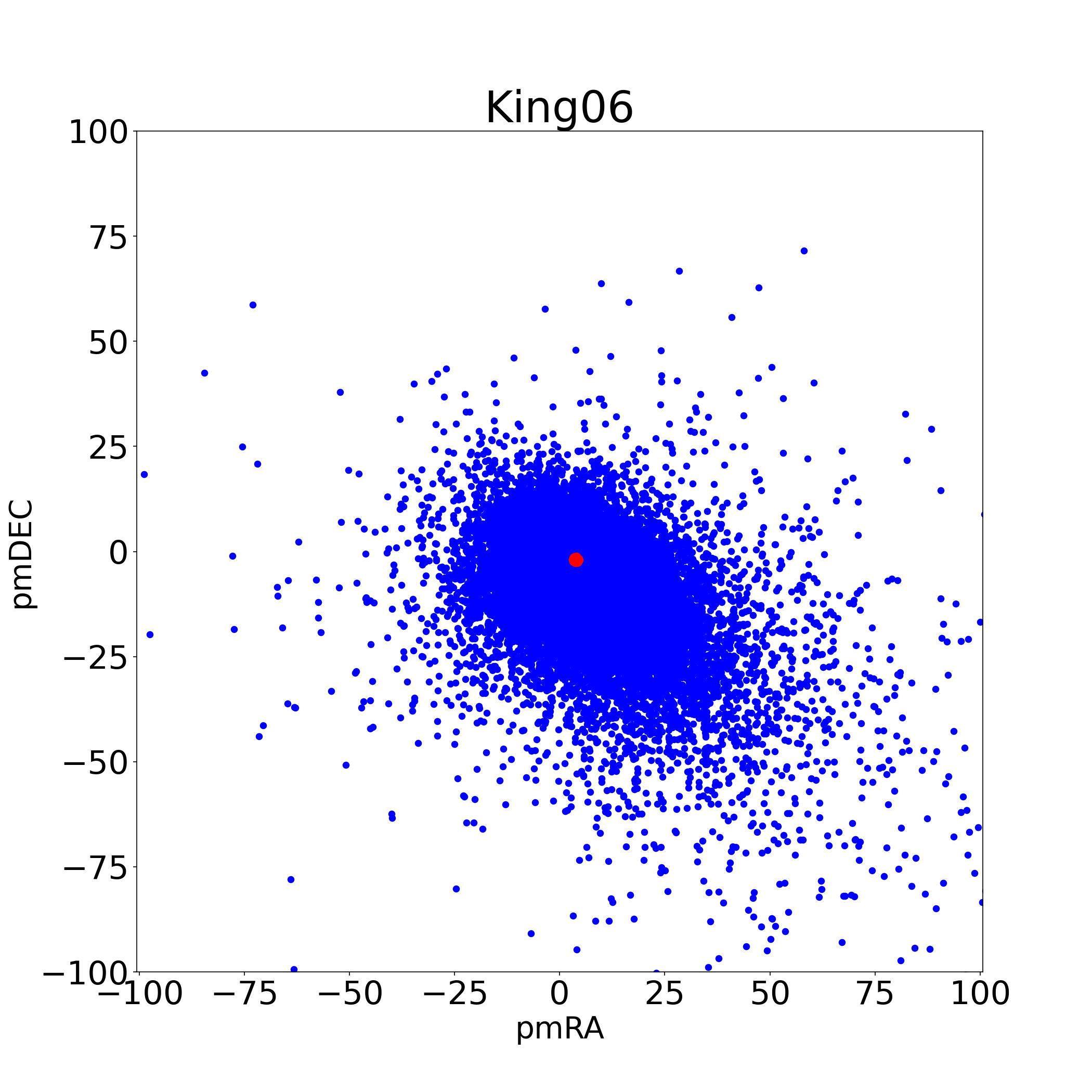}

        \end{subfigure}
        \begin{subfigure}{0.44\textwidth}
                \centering

                \includegraphics[width=\textwidth]{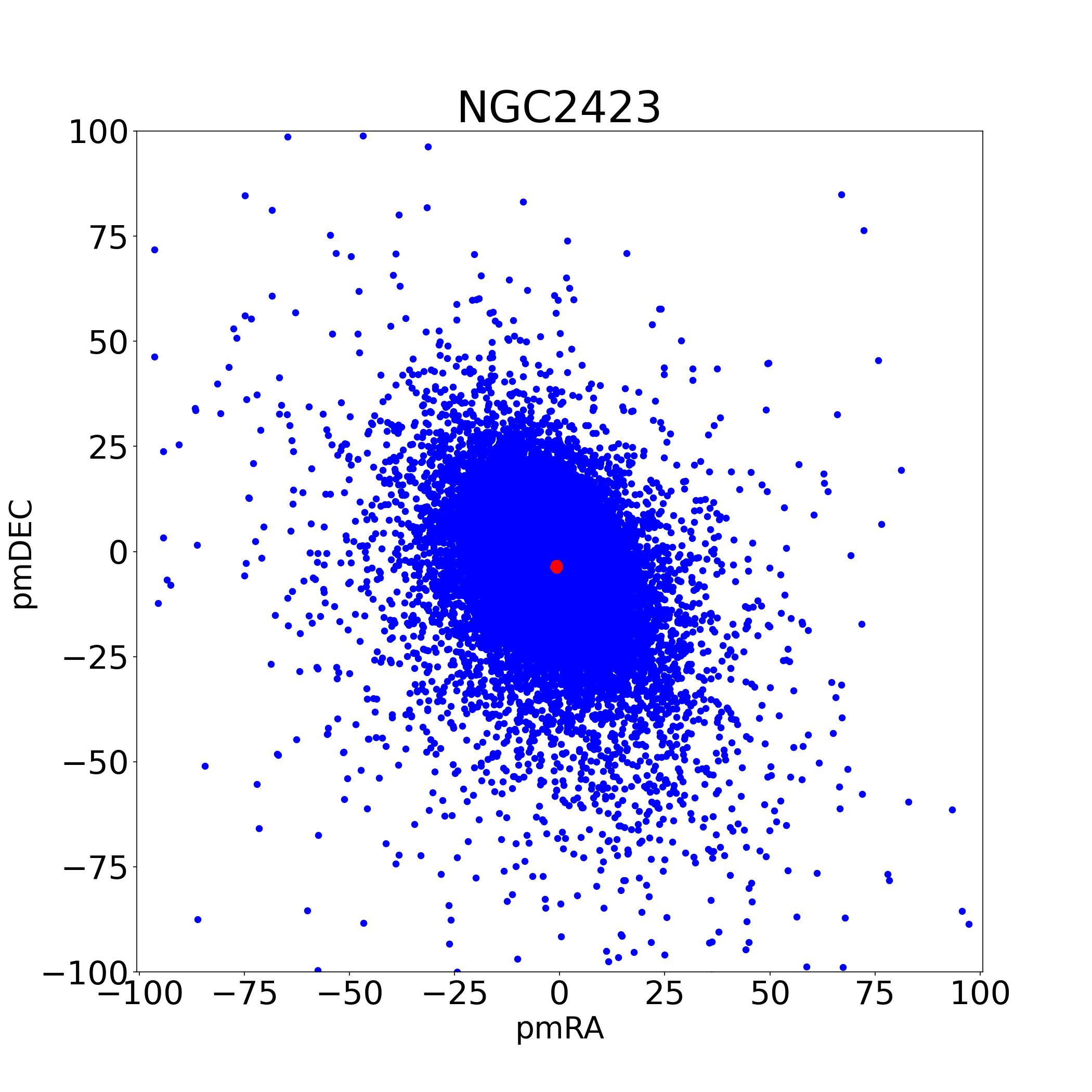}

        \end{subfigure}
        \begin{subfigure}{0.44\textwidth}
                \centering

                \includegraphics[width=\textwidth]{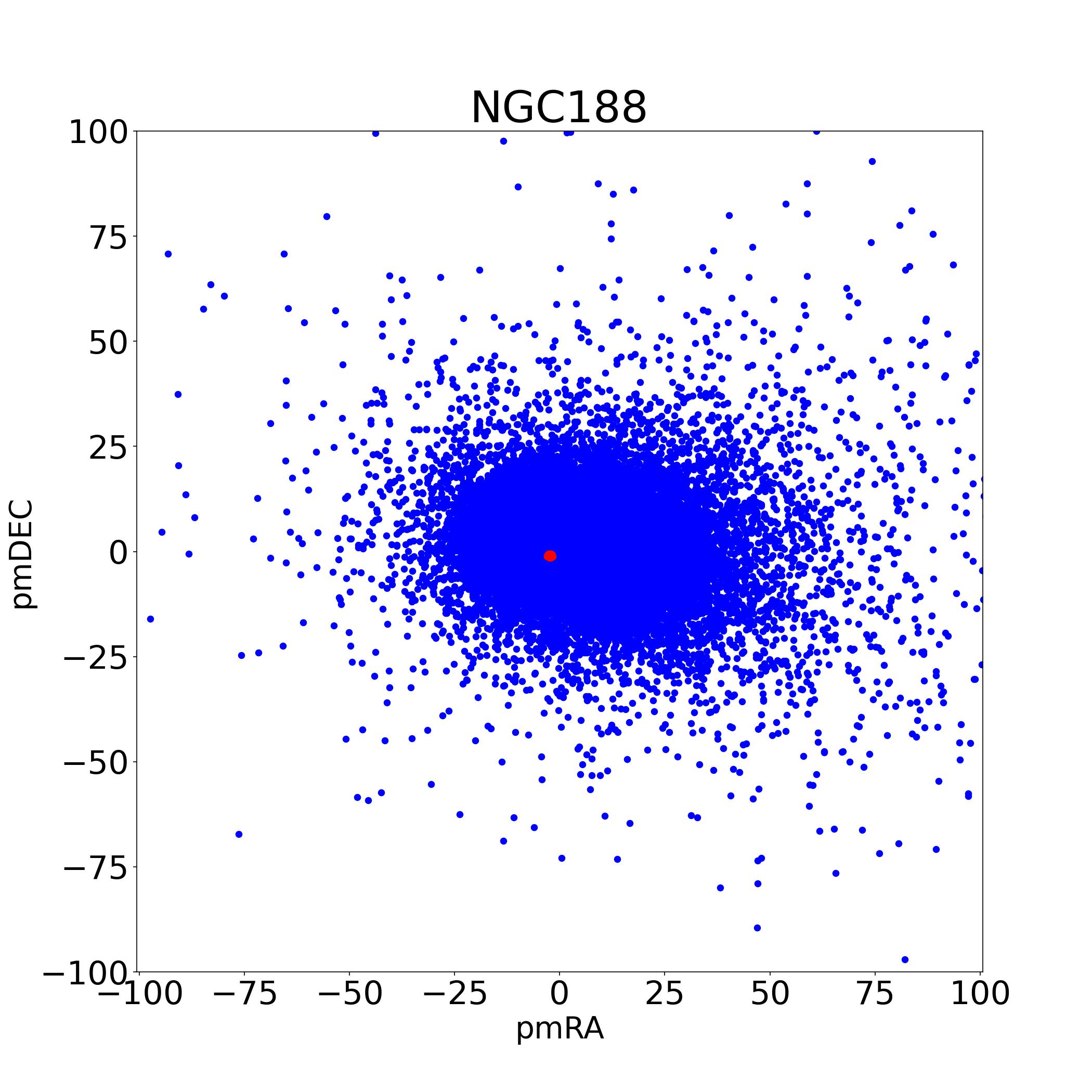}

        \end{subfigure}

  \caption{The proper motion of the DBSCAN-selected members among field stars. Grey dots show the field stars and blue dots show stars that were selected by DBSCAN}
  \label{proper motion of dbscan.fig}
\end{figure}

\begin{figure}
  \centering
  \captionsetup[subfigure]{labelformat=empty}
        \begin{subfigure}{0.43\textwidth}
        \centering

                \includegraphics[width=\textwidth]{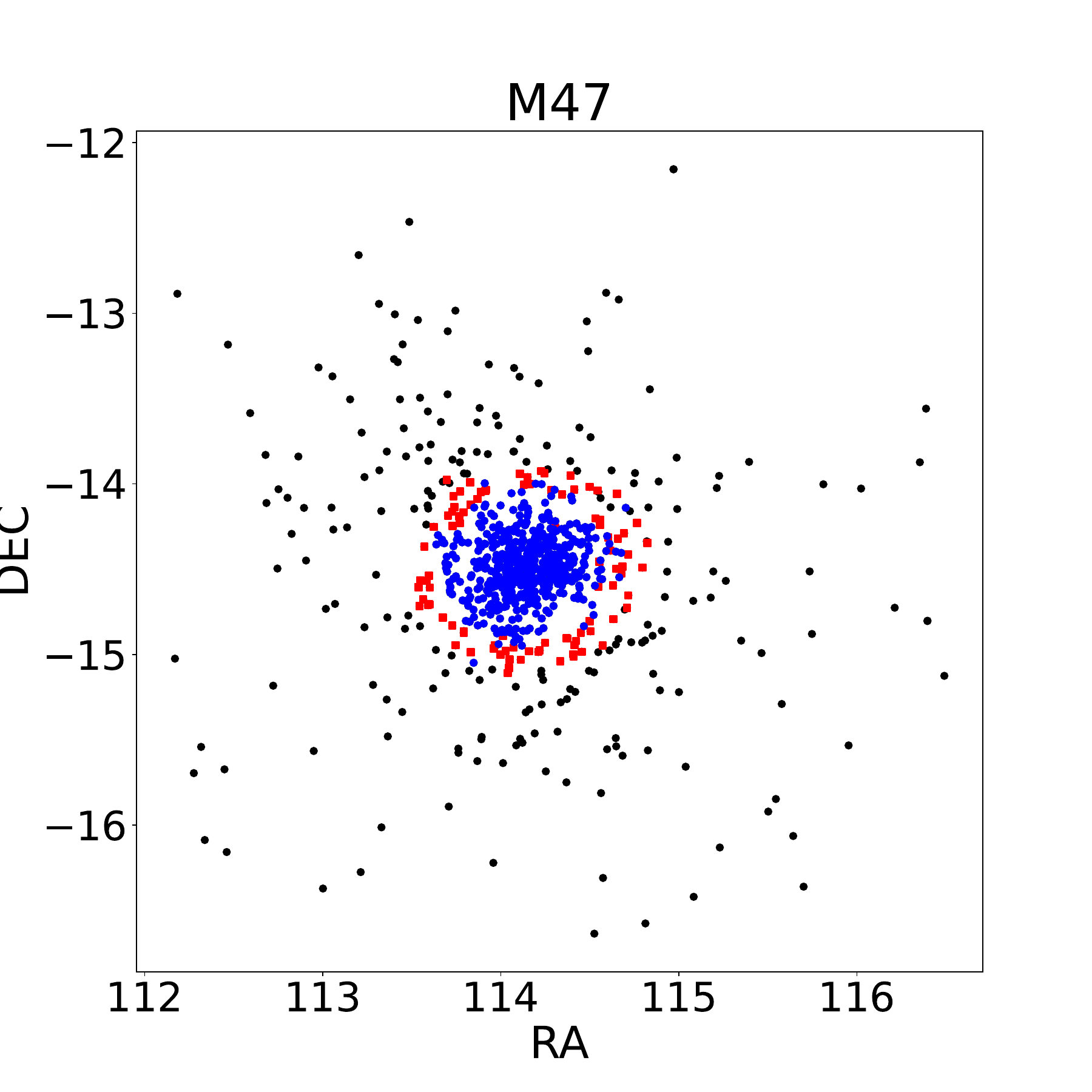}
        \end{subfigure}
        \begin{subfigure}{0.43\textwidth}

                \centering
                \includegraphics[width=\textwidth]{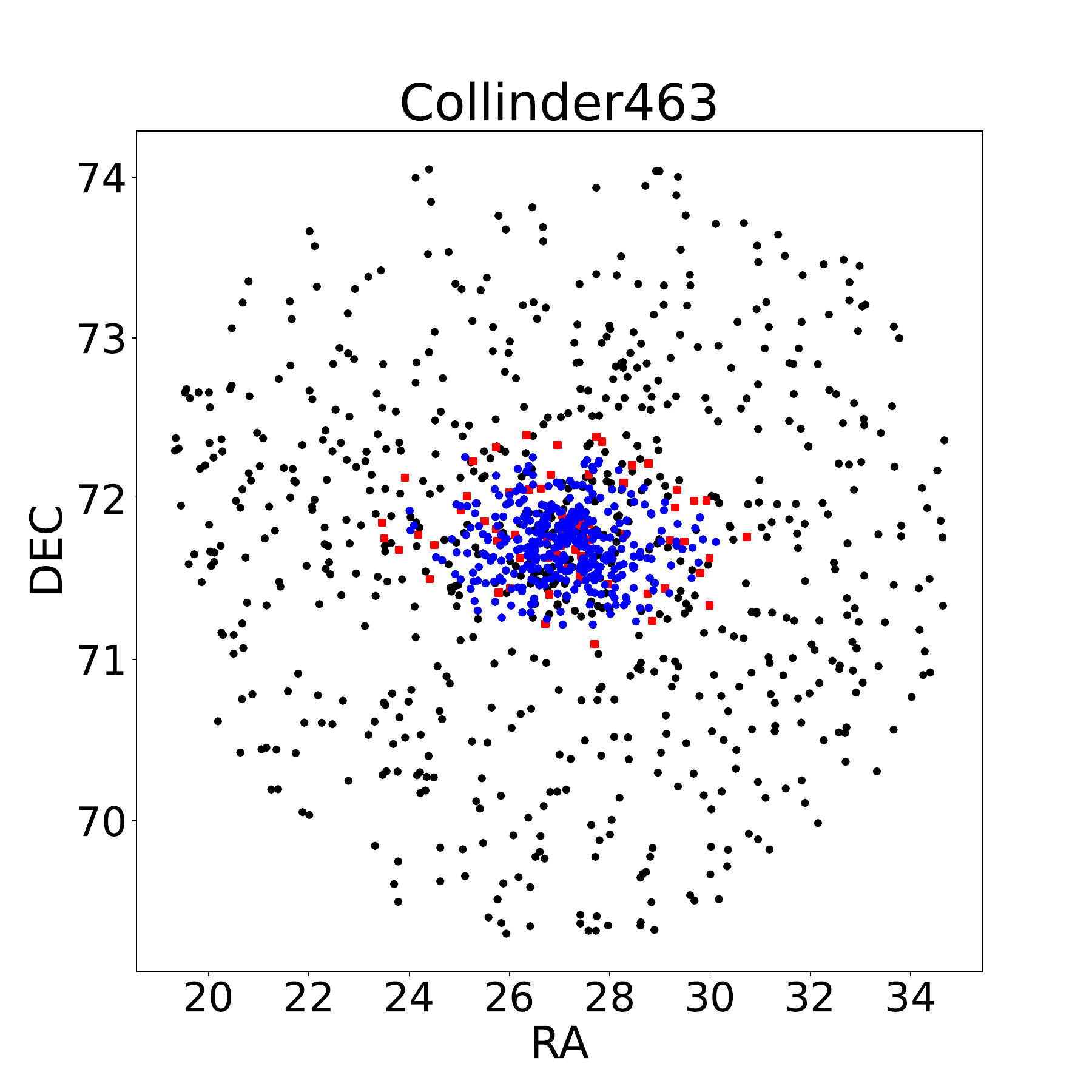}

        \end{subfigure}

  \begin{subfigure}{0.43\textwidth}
                \centering

                \includegraphics[width=\textwidth]{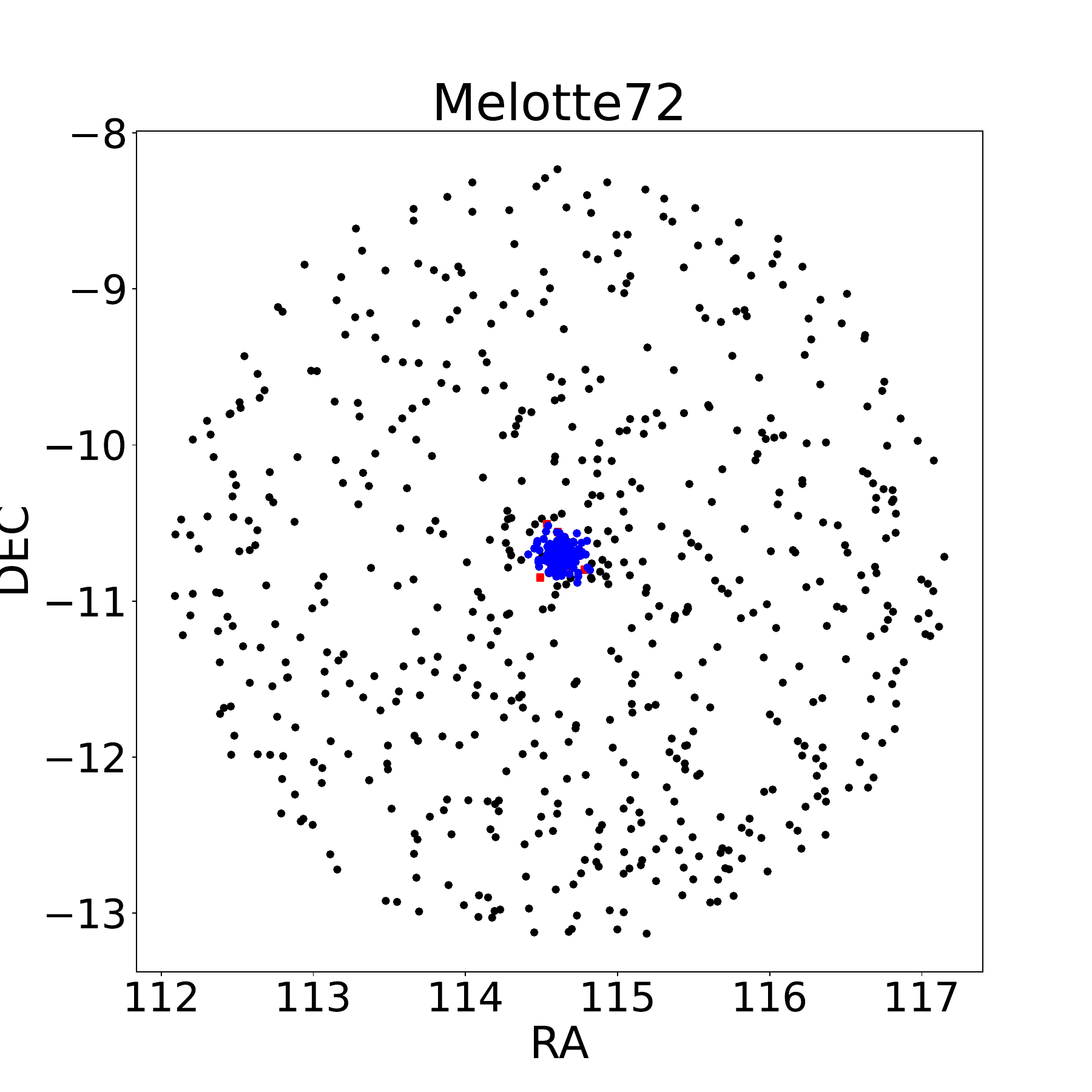}
        \end{subfigure}
        \begin{subfigure}{0.43\textwidth}
                \centering

                \includegraphics[width=\textwidth]{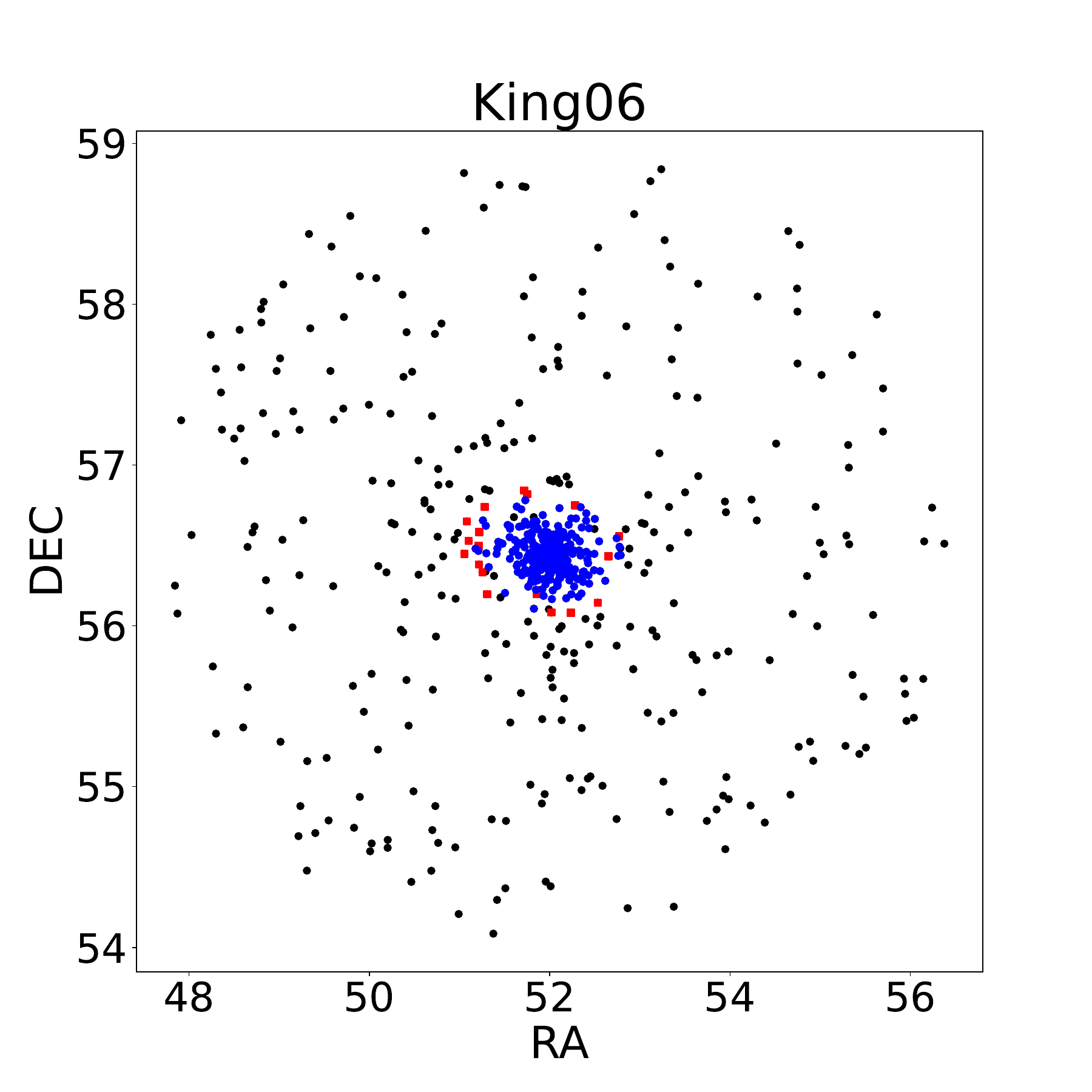}

        \end{subfigure}
        \begin{subfigure}{0.43\textwidth}
                \centering

                \includegraphics[width=\textwidth]{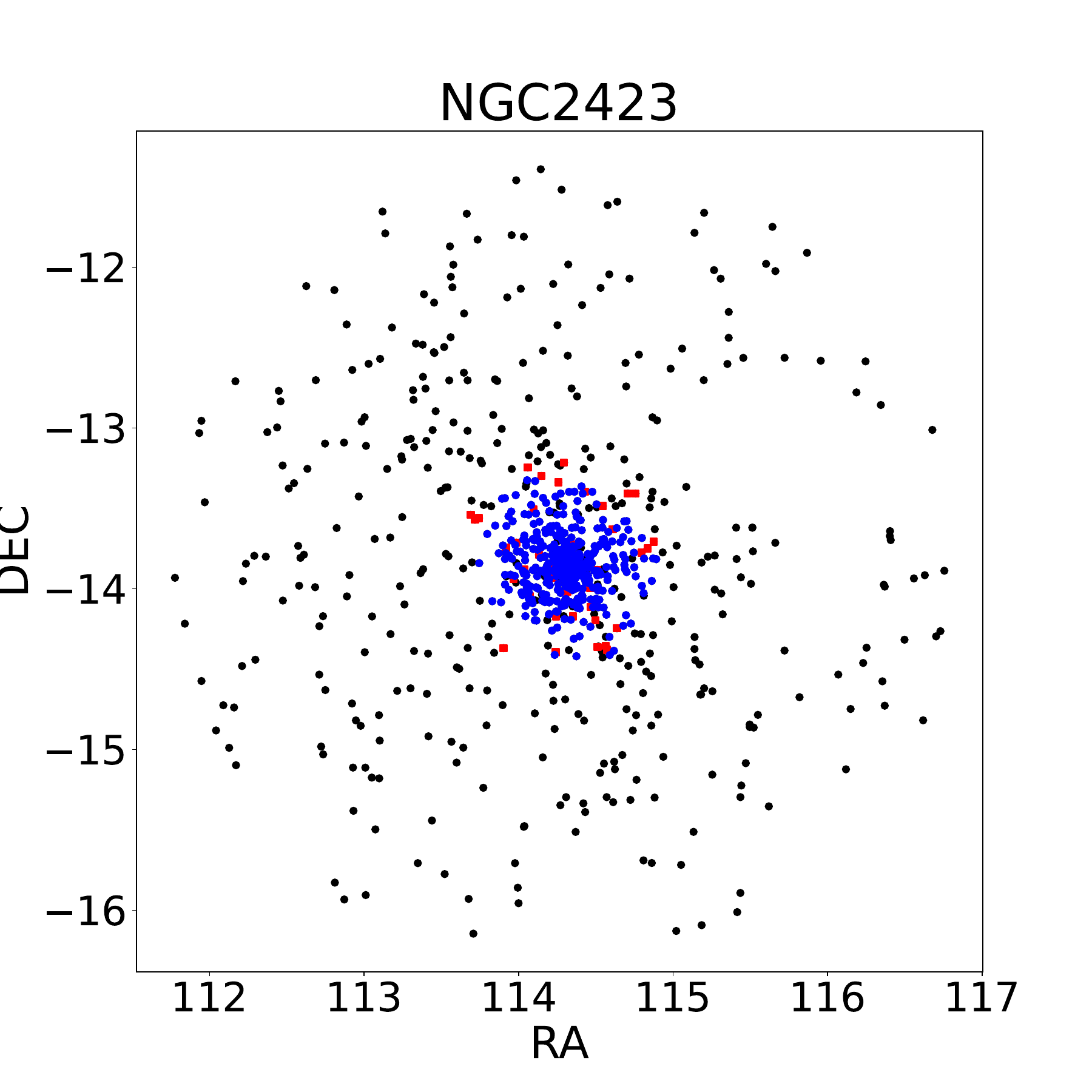}

        \end{subfigure}
        \begin{subfigure}{0.43\textwidth}
                \centering

                \includegraphics[width=\textwidth]{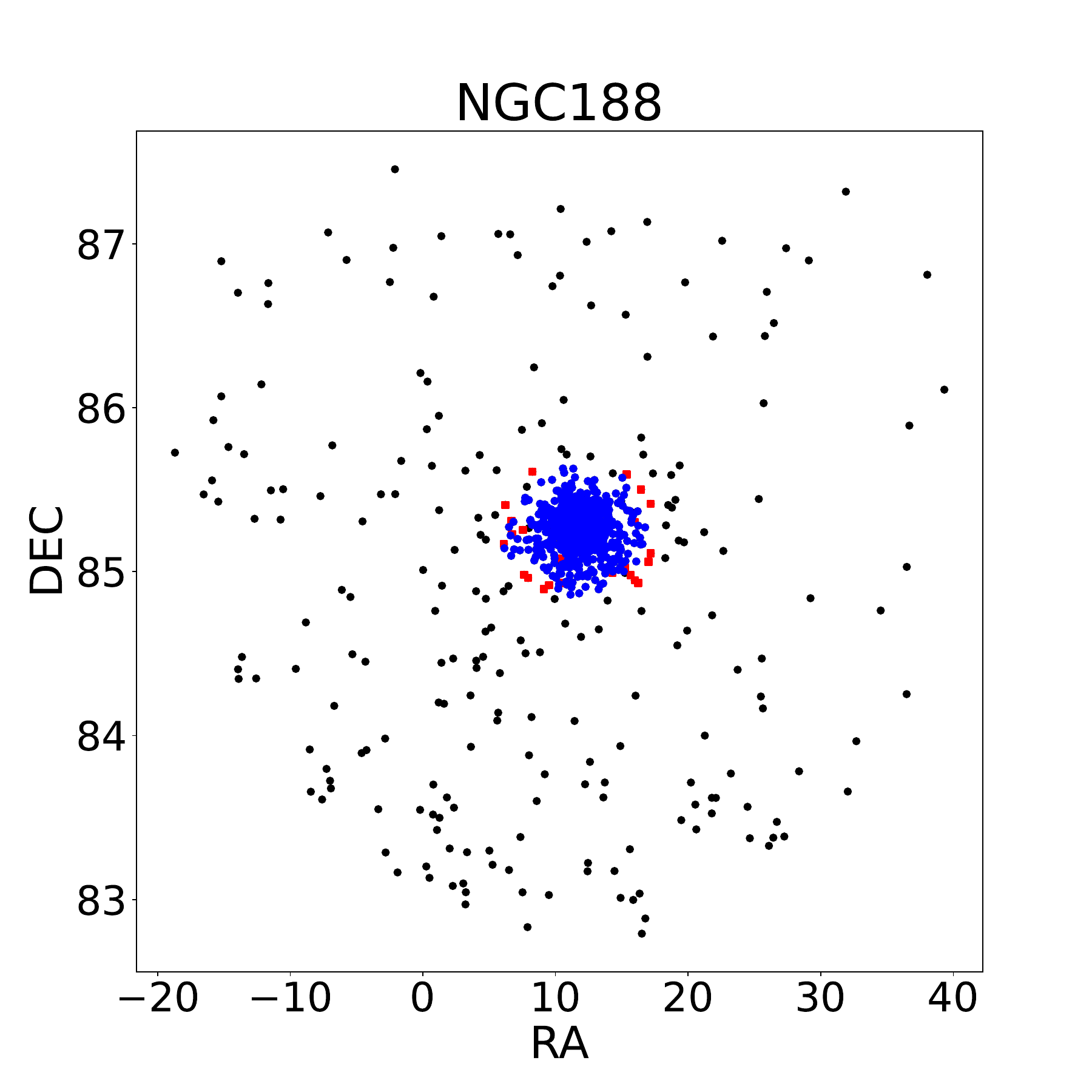}

        \end{subfigure}
  \caption{The position of stars selected by DBSCAN and GMM is as follows: Grey dots represent stars chosen by DBSCAN but not selected by the GMM algorithm. Red dots indicate stars selected by the GMM algorithm, with membership probabilities ranging between 0.5 and 0.8. Blue dots correspond to stars selected by the GMM algorithm, with a probability greater than 0.8.}
  \label{position of dbscan and GMM.fig}
\end{figure}

\begin{figure}
  \centering
  \captionsetup[subfigure]{labelformat=empty}
        \begin{subfigure}{0.43\textwidth}
        \centering

                \includegraphics[width=\textwidth]{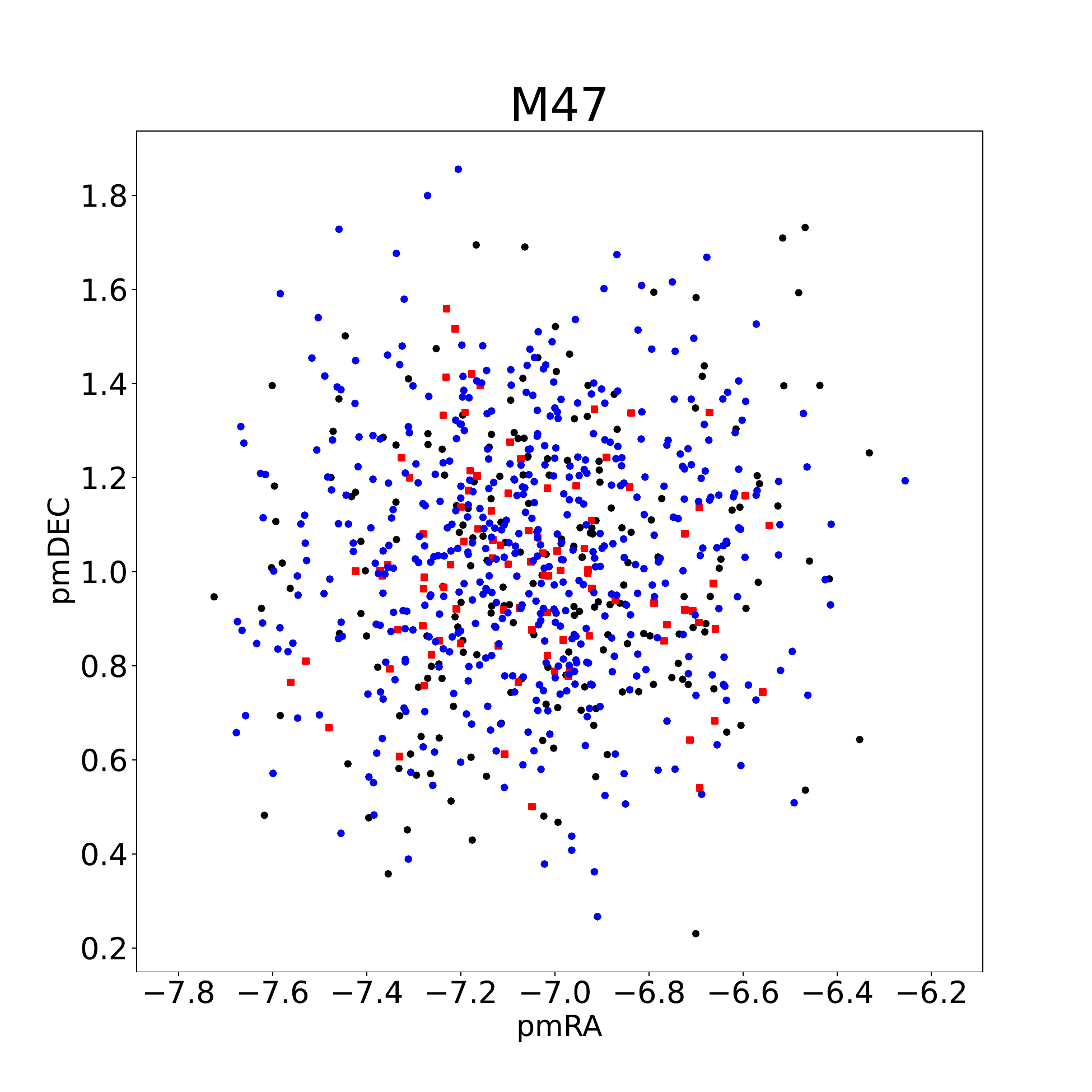}
        \end{subfigure}
        \begin{subfigure}{0.43\textwidth}

                \centering
                \includegraphics[width=\textwidth]{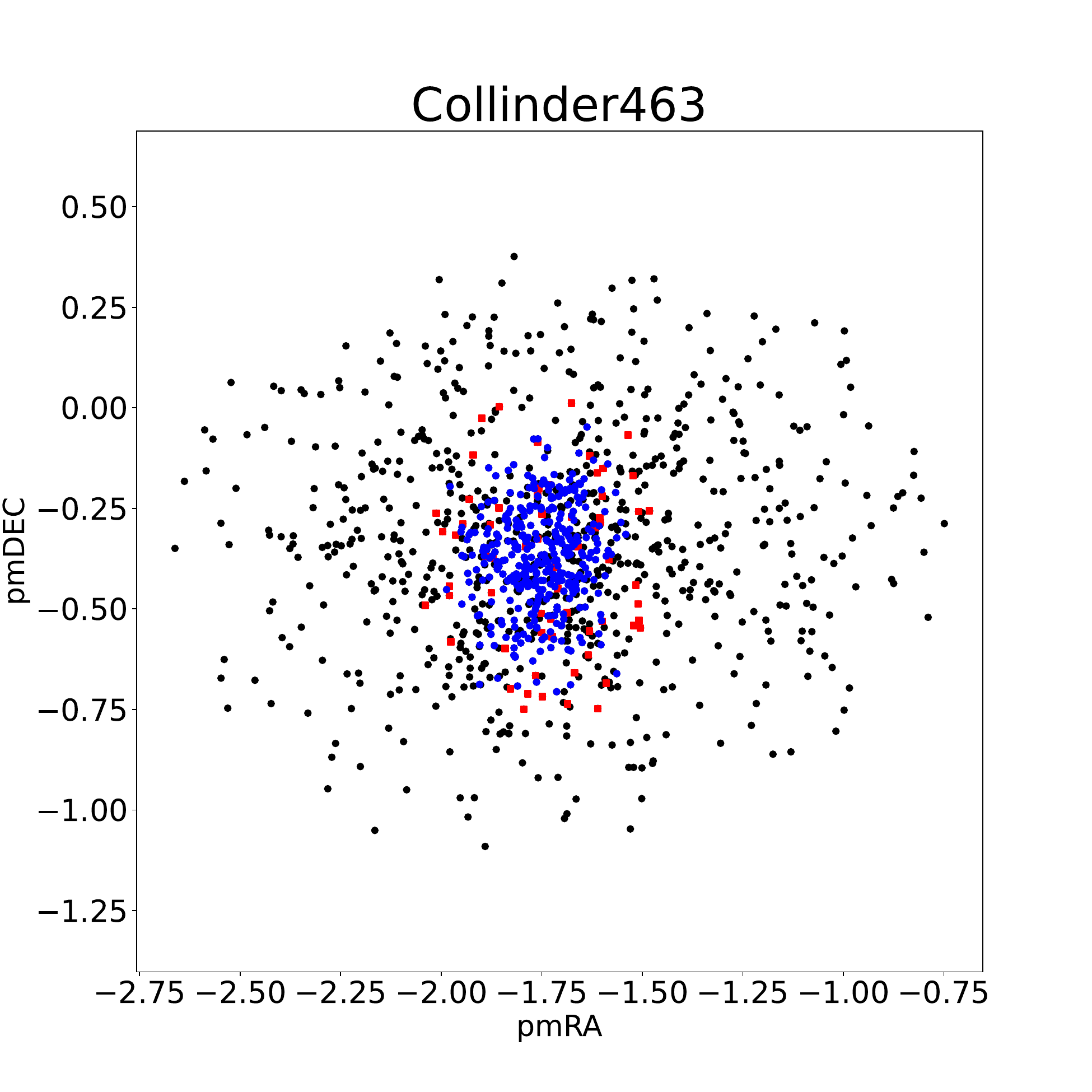}

        \end{subfigure}
  \begin{subfigure}{0.43\textwidth}
                \centering

                \includegraphics[width=\textwidth]{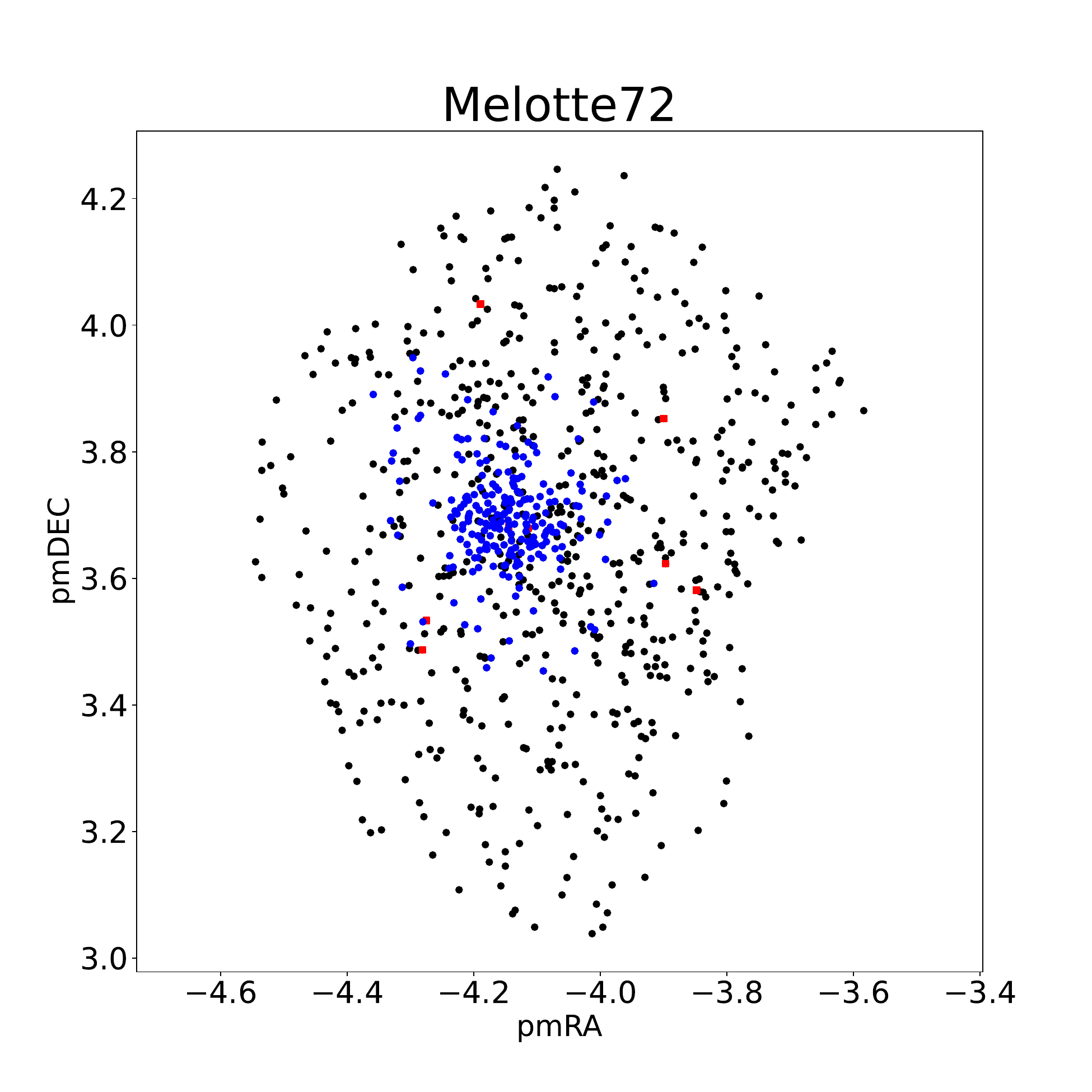}
        \end{subfigure}
        \begin{subfigure}{0.43\textwidth}
                \centering

                \includegraphics[width=\textwidth]{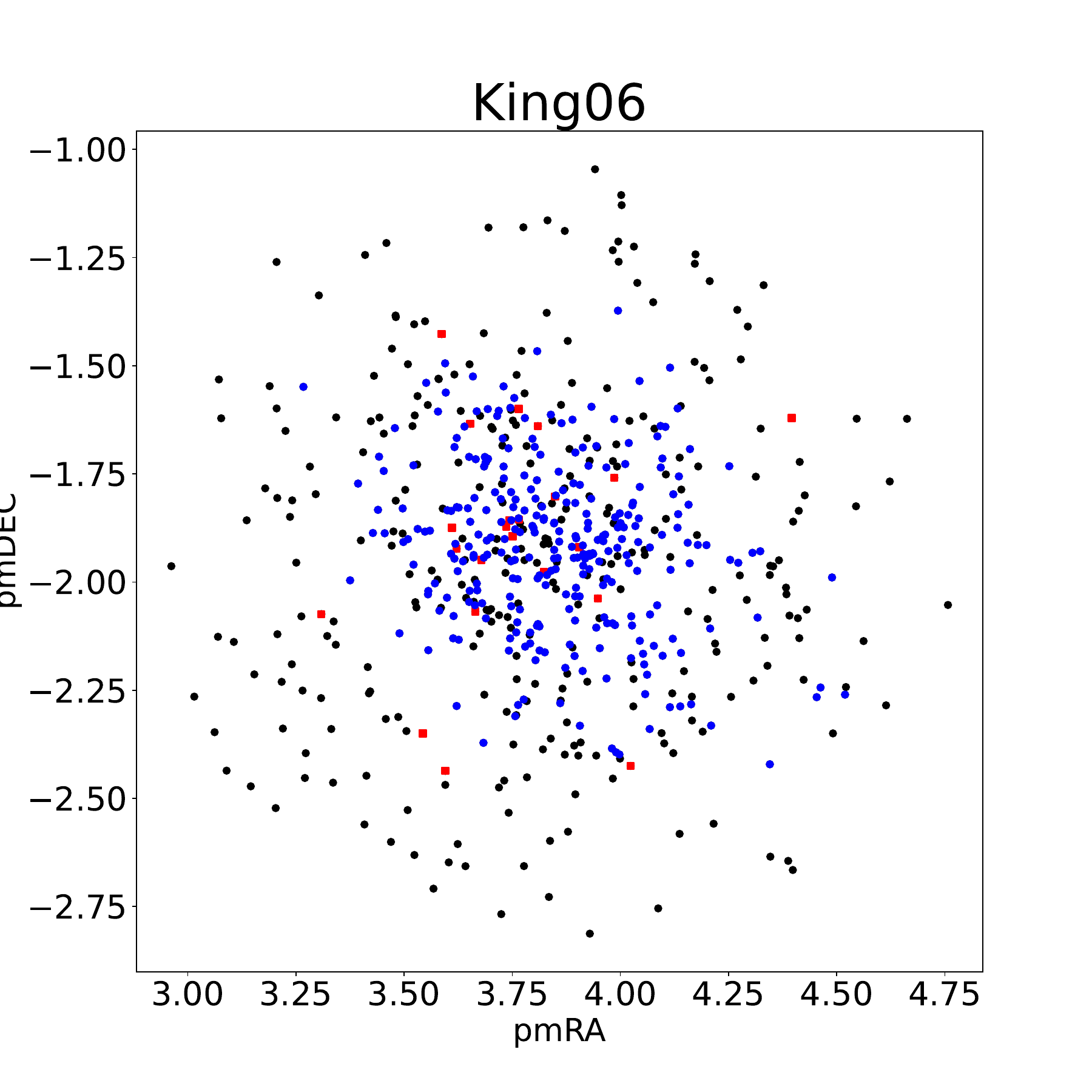}

        \end{subfigure}
        \begin{subfigure}{0.43\textwidth}
                \centering

                \includegraphics[width=\textwidth]{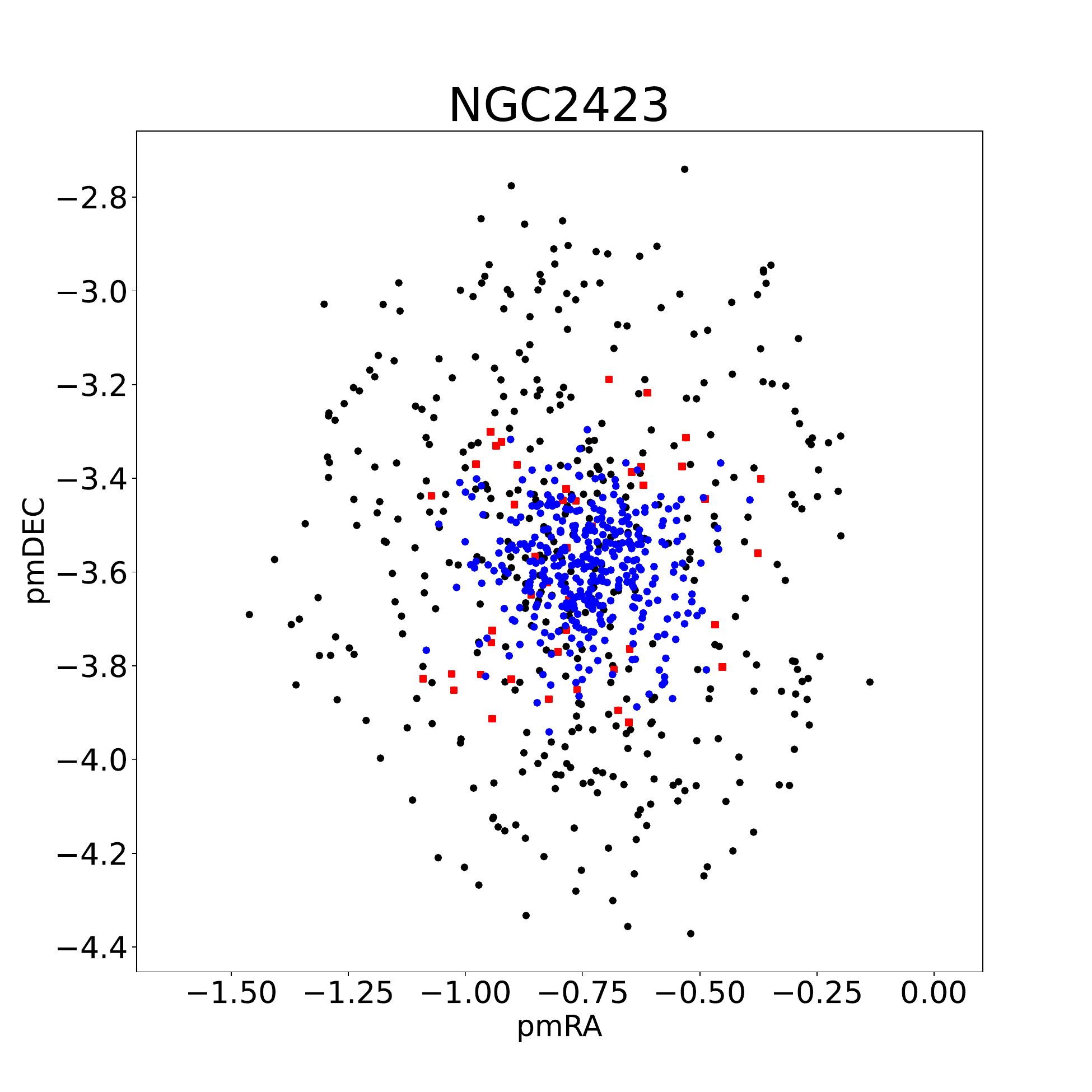}

        \end{subfigure}
        \begin{subfigure}{0.43\textwidth}
                \centering

                \includegraphics[width=\textwidth]{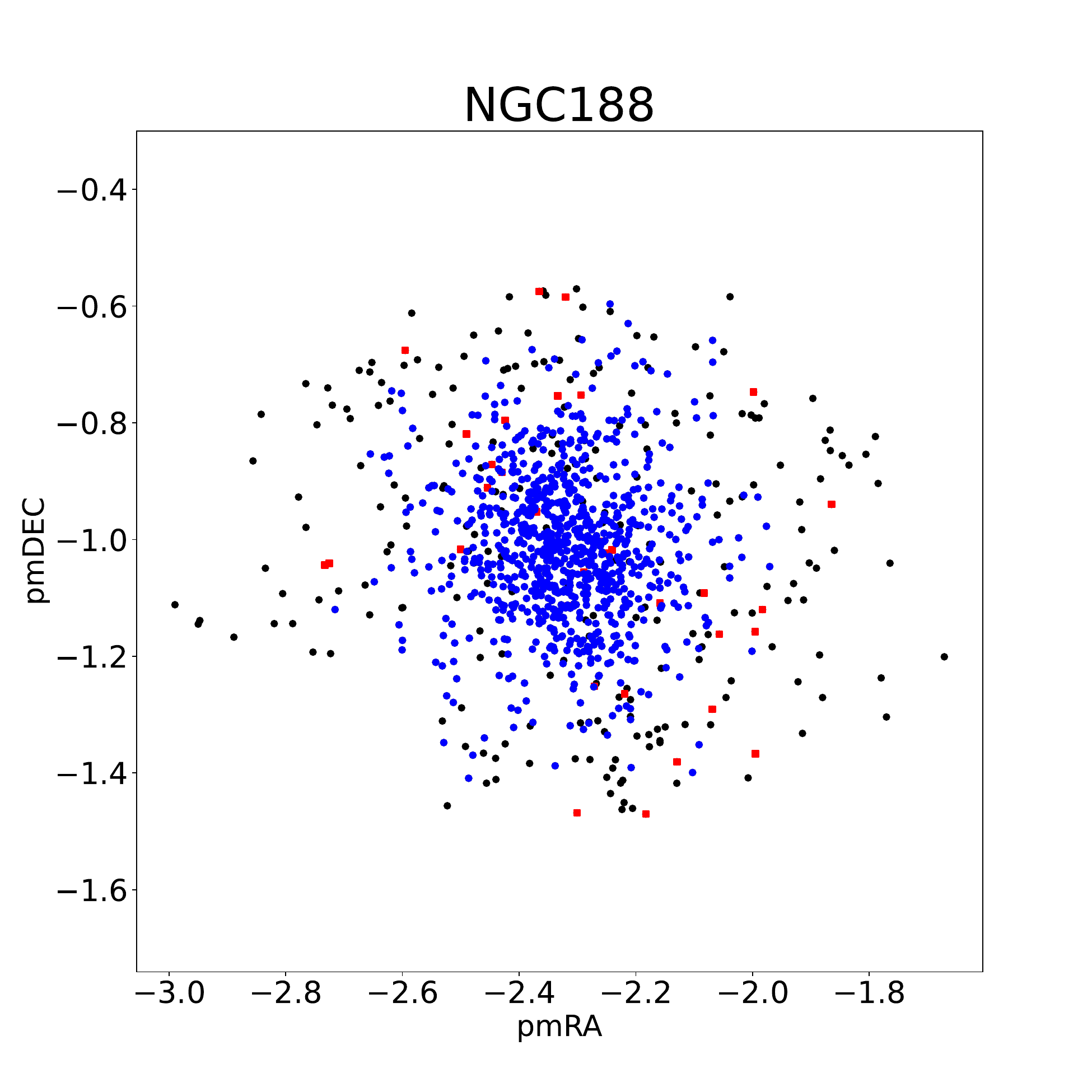}

        \end{subfigure}
  \caption{The proper motion of selected cluster members is determined by the GMM algorithm. Grey dots represent stars chosen by DBSCAN with a membership probability less than 0.5. Red dots indicate stars selected by the GMM algorithm, with membership probabilities ranging between 0.5 and 0.8. Blue dots correspond to stars selected by the GMM algorithm with a probability greater than 0.8.}
  \label{proper motion of dbscan and GMM.fig}
\end{figure}

\begin{figure}
  \centering
  \captionsetup[subfigure]{labelformat=empty}
        \begin{subfigure}{0.43\textwidth}
        \centering

                \includegraphics[width=\textwidth]{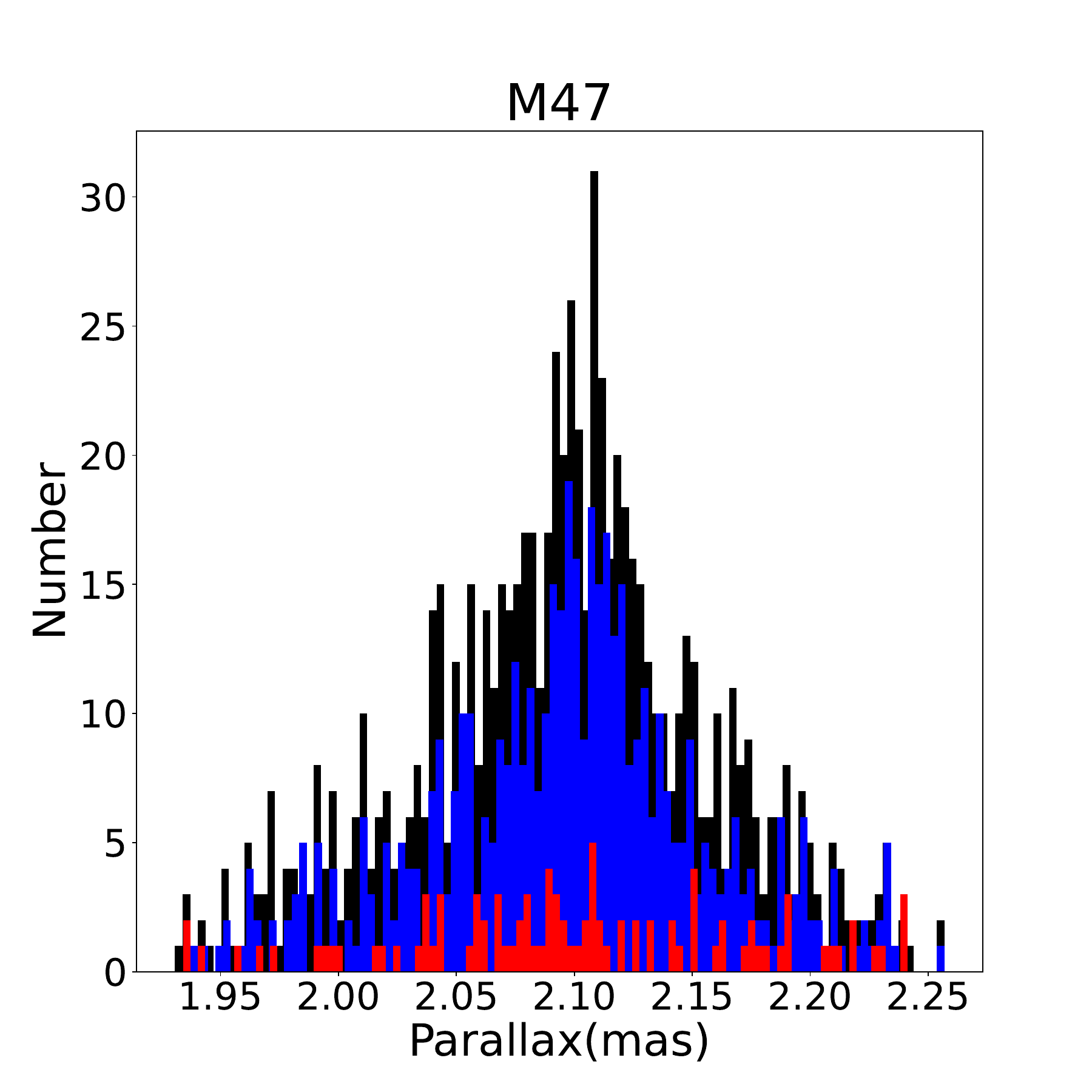}
        \end{subfigure}
        \begin{subfigure}{0.43\textwidth}

                \centering
                \includegraphics[width=\textwidth]{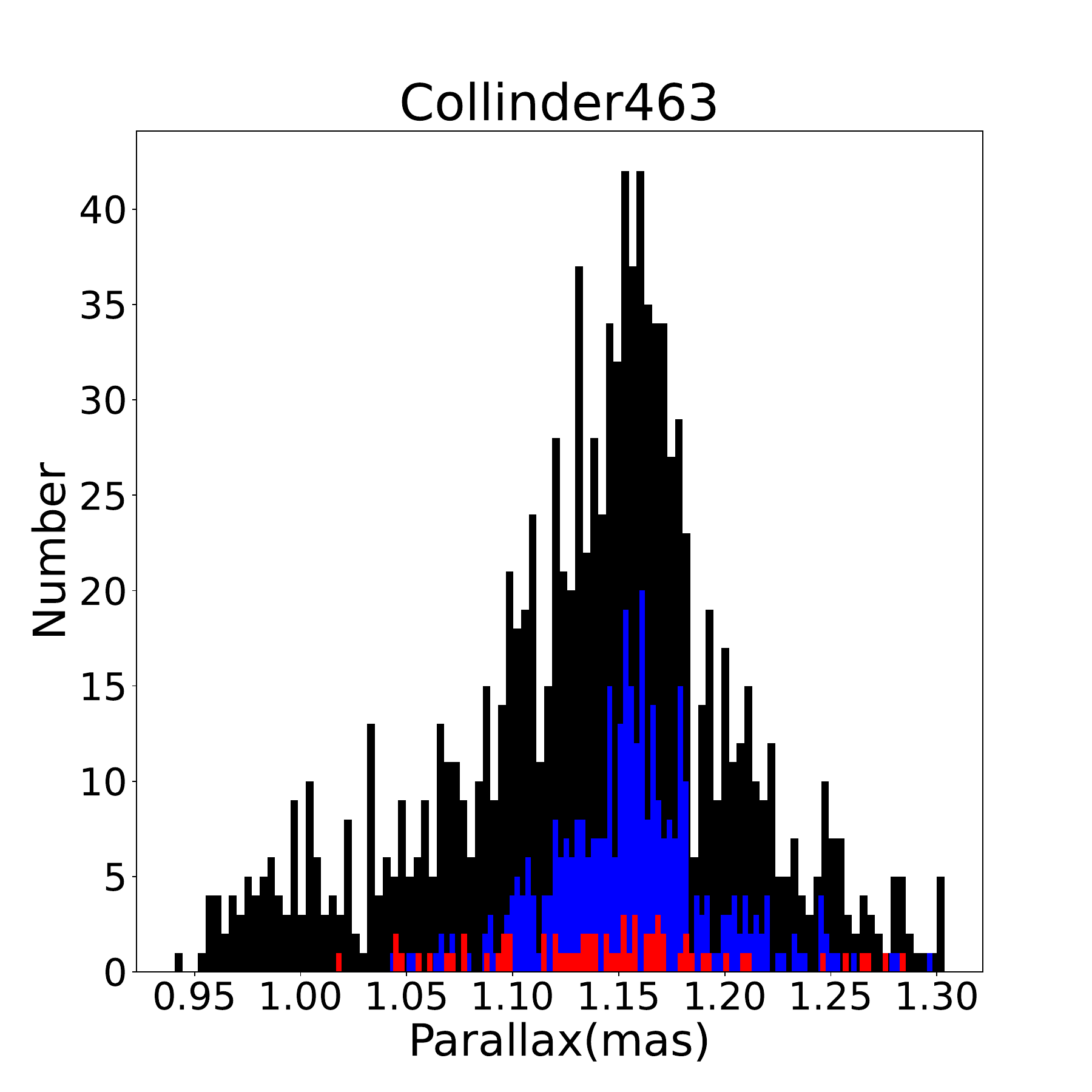}

        \end{subfigure}
        \begin{subfigure}{0.43\textwidth}
                \centering

                \includegraphics[width=\textwidth]{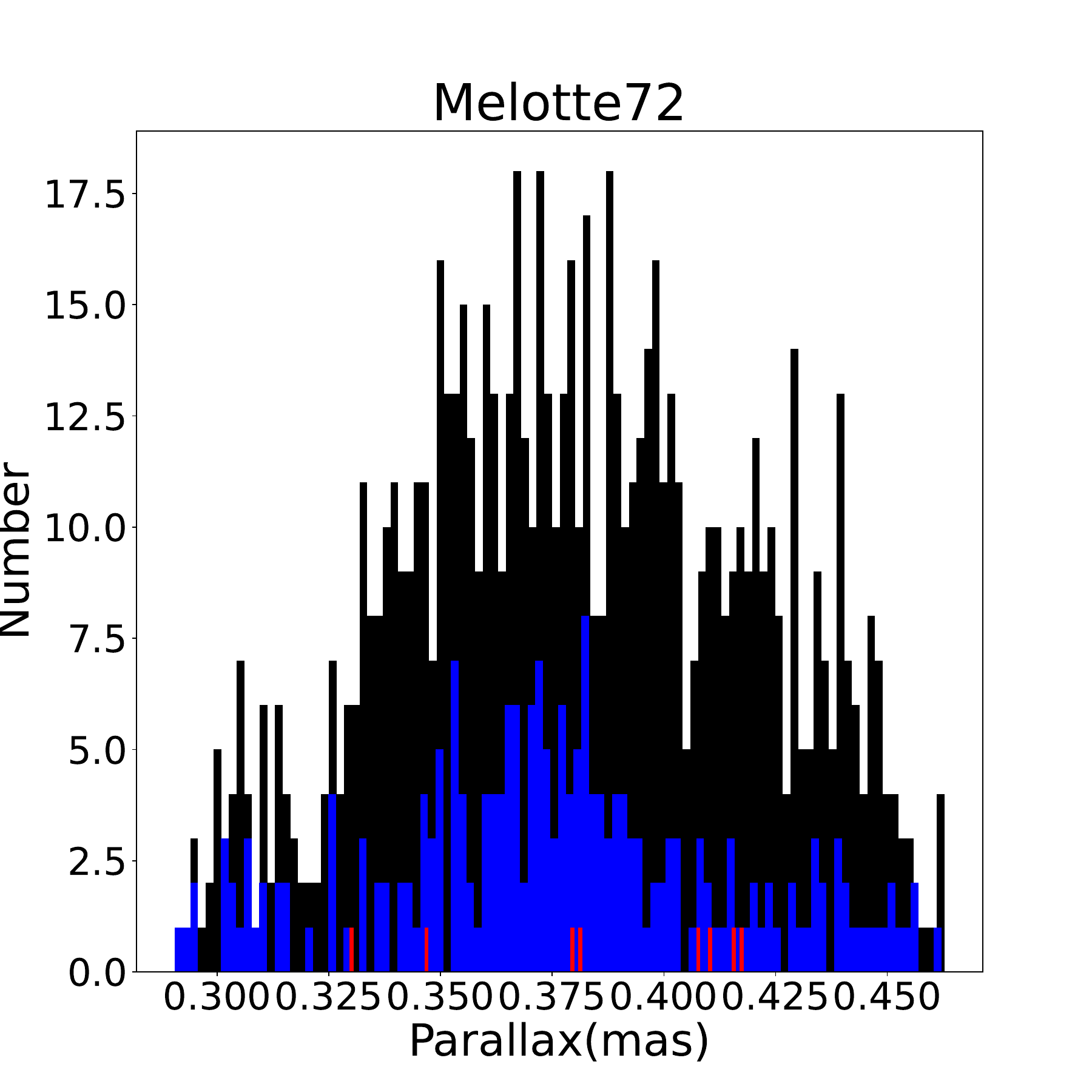}
        \end{subfigure}
        \begin{subfigure}{0.43\textwidth}
                \centering

                \includegraphics[width=\textwidth]{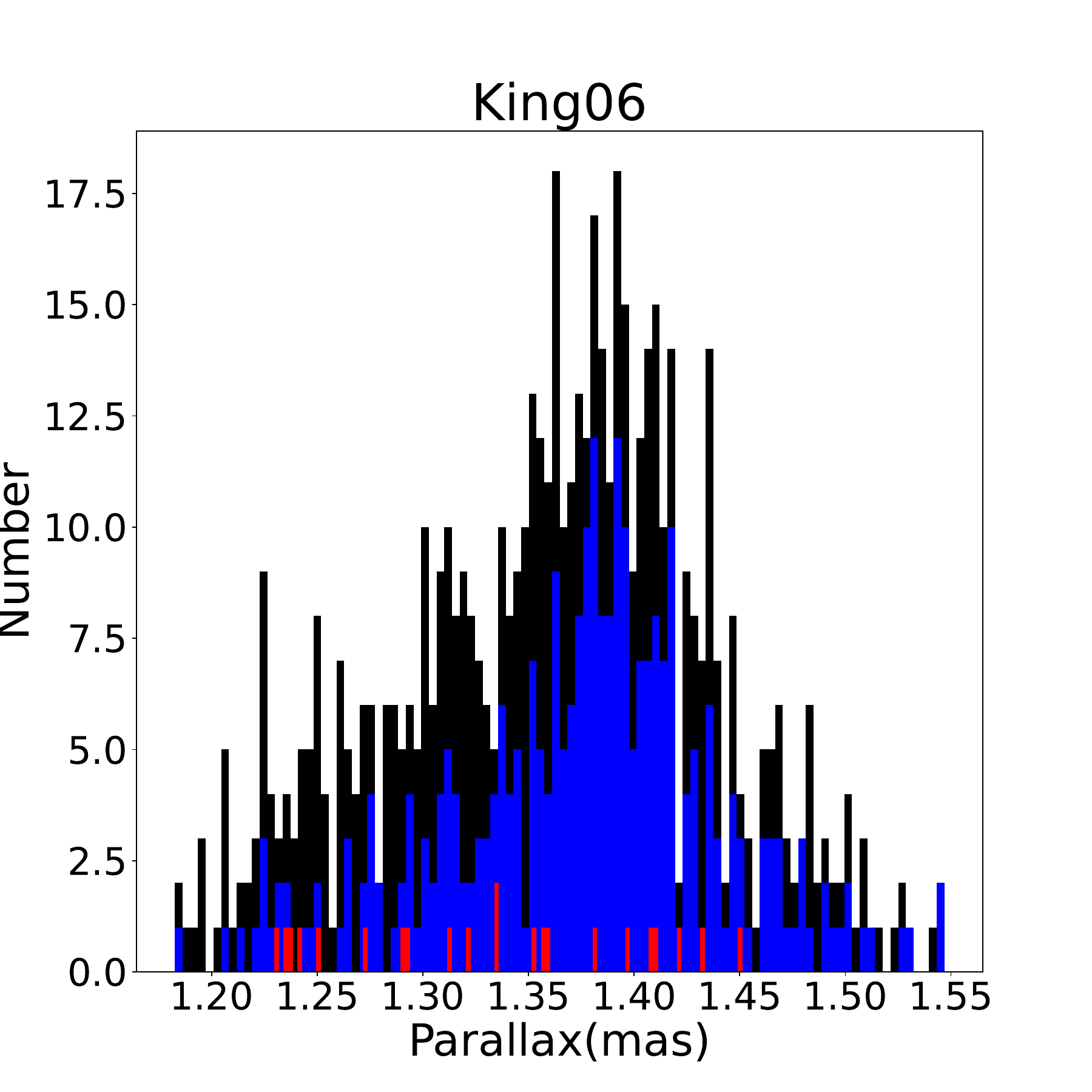}

        \end{subfigure}
        \begin{subfigure}{0.43\textwidth}
                \centering

                \includegraphics[width=\textwidth]{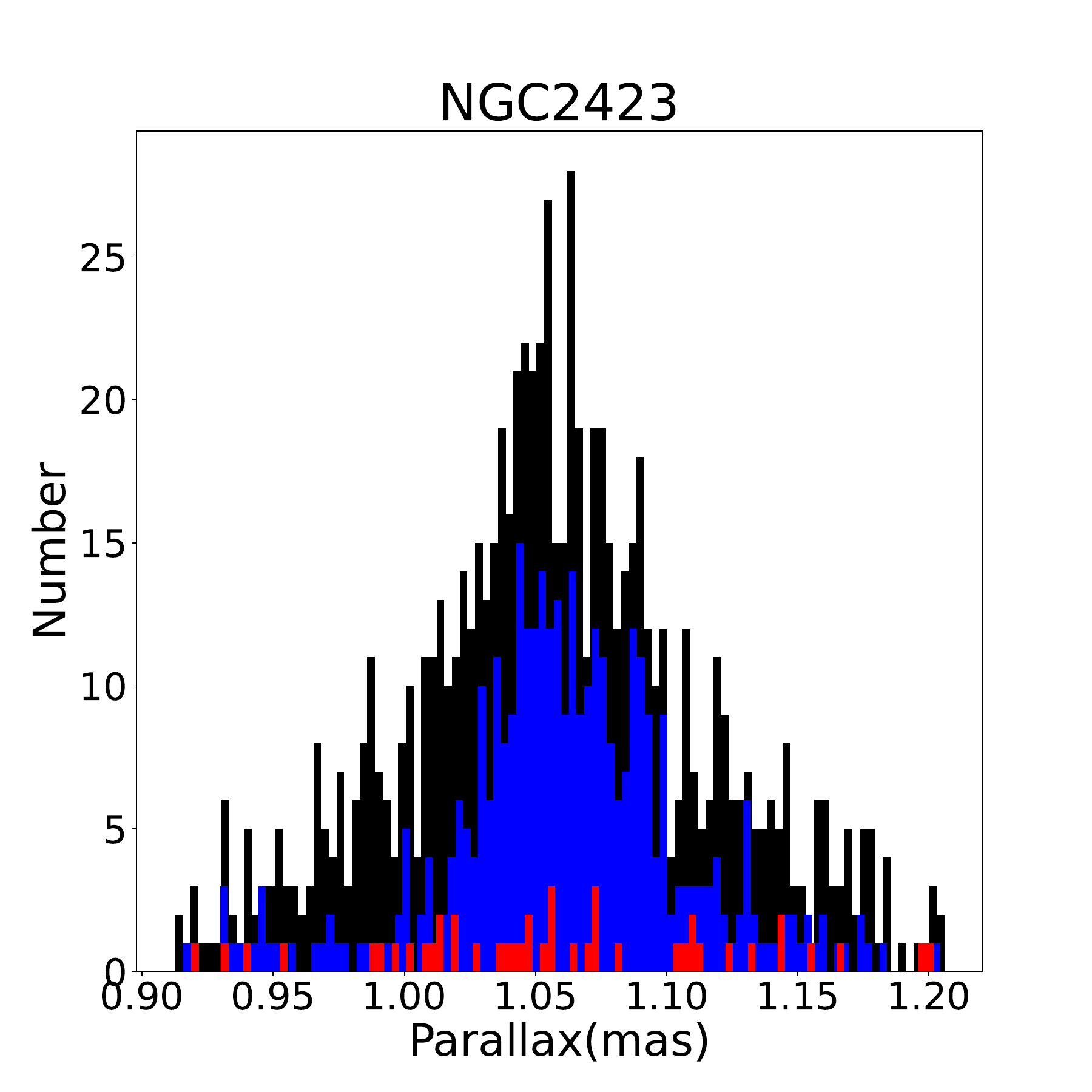}

        \end{subfigure}
        \begin{subfigure}{0.43\textwidth}
                \centering

                \includegraphics[width=\textwidth]{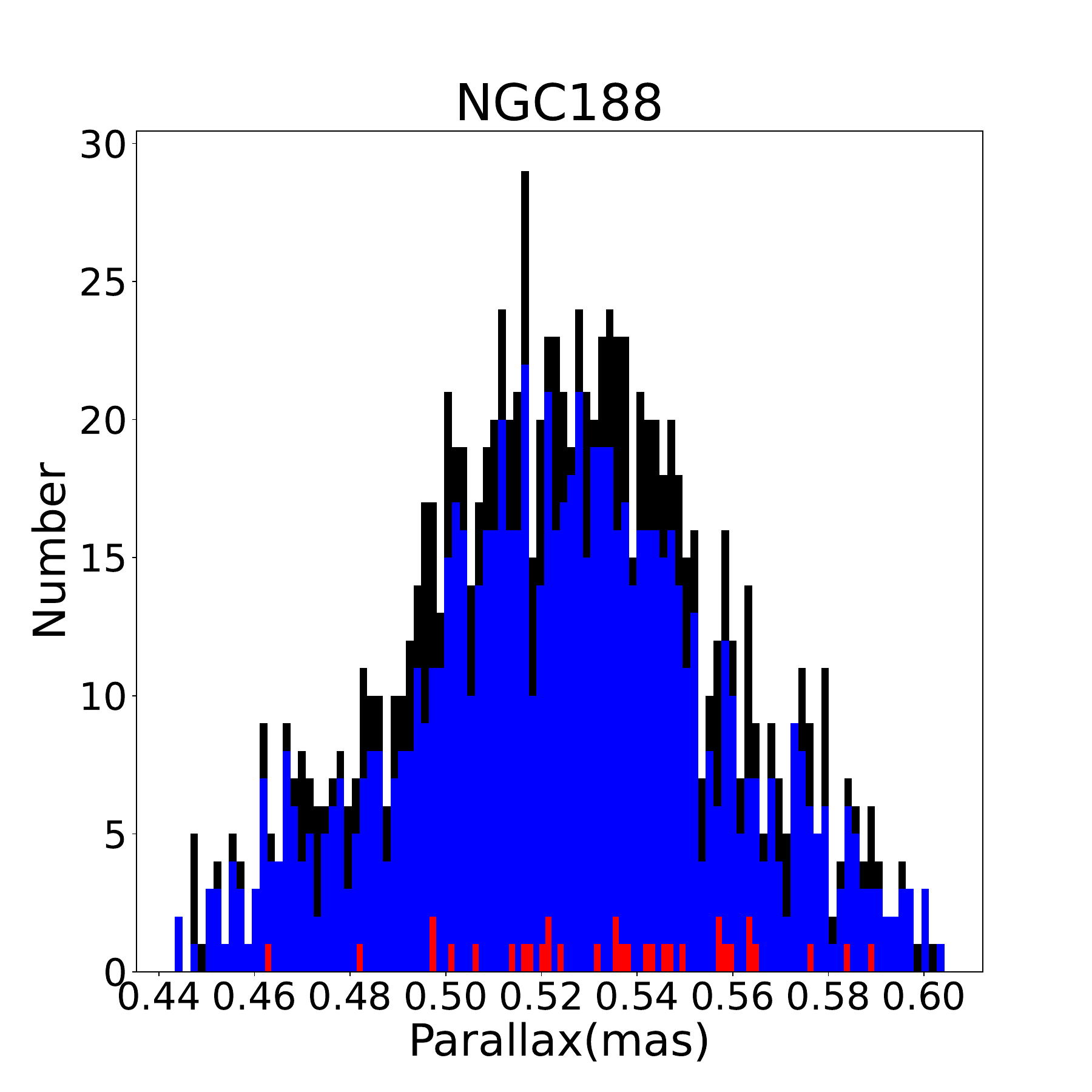}

        \end{subfigure}
  \caption{The parallax of selected cluster members is determined by the GMM algorithm. Grey lines represent stars chosen by DBSCAN with a membership probability less than 0.5. Red lines indicate stars selected by the GMM algorithm, with membership probabilities ranging between 0.5 and 0.8. Blue lines correspond to stars selected by the GMM algorithm with a probability greater than 0.8}
  \label{parallax of dbscan and GMM.fig}
\end{figure}

\begin{figure}
  \centering
  \captionsetup[subfigure]{labelformat=empty}
        \begin{subfigure}{0.43\textwidth}
        \centering

                \includegraphics[width=\textwidth]{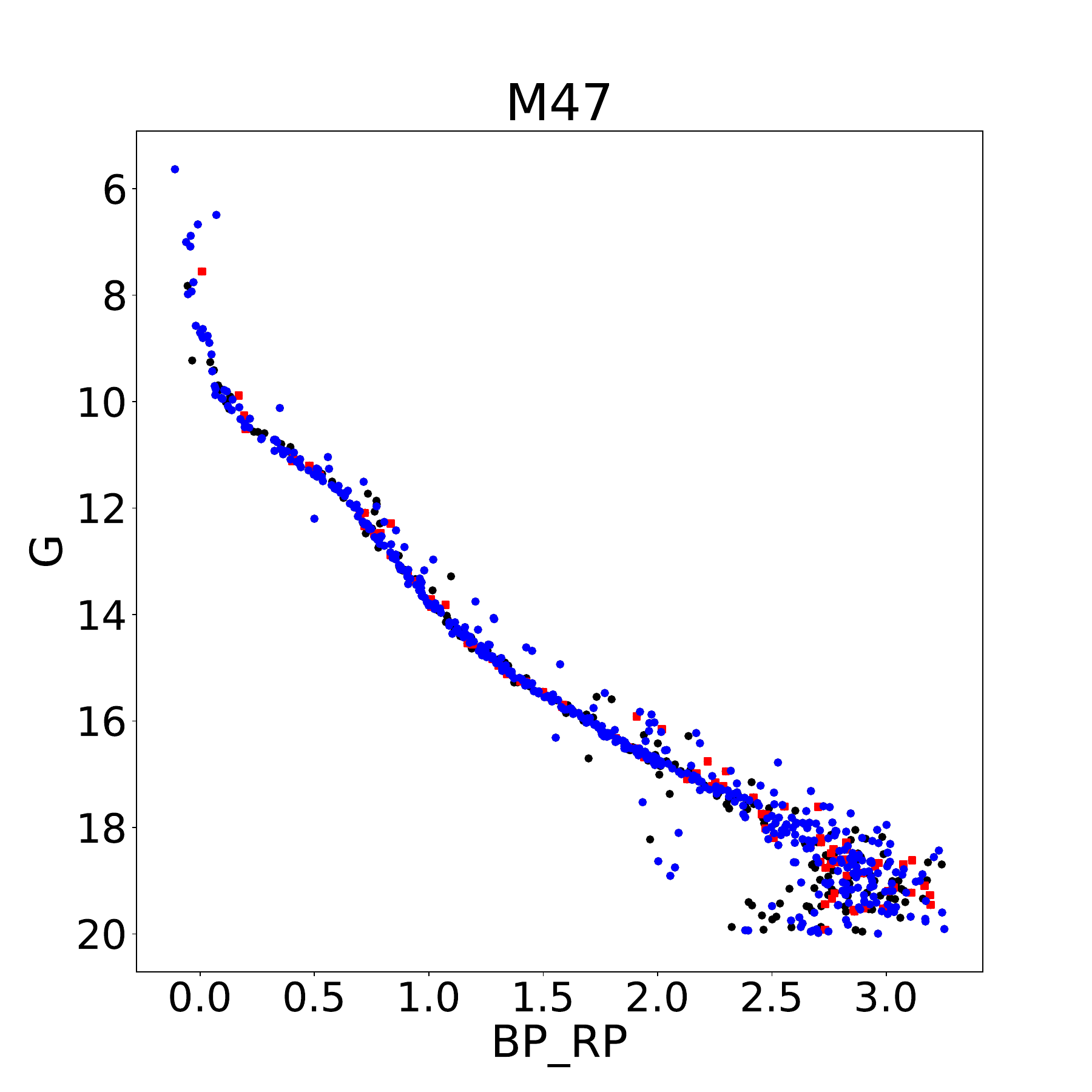}
        \end{subfigure}
        \begin{subfigure}{0.43\textwidth}

                \centering
                \includegraphics[width=\textwidth]{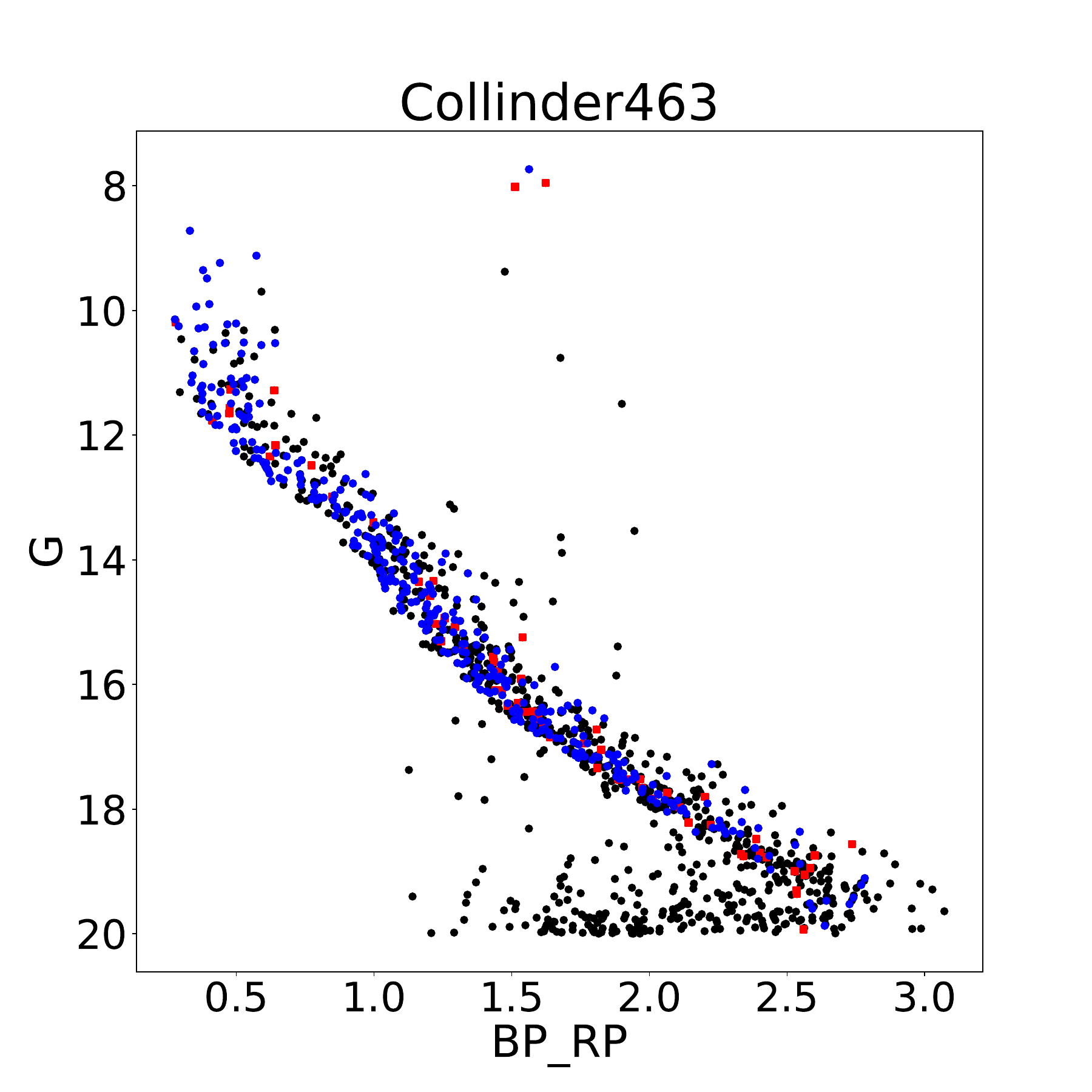}

        \end{subfigure}
        \begin{subfigure}{0.43\textwidth}
                \centering

                \includegraphics[width=\textwidth]{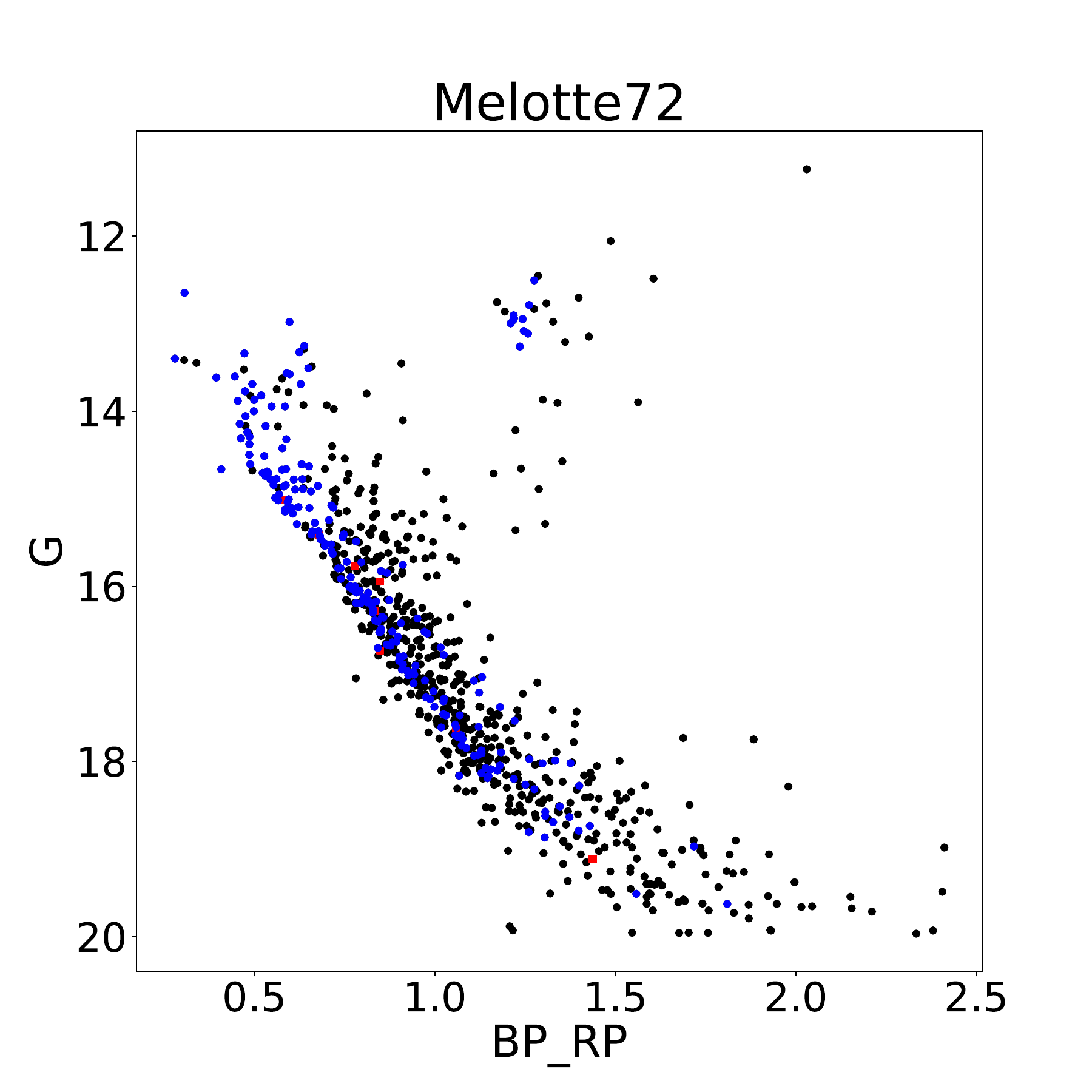}
        \end{subfigure}
        \begin{subfigure}{0.43\textwidth}
                \centering

                \includegraphics[width=\textwidth]{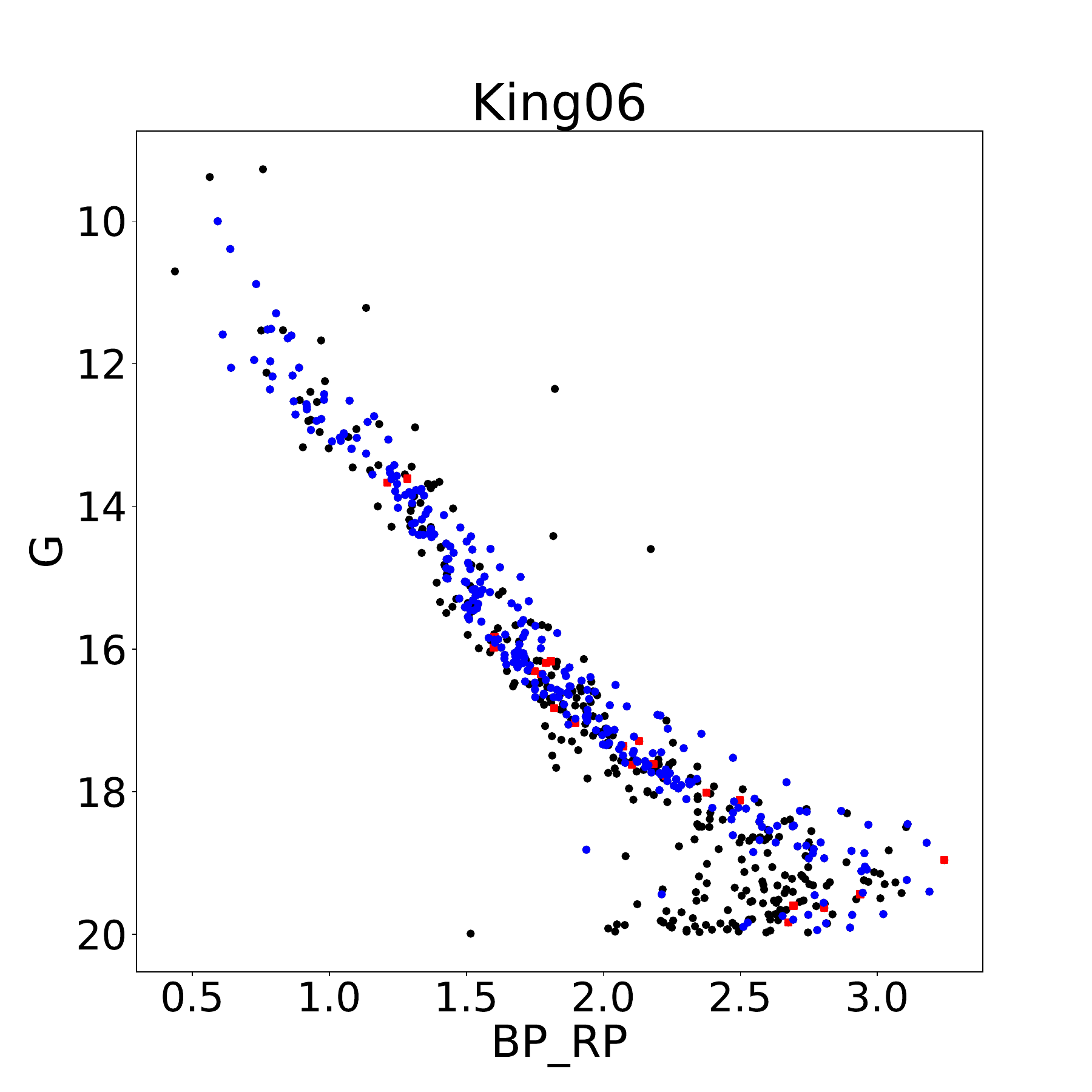}

        \end{subfigure}
        \begin{subfigure}{0.43\textwidth}
                \centering

                \includegraphics[width=\textwidth]{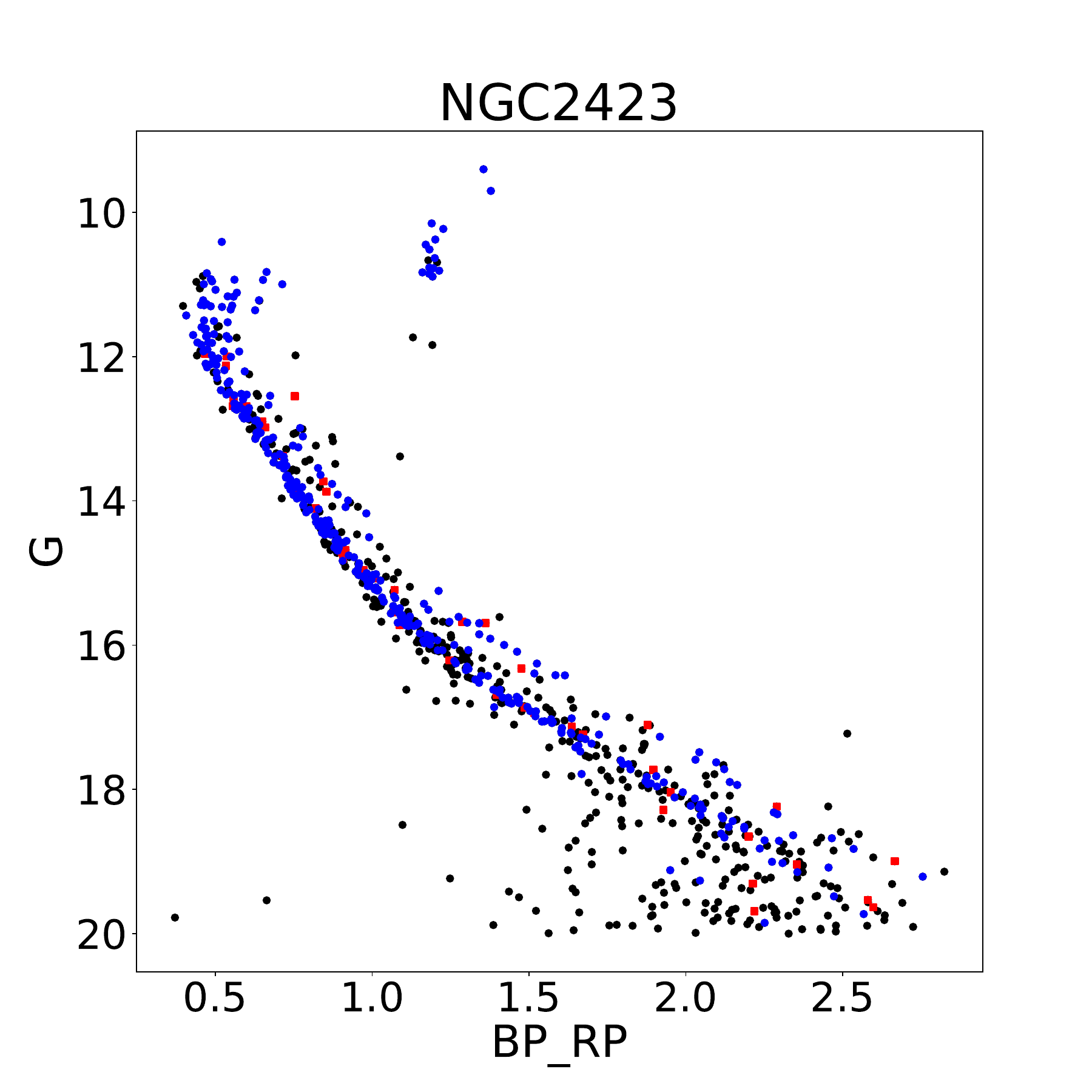}

        \end{subfigure}
        \begin{subfigure}{0.43\textwidth}
                \centering

                \includegraphics[width=\textwidth]{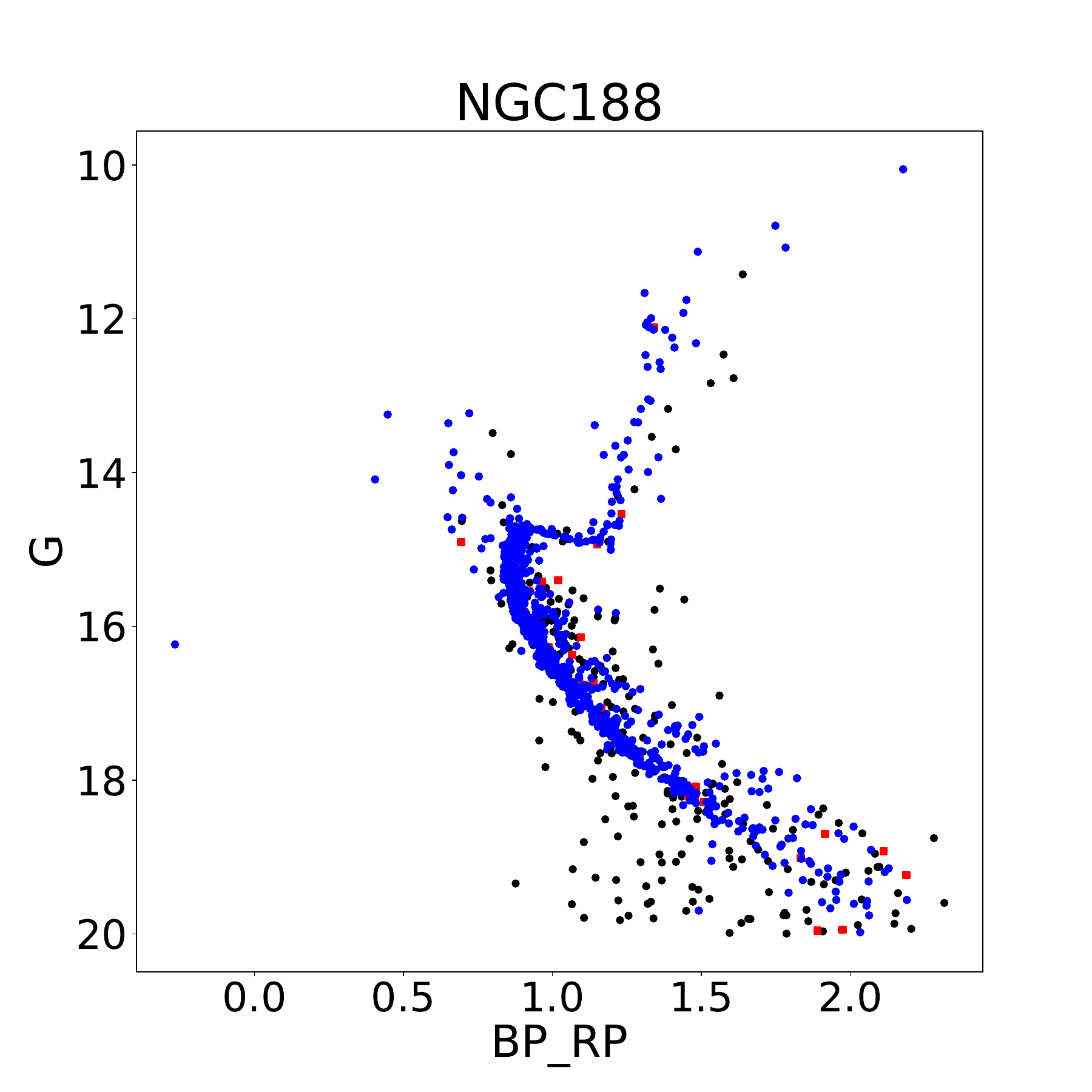}

        \end{subfigure}
  \caption{The CMD plot displays selected cluster members identified by the GMM algorithm. Grey dots represent stars chosen by DBSCAN with a membership probability less than 0.5. Red dots indicate stars selected by the GMM algorithm, with membership probabilities ranging between 0.5 and 0.8. Blue dots correspond to cluster members with probabilities greater than 0.8.}
  \label{CMD of dbscan and GMM.fig}
\end{figure}

\begin{figure}
\centering
\captionsetup[subfigure]{labelformat=empty}
\begin{subfigure}{0.25\textwidth}
        \centering
           \includegraphics[width=\textwidth]{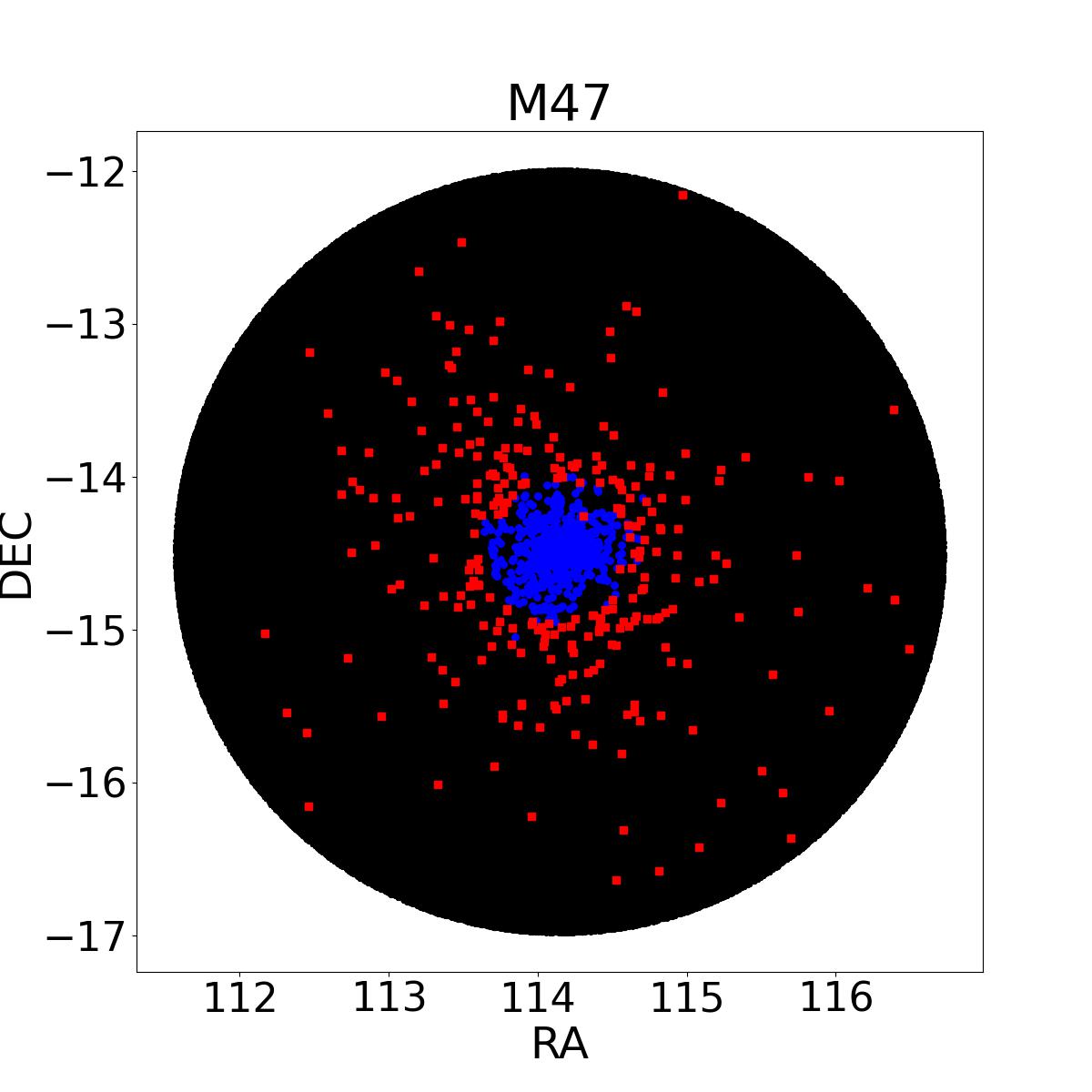}

        \end{subfigure}
        \begin{subfigure}{0.25\textwidth}

                \centering
                \includegraphics[width=\textwidth]{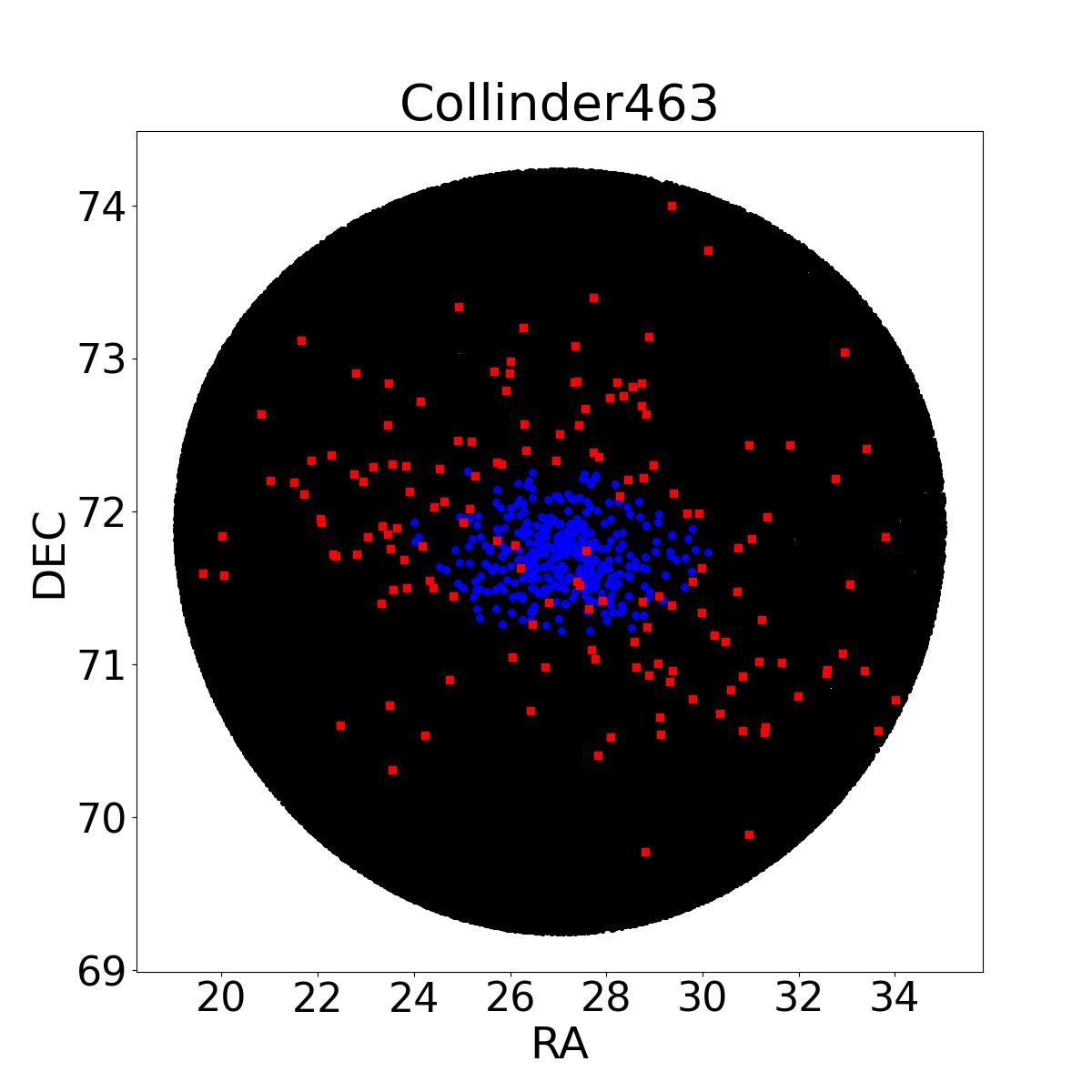}

        \end{subfigure}
        \begin{subfigure}{0.25\textwidth}
                \centering
           \includegraphics[width=\textwidth]{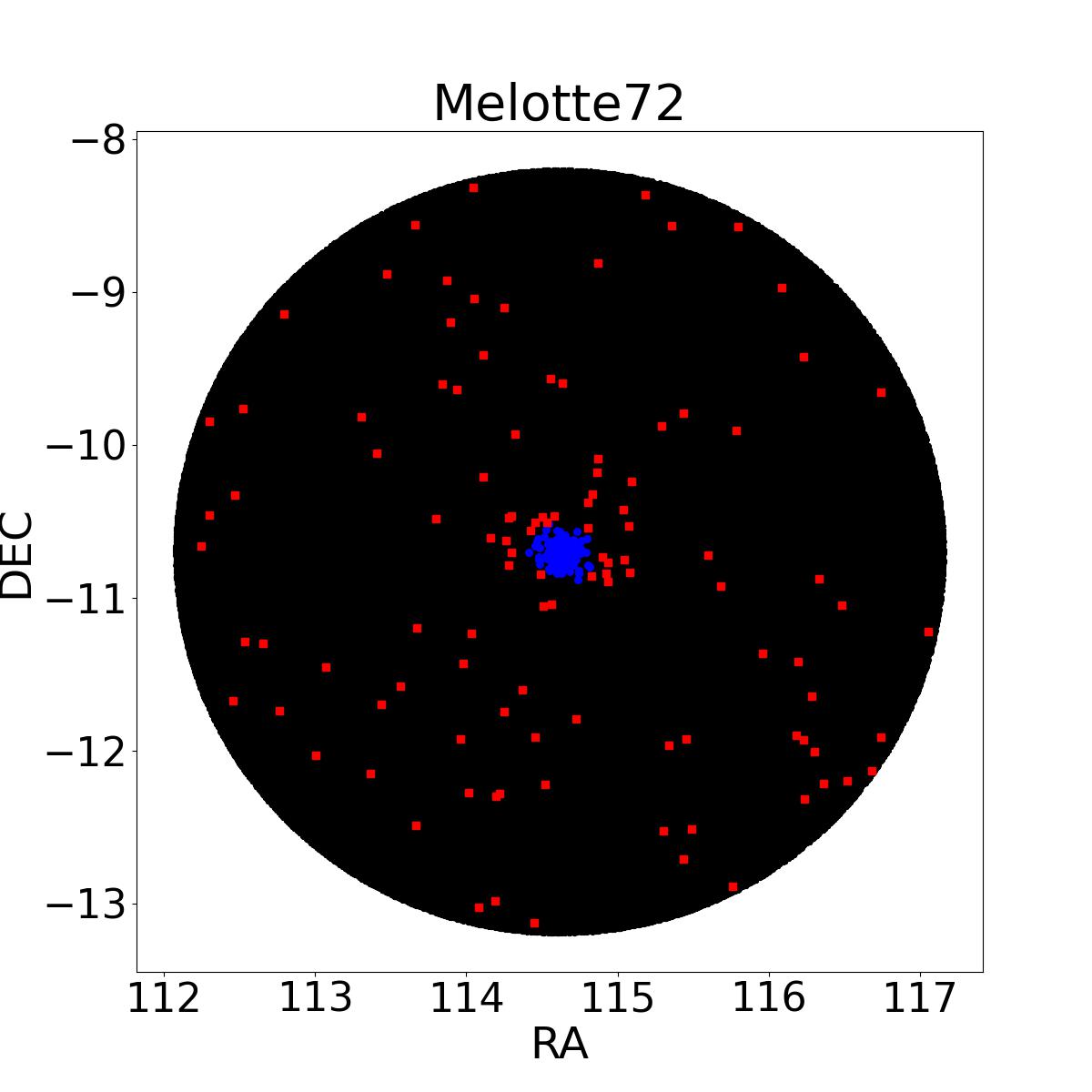}

        \end{subfigure}
        \begin{subfigure}{0.25\textwidth}
                \centering

                \includegraphics[width=\textwidth]{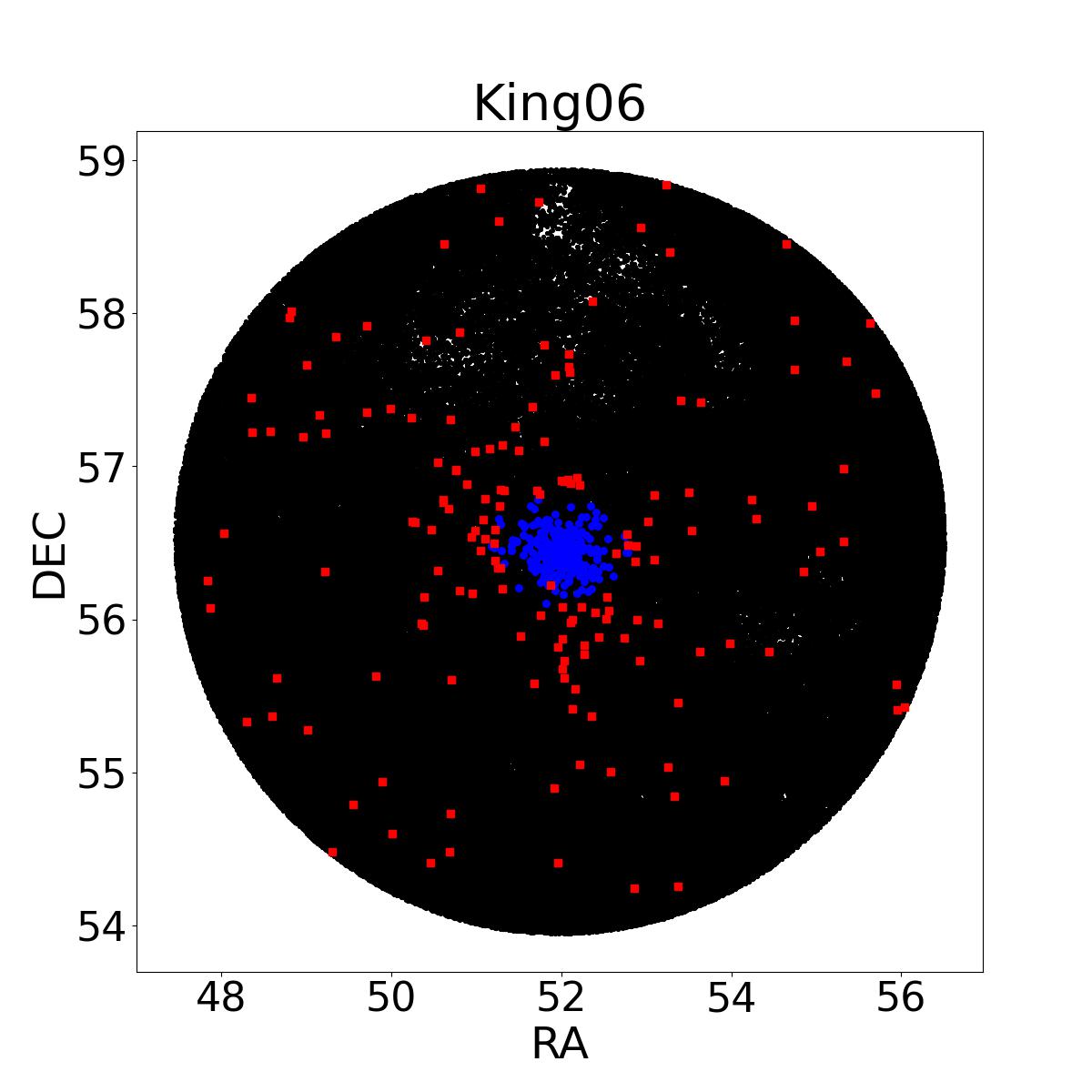}

        \end{subfigure}
        \begin{subfigure}{0.25\textwidth}
                \centering

                \includegraphics[width=\textwidth]{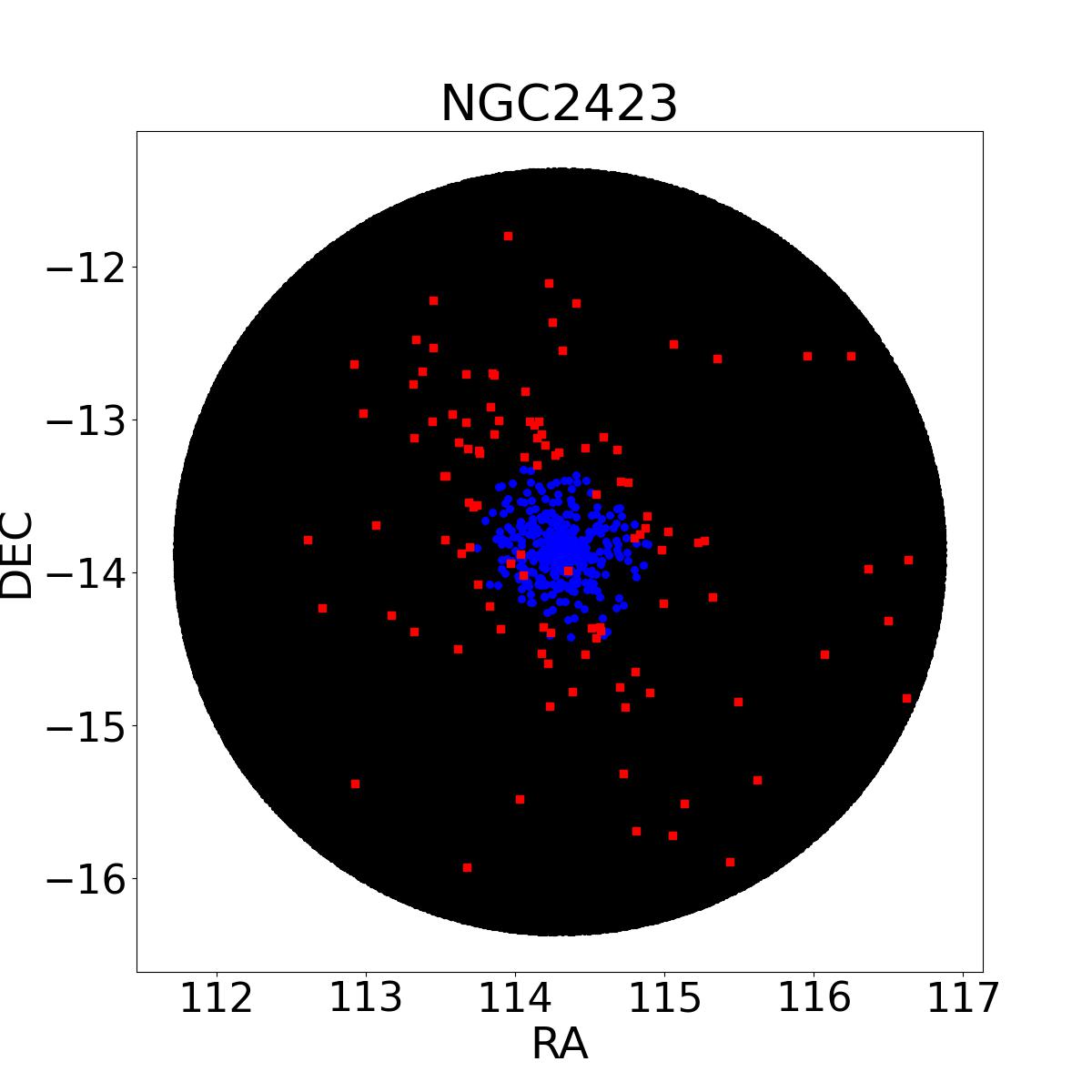}

        \end{subfigure}
        \begin{subfigure}{0.25\textwidth}
                \centering

                \includegraphics[width=\textwidth]{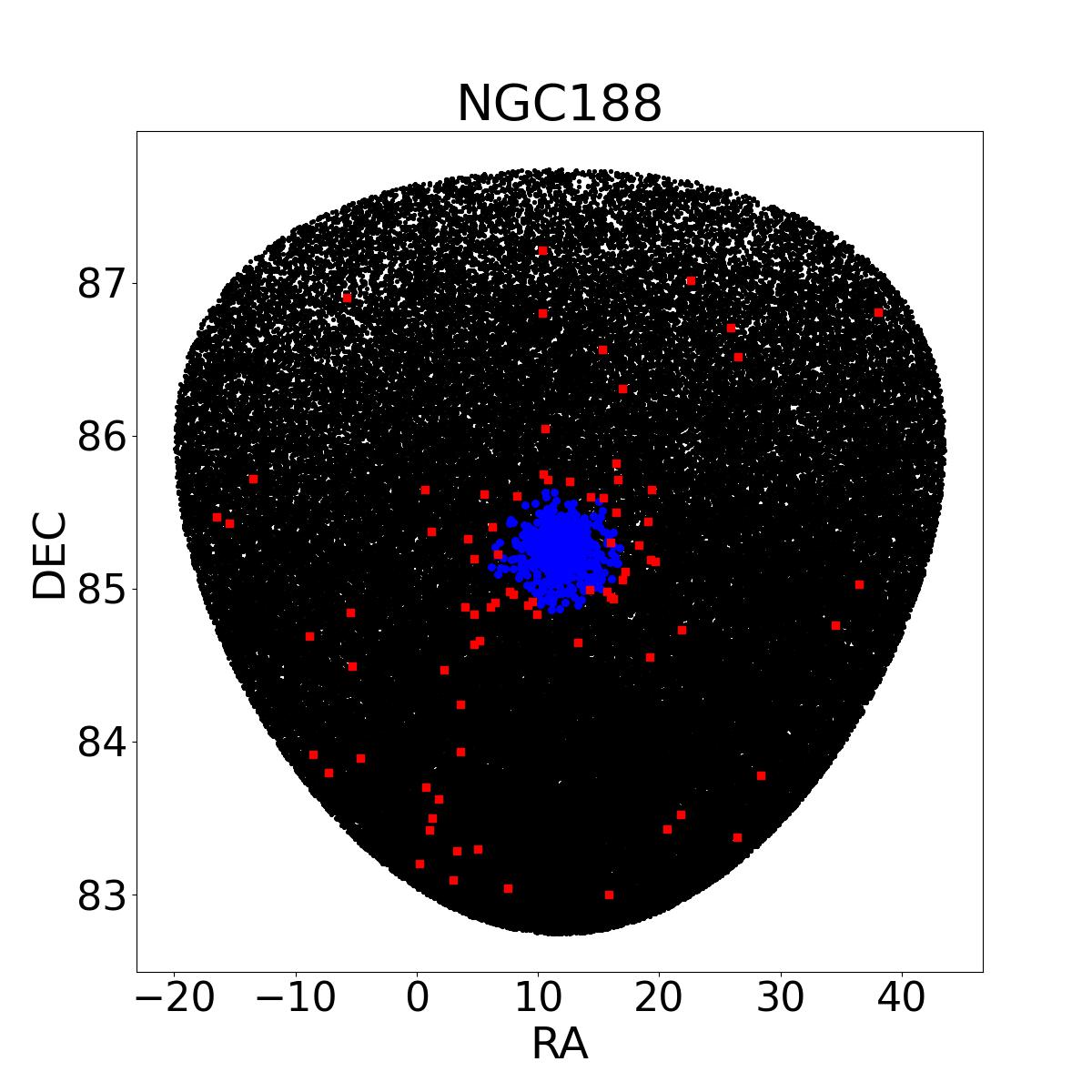}

        \end{subfigure}
        \begin{subfigure}{0.25\textwidth}
        \centering
           \includegraphics[width=\textwidth]{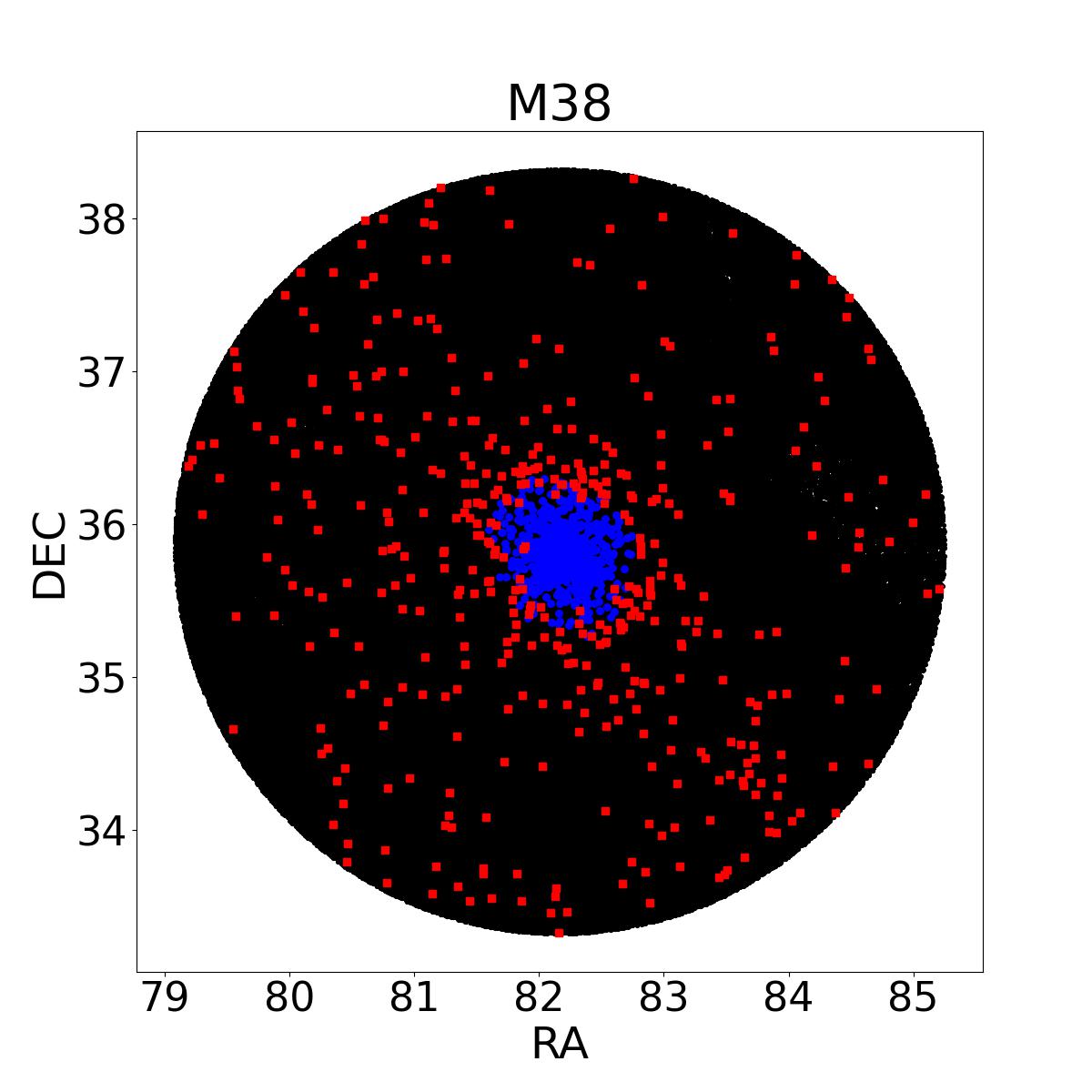}

        \end{subfigure}
        \begin{subfigure}{0.25\textwidth}
        \centering
           \includegraphics[width=\textwidth]{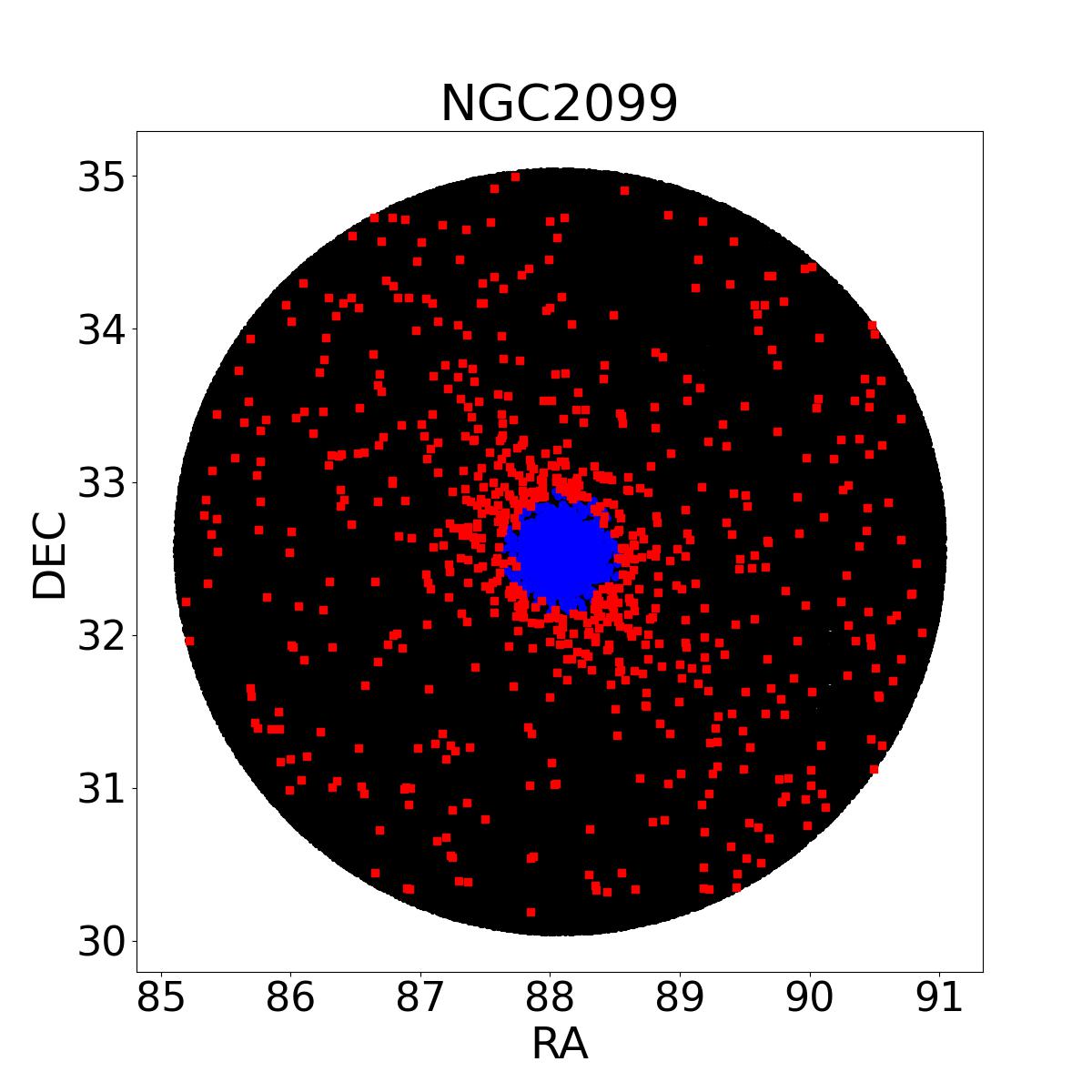}

        \end{subfigure}
        \begin{subfigure}{0.25\textwidth}
        \centering
           \includegraphics[width=\textwidth]{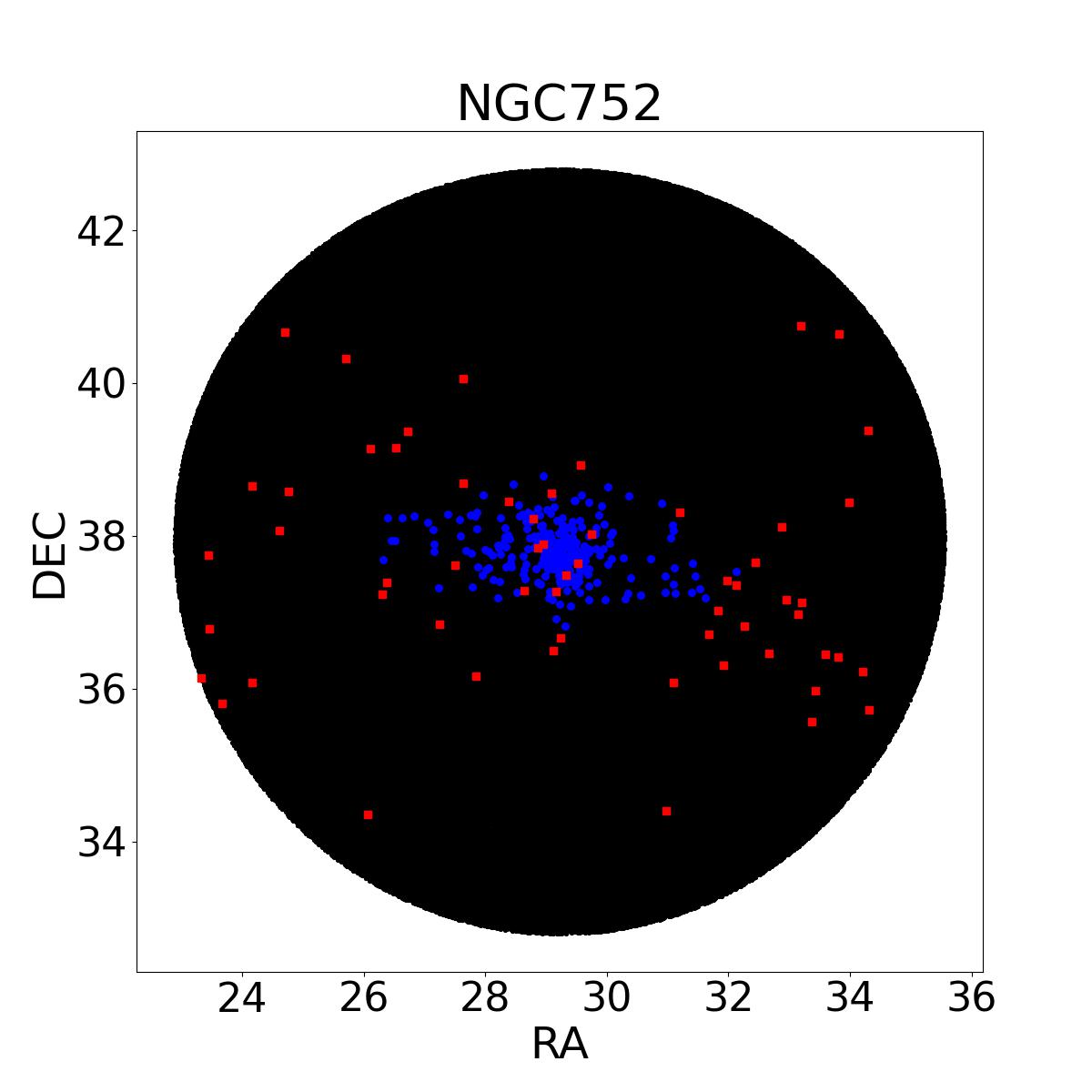}

        \end{subfigure}
        \begin{subfigure}{0.25\textwidth}
        \centering
           \includegraphics[width=\textwidth]{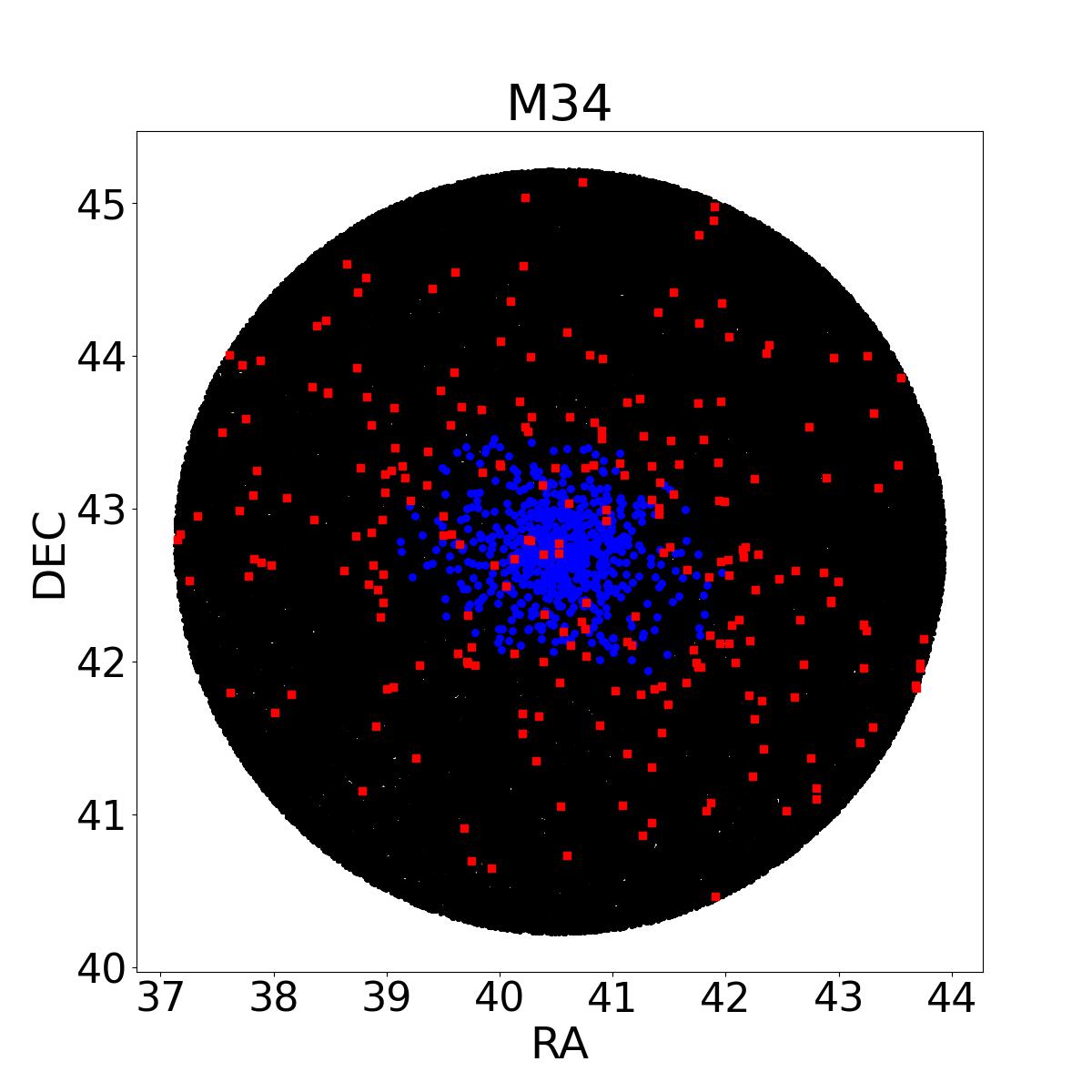}

        \end{subfigure}
        \begin{subfigure}{0.25\textwidth}
        \centering
           \includegraphics[width=\textwidth]{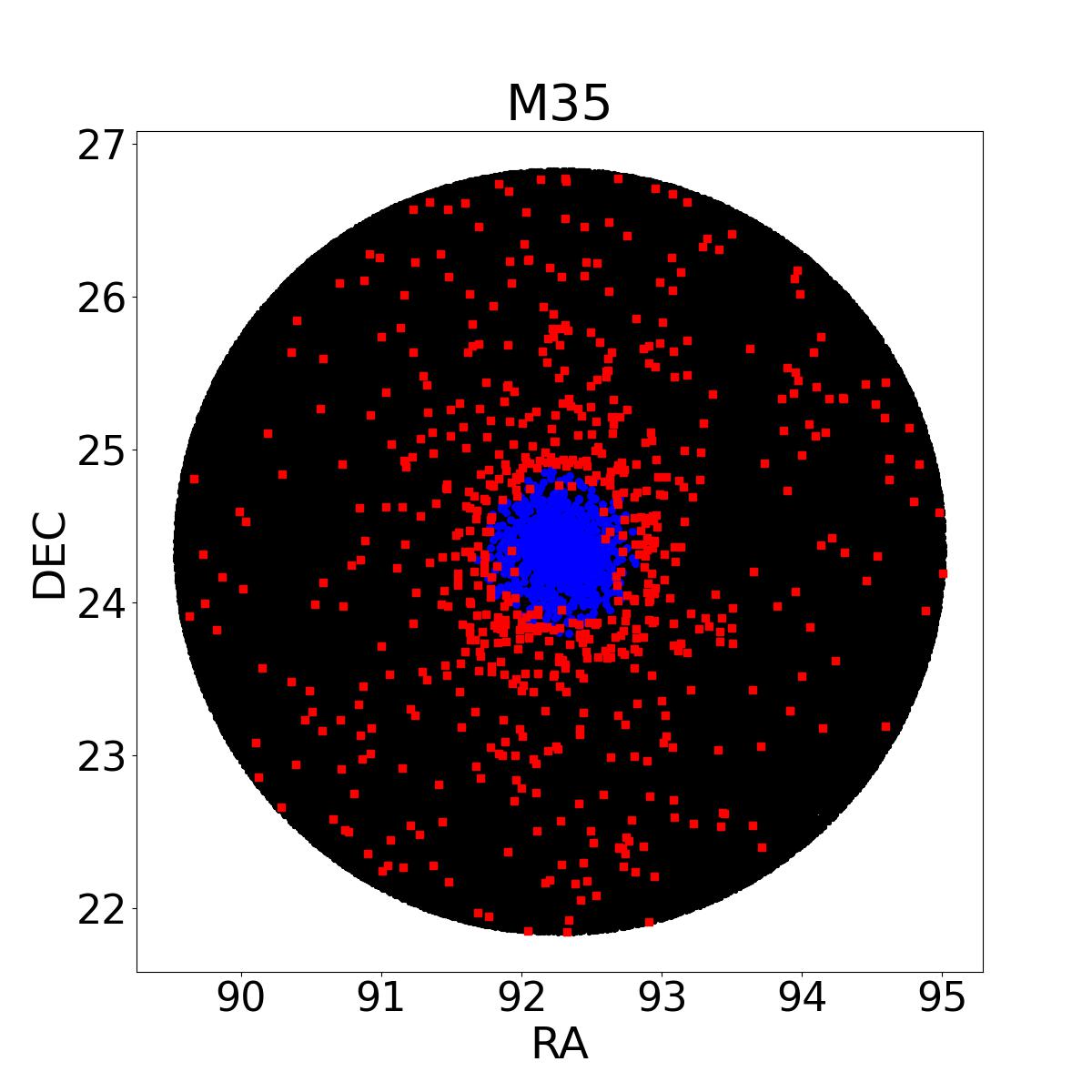}

        \end{subfigure}
        \begin{subfigure}{0.25\textwidth}
        \centering
           \includegraphics[width=\textwidth]{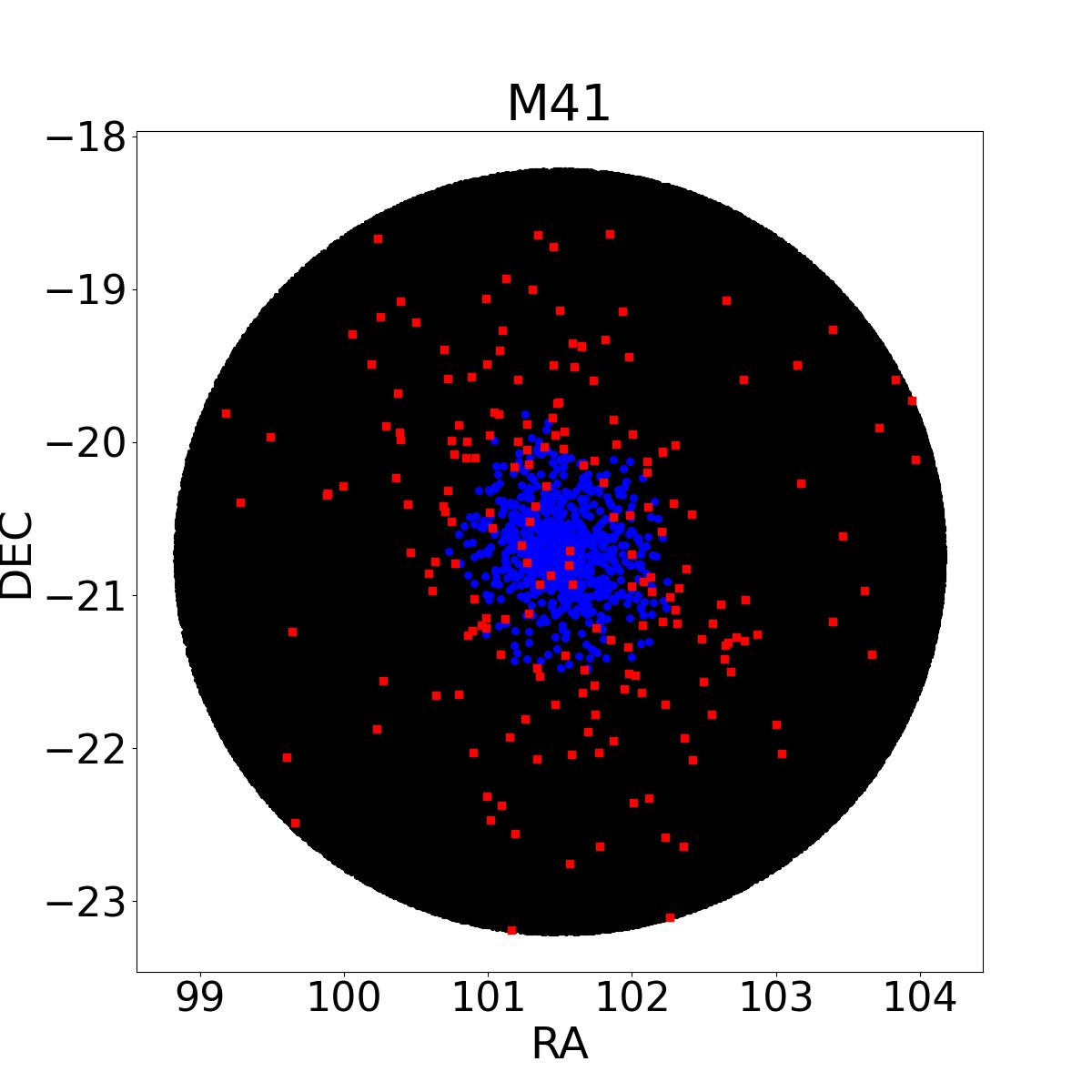}

        \end{subfigure}
        \begin{subfigure}{0.25\textwidth}
        \centering
           \includegraphics[width=\textwidth]{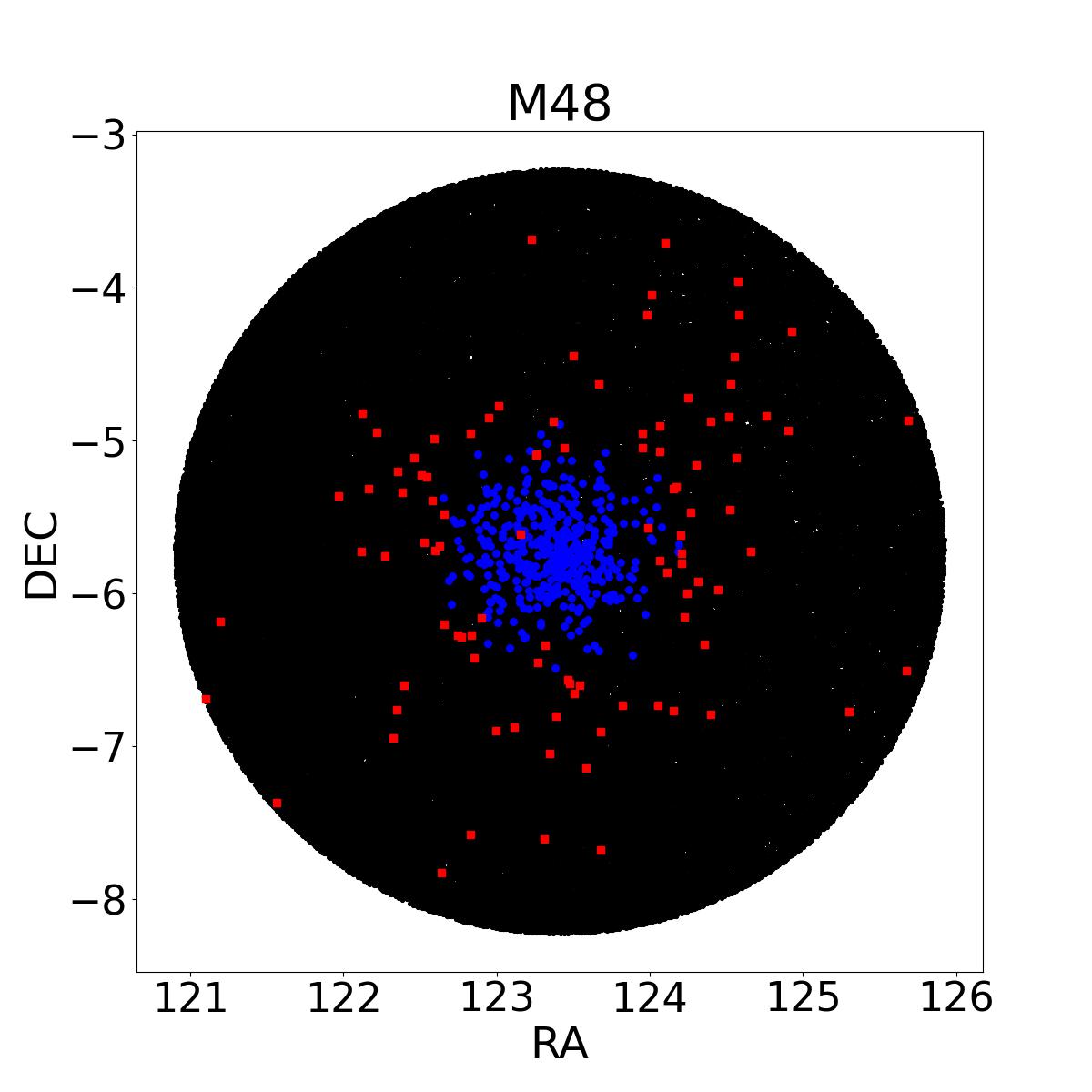}

        \end{subfigure}
        \begin{subfigure}{0.25\textwidth}
        \centering
           \includegraphics[width=\textwidth]{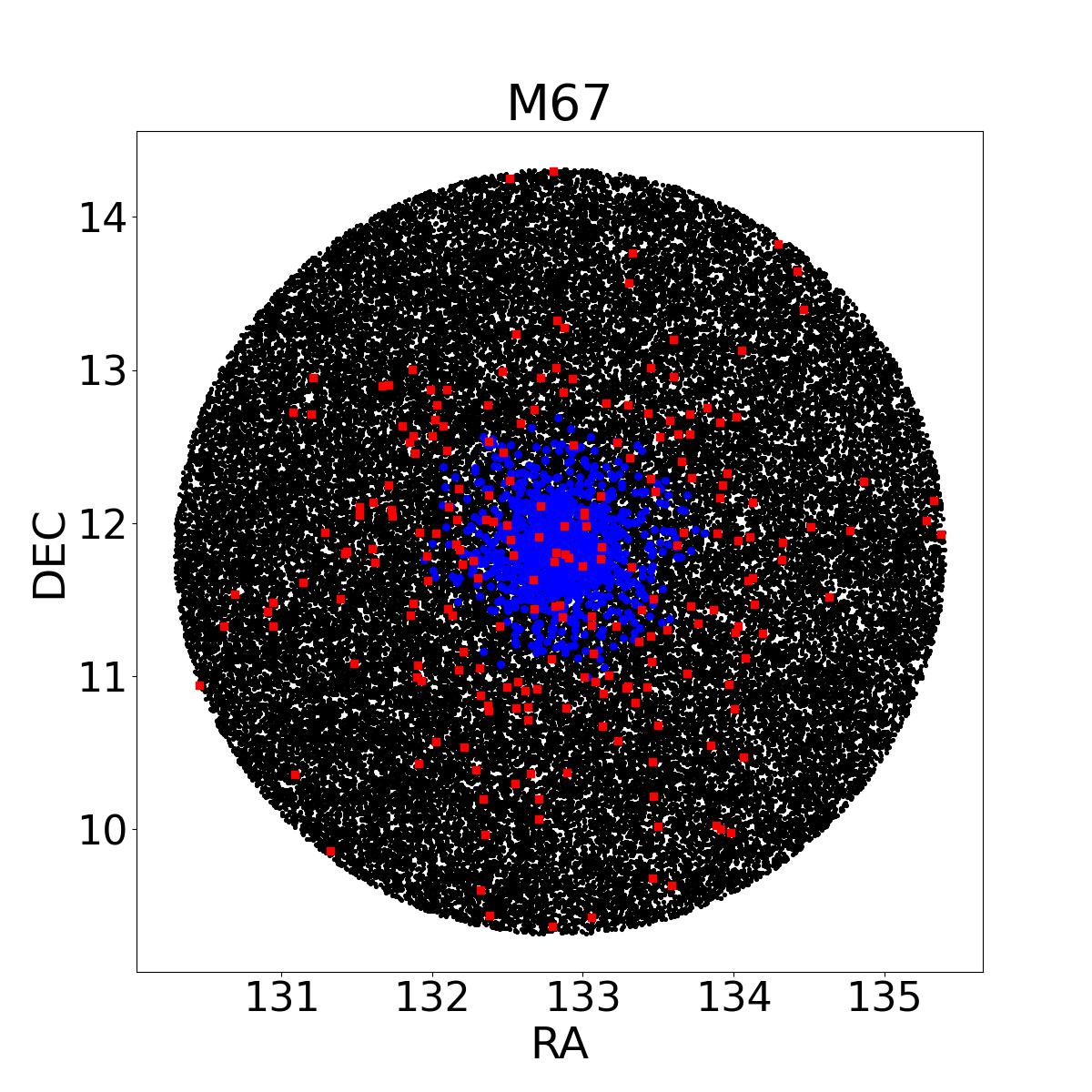}

        \end{subfigure}
        \begin{subfigure}{0.25\textwidth}
        \centering
           \includegraphics[width=\textwidth]{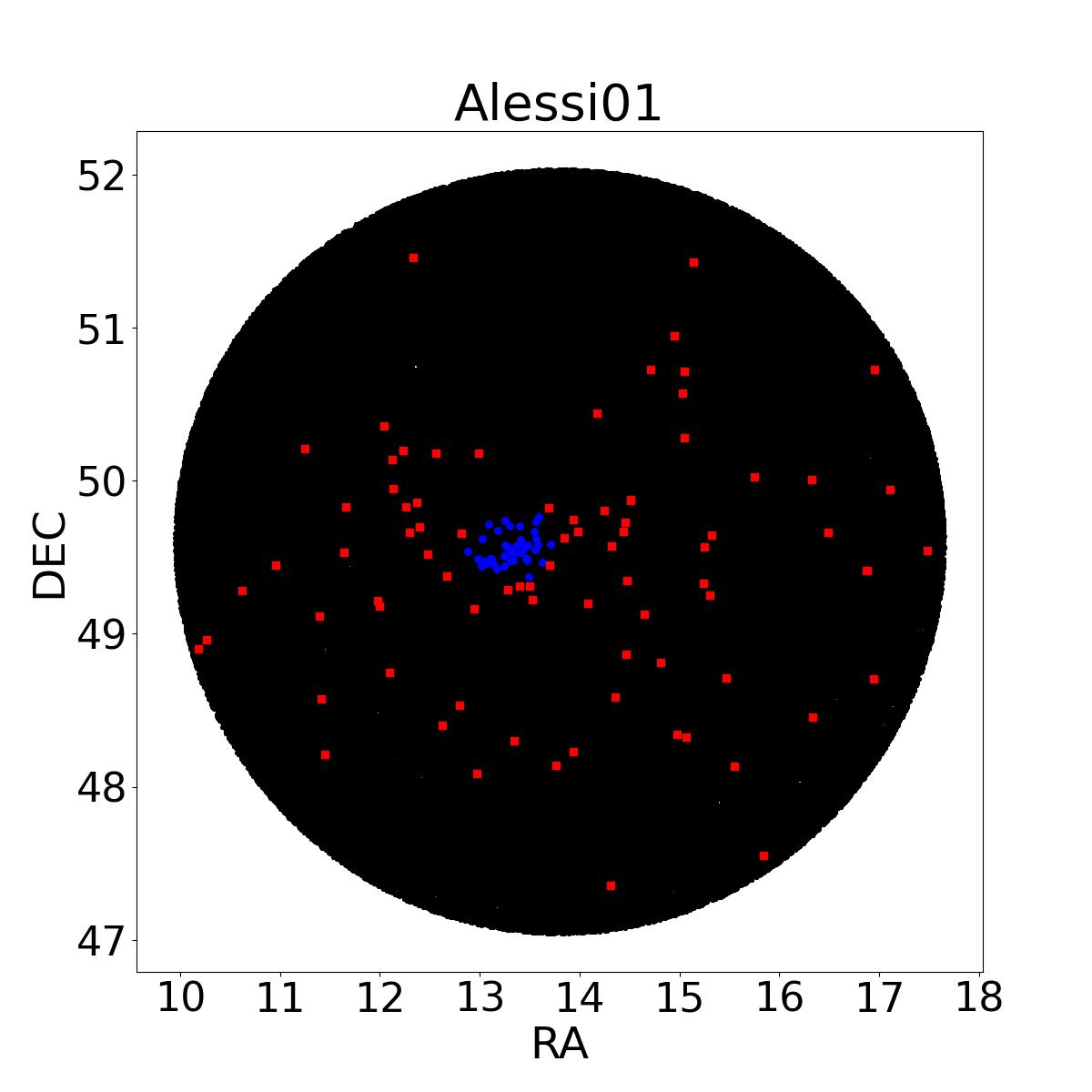}

        \end{subfigure}

  \caption{The Position of the Cluster's Inner and Outer Members Among the Field Stars: Black dots represent the field stars, while red dots indicate stars that were selected by Random Forest algorithm, and blue dots show stars that were selected by the GMM algorithm with a probability higher than 0.8.}
  \label{position dgr.fig}
\end{figure}

\begin{figure}
  \centering
  \captionsetup[subfigure]{labelformat=empty}
        \begin{subfigure}{0.25\textwidth}
        \centering
           \includegraphics[width=\textwidth]{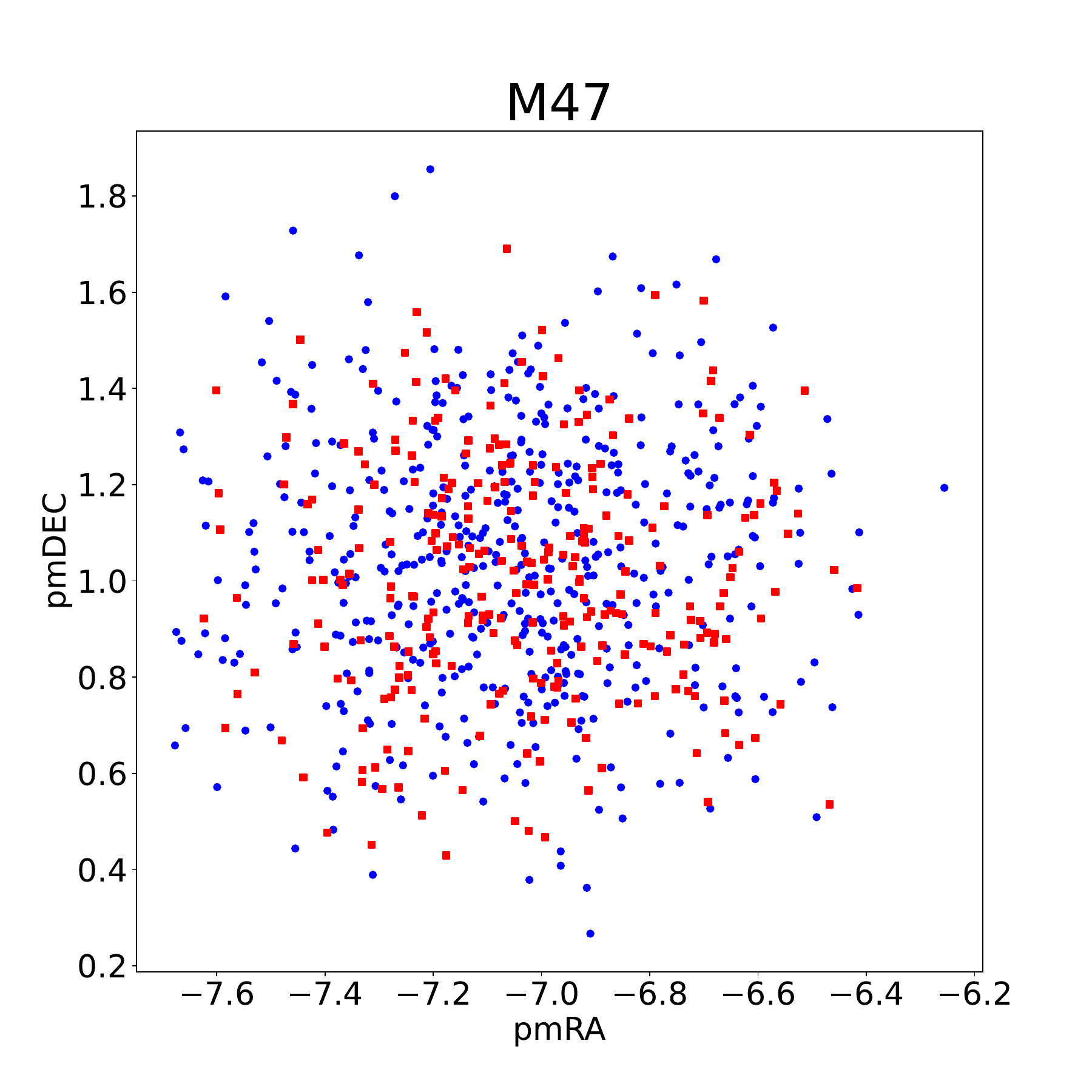}

        \end{subfigure}
        \begin{subfigure}{0.25\textwidth}

                \centering
                \includegraphics[width=\textwidth]{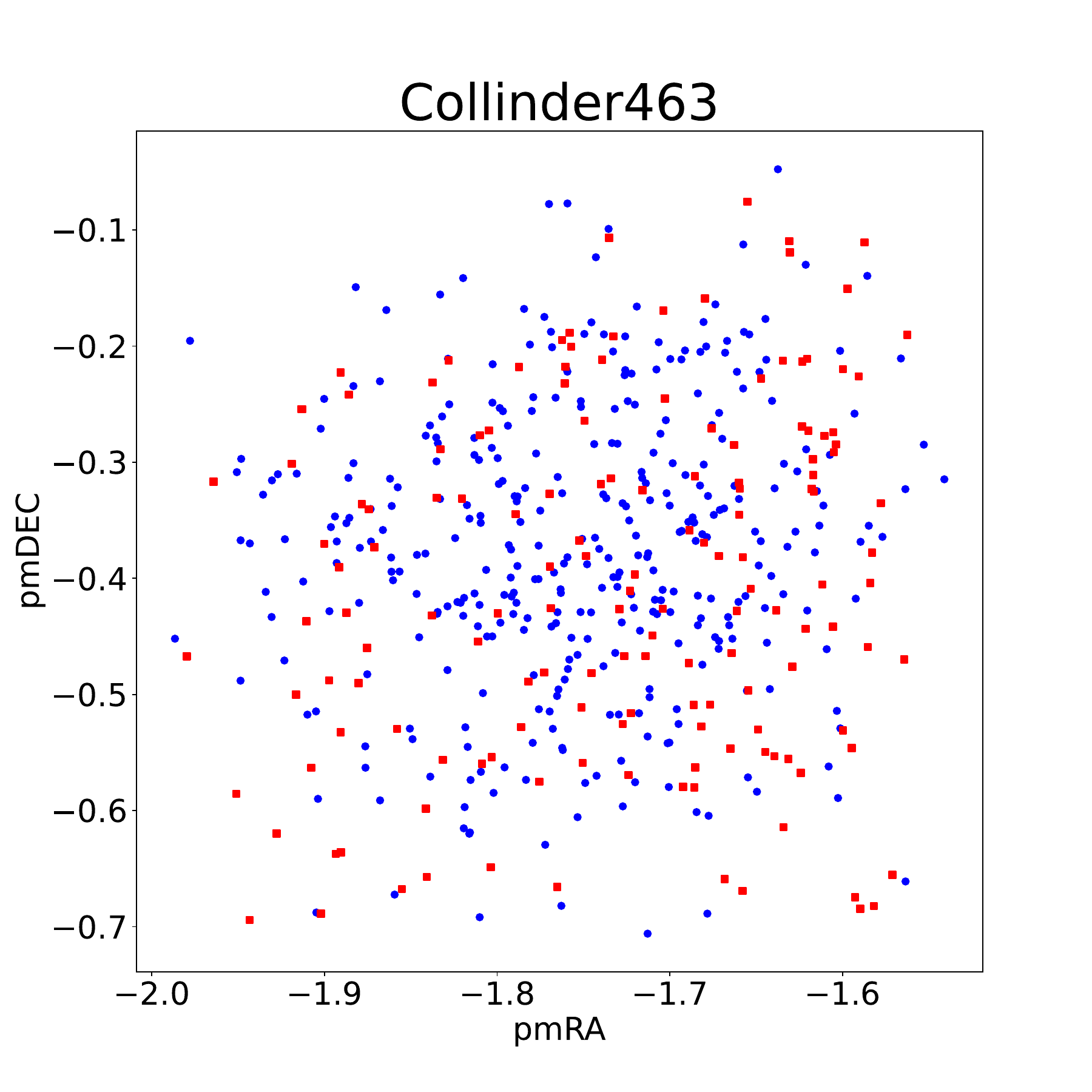}

        \end{subfigure}
        \begin{subfigure}{0.25\textwidth}
                \centering
           \includegraphics[width=\textwidth]{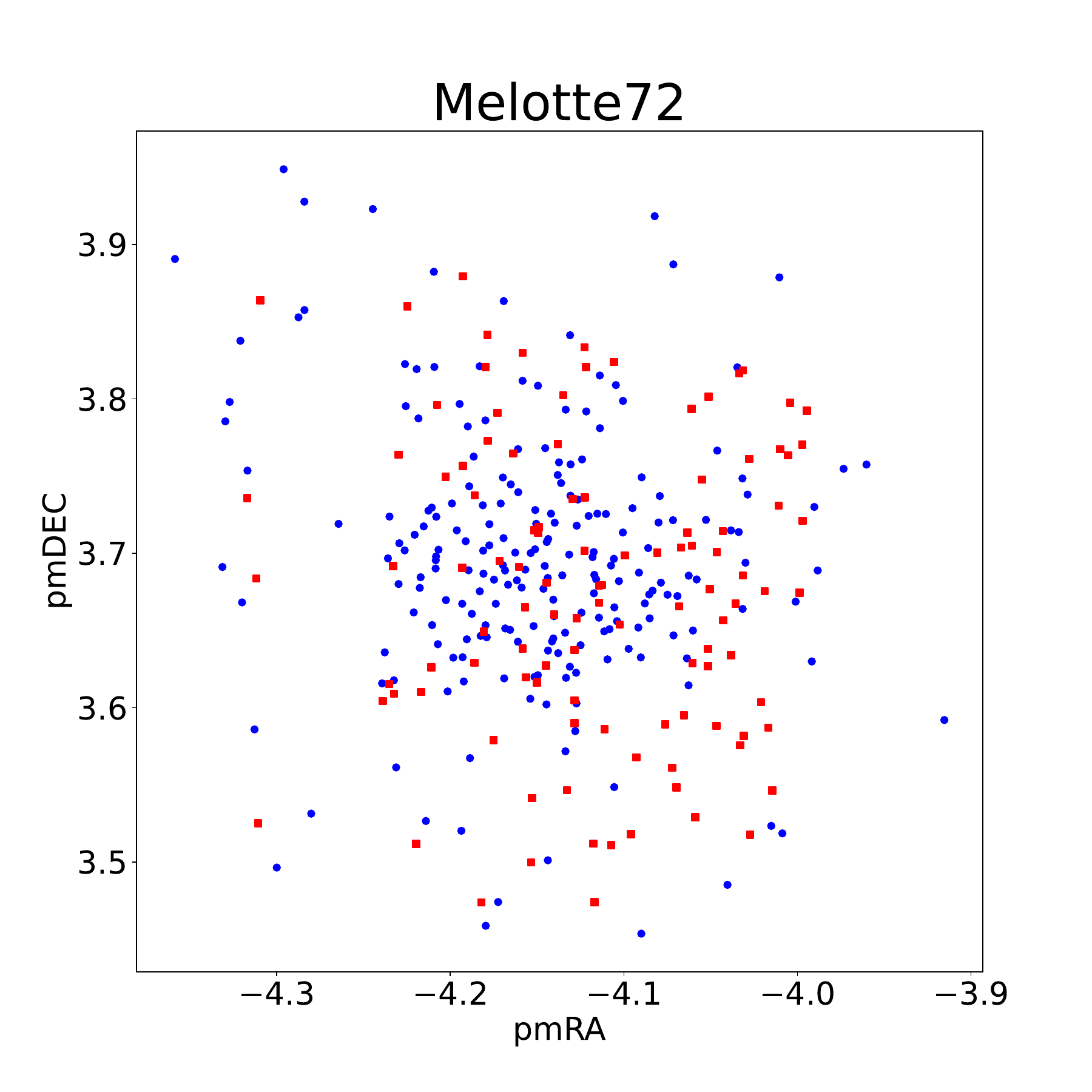}

        \end{subfigure}
        \begin{subfigure}{0.25\textwidth}
                \centering

                \includegraphics[width=\textwidth]{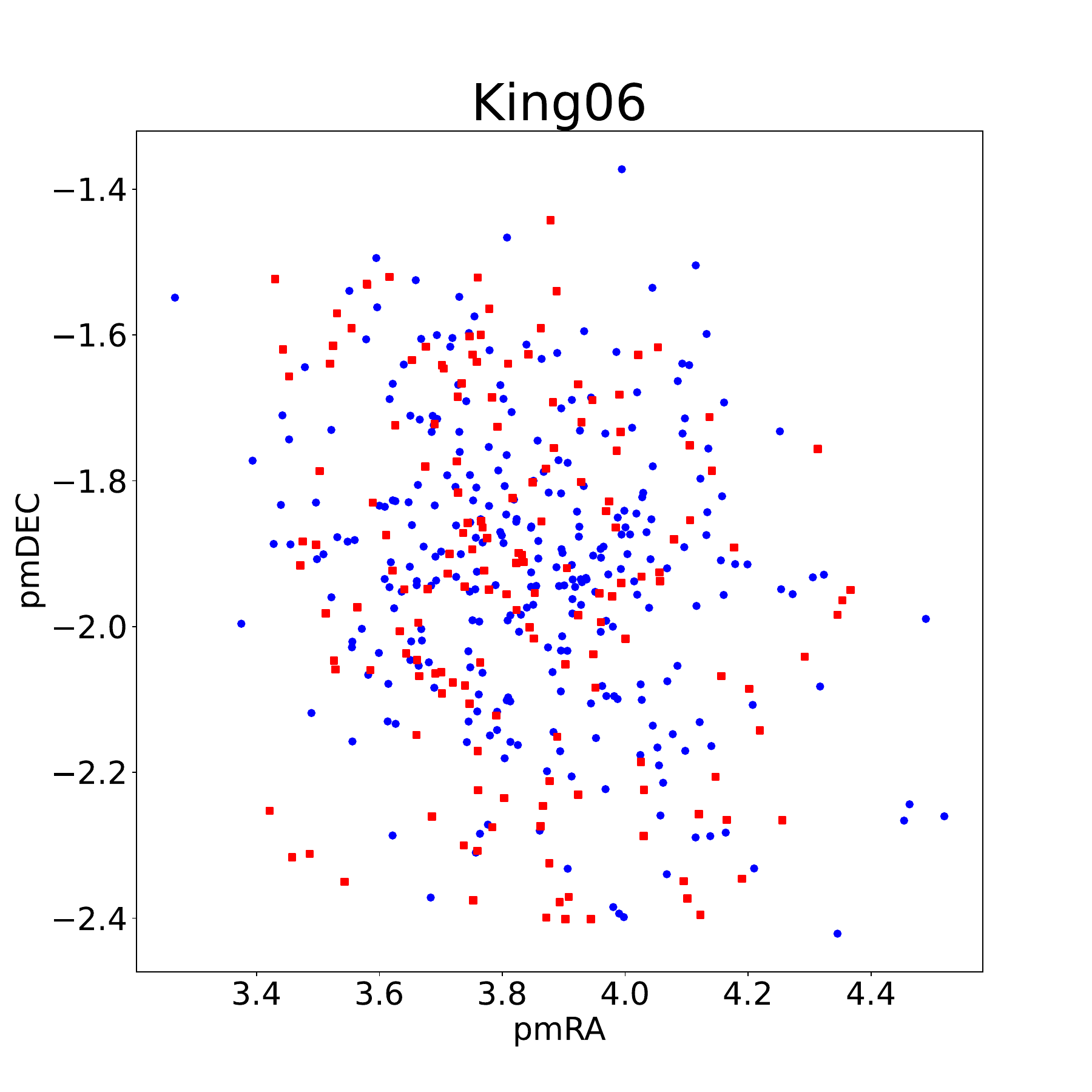}

        \end{subfigure}
        \begin{subfigure}{0.25\textwidth}
                \centering

                \includegraphics[width=\textwidth]{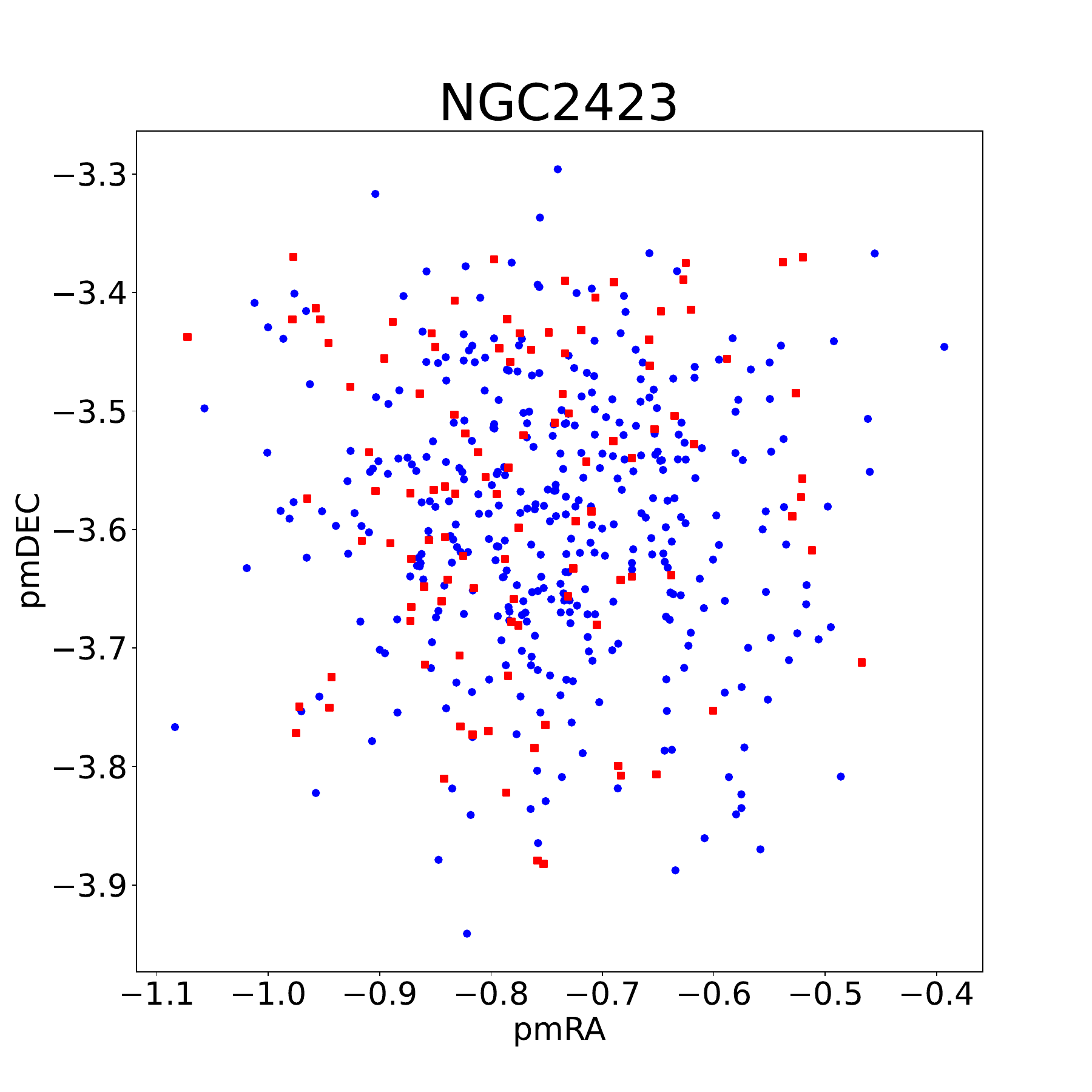}

        \end{subfigure}
        \begin{subfigure}{0.25\textwidth}
                \centering

                \includegraphics[width=\textwidth]{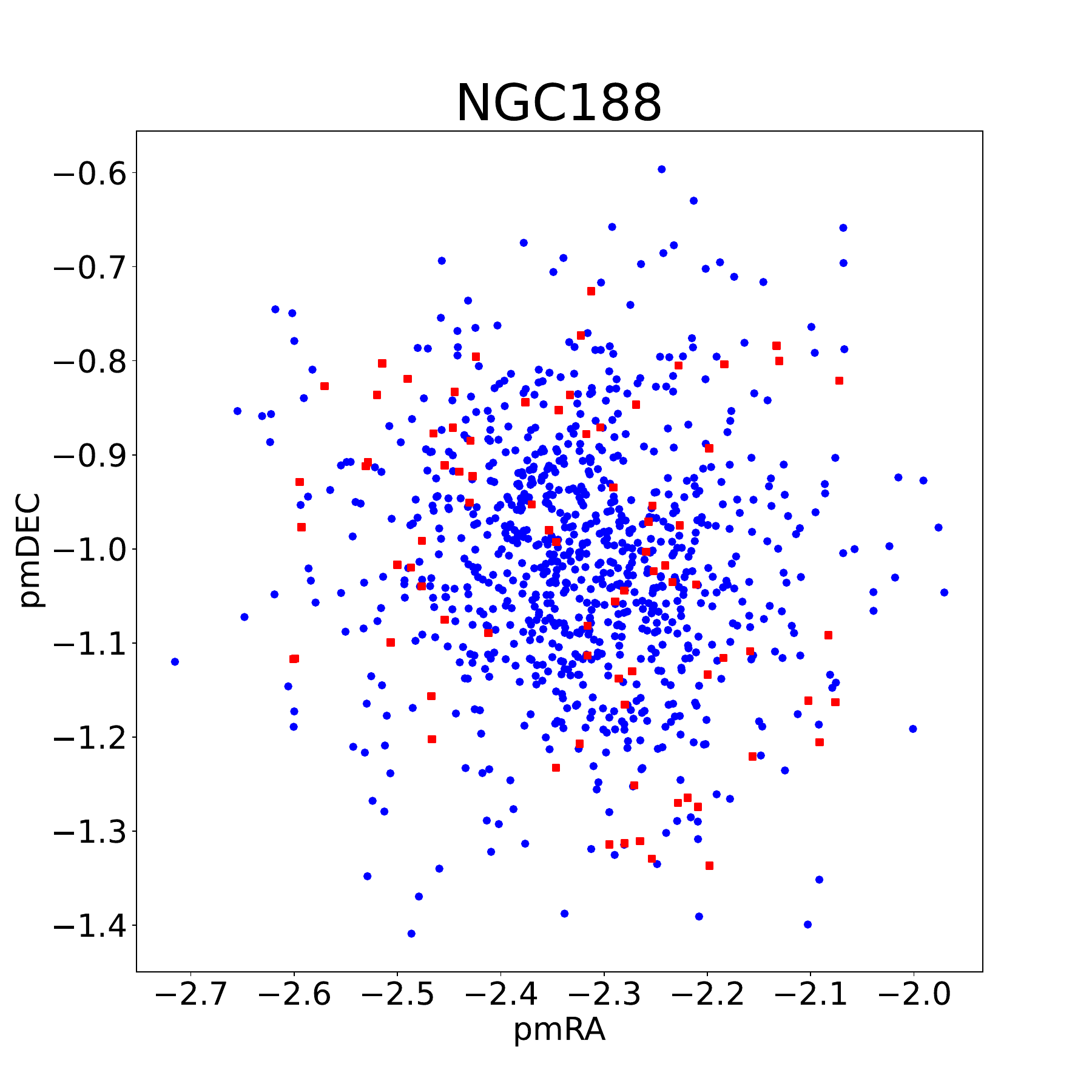}

        \end{subfigure}
        \begin{subfigure}{0.25\textwidth}
        \centering
           \includegraphics[width=\textwidth]{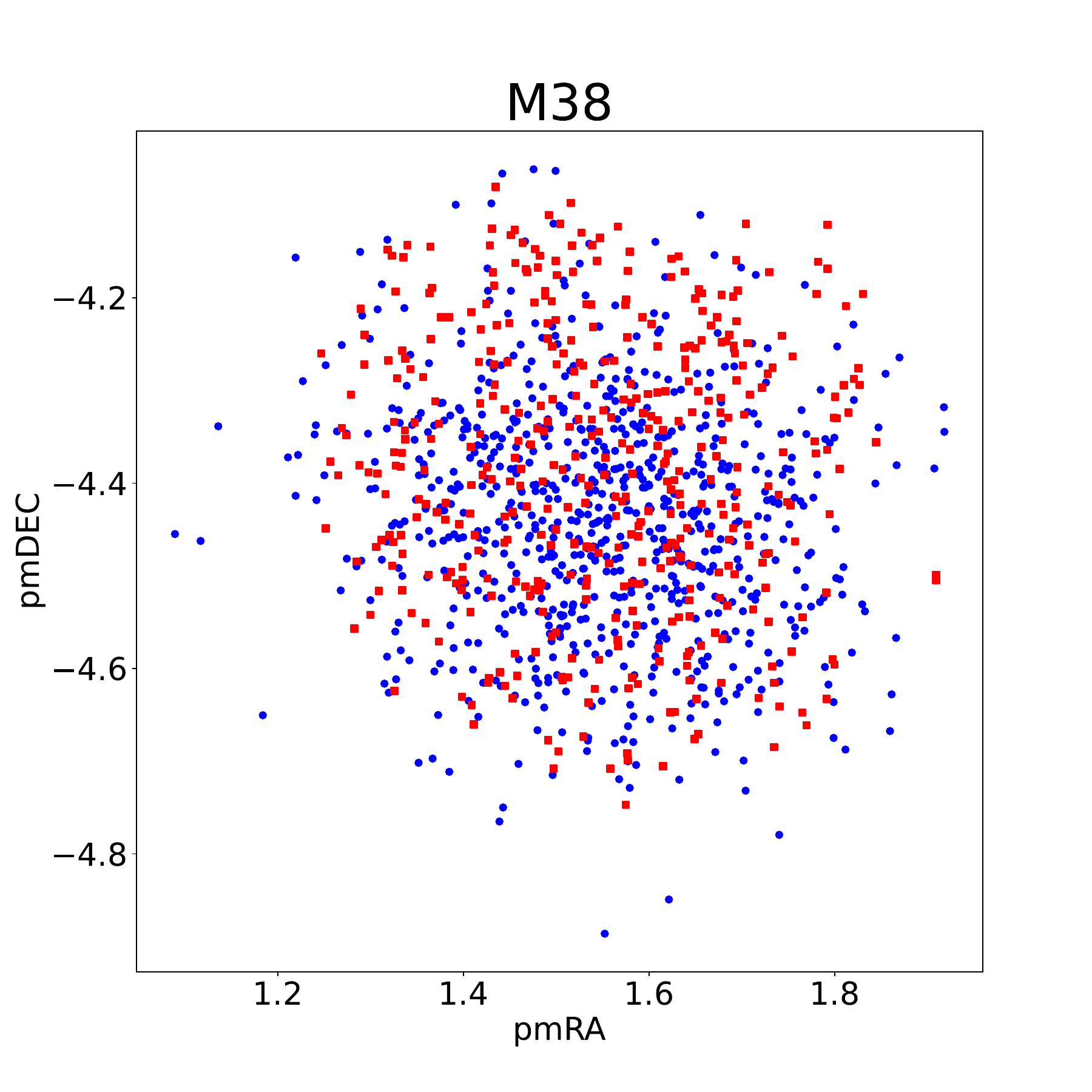}

        \end{subfigure}
        \begin{subfigure}{0.25\textwidth}
        \centering
           \includegraphics[width=\textwidth]{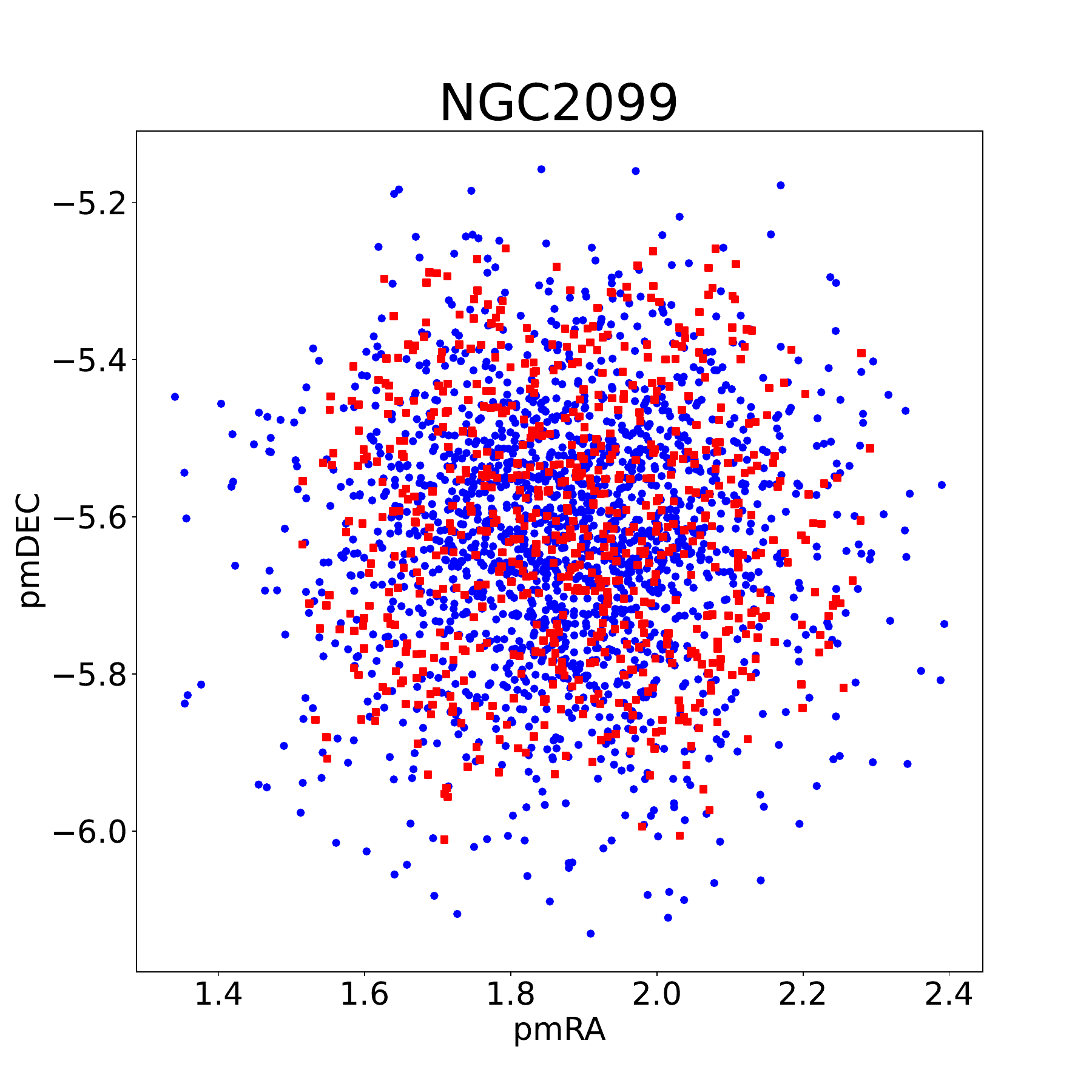}

        \end{subfigure}
        \begin{subfigure}{0.25\textwidth}
        \centering
           \includegraphics[width=\textwidth]{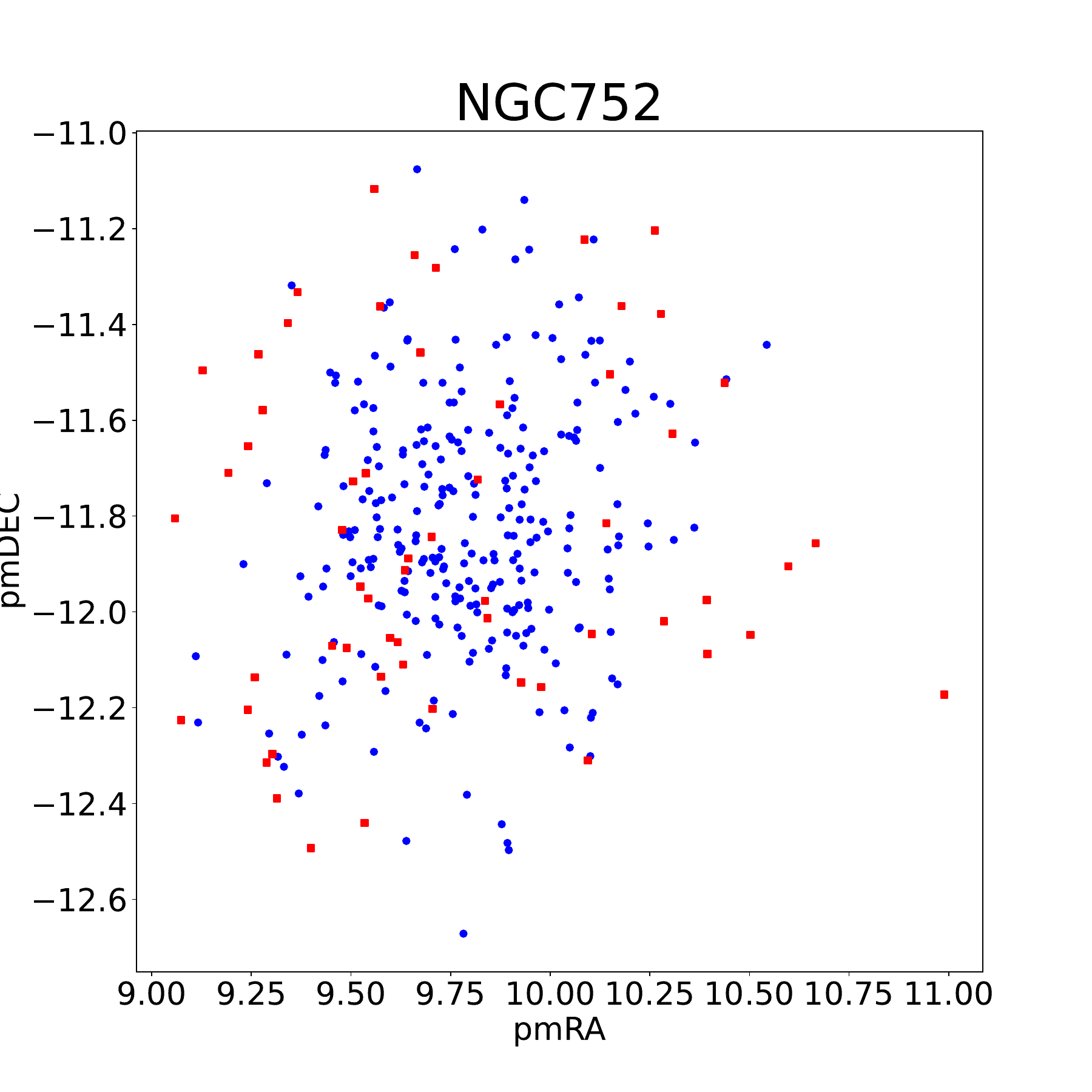}

        \end{subfigure}
        \begin{subfigure}{0.25\textwidth}
        \centering
           \includegraphics[width=\textwidth]{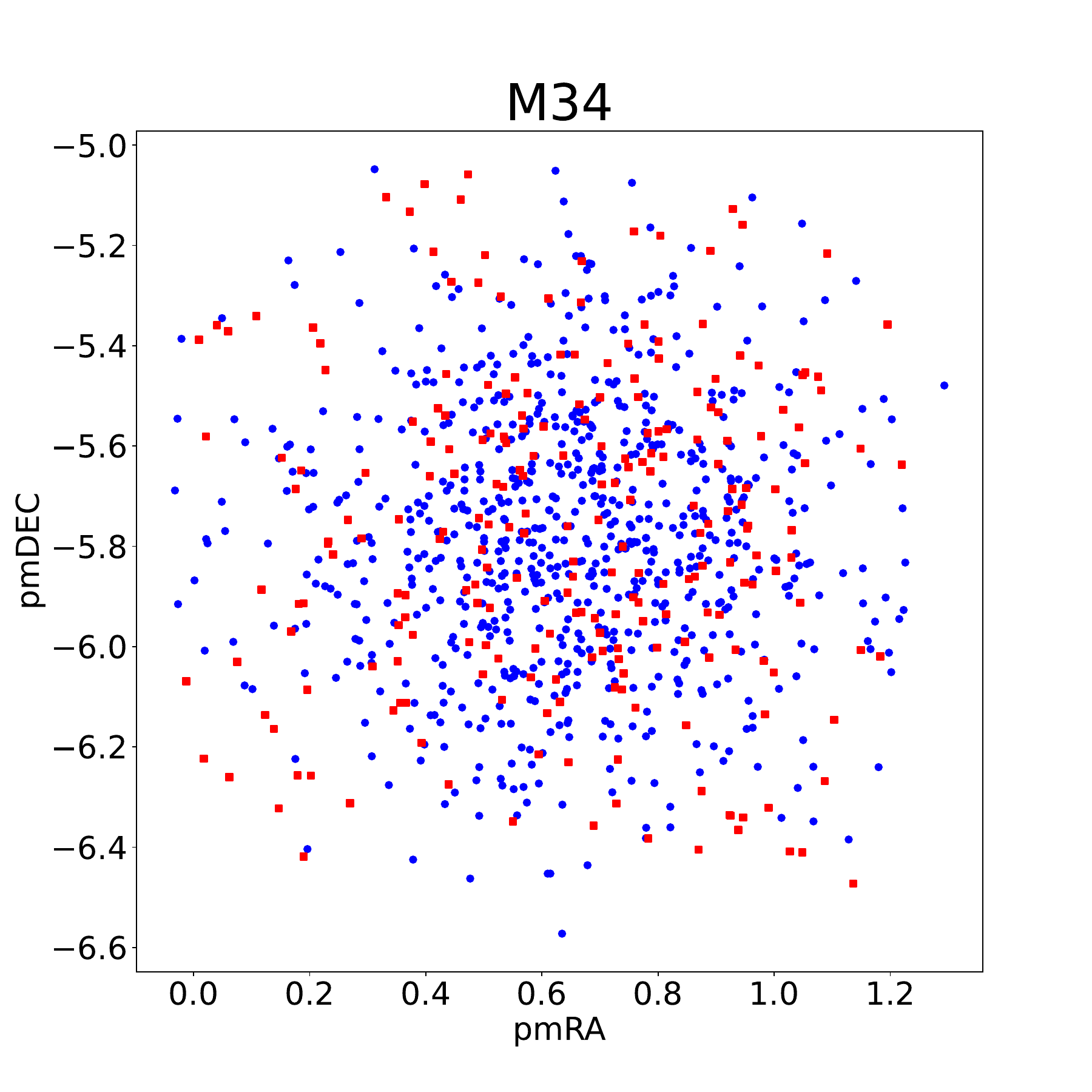}

        \end{subfigure}
        \begin{subfigure}{0.25\textwidth}
        \centering
           \includegraphics[width=\textwidth]{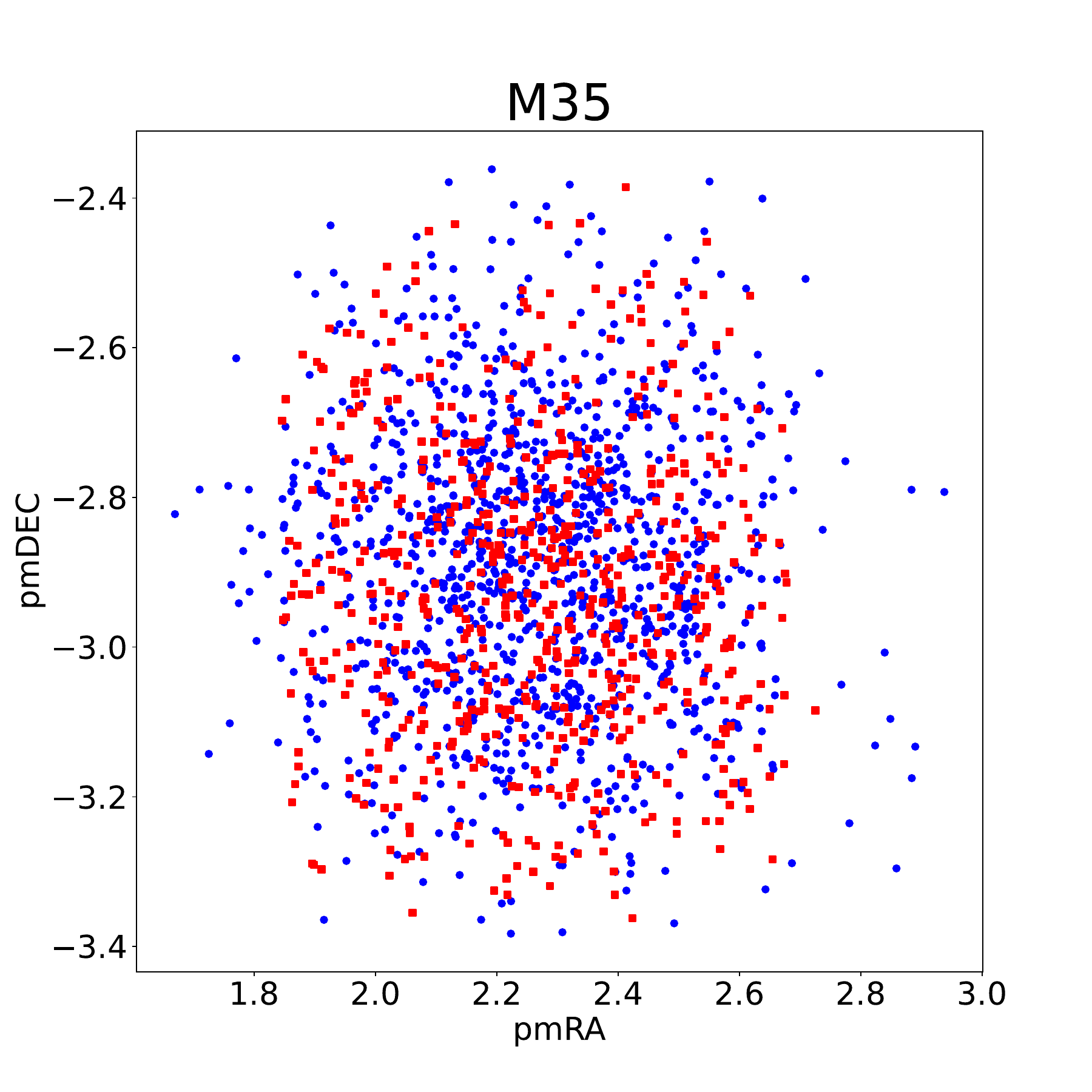}

        \end{subfigure}
        \begin{subfigure}{0.25\textwidth}
        \centering
           \includegraphics[width=\textwidth]{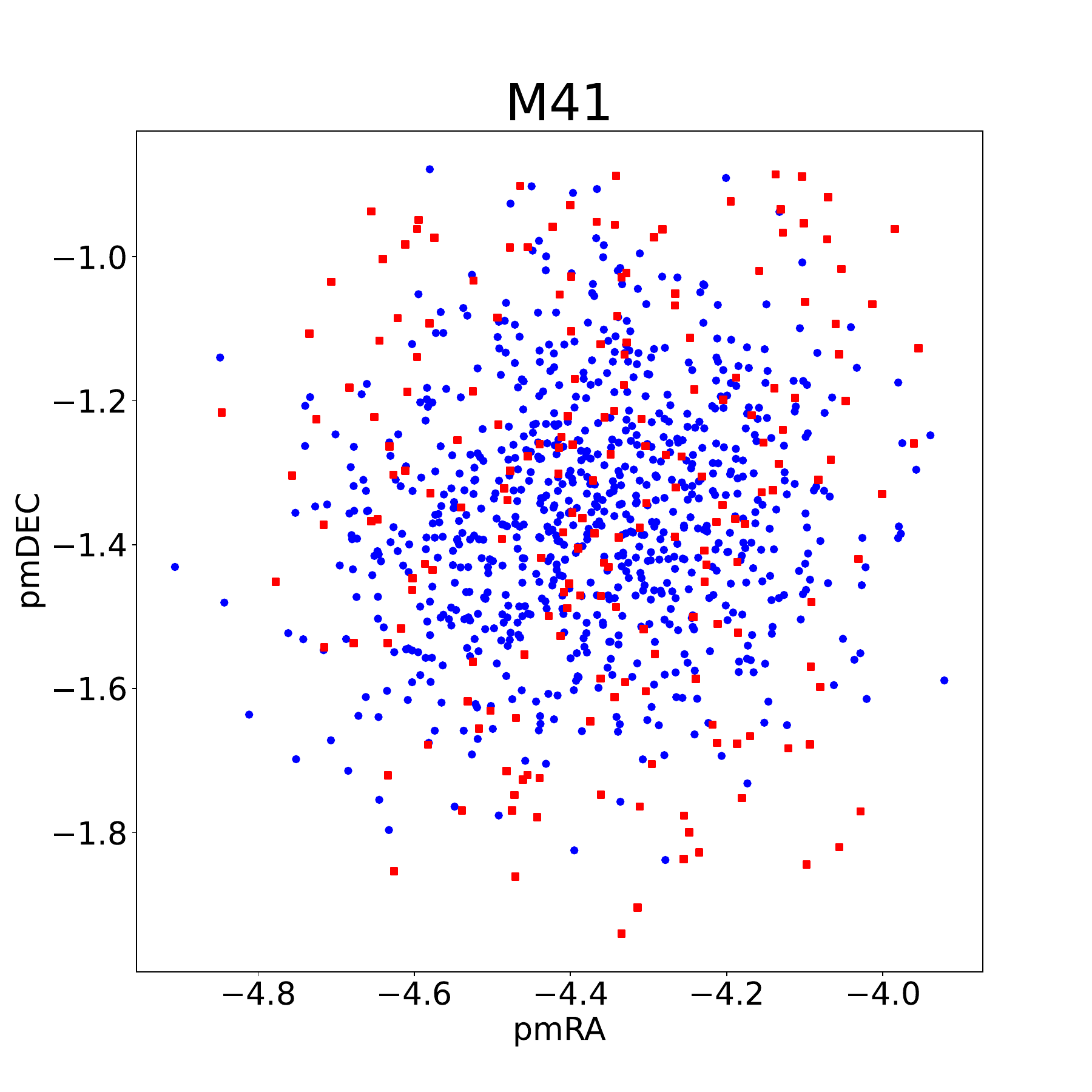}

        \end{subfigure}
        \begin{subfigure}{0.25\textwidth}
        \centering
           \includegraphics[width=\textwidth]{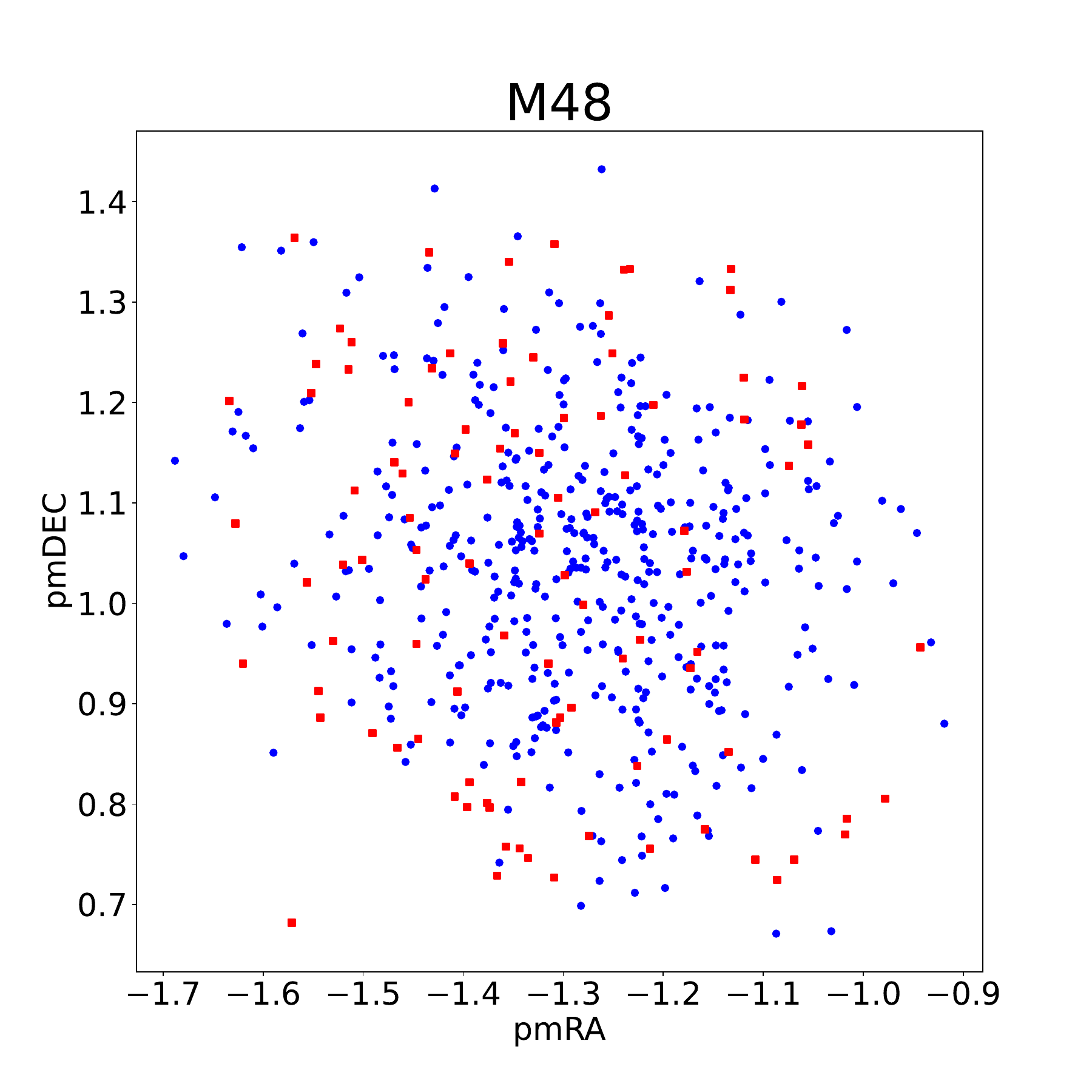}

        \end{subfigure}
        \begin{subfigure}{0.25\textwidth}
        \centering
           \includegraphics[width=\textwidth]{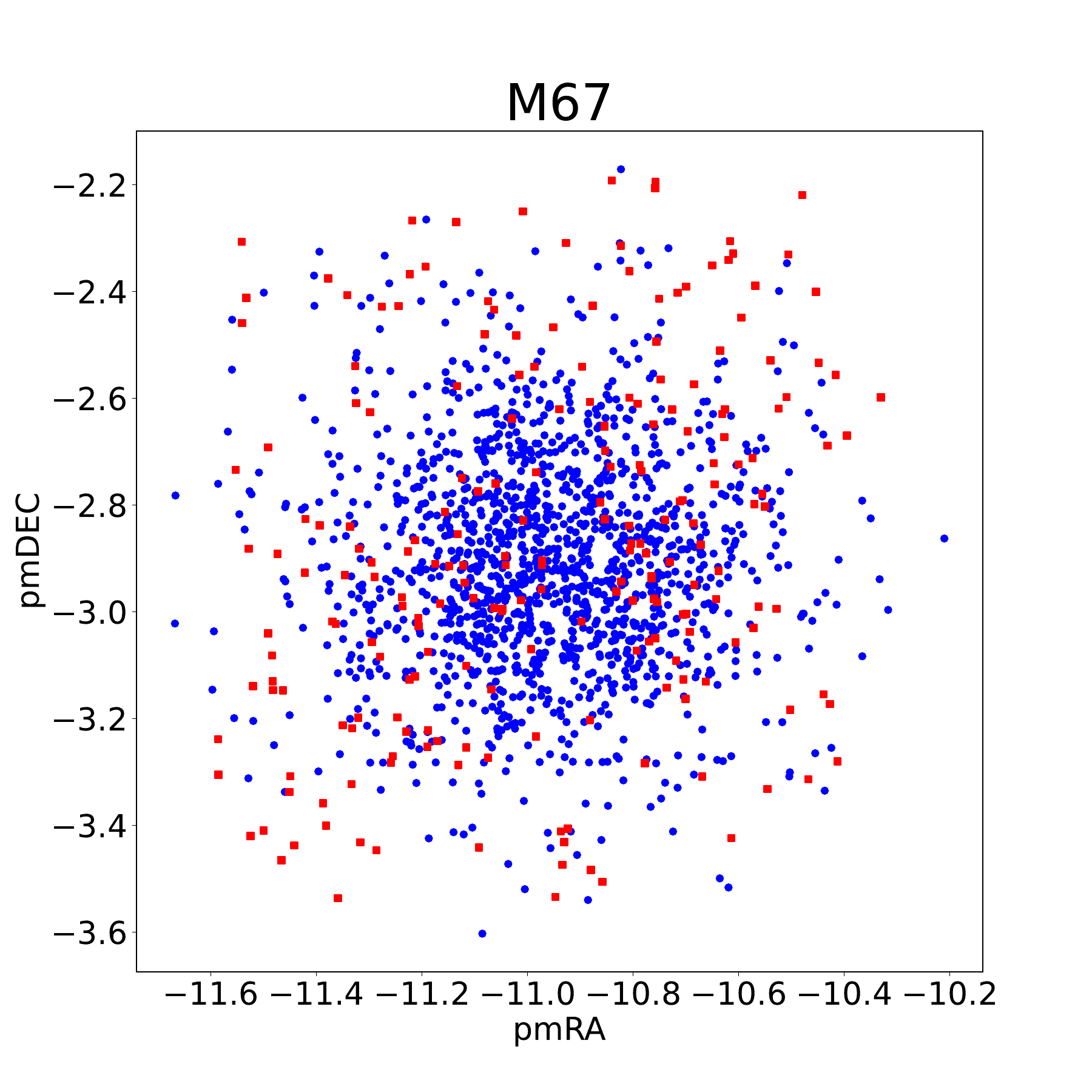}

        \end{subfigure}
        \begin{subfigure}{0.25\textwidth}
        \centering
           \includegraphics[width=\textwidth]{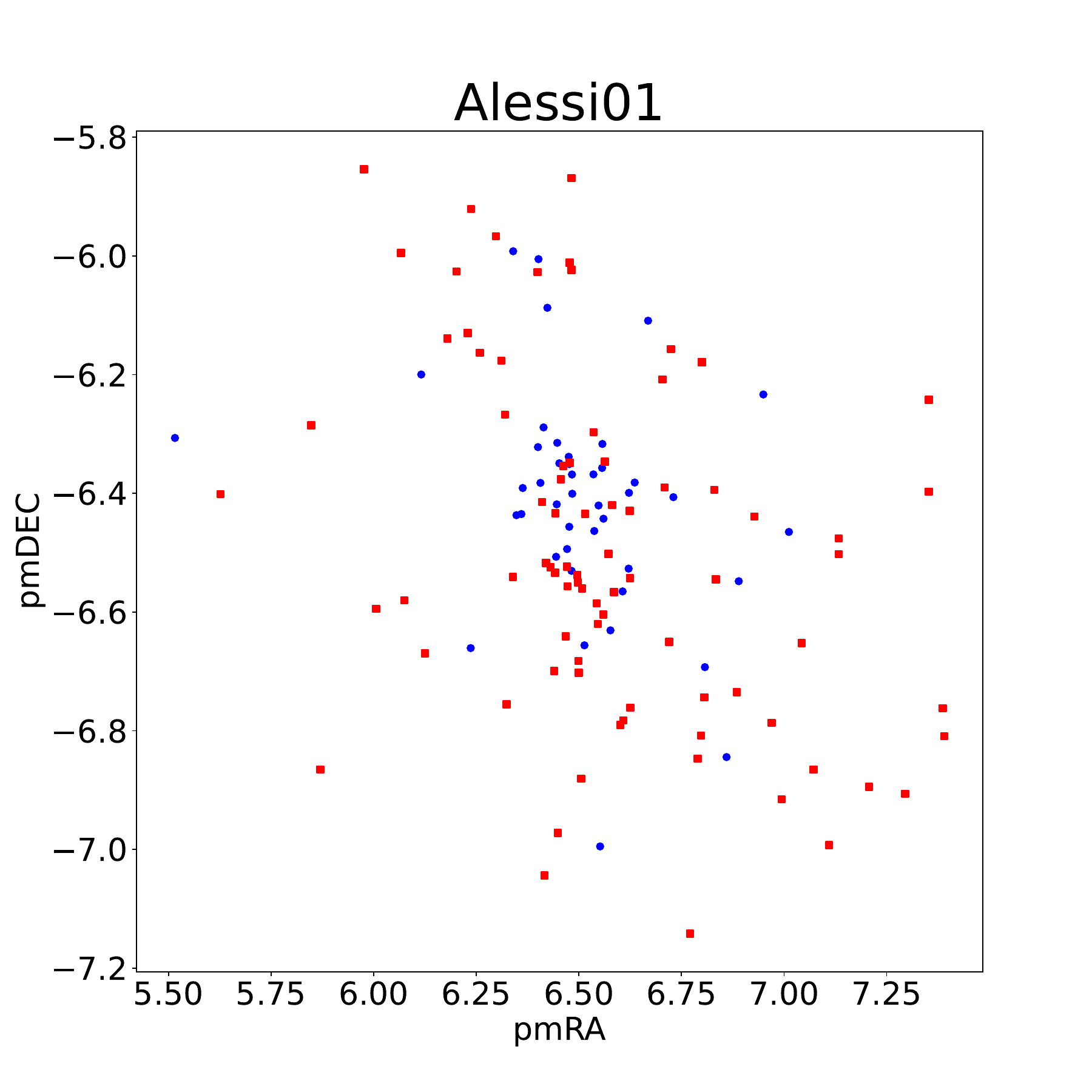}

        \end{subfigure}
  \caption{The proper motion of the clusters inner and outer members. Red dots represent stars that were selected by Random Forest, while blue dots indicate stars that were selected by GMM with a probability higher than 0.8.}
  \label{proper motion dgr.fig}
\end{figure}

\begin{figure}
  \centering
  \captionsetup[subfigure]{labelformat=empty}
        \begin{subfigure}{0.25\textwidth}
        \centering
           \includegraphics[width=\textwidth]{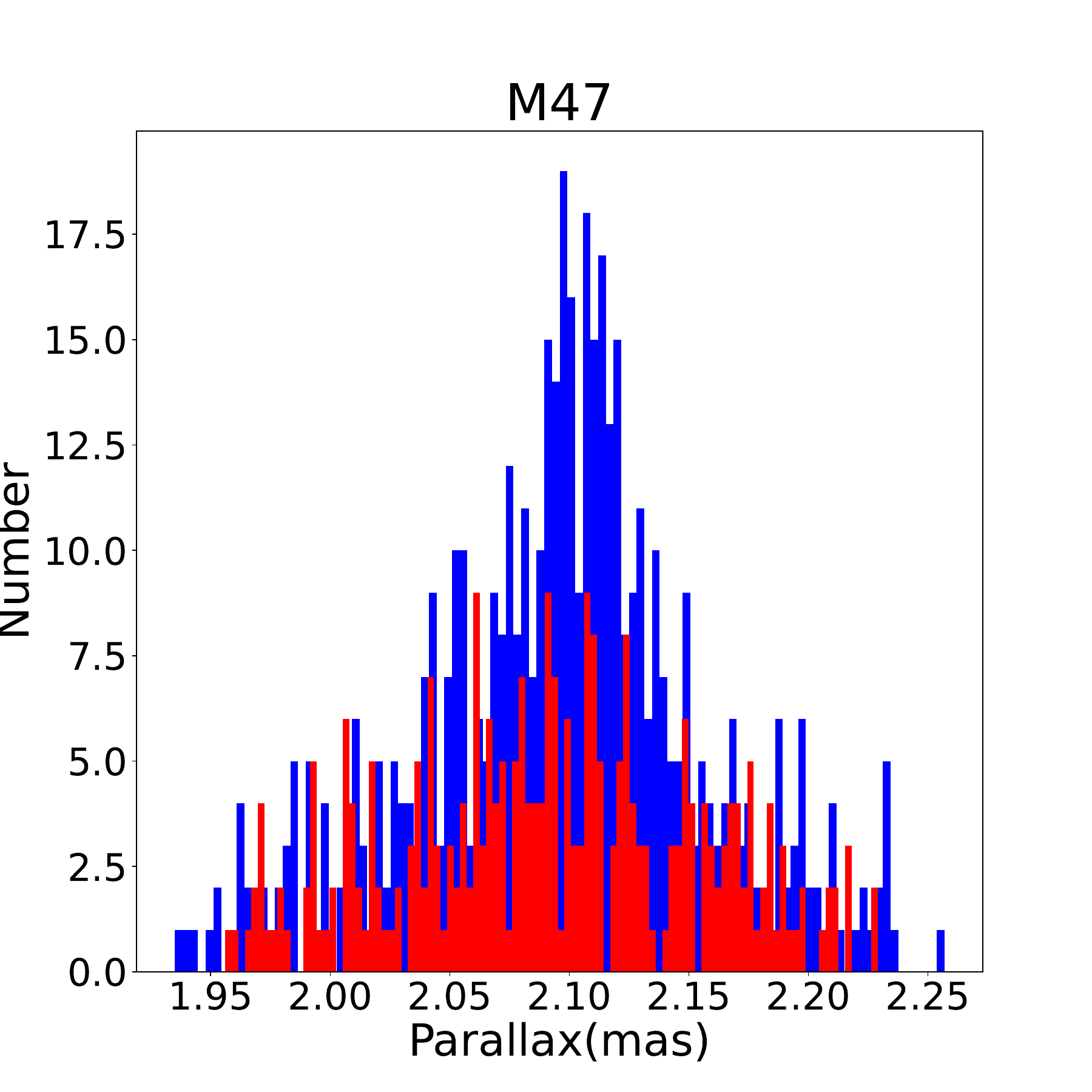}

        \end{subfigure}
        \begin{subfigure}{0.25\textwidth}

                \centering
                \includegraphics[width=\textwidth]{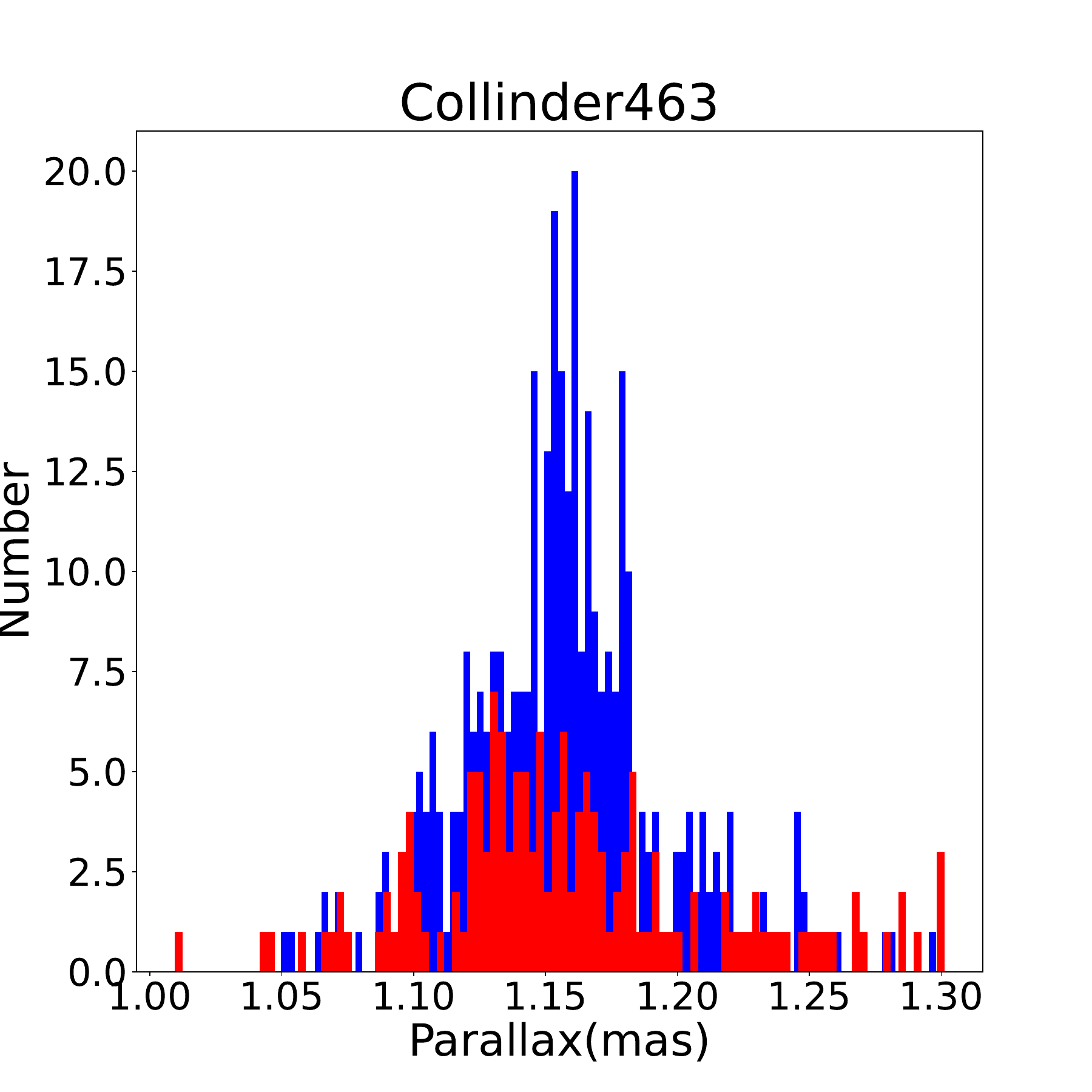}

        \end{subfigure}
        \begin{subfigure}{0.25\textwidth}
                \centering
           \includegraphics[width=\textwidth]{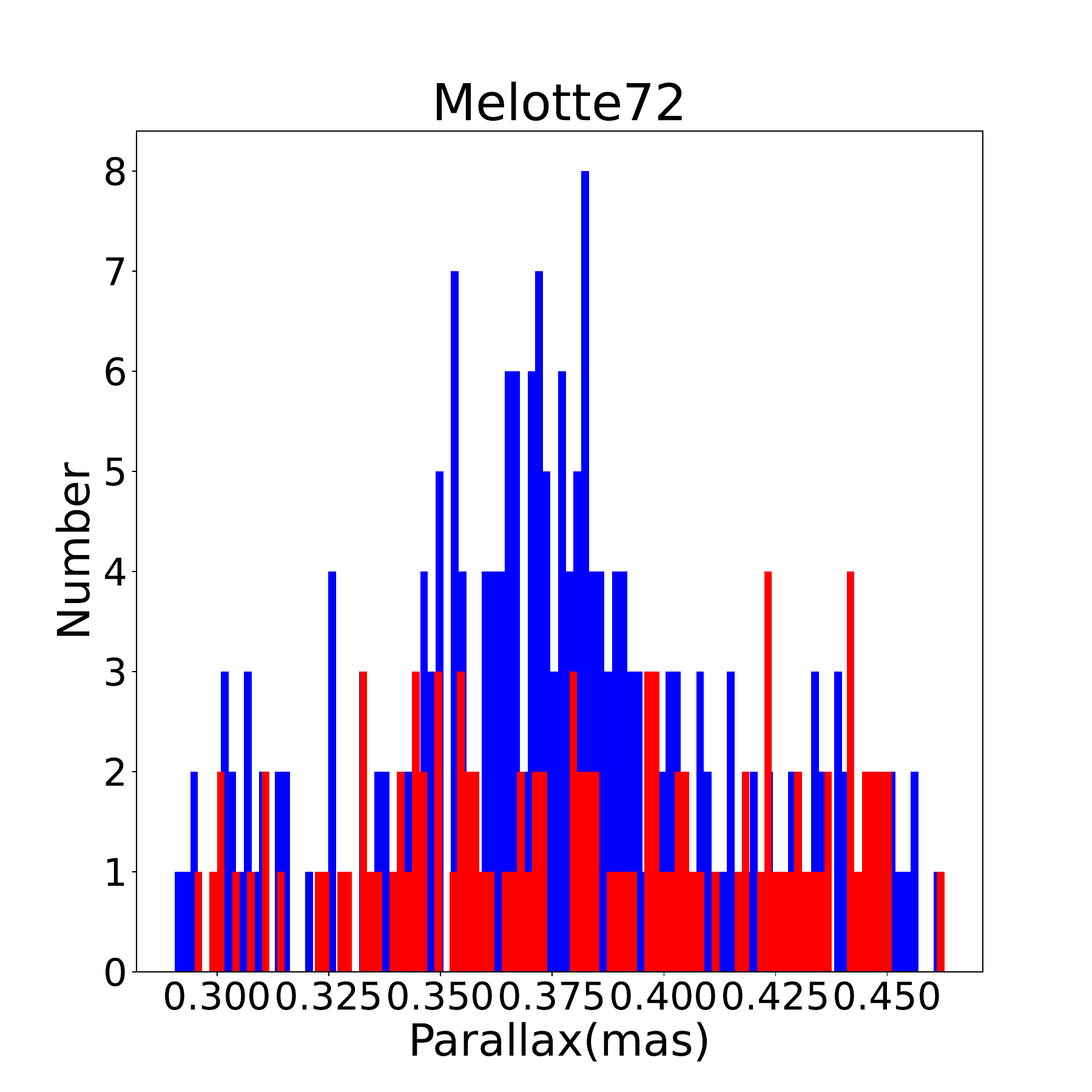}

        \end{subfigure}
        \begin{subfigure}{0.25\textwidth}
                \centering

                \includegraphics[width=\textwidth]{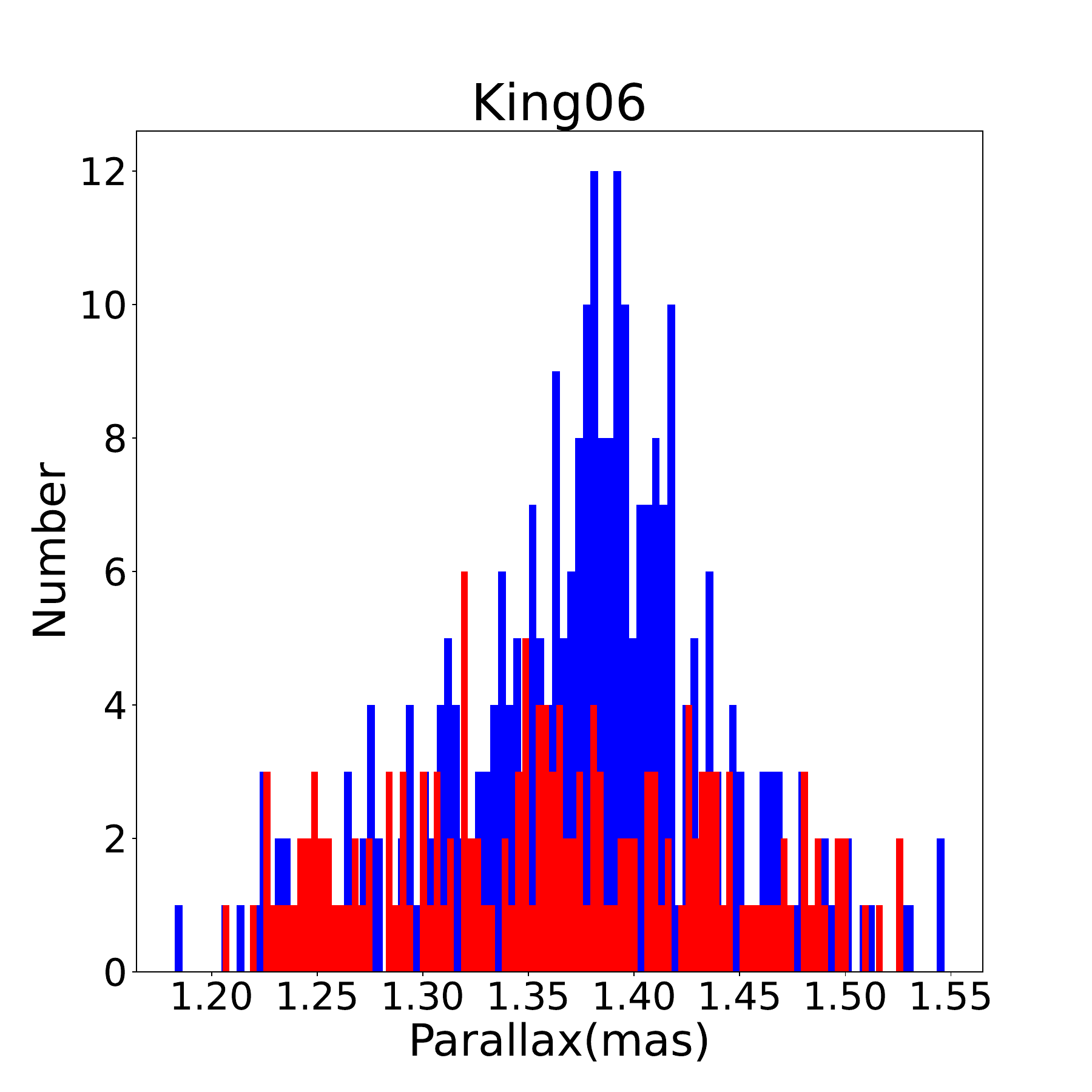}

        \end{subfigure}
        \begin{subfigure}{0.25\textwidth}
                \centering

                \includegraphics[width=\textwidth]{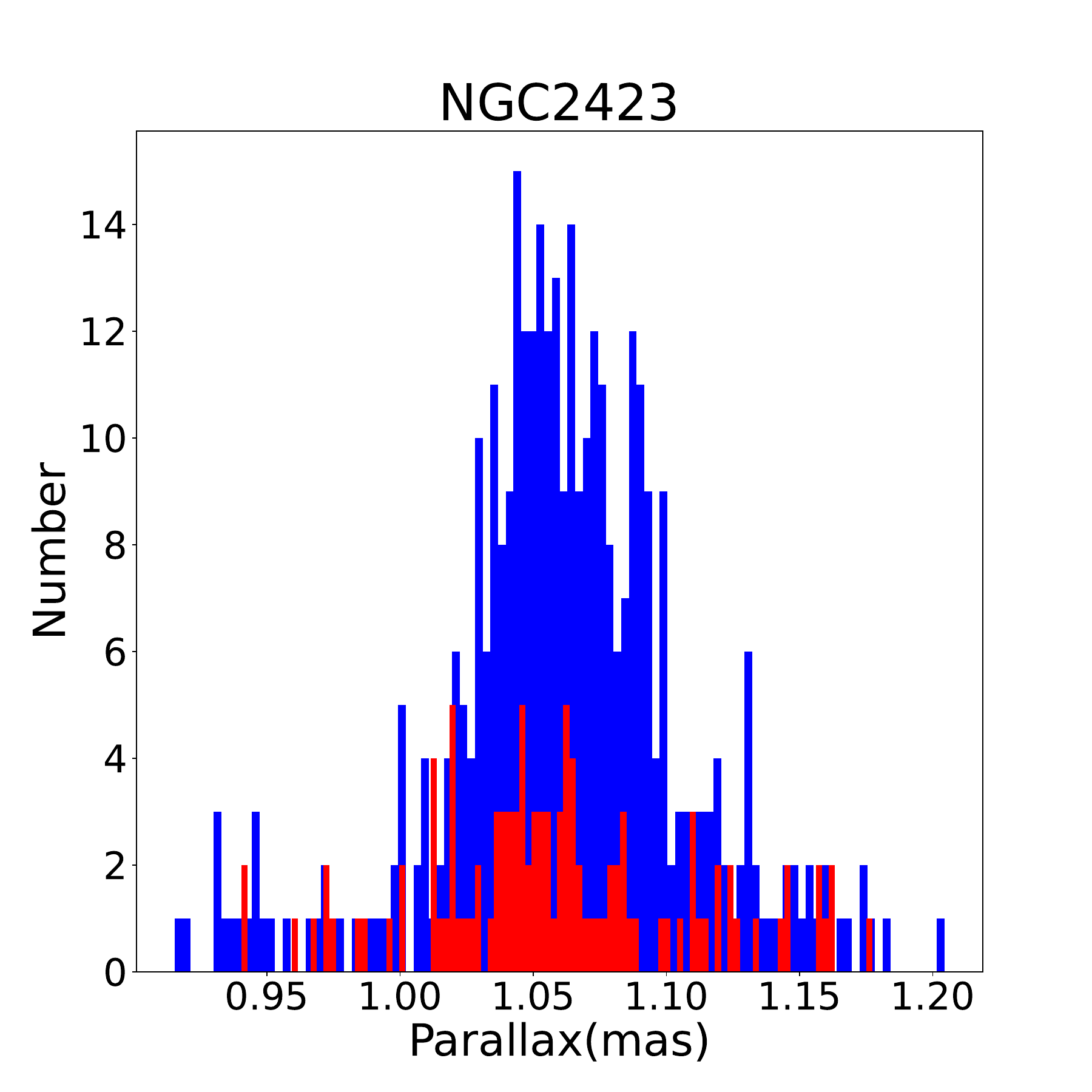}

        \end{subfigure}
        \begin{subfigure}{0.25\textwidth}
                \centering

                \includegraphics[width=\textwidth]{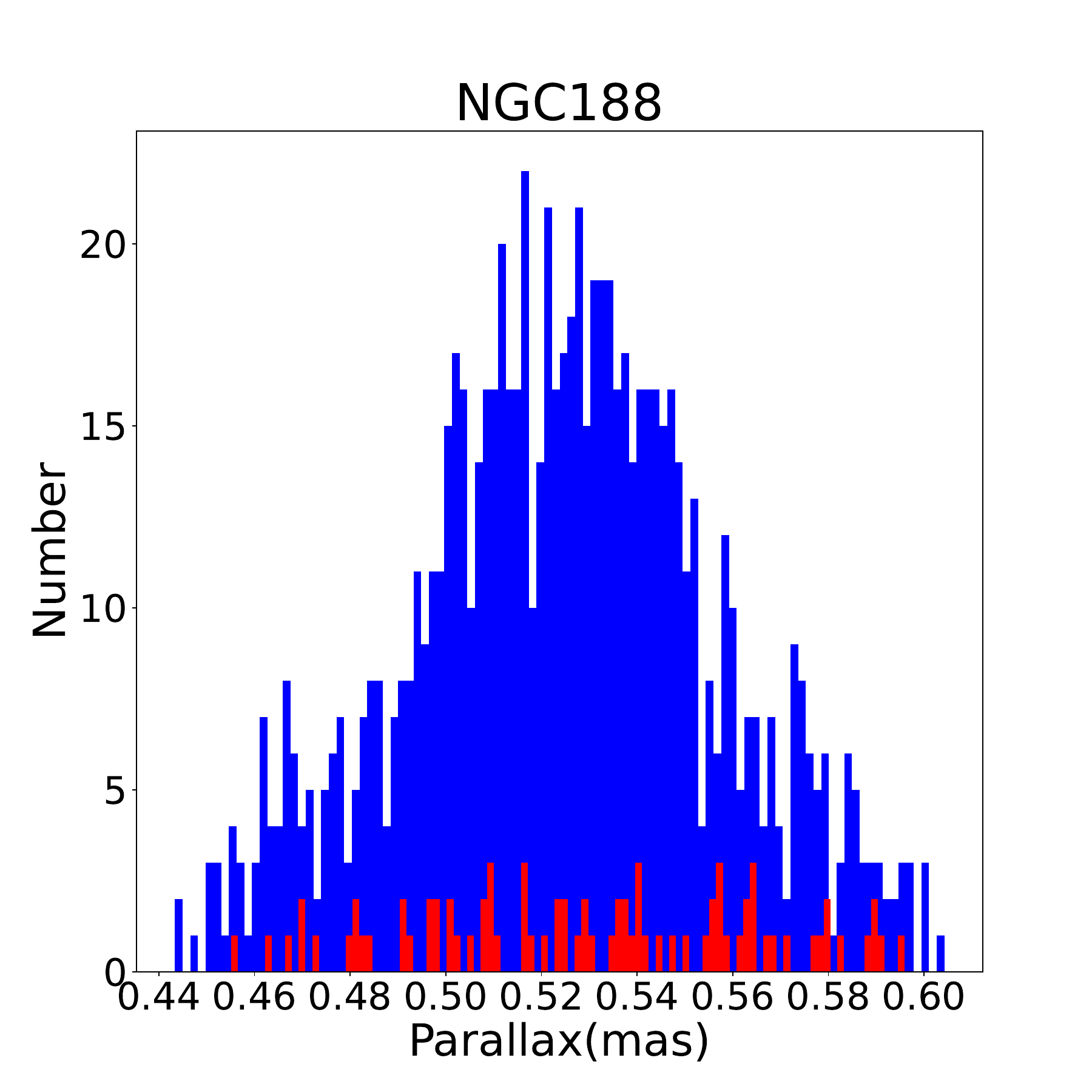}

        \end{subfigure}
        \begin{subfigure}{0.25\textwidth}
        \centering
           \includegraphics[width=\textwidth]{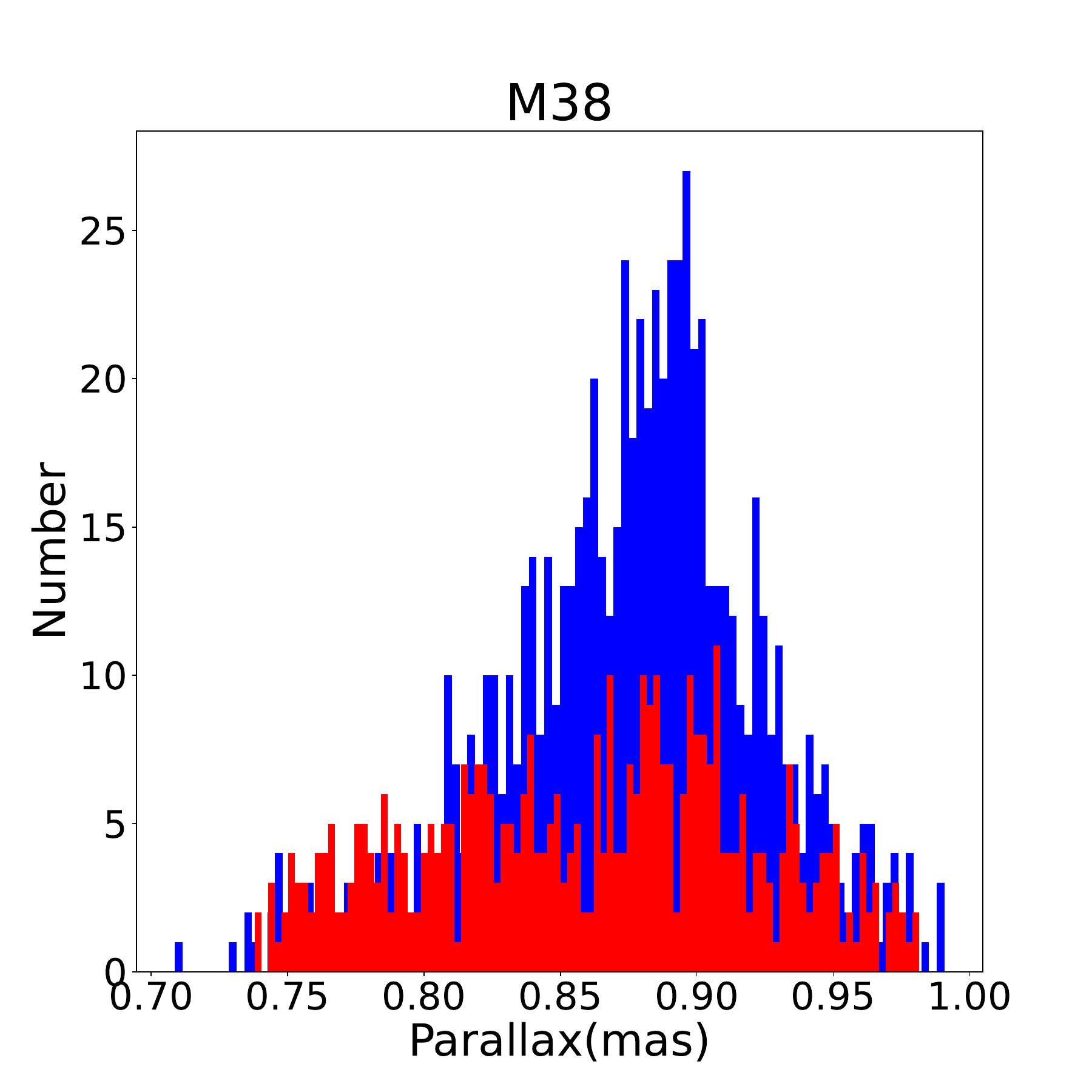}

        \end{subfigure}
        \begin{subfigure}{0.25\textwidth}
        \centering
           \includegraphics[width=\textwidth]{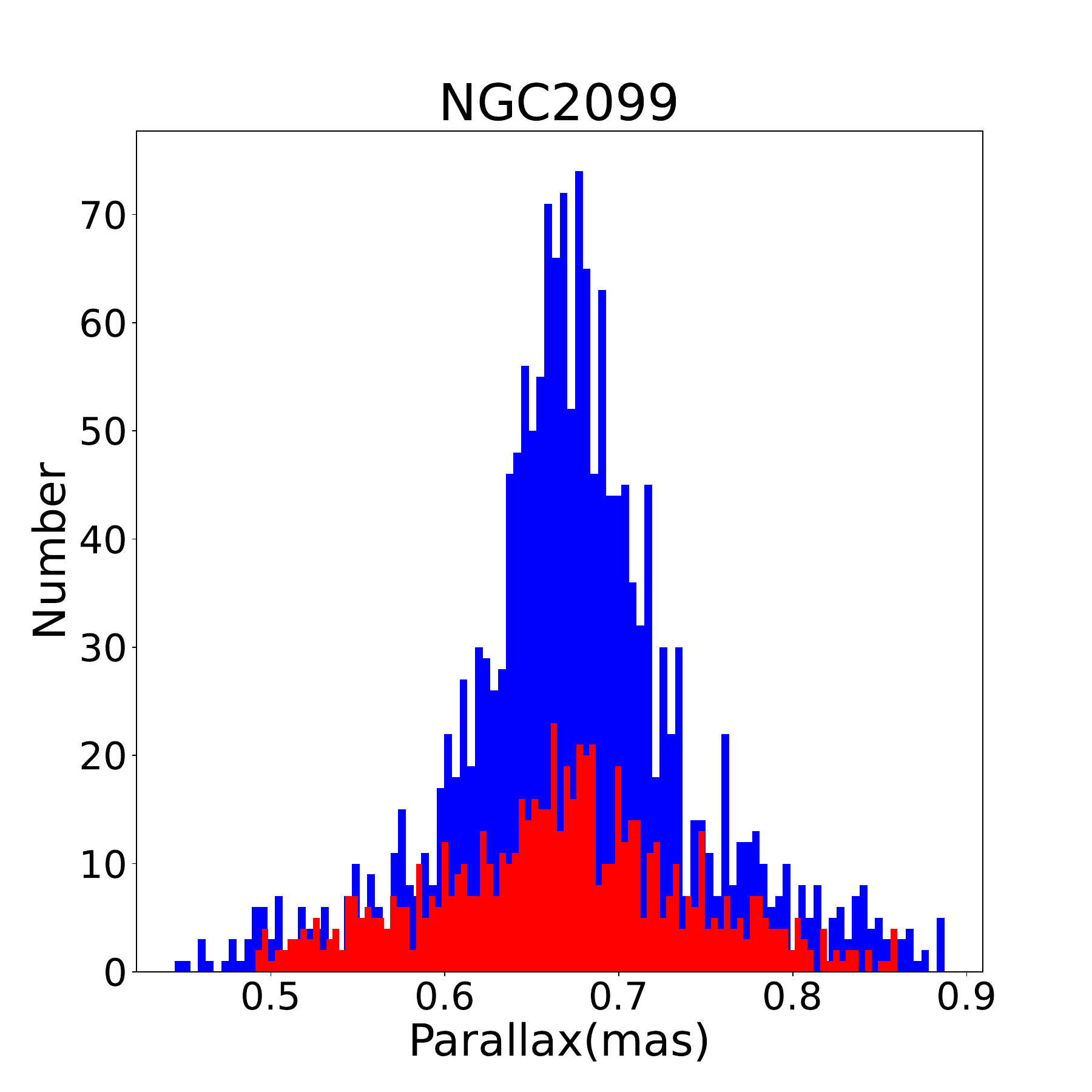}

        \end{subfigure}
        \begin{subfigure}{0.25\textwidth}
        \centering
           \includegraphics[width=\textwidth]{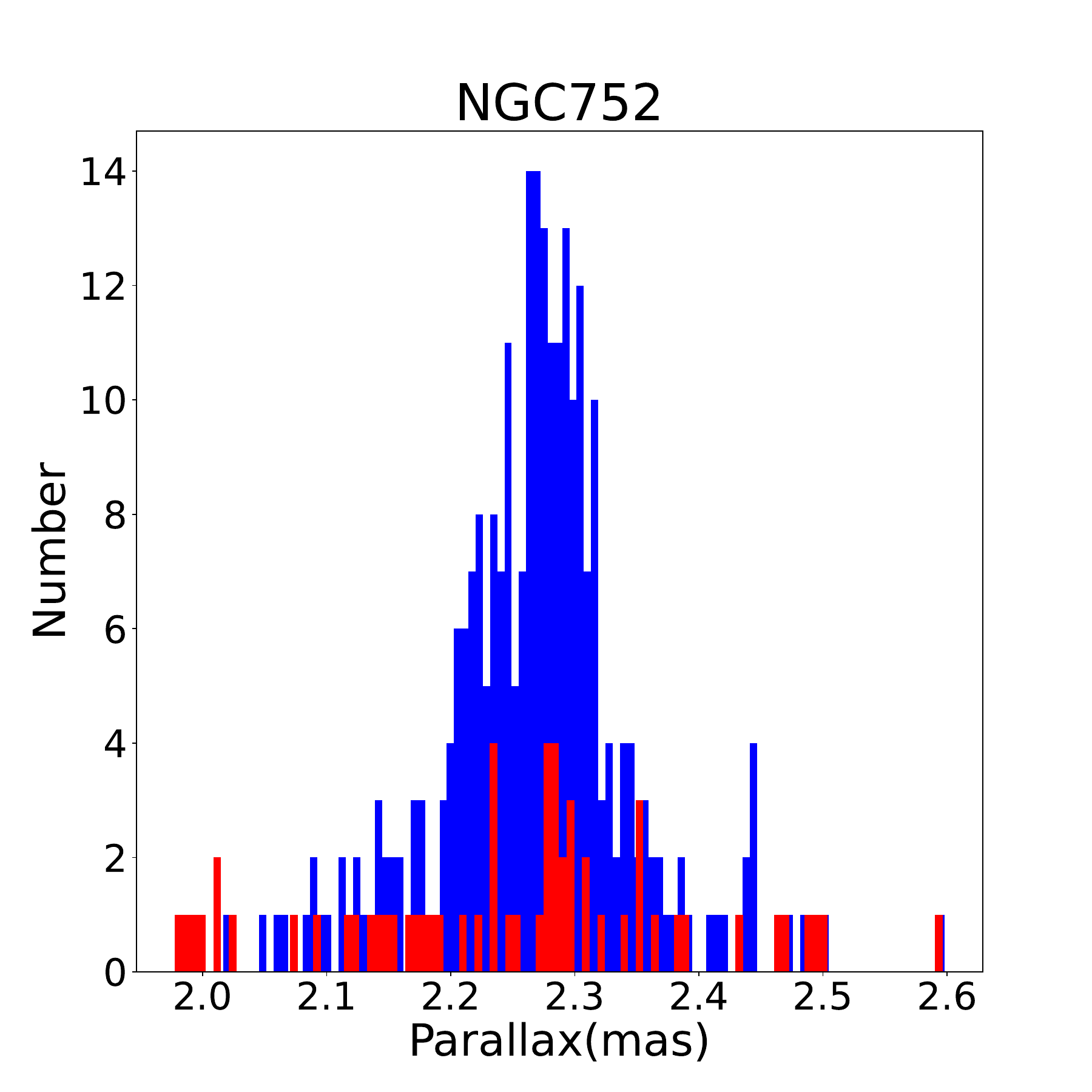}

        \end{subfigure}
        \begin{subfigure}{0.25\textwidth}
        \centering
           \includegraphics[width=\textwidth]{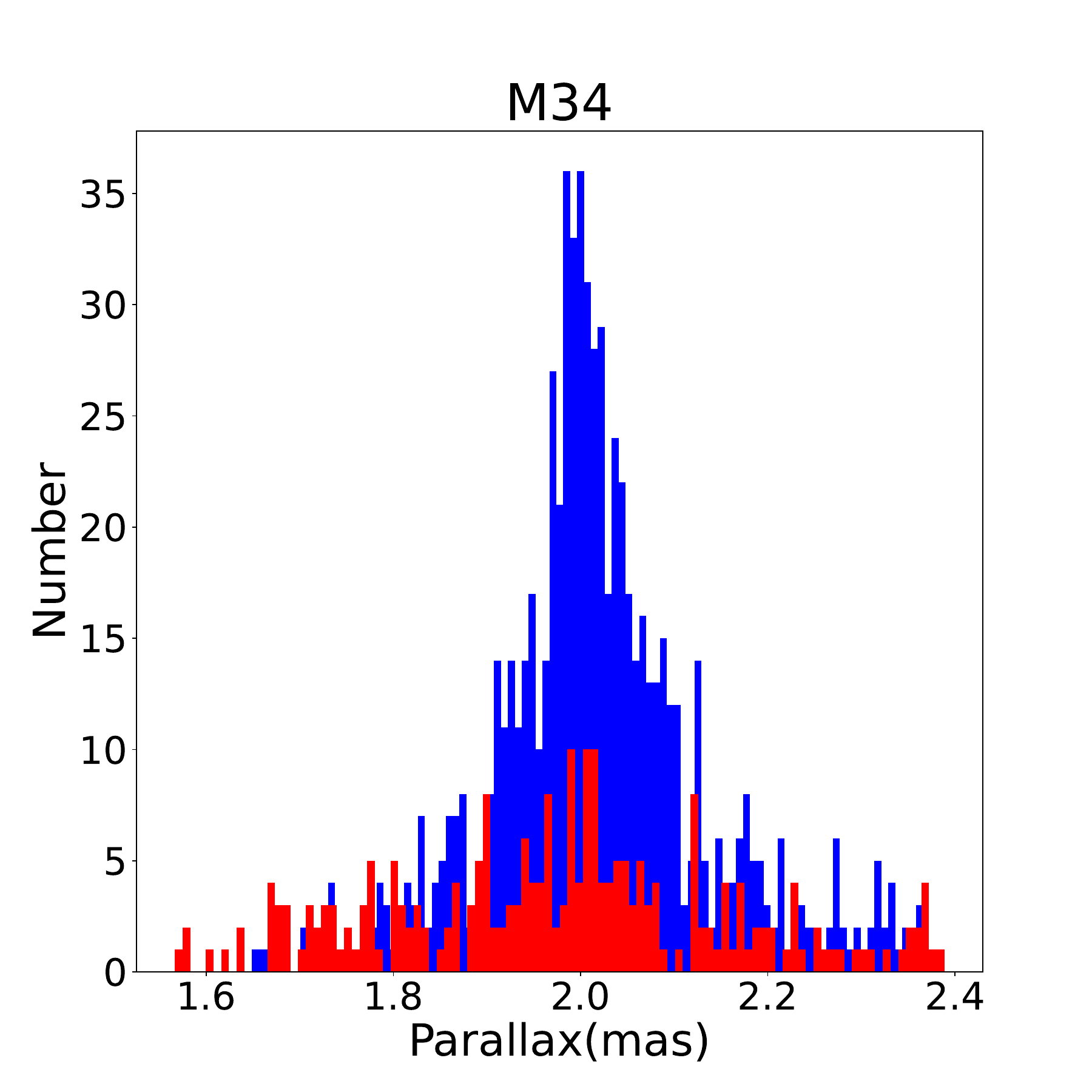}

        \end{subfigure}
        \begin{subfigure}{0.25\textwidth}
        \centering
           \includegraphics[width=\textwidth]{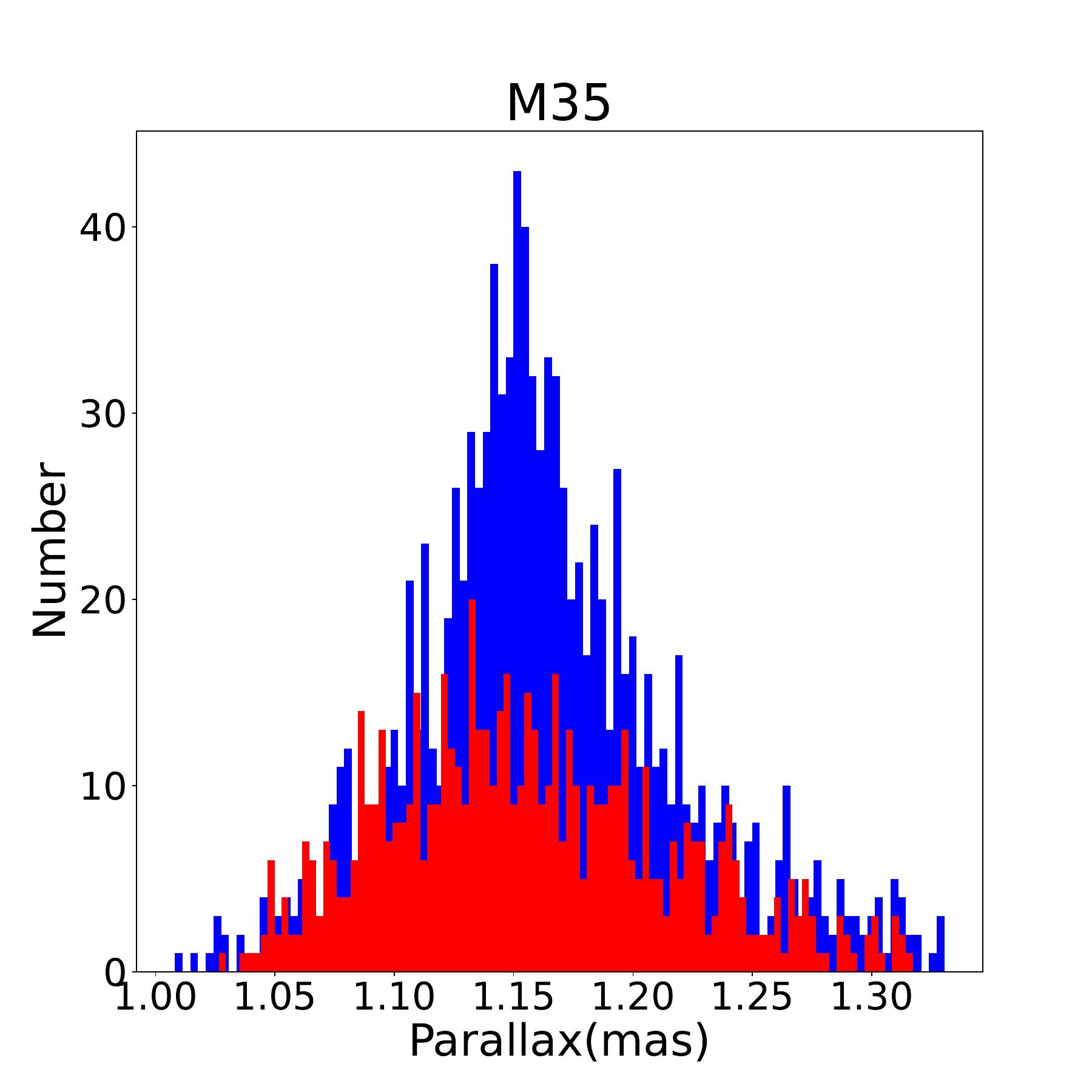}

        \end{subfigure}
        \begin{subfigure}{0.25\textwidth}
        \centering
           \includegraphics[width=\textwidth]{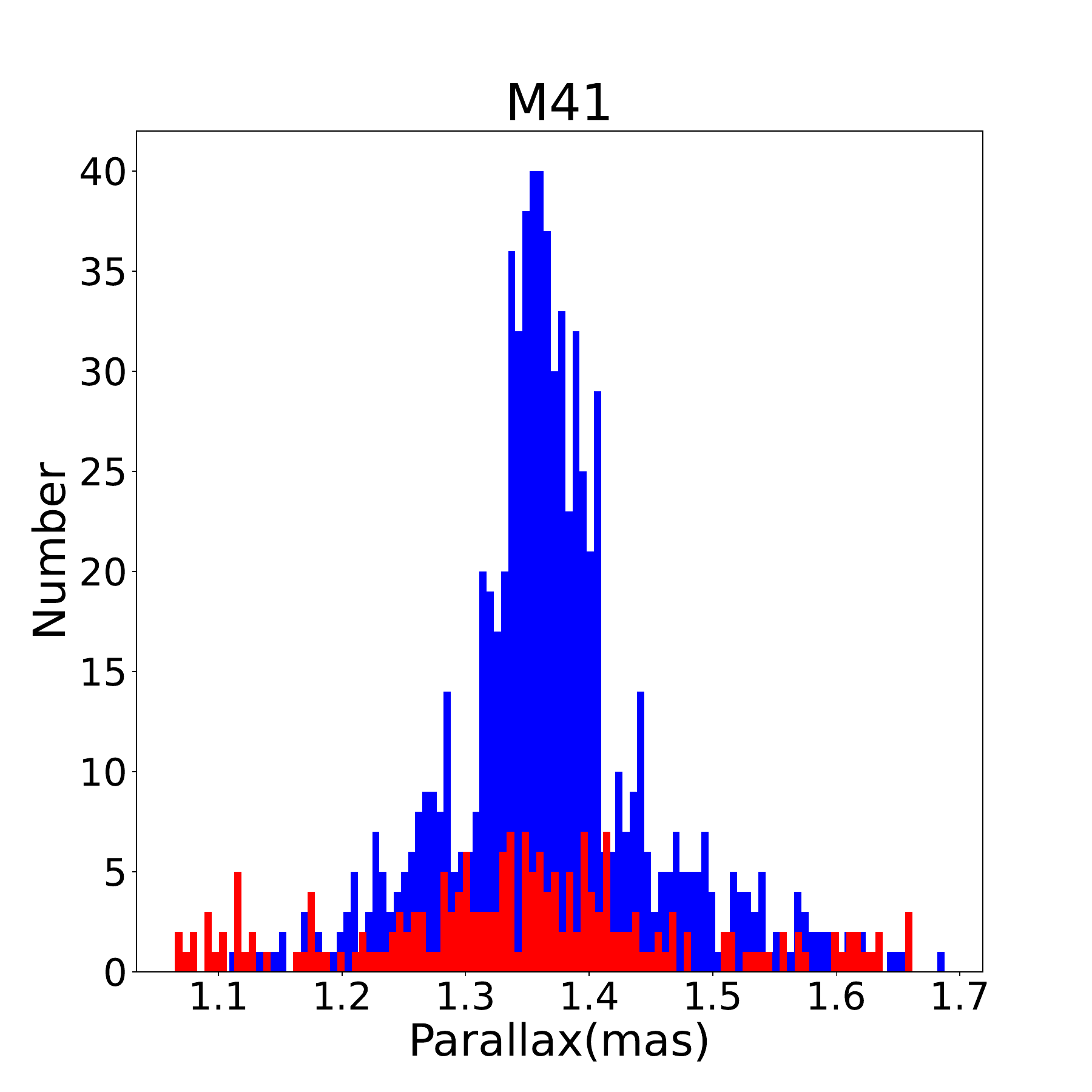}

        \end{subfigure}
        \begin{subfigure}{0.25\textwidth}
        \centering
           \includegraphics[width=\textwidth]{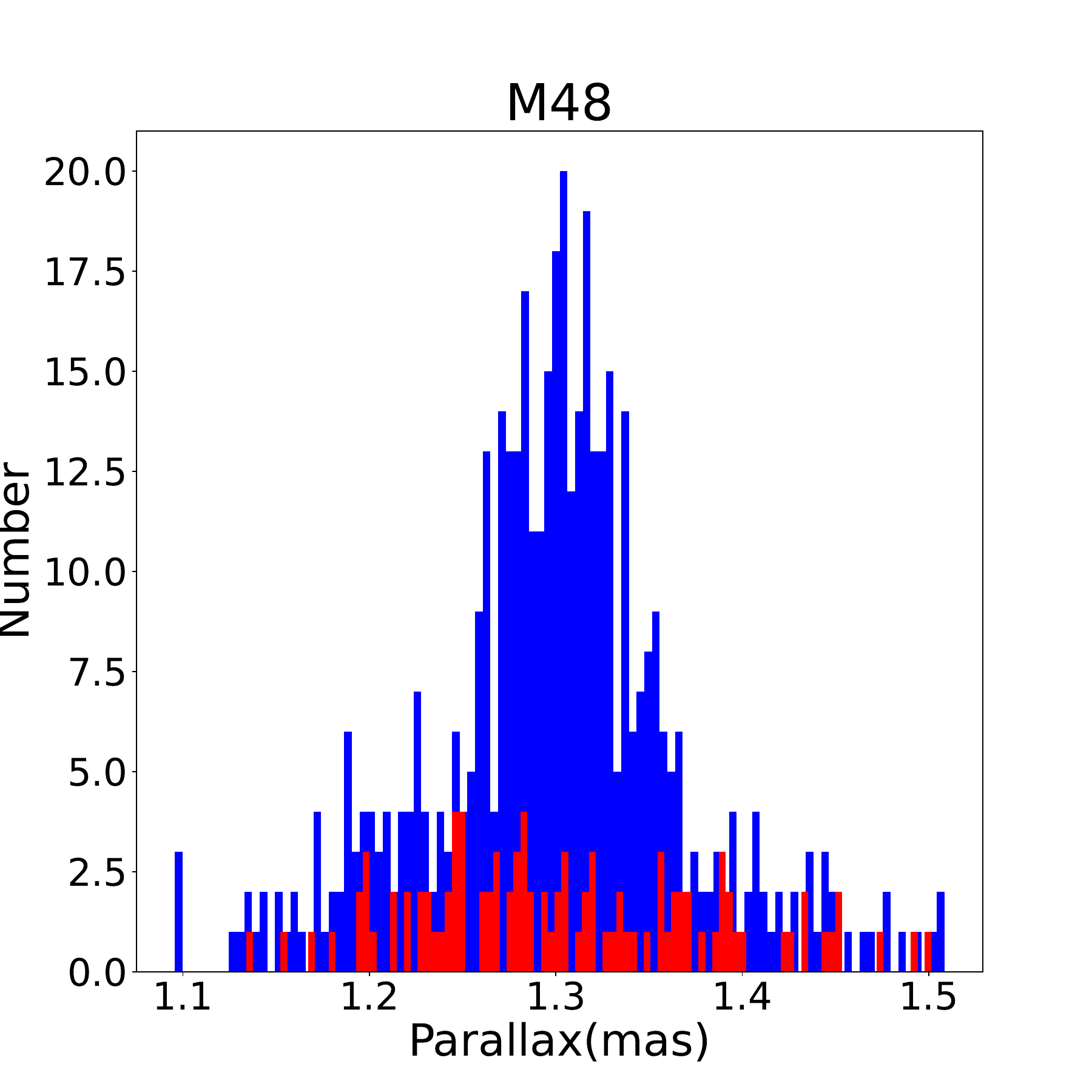}

        \end{subfigure}
        \begin{subfigure}{0.25\textwidth}
        \centering
           \includegraphics[width=\textwidth]{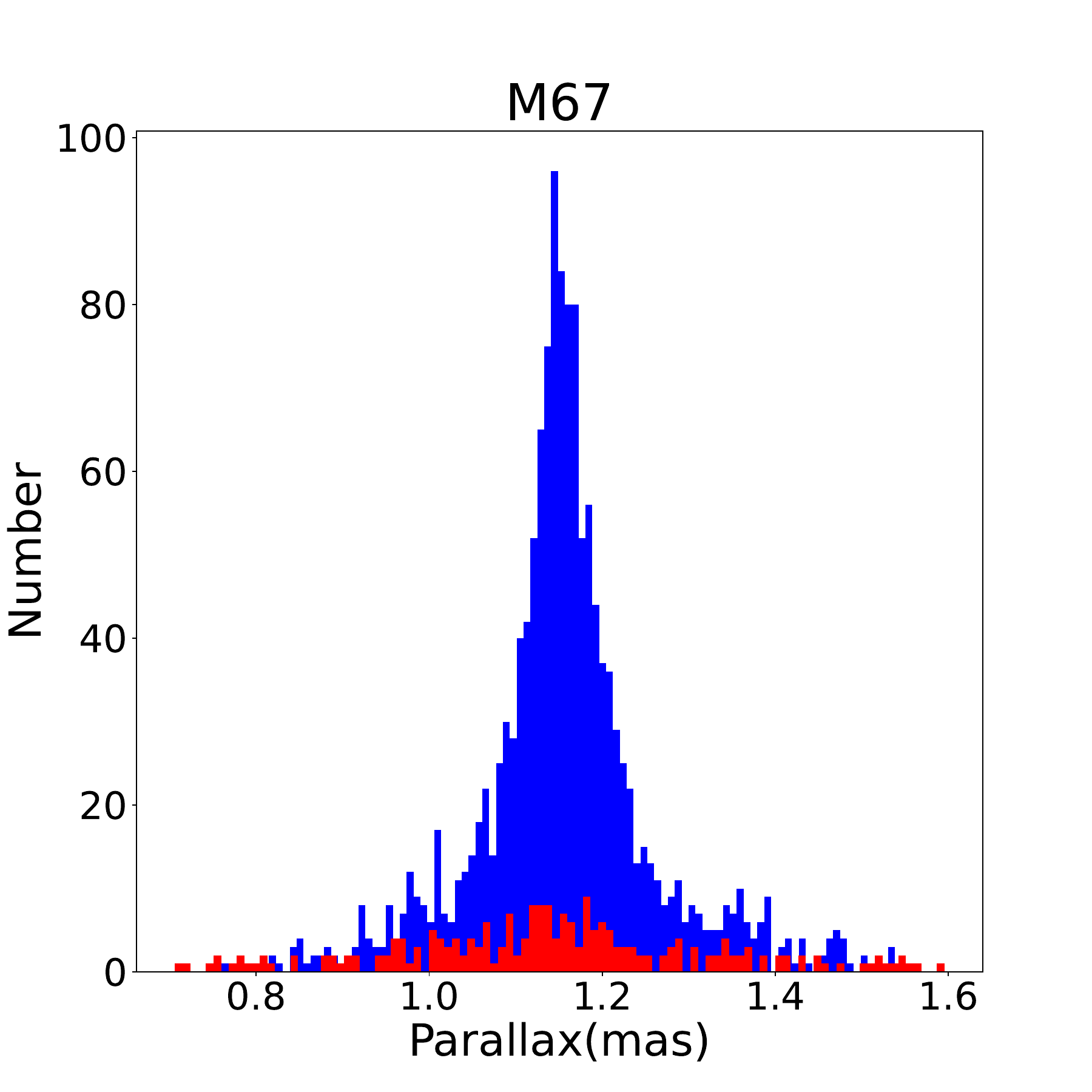}

        \end{subfigure}
        \begin{subfigure}{0.25\textwidth}
        \centering
           \includegraphics[width=\textwidth]{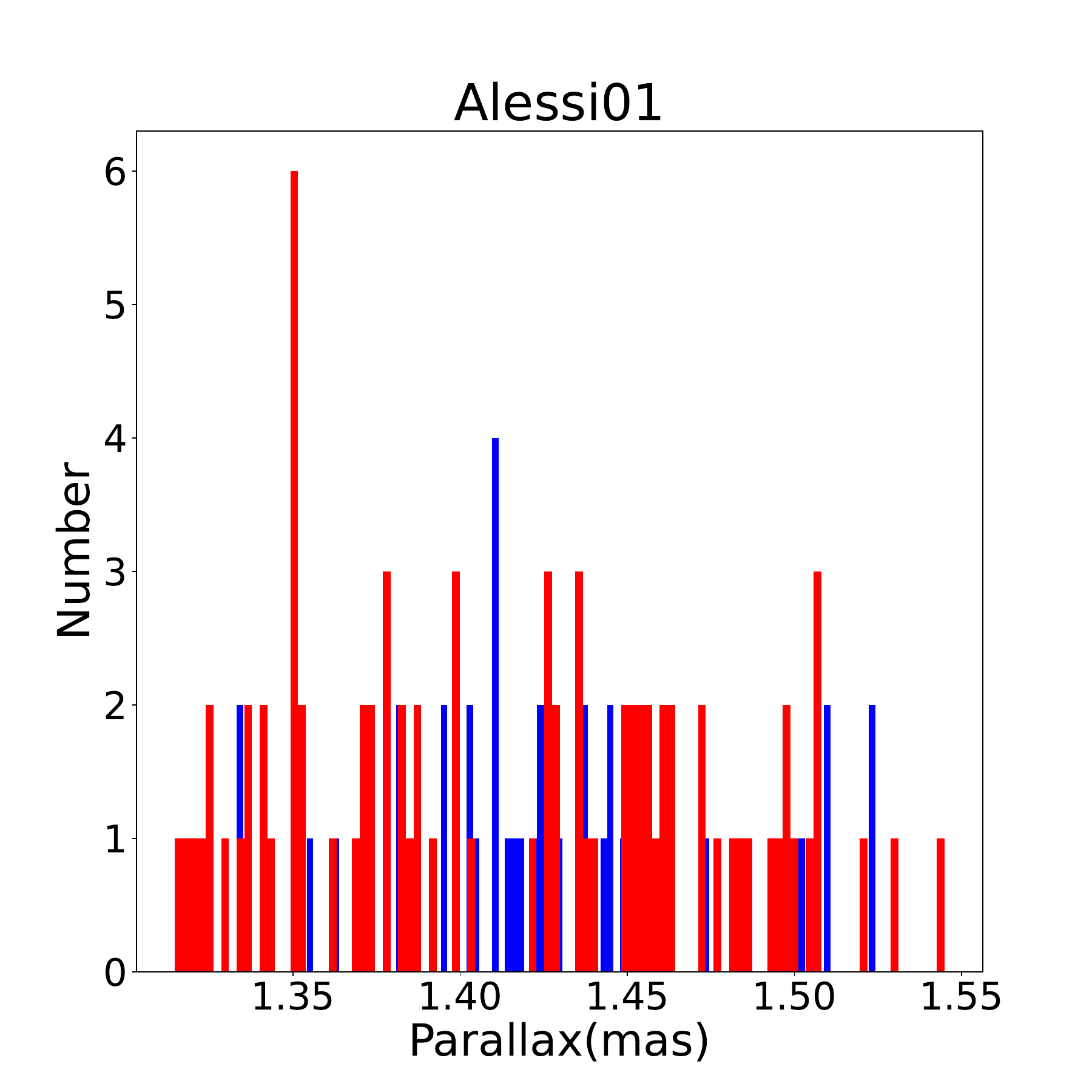}

        \end{subfigure}
  \caption{The parallax of the clusters inner and outer members. Red lines represent stars that were selected by Random Forest, while blue lines indicate stars that were selected by GMM with a probability higher than 0.8.}
  \label{parallax dgr.fig}
\end{figure}

\begin{figure}
  \centering
  \captionsetup[subfigure]{labelformat=empty}
        \begin{subfigure}{0.25\textwidth}
        \centering
           \includegraphics[width=\textwidth]{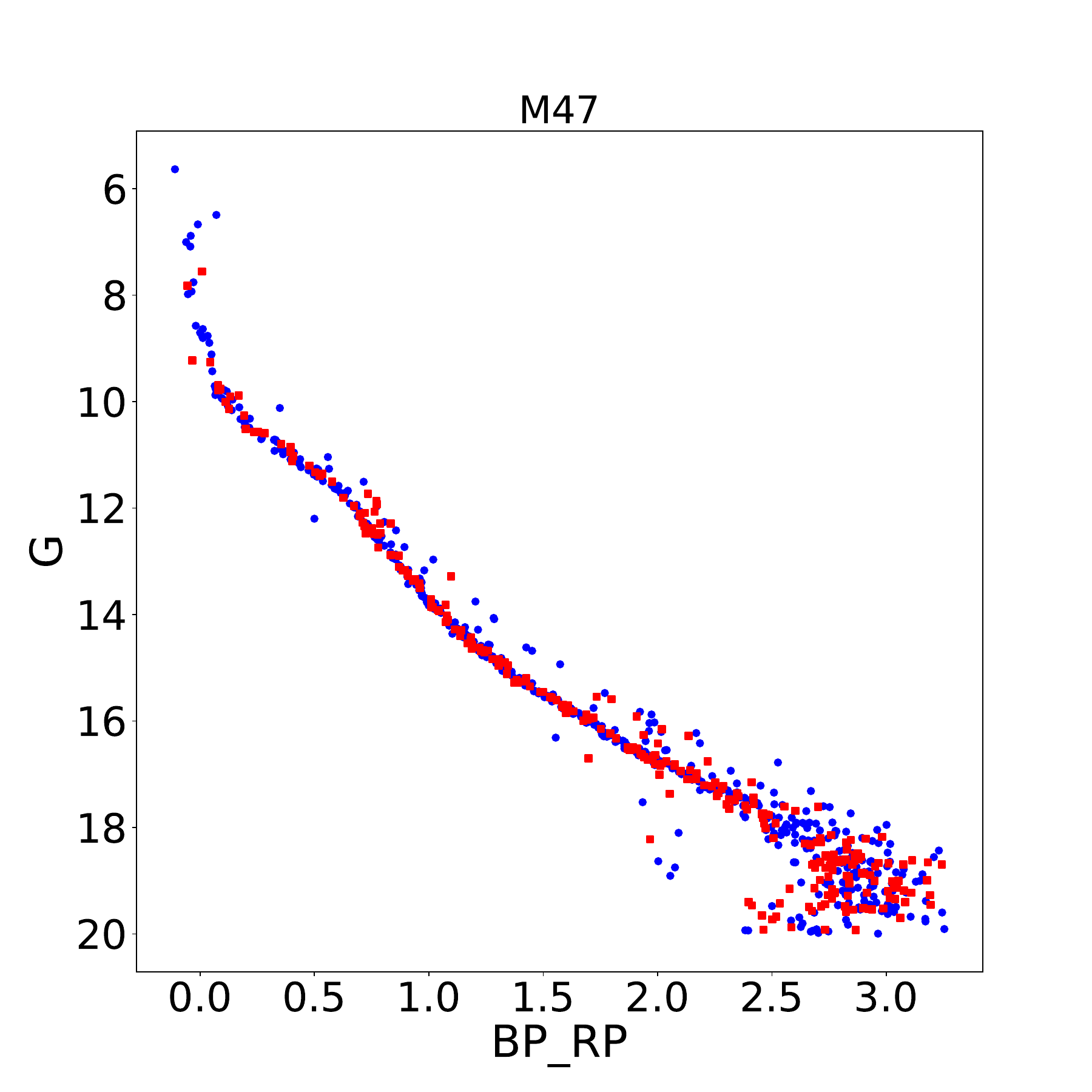}

        \end{subfigure}
        \begin{subfigure}{0.25\textwidth}

                \centering
                \includegraphics[width=\textwidth]{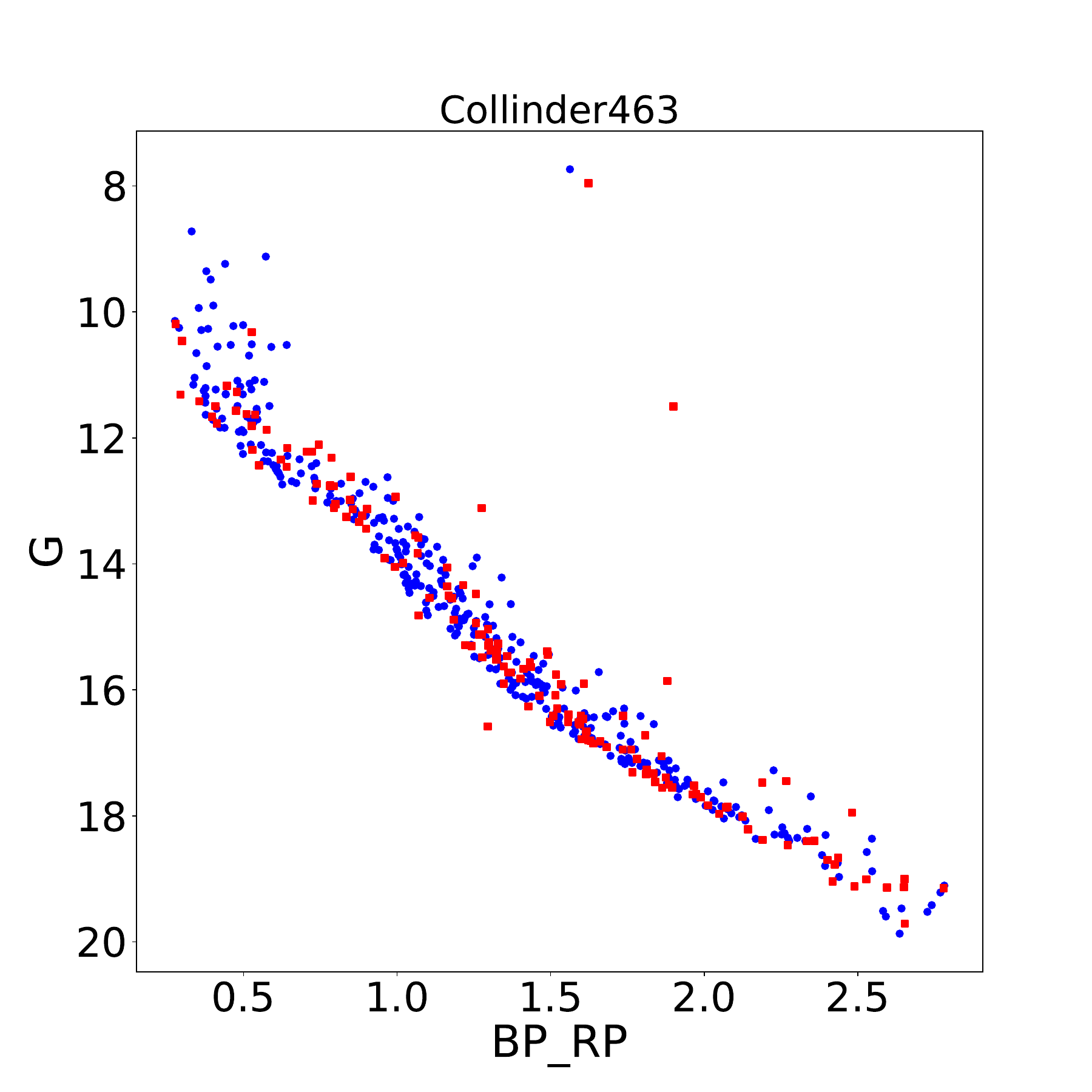}

        \end{subfigure}
        \begin{subfigure}{0.25\textwidth}
                \centering
           \includegraphics[width=\textwidth]{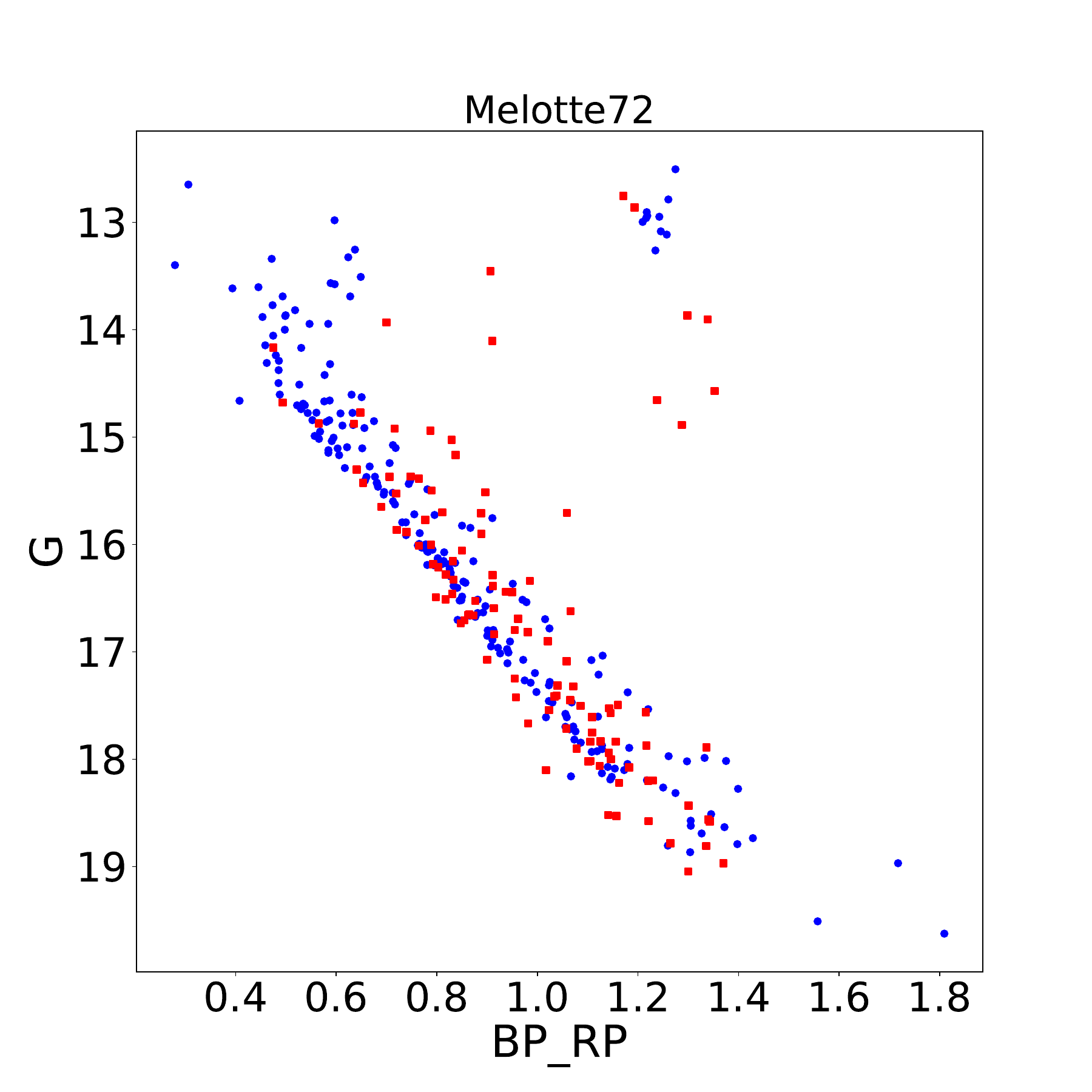}

        \end{subfigure}
        \begin{subfigure}{0.25\textwidth}
                \centering

                \includegraphics[width=\textwidth]{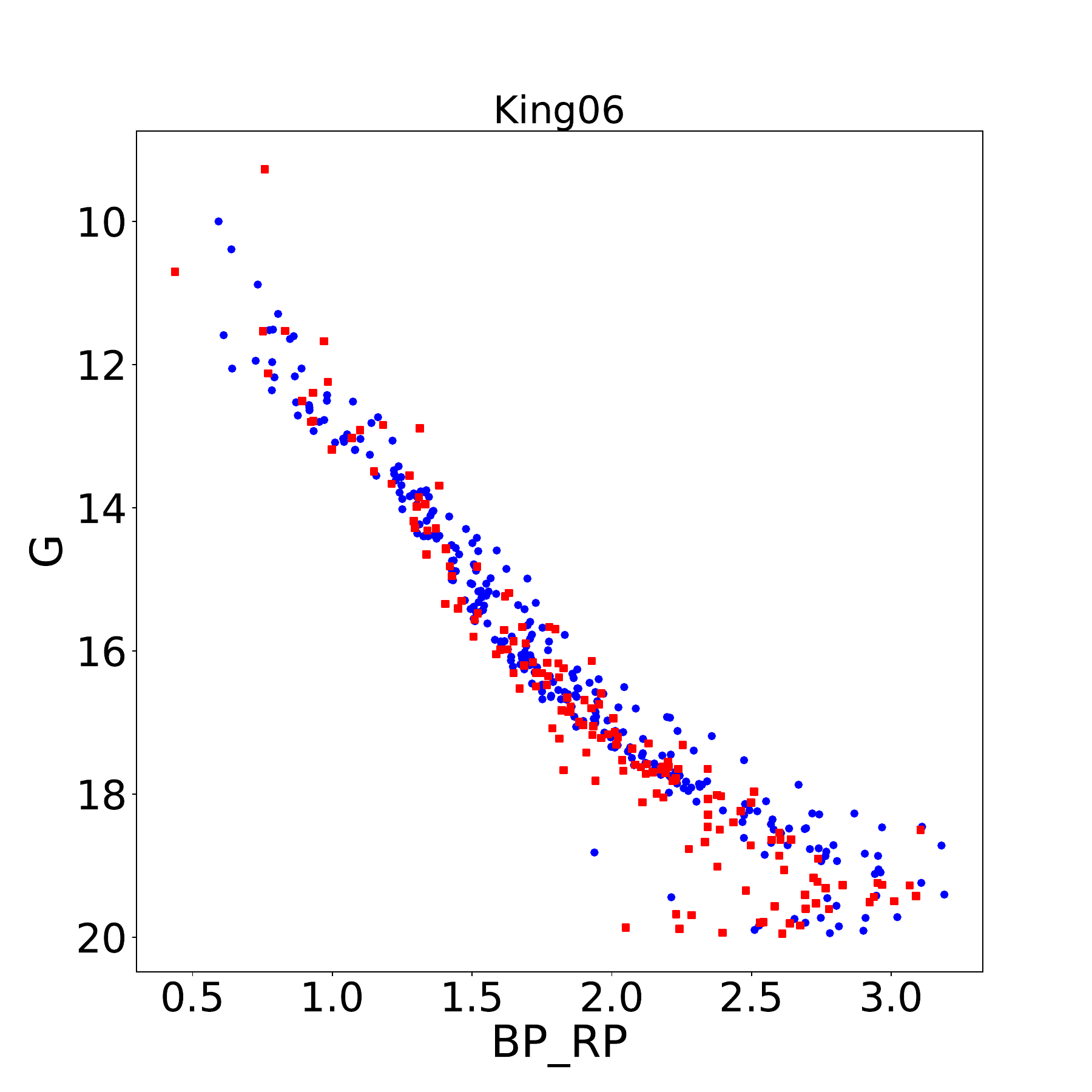}

        \end{subfigure}
        \begin{subfigure}{0.25\textwidth}
                \centering

                \includegraphics[width=\textwidth]{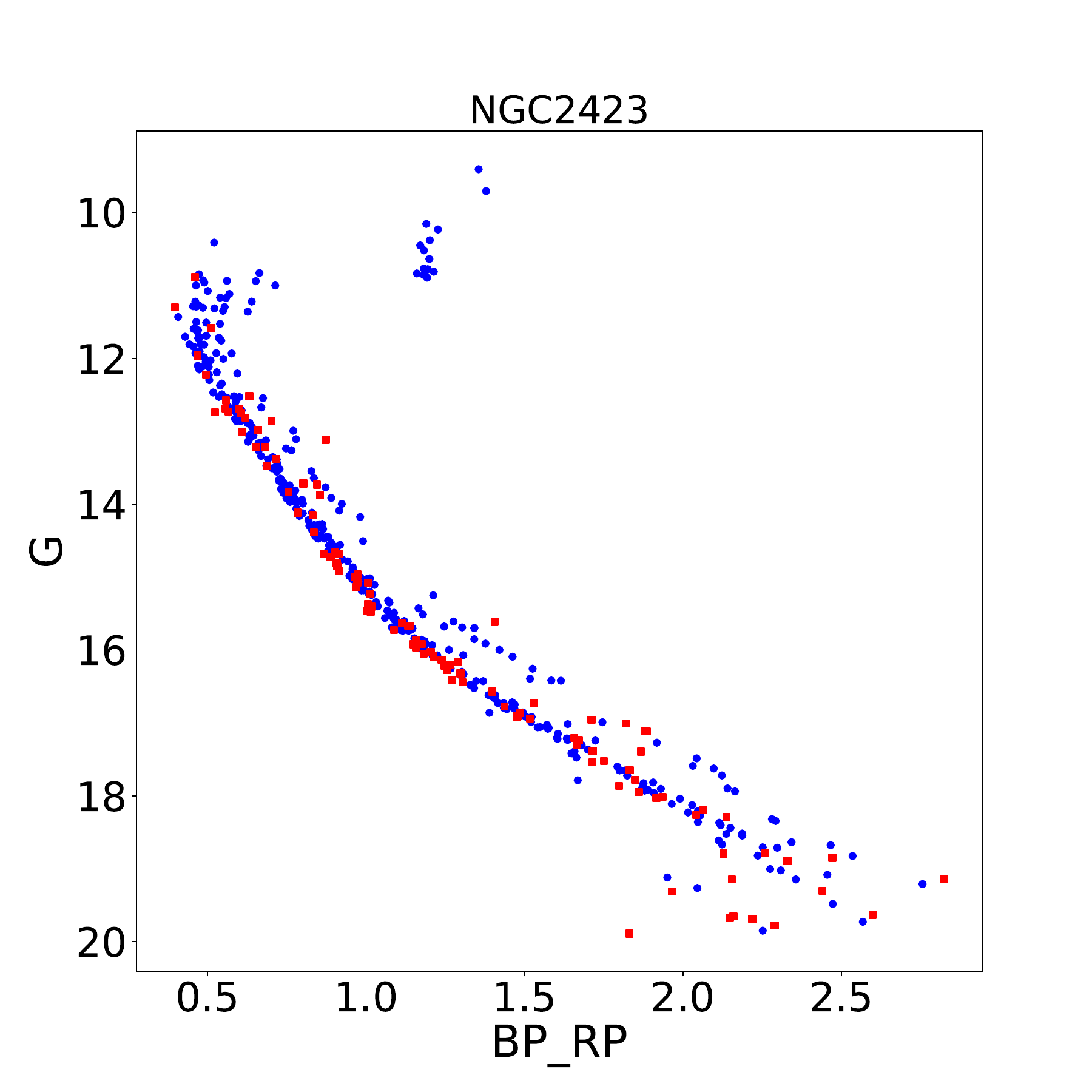}

        \end{subfigure}
        \begin{subfigure}{0.25\textwidth}
                \centering

                \includegraphics[width=\textwidth]{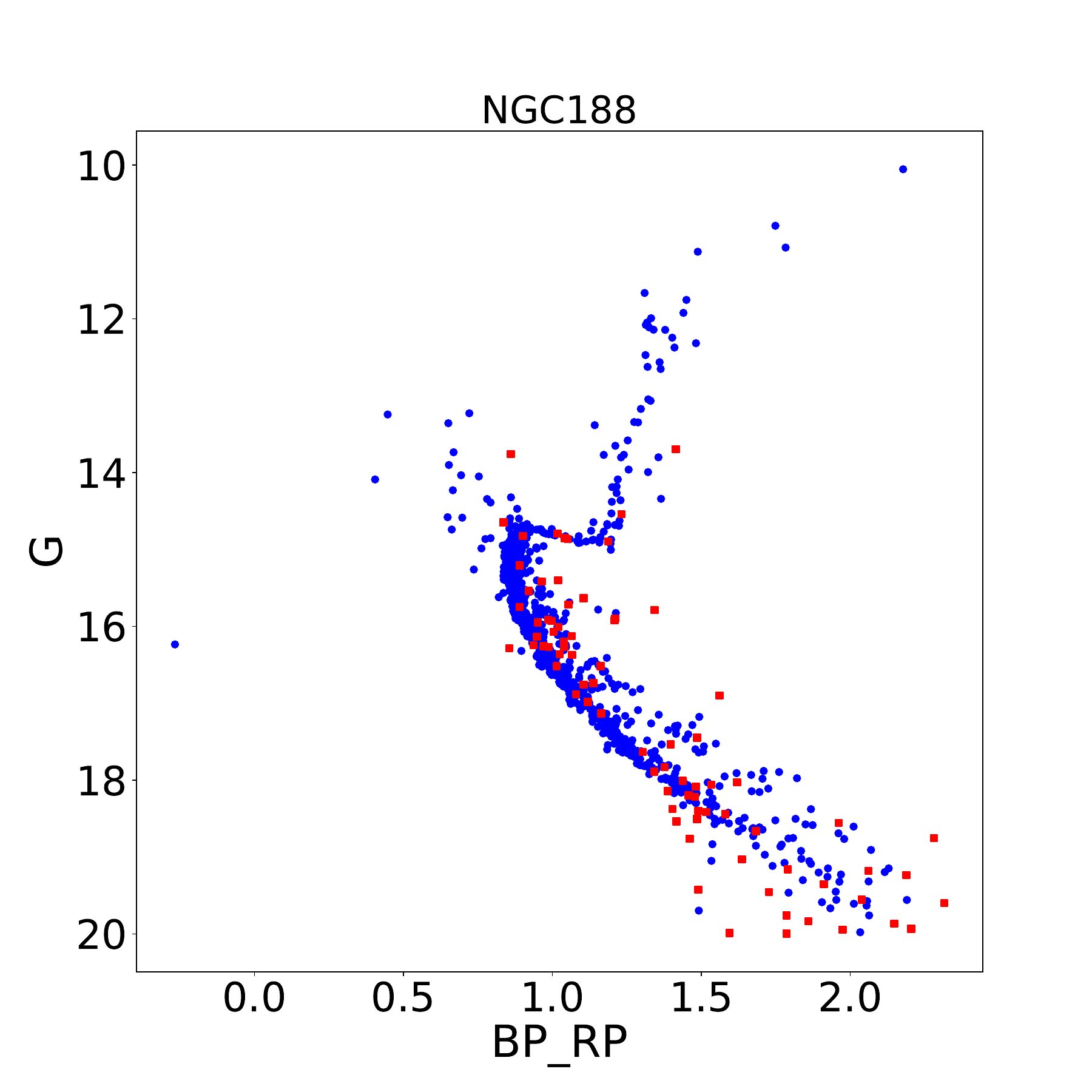}

        \end{subfigure}
        \begin{subfigure}{0.25\textwidth}
        \centering
           \includegraphics[width=\textwidth]{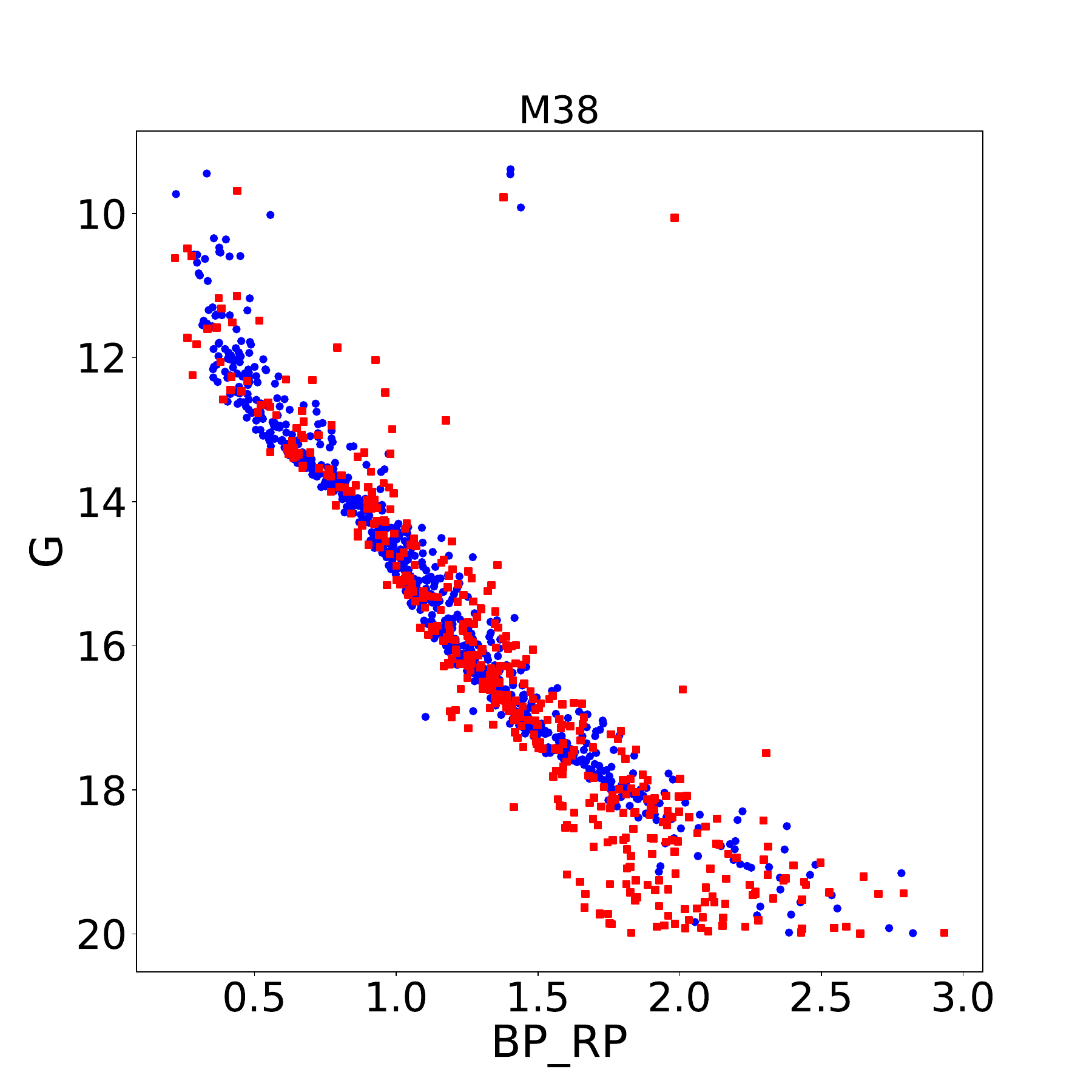}

        \end{subfigure}
        \begin{subfigure}{0.25\textwidth}
        \centering
           \includegraphics[width=\textwidth]{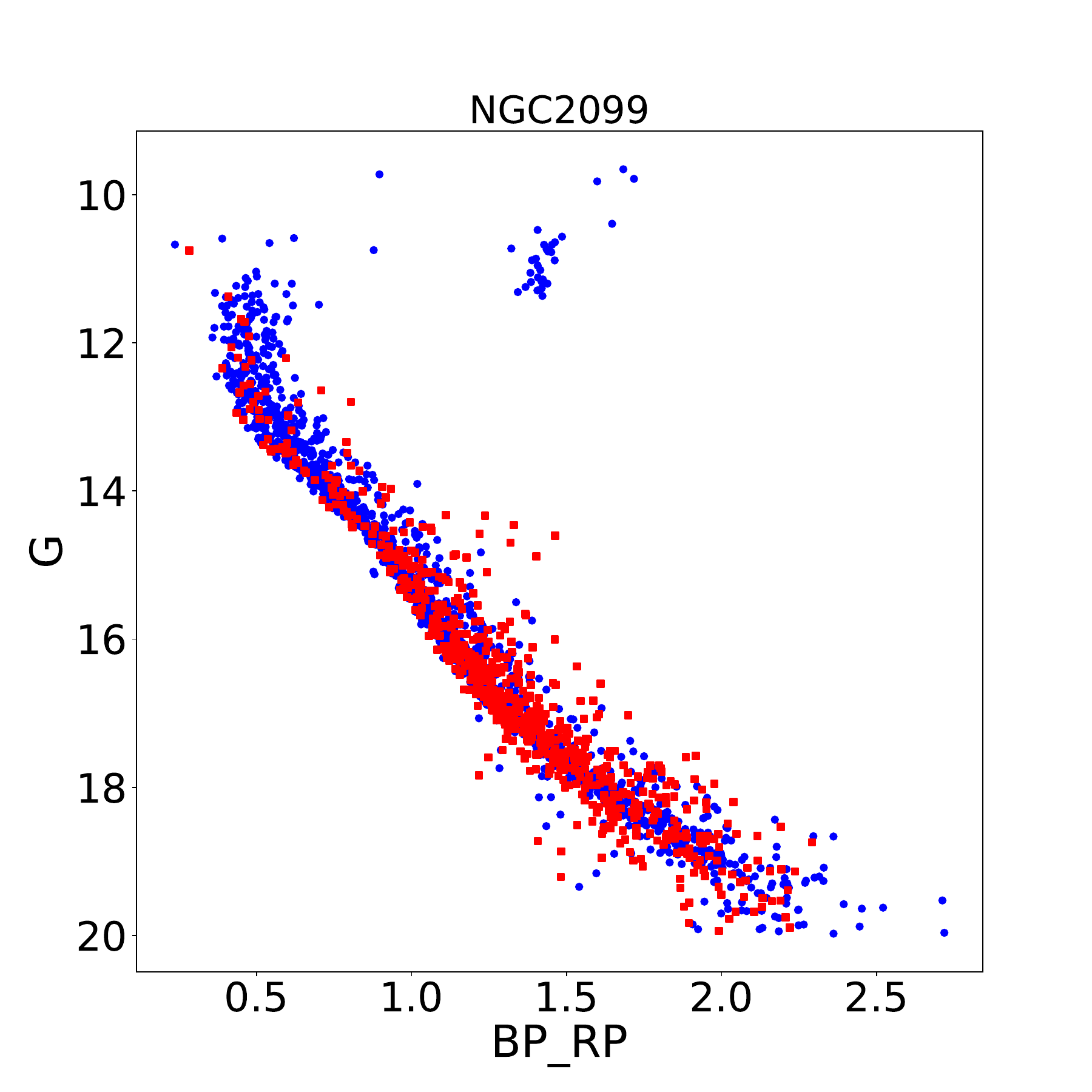}

        \end{subfigure}
        \begin{subfigure}{0.25\textwidth}
        \centering
           \includegraphics[width=\textwidth]{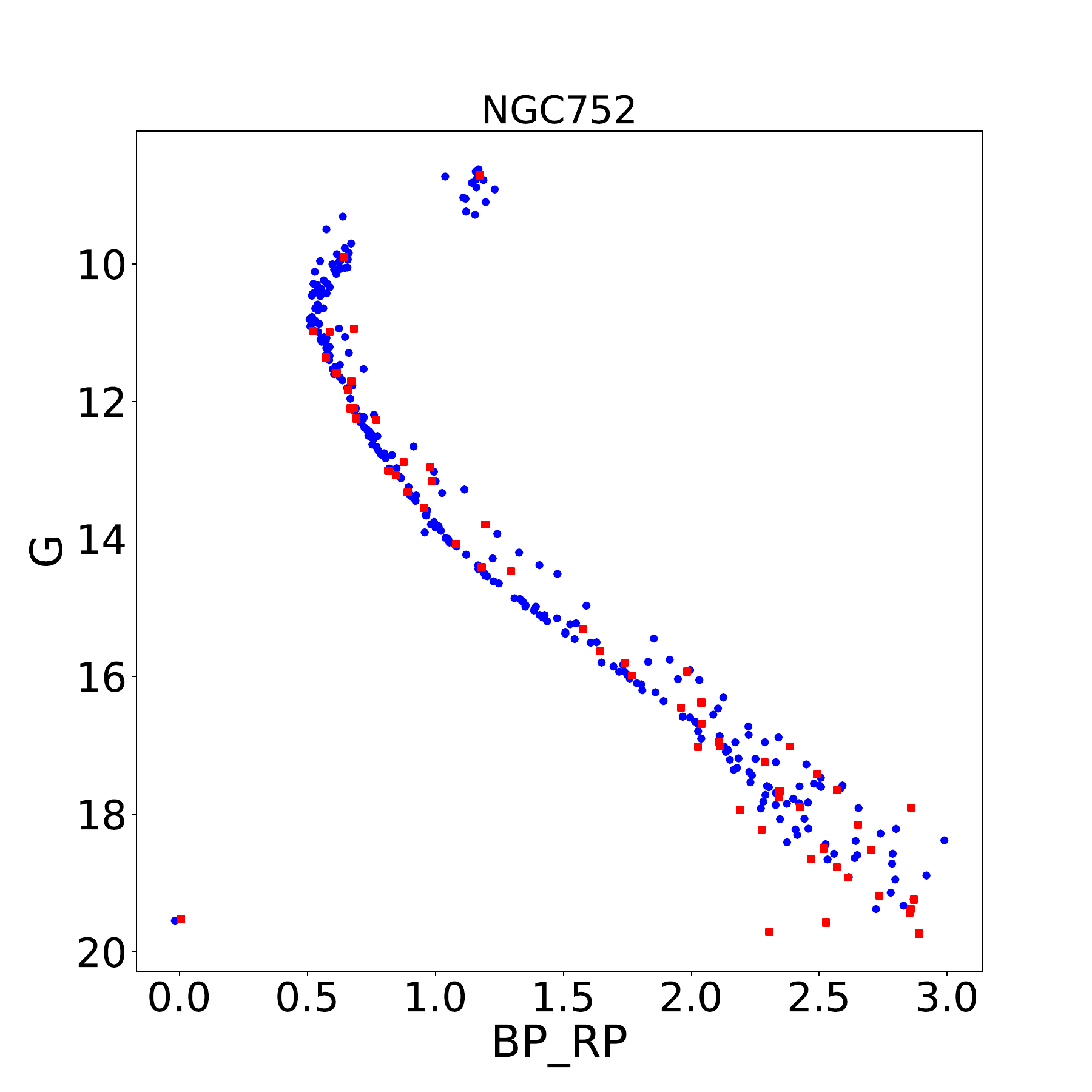}

        \end{subfigure}
        \begin{subfigure}{0.25\textwidth}
        \centering
           \includegraphics[width=\textwidth]{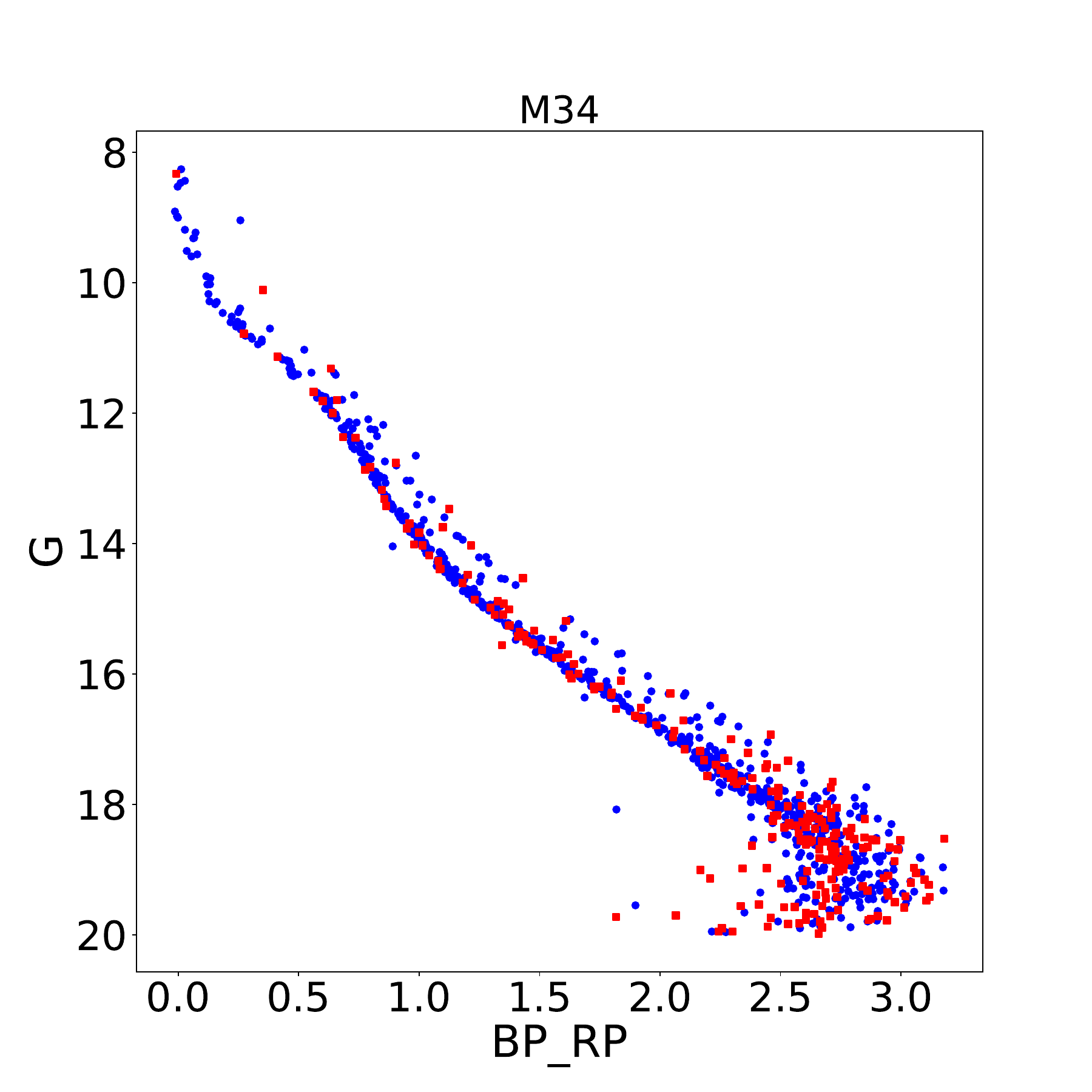}

        \end{subfigure}
        \begin{subfigure}{0.25\textwidth}
        \centering
           \includegraphics[width=\textwidth]{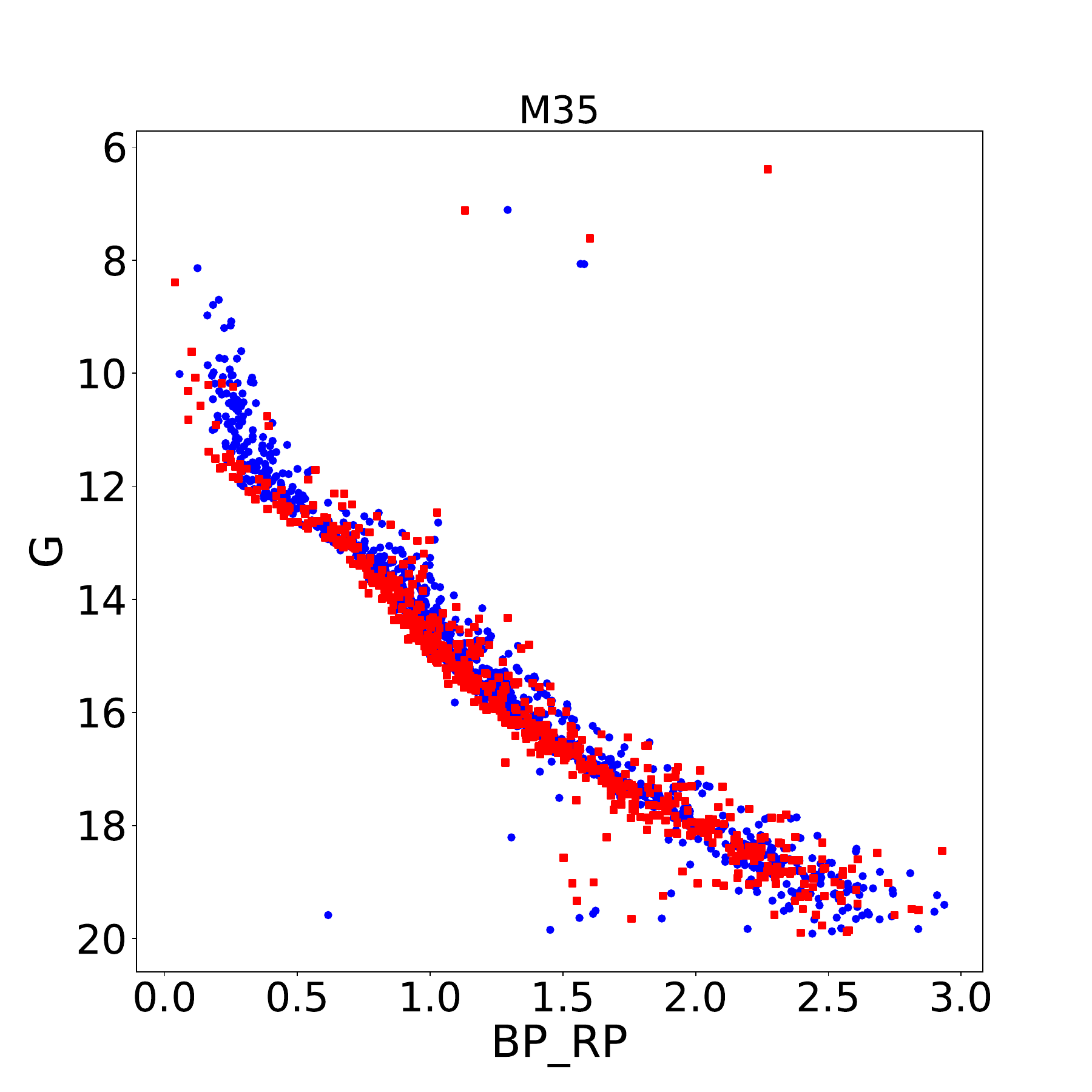}

        \end{subfigure}
        \begin{subfigure}{0.25\textwidth}
        \centering
           \includegraphics[width=\textwidth]{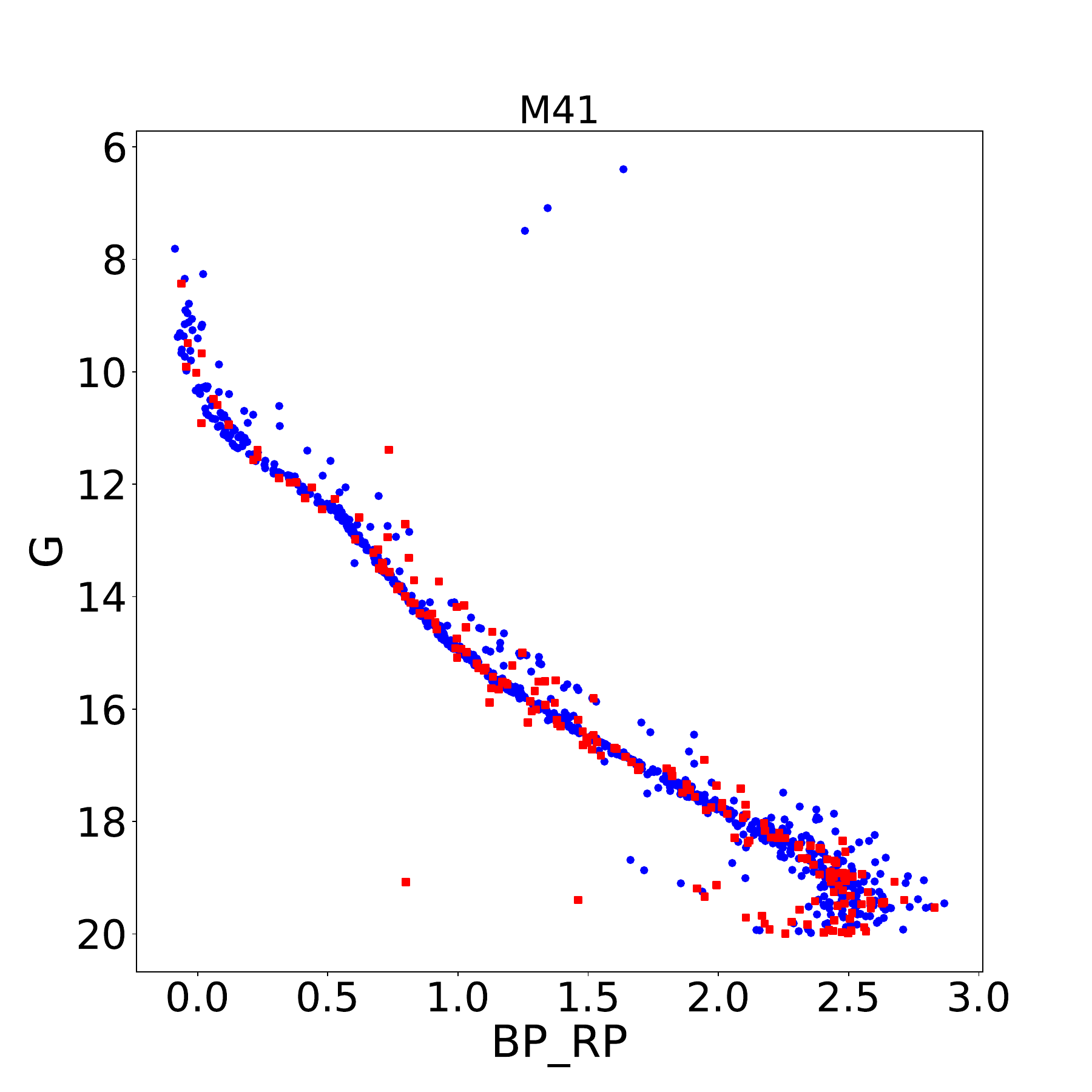}

        \end{subfigure}
        \begin{subfigure}{0.25\textwidth}
        \centering
           \includegraphics[width=\textwidth]{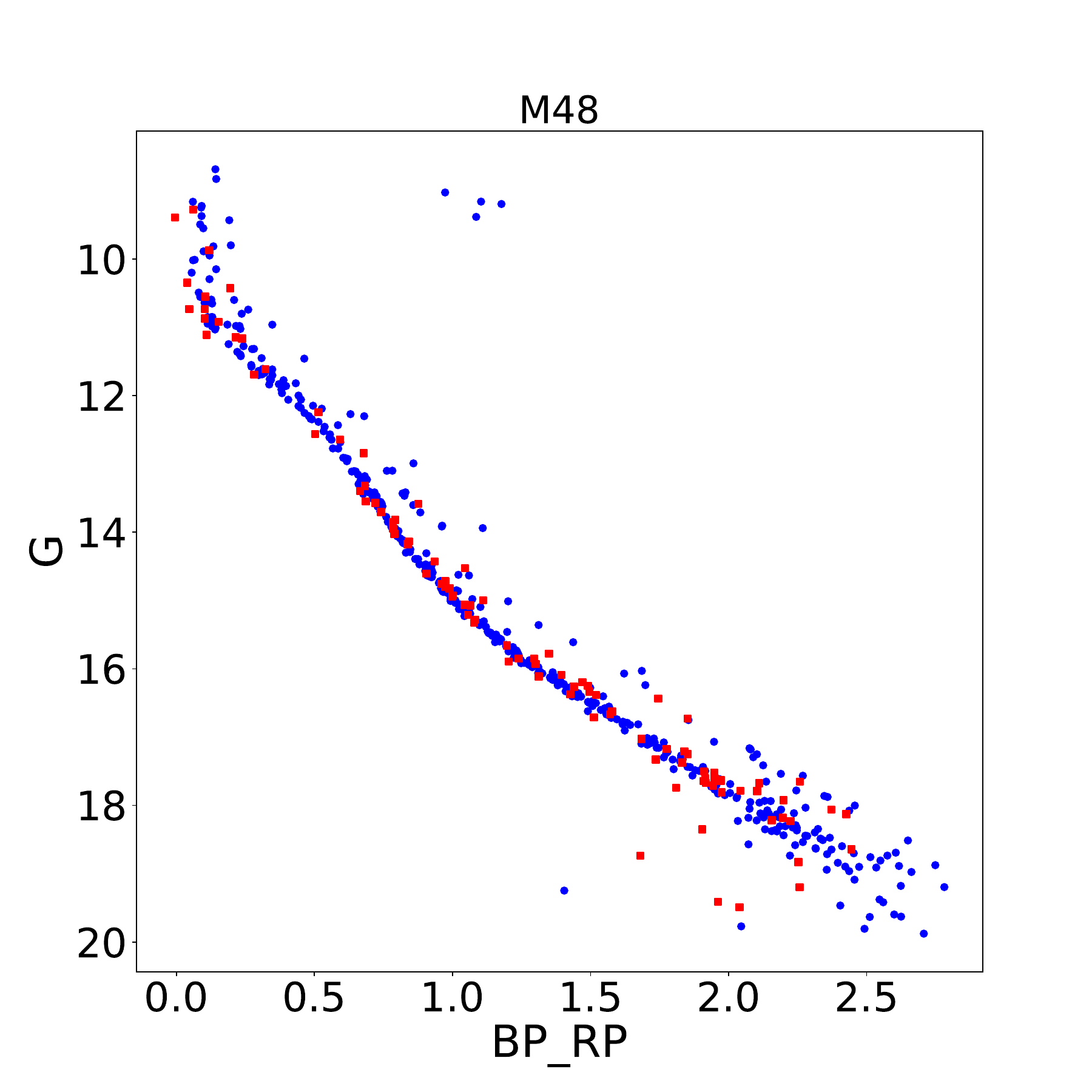}

        \end{subfigure}
        \begin{subfigure}{0.25\textwidth}
        \centering
           \includegraphics[width=\textwidth]{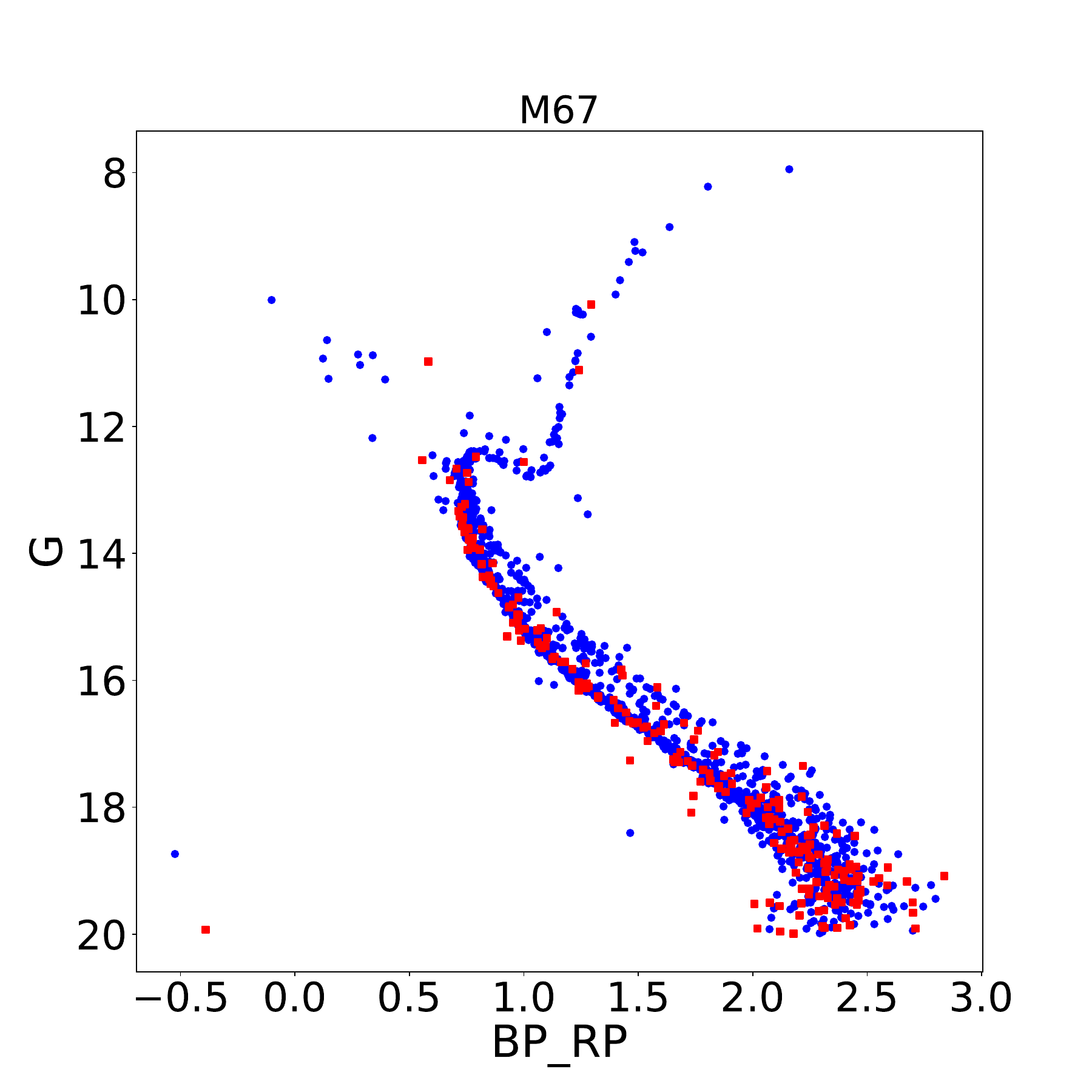}

        \end{subfigure}
        \begin{subfigure}{0.25\textwidth}
        \centering
           \includegraphics[width=\textwidth]{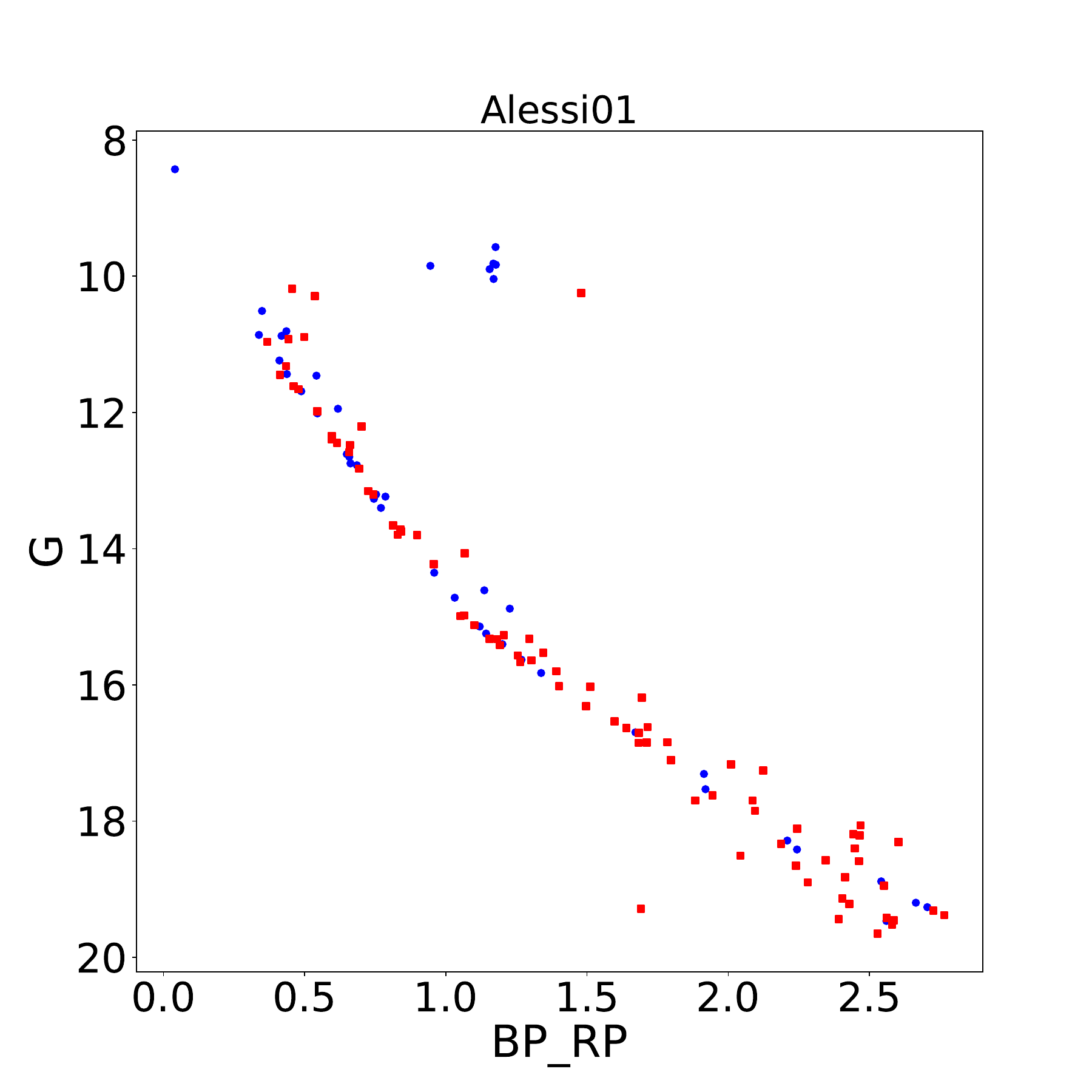}

        \end{subfigure}
  \caption{ The color-magnitude diagram (CMD) of the cluster includes stars both inside and outside the tidal radius. Red dots represent stars that were selected by Random Forest, while blue dots indicate stars that were selected by GMM with a probability higher than 0.8.}
  \label{CMD dgr.fig}
\end{figure}

\begin{table}
\caption{Results in every step. (1): Cluster name. (2): The number of sample sources from each cluster when filtered by photometric and astrometric conditions in this work. (3): Stars that DBSCAN selected among sample sources. (4) and (5): Two free parameters of the DBSCAN algorithm (refer to the text for more details). (6): Number of stars that are selected by the GMM algorithm as cluster members with a membership probability higher than 0.5. (7): Cluster members with a membership probability higher than 0.8. (8): Cluster members detected by Random Forest. (9): The evaluated F1 score value for Random Forest. (10): Range of parallax of field stars for training Random Forest. (11): Number of field stars for training Random Forest.}

  \centering
  \begin{tabular}{ccccccccccc}
    \hline
    \hline
    (1) & (2) & (3) & (4) & (5) &(6) &(7) &(8) &(9) &(10) &(11) \\
    Name & Sa & DB & Eps & MinPts & GMM & CM$>0.8$ & RF & $f1_-score$ & Range & N  \\
    \hline
    M\,38 & 444118 & 1584 & 0.05 & 63 & 853 & 753 & 431 & $[0.95852535, 0.99965926]$ & [0.65,1.1] & 88011 \\
    NGC\,2099 & 526909 & 4750 & 0.1 & 225 & 1878 & 1727 & 713 & $[0.98231827, 0.99991931]$ & [0.2,1.1] & 371754\\
    Collinder\,463 & 396681 & 1134 & 0.08 & 90 & 422 & 357 & 152 & $[0.99065421, 0.99983108]$ & [1,1.3] & 19731 \\
    NGC\,752 & 475490 & 736 & 0.15 & 80 & 290 & 269 & 59 & $[0.98113208, 0.99982608]$ & [1.8,2.7] & 28743\\
    M\,67 & 64544 & 2579 & 0.15 & 280 & 1496 & 1425 & 220 & $[0.98937426, 0.99954042]$ & [0.5,1.7] & 32624 \\
    NGC\,188 & 130483 & 1121 & 0.034 & 200 & 891 & 858 & 80 & $[0.97637795, 0.99946048]$ & [0.4,0.7] & 37057\\
    NGC\,2423 & 718872 & 808 & 0.1 & 90 & 412 & 368 & 108 & $[0.93203883, 0.99989088]$ & [0.5,1.5] & 213800\\
    Melotte\,72 & 596210 & 822 & 0.1 & 440 & 223 & 215 & 109 & $[0.93442623, 0.99994752]$ & [0.2,0.5] & 254063\\
    King\,06 & 297121 & 610 & 0.09 & 100 & 310 & 288 & 161 & $[0.95121951, 0.99949755]$ & [1.1,1.6] & 26521\\
    Alessi\,01 & 290461 & 152 & 0.1 & 65 & 52 & 43 & 82 & $[1., 1.]$ & [1.4,1.43] & 1171\\
    M\,34 & 219779 & 1700 & 0.18 & 223 & 851 & 742 & 241 & $[0.9955157 , 0.99970078]$ & [1.6,2.4] & 11139\\
    M\,35 & 514211 & 3626 & 0.11 & 320 & 1353 & 1107 & 642 & $[0.97846154, 0.99951128]$  & [0.9,1.4] & 47718\\
    M\,41 & 456358 & 1275 & 0.12 & 52 & 873 & 781 & 197 & $[0.99785867, 0.99995294]$  & [1,1.72] & 35415\\
    M\,47 & 732791 & 792 & 0.1 & 62 & 585 & 490 & 280 & $[0.96527778, 0.99896502]$  & [1.6,2.5] & 16092\\
    M\,48 & 214421 & 1723 & 0.15 & 480 & 516 & 454 & 99 & $[0.98507463, 0.99978658]$ & [0.9,1.6] & 31229\\
    \hline
  \end{tabular}
  \label{results.tab}
\end{table}

\begin{table}
\centering
\caption{Physical parameters of clusters}
\subcaption*{This work}
\begin{tabular}{c c c c c c}
    \hline
    \hline
    Name & Parallax~(mas) & pmRA~(mas\,yr$^{-1}$) & pmDEC~(mas\,yr$^{-1}$) & D~(pc) \\
    \hline
    M\,38 & $0.87\pm0.09$ & $1.54\pm0.10$ & $-4.41\pm0.07$ & $1252.49$  \\
    NGC\,2099 & $0.67\pm0.08$ & $1.87\pm0.09$ & $-5.62\pm0.06$ & $1604.09$ \\
    King\,06 & $1.37\pm0.10$ & $3.84\pm0.10$ & $-1.92\pm0.10$ & $768.11$ \\
    NGC\,752 & $2.26\pm0.07$ & $9.78\pm0.07$ & $-11.82\pm0.08$ & $439.31$ \\
    M\,67 & $1.15\pm0.10$ & $-10.96\pm0.10$ & $-2.90\pm0.07$ & $870.81$ \\
    NGC\,188 & $0.52\pm0.05$ & $-2.32\pm0.06$ & $-1.01\pm0.06$ & $1840.81$ \\
    NGC\,2423 & $1.06\pm0.05$ & $-0.74\pm0.05$ & $-3.58\pm0.05$ & $993.17$ \\
    Melotte\,72 & $0.37\pm0.06$ & $-4.13\pm0.05$ & $3.68\pm0.05$ & $2606.93$ \\
    Alessi\,01 & $1.41\pm0.08$ & $6.55\pm0.07$ & $-6.47\pm0.06$ & $712.14$ \\
    Collinder\,463 & $1.15\pm0.04$ & $-1.74\pm0.04$ & $-0.38\pm0.05$ & $852.68$ \\
    M\,34 & $2.00\pm0.14$ & $0.64\pm0.15$ & $-5.78\pm0.14$ & $512.73$ \\
    M\,35 & $1.16\pm0.08$ & $2.26\pm0.09$ & $-2.89\pm0.07$ & $908.03$ \\
    M\,41 & $1.36\pm0.09$ & $-4.37\pm0.05$ & $-1.35\pm0.08$ & $762.80$ \\
    M\,47 & $2.09\pm0.10$ & $-7.05\pm0.09$ & $1.02\pm0.09$ & $484.56$ \\
    M\,48 & $1.29\pm0.07$ & $-1.29\pm0.07$ & $1.03\pm0.06$ & $768.34$ \\
    \hline
  \end{tabular}
\bigskip
\subcaption*{Physical parameters in~\cite{cg2018}. }
\begin{tabular}{c c c c c c}
    \hline
    \hline
    Name & Parallax~(mas) & pmRA~(mas\,yr$^{-1}$) & pmDEC~(mas\,yr$^{-1}$) & D~(pc) \\
    \hline
    M\,38 & $0.87$ & $1.58$ & $-4.42$ & $1107.4$  \\
    NGC\,2099 & $0.66$ & $1.92$ & $-5.64$ & $1438.1$ \\
    King\,06 & $1.34$ & $3.86$ & $-1.81$ & $727.3$ \\
    NGC\,752 & $2.23$ & $9.81$ & $-11.71$ & $441.0$ \\
    M\,67 & $1.13$ & $-10.98$ & $-2.96$ & $859.1$ \\
    NGC\,188 & $0.50$ & $-2.30$ & $-0.96$ & $1864.3$ \\
    NGC\,2423 & $1.04$ & $-0.73$ & $-3.63$ & $930.7$ \\
    Melotte\,72 & $0.36$ & $-4.15$ & $3.68$ & $2514.7$ \\
    Alessi\,01 & $1.39$ & $6.53$ & $-6.24$ & $704.8$ \\
    Collinder\,463 & $1.13$ & $-1.71$ & $-0.30$ & $857.5$ \\
    M\,34 & $1.94$ & $0.72$ & $-5.68$ & $505.5$ \\
    M\,35 & $1.13$ & $2.30$ & $-2.90$ & $862.4$ \\
    M\,41 & $1.36$ & $-4.33$ & $-1.38$ & $720.1$ \\
    M\,47 & $2.07$ & $-7.05$ & $0.99$ & $476.4$ \\
    M\,48 & $1.28$ & $-1.31$ & $1.02$ & $758.8$ \\
    \hline
  \end{tabular}
  \label{astronomy.tab}
\end{table}
  
\twocolumn
\noindent
and Melotte\,72, higher than 0.7 for Collinder\,463, M\,34, M\,48, higher than 0.8 for M\,35, and NGC\,188 and higher than 0.9 for NGC\,2099.
\section{Results} \label{result.Method}
Fig~\ref{proper motion of dbscan.fig}  shows the distribution of member candidates among field stars for six clusters in two parameters: pmRA, and pmDEC. As seen in Fig~\ref{proper motion of dbscan.fig} , the DBSCAN selection data reveal a dense distribution among the sample sources. This indicates that DBSCAN can detect data between huge sample sources using just two filters: positive parallax and stars brighter than 20 mag.\\ 
Fig~\ref{position of dbscan and GMM.fig} to~\ref{CMD of dbscan and GMM.fig} show stars that were selected by the GMM algorithm in five parameters (RA, DEC, pmRA, pmDEC, and Parallax). In the Gaussian Mixture Model (GMM), we selected a cluster number equal to 2, which corresponded to the cluster and field excluding Melotte 72. To distinguish members of Melotte 72 from field stars, we utilized three different values for the GMM cluster number. In the case of this specific cluster, all data points within the other two GMM clusters were considered as suspect data, subject to a decision by the Random Forest algorithm in the final step. If these stars are indeed members of the cluster, they were identified by the Random Forest in the last stage. The confusion matrix is shown in Fig~\ref{cmatrix.fig}. \\
In this work, stars that have a probability higher than 0.8 are considered as cluster members. As seen in Fig~\ref{position of dbscan and GMM.fig}, members that are in the outer radius from the cluster center (cluster dense region) have a probability lower than 0.8, nevertheless, some of them can be selected as escape members. Fig~\ref{CMD of dbscan and GMM.fig} shows a clear main sequence and for older clusters, a red giant branch. Fig~\ref{position dgr.fig} shows data that were detected with the Random Forest algorithm among GMM detection members and field stars based on five parameters (pmRA, pmDEC, Parallax, G magnitude, and Bp-Rp color index). Position parameters were not applied in the Random Forest model to obtain the best view of cluster morphology.  As seen in Fig~\ref{position dgr.fig}, stars selected by Random Forest are in the outer layer than the cluster center but these members are in the range of proper motion, parallax, and CMD of members that were selected by GMM with a probability higher than 0.8 as are shown in Fig~\ref{proper motion dgr.fig} to Fig~\ref{CMD dgr.fig}.\\ 
As shown in Fig~\ref{position dgr.fig}, the morphology of the clusters can be observed in detail, including their members, corona, and tidal tails. This method can detect members of the smallest cluster, even those far away from the center of the clusters, such as Alessi\,01 and Melotte\,72. In the case of Melotte\,72, which is at a high distance from Earth, as depicted in Fig~\ref{position dgr.fig}, the candidate members fall within a large distance range of approximately 100 parsecs from the cluster. In Table~\ref{results.tab} , we present the selection data at each step. Notably, for the richness cluster (M\,35 and NGC\,2099), the Random Forest algorithm detected more members compared to other clusters. Moving on to Table~\ref{astronomy.tab}, it displays the physical parameters for the selected members using the Gaussian Mixture Model (GMM) with a probability higher than 0.8, and those identified by the Random Forest and comparison with \cite{cg2018}. In this method, the selected parameters correspond with the physical characteristics detected by GMM with a probability higher than 0.8. In the case of the oldest cluster in this study (NGC\,188), only a few data members were detected using Random Forest. This observation could be attributed to its age and dense shape.\\
As seen in Fig~\ref{CMD dgr.fig}, stars detected by Random Forest lie in the main sequence, red giant branch, and also the binary region. These stars could be studied in other works to discuss star formation theory, check simulation codes related to cluster star evolution, survey the chemical elements of clusters, study cluster morphology, calculate the gravitational effect from the Galaxy to the cluster, estimate the cluster’s initial mass, and determine a reliable value for the clusters age.\\
By viewing the region of members detected by GMM in Fig~\ref{position dgr.fig}, Random Forest selected only a few members for cluster-dense regions that were detected by GMM. This could indicate that GMM algorithms detected cluster members in five dimensions in the cluster-dense regions very well. The distance of the clusters is obtained from \cite{Bailer-Jones}.

\section{Discussion} \label{descotion.Method}

\begin{table}
\caption{The distribution of stars in clusters is as follows: \\
(1): cluster name . (2): The radius that we found cluster members candidate. (3): Tidal radius. (4): Number of stars inside the tidal radius. (5): Number of stars outside the tidal radius. (6): Number of stars inside the core radius. (7): the visual extinction astronomy}  [1]~\cite{mass-extended}, [2]~\cite{ngc2099red}, [3]~\cite{collinder469-semionov}, [4]~\cite{ngc752red}, [5]~\cite{ngc188red}, [6]~\cite{m34red}, [7]~\cite{m35red}, [8]~\cite{m47-Prisinzano}
  \centering
  \begin{tabular}{ccccccc}
    \hline
    \hline
    (1) & (2) & (3) & (4) & (5) & (6) & (7)\\
    Name & $R[pc]$ & $R_t[pc]$ & $N_{tin}$ & $N_{tout}$ &  $N_{c}$ & $A_v[mag]$  \\
    \hline
    M\,38 & $55.30$ & $21.52$ & $930$ & $254$ & $213$ & $0.998$~[1] \\
    NGC\,2099 & $71.17$ & $28.92$ & $2073$ & $367$ & $356$ & $0.93$~[2] \\
    Collinder\,463 & $35.92$ & $15.78$ & $415$ & $94$ & $158$ & $0.836$~[3] \\
    NGC\,752 & $37.45$ & $12.11$ & $266$ & $62$ & $59$ & $0.1085$~[4] \\
    M\,67 & $38.07$ & $16.01$ & $1517$ & $128$ & $211$ & $0.164$~[1] \\
    NGC\,188 & $87.72$ & $30.27$ & $899$ & $39$ & $287$ & $0.2697$~[5] \\
    NGC\,2423 & $42.70$ & $14.80$ & $412$ & $64$ & $75$ & $0.353$~[1] \\
    Alessi\,01 & $33.52$ & $8.19$ & $60$ & $65$ & $12$ &  \\
    Melotte\,72 & $112.95$ & $20.20$ & $236$ & $88$ & $64$ & $0.230$~[1] \\
    King\,06 & $33.95$ & $8.22$ & $331$ & $118$ & $27$ & $1.909$~[1] \\
    M\,34 & $22.54$ & $12.33$ & $875$ & $108$ & $169$ & $0.217$~[6] \\
    M\,35 & $39.88$ & $22.76$ & $1523$ & $226$ & $318$ & $0.682$~[7] \\
    M\,41 & $33.59$ & $16.11$ & $889$ & $89$ & $254$ & $0.225$~[1] \\
    M\,47 & $21.05$ & $9.25$ & $684$ & $86$ & $172$ & $0.273$~[8] \\
    M\,48 & $33.07$ & $20.38$ & $530$ & $23$ & $156$ & $0.127$~[1] \\
    \hline
  \end{tabular}
  \label{tab_discussion}
\end{table}

\begin{figure}
  \centering
  \captionsetup[subfigure]{labelformat=empty}
        \begin{subfigure}{0.25\textwidth}
        \centering
           \includegraphics[width=\textwidth]{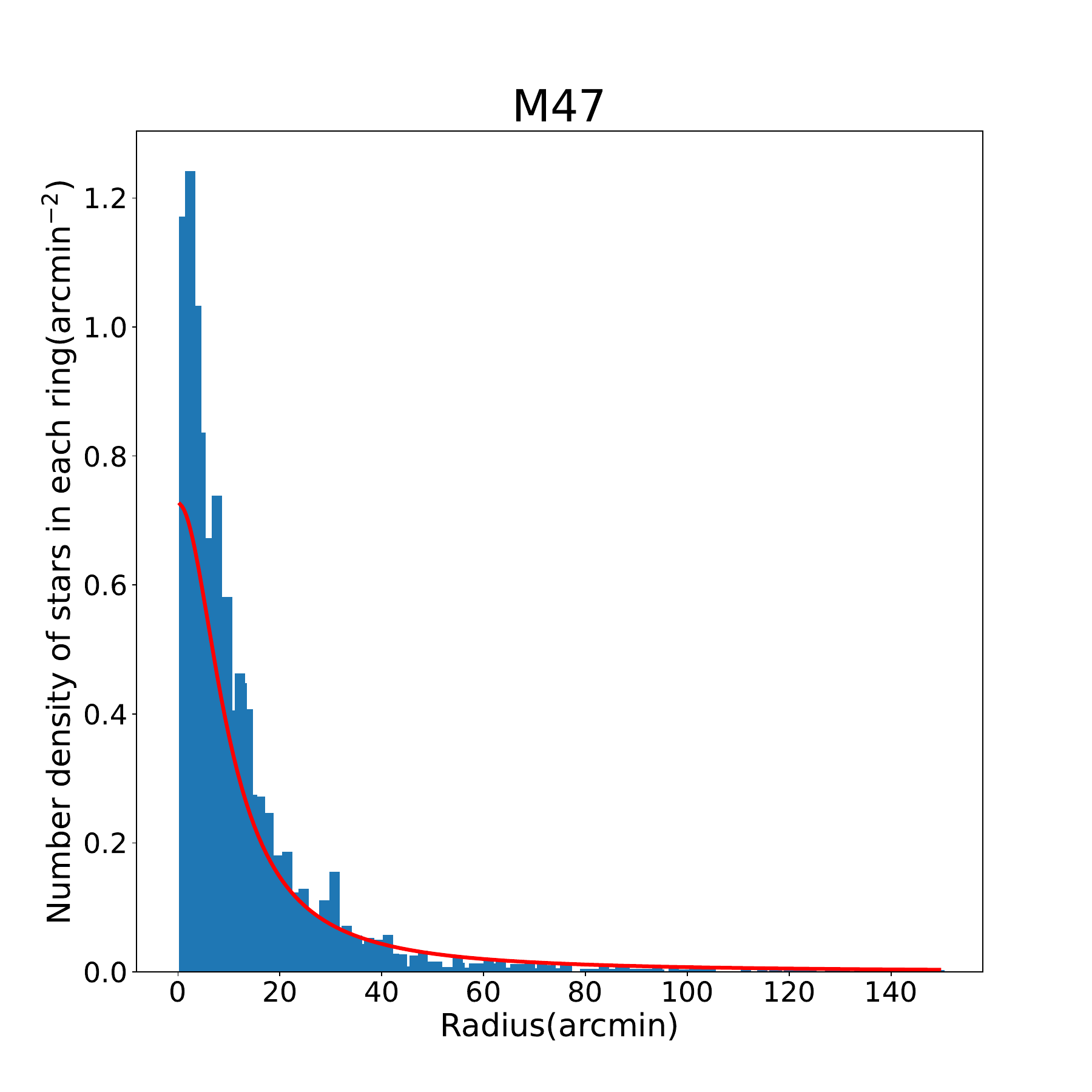}

        \end{subfigure}
        \begin{subfigure}{0.25\textwidth}

                \centering
                \includegraphics[width=\textwidth]{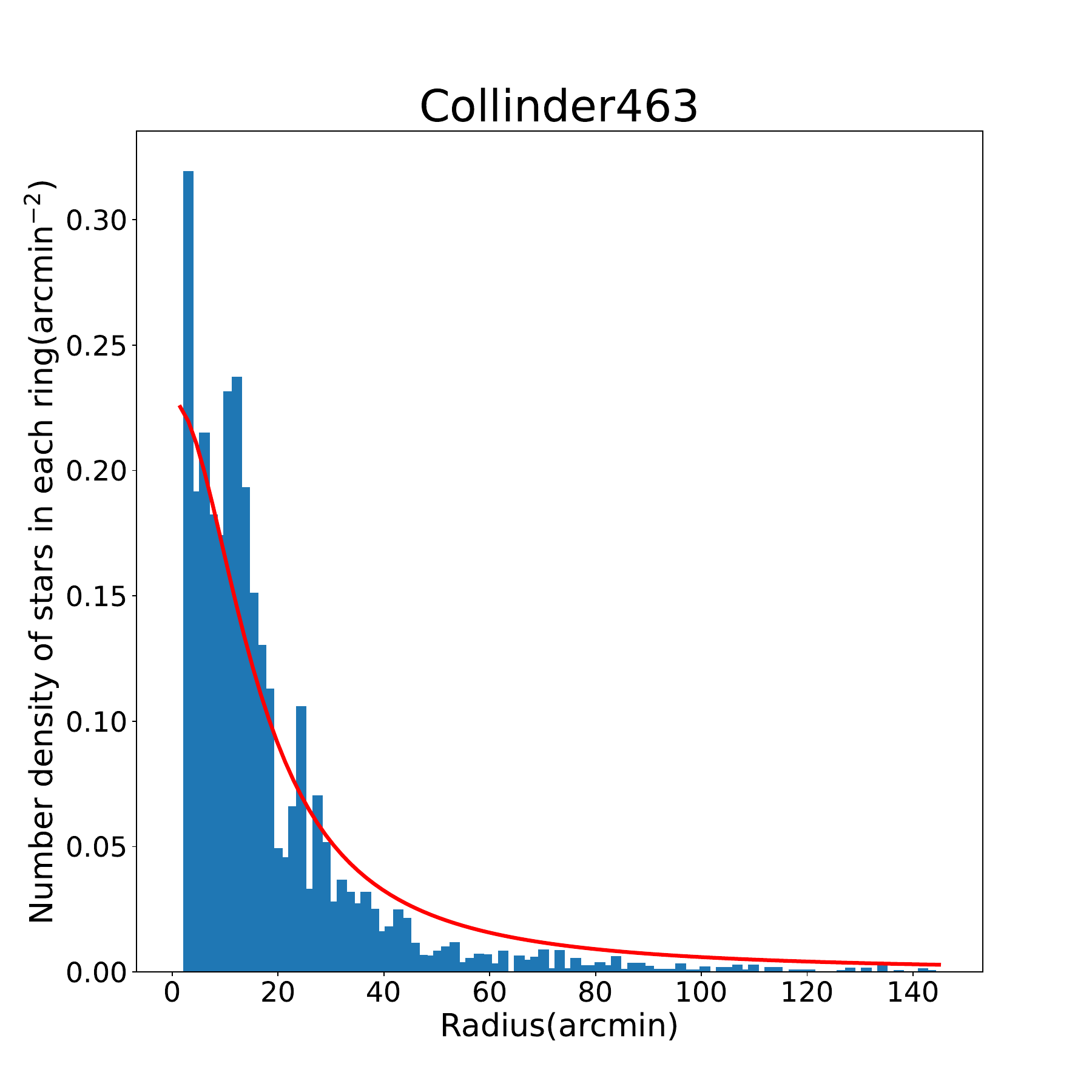}

        \end{subfigure}
        \begin{subfigure}{0.25\textwidth}
                \centering
           \includegraphics[width=\textwidth]{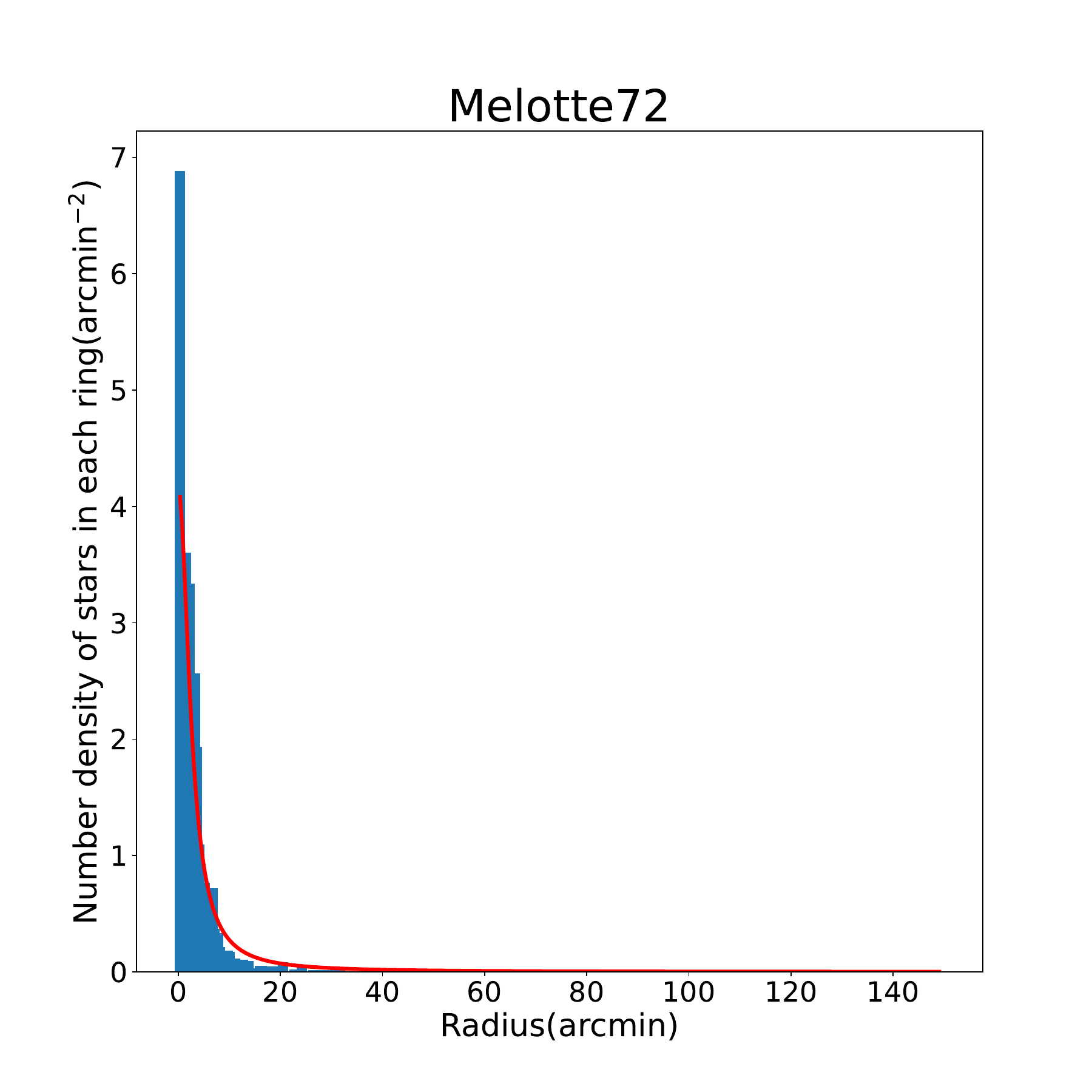}

        \end{subfigure}
        \begin{subfigure}{0.25\textwidth}
                \centering

                \includegraphics[width=\textwidth]{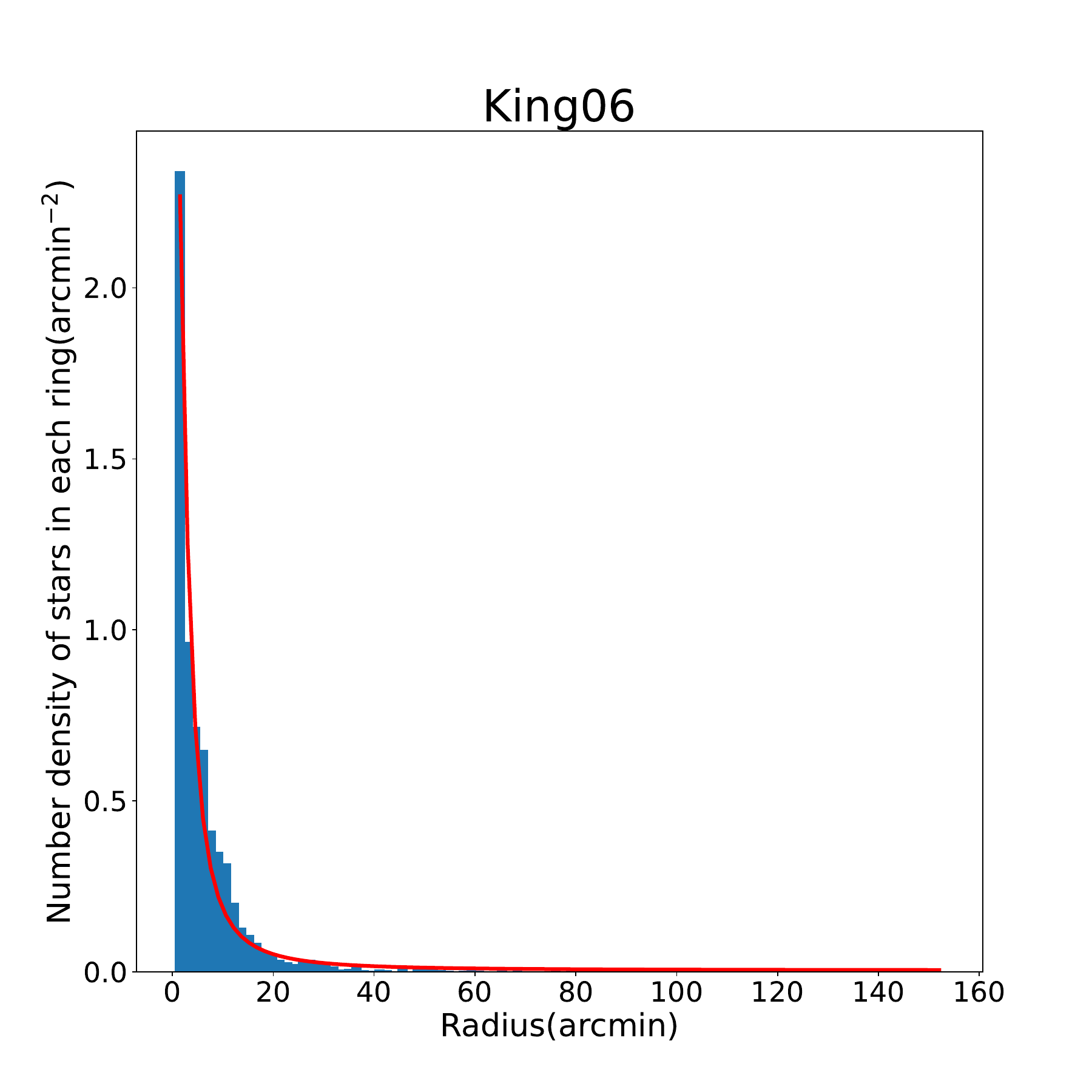}

        \end{subfigure}
        \begin{subfigure}{0.25\textwidth}
                \centering

                \includegraphics[width=\textwidth]{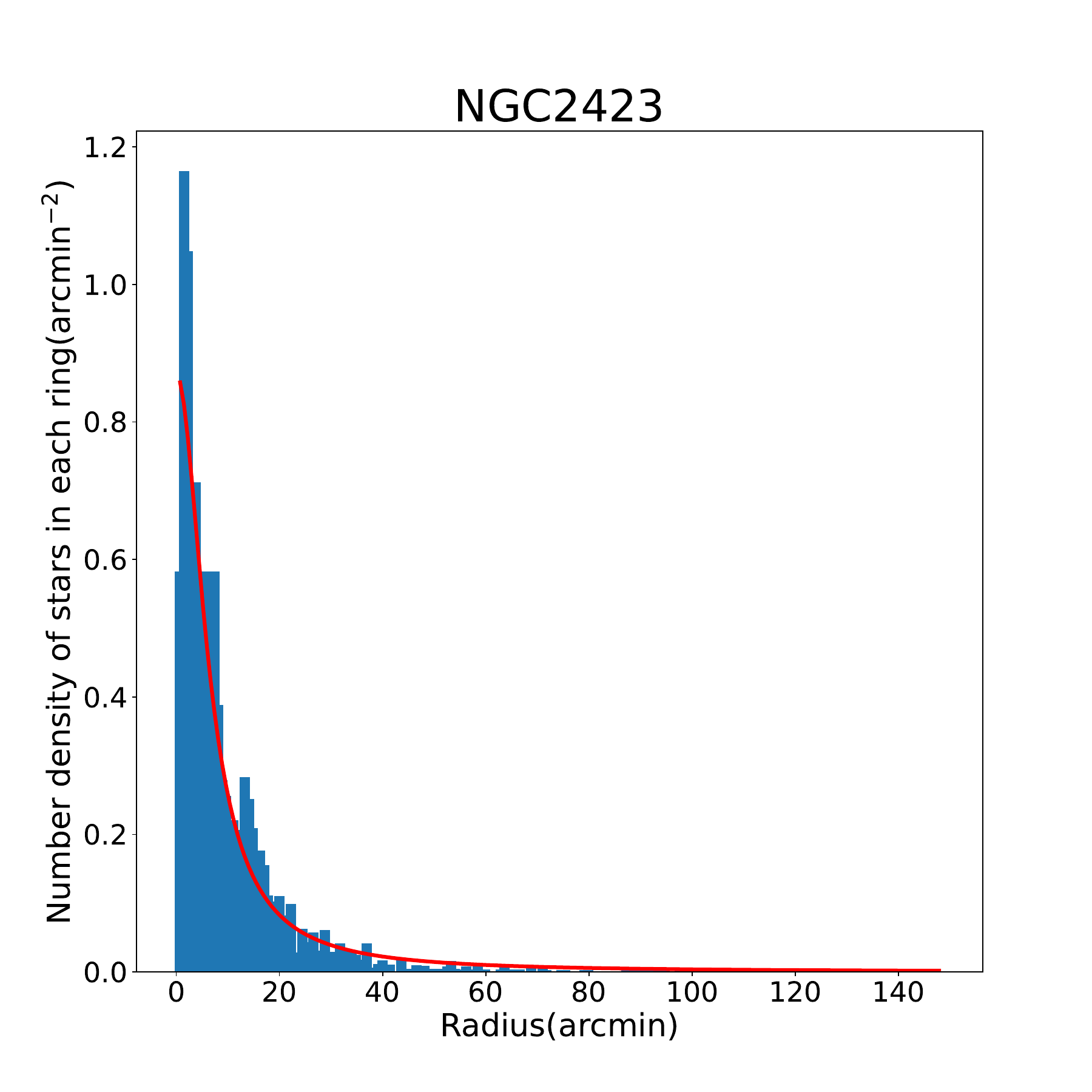}

        \end{subfigure}
        \begin{subfigure}{0.25\textwidth}
                \centering

                \includegraphics[width=\textwidth]{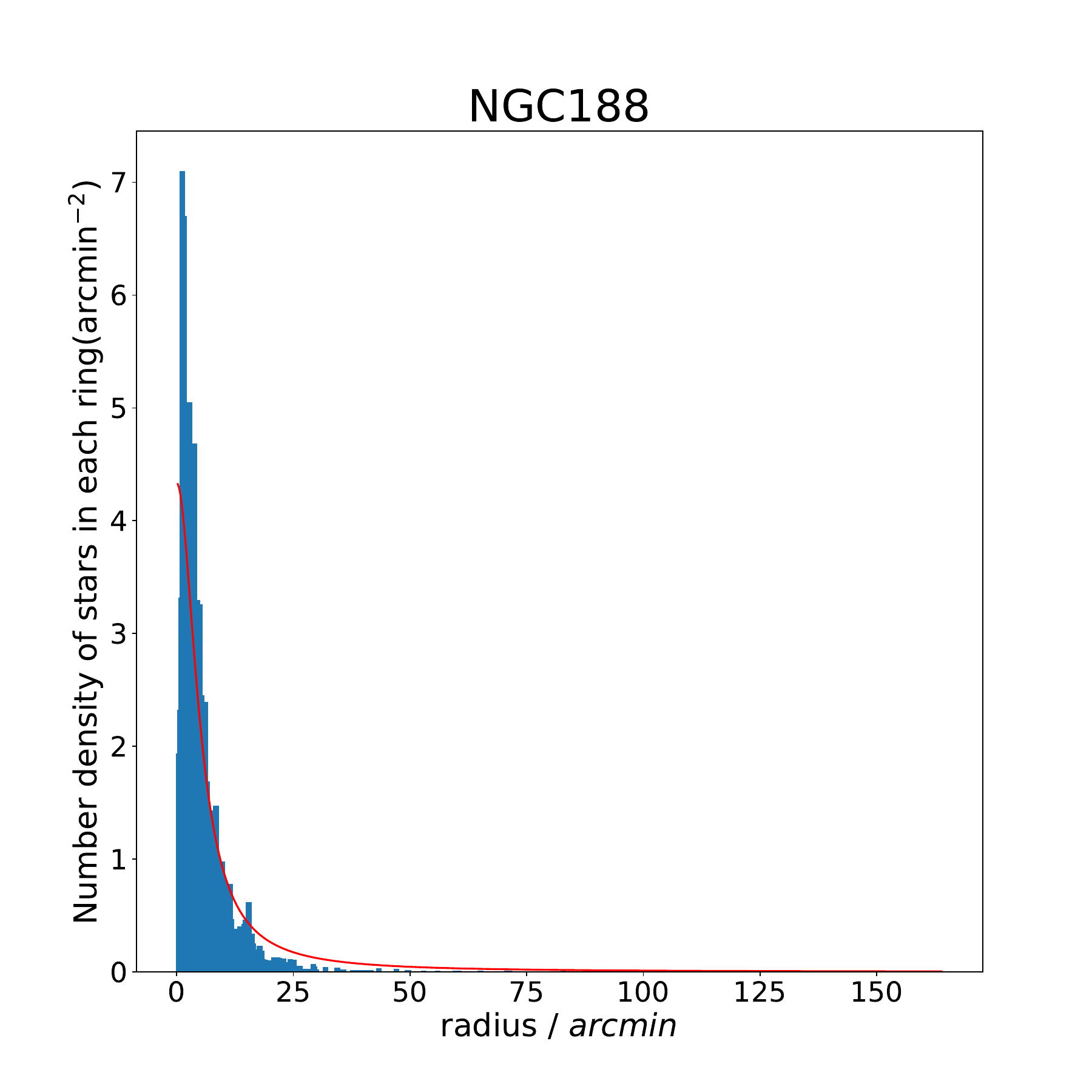}

        \end{subfigure}
        \begin{subfigure}{0.25\textwidth}
        \centering
           \includegraphics[width=\textwidth]{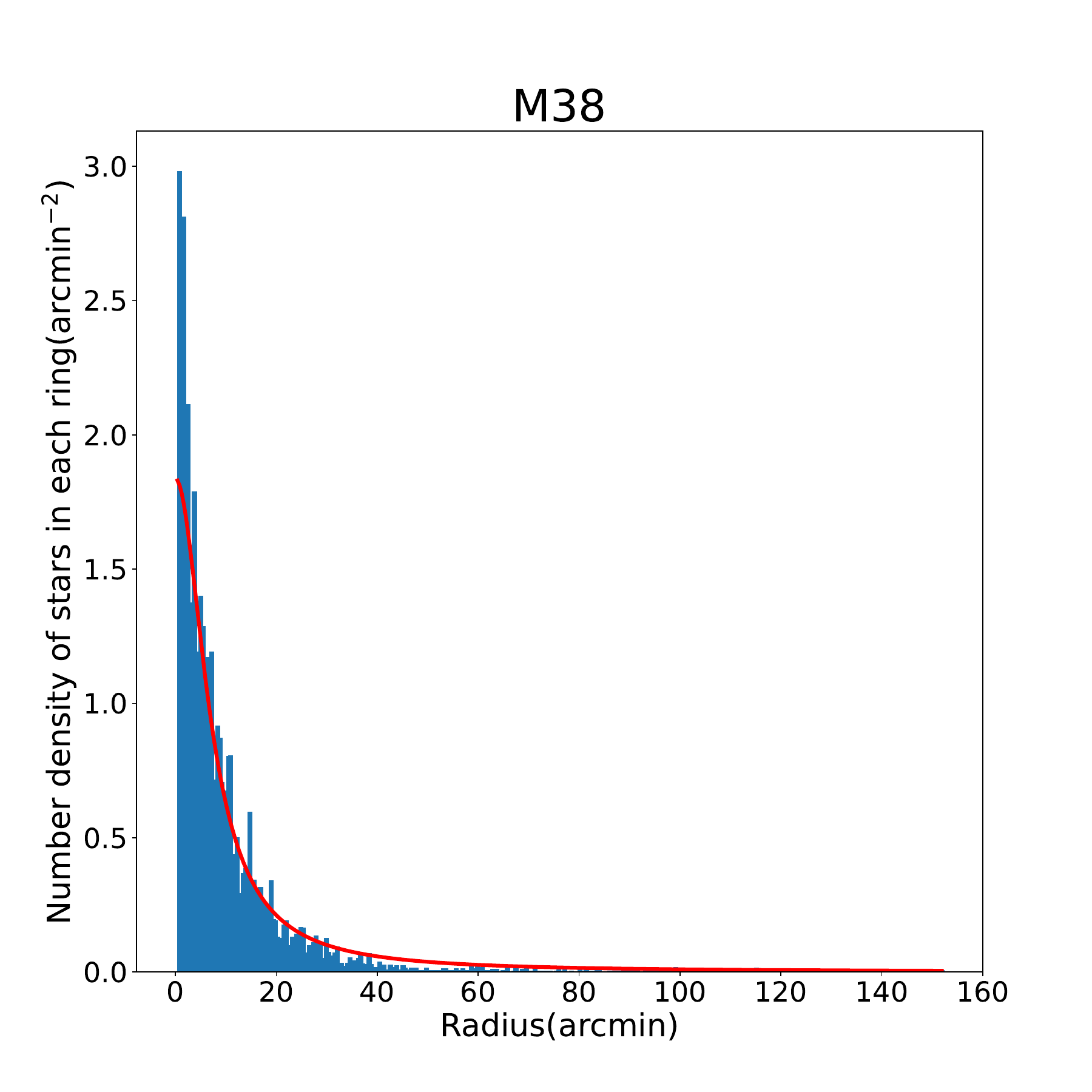}

        \end{subfigure}
        \begin{subfigure}{0.25\textwidth}
        \centering
           \includegraphics[width=\textwidth]{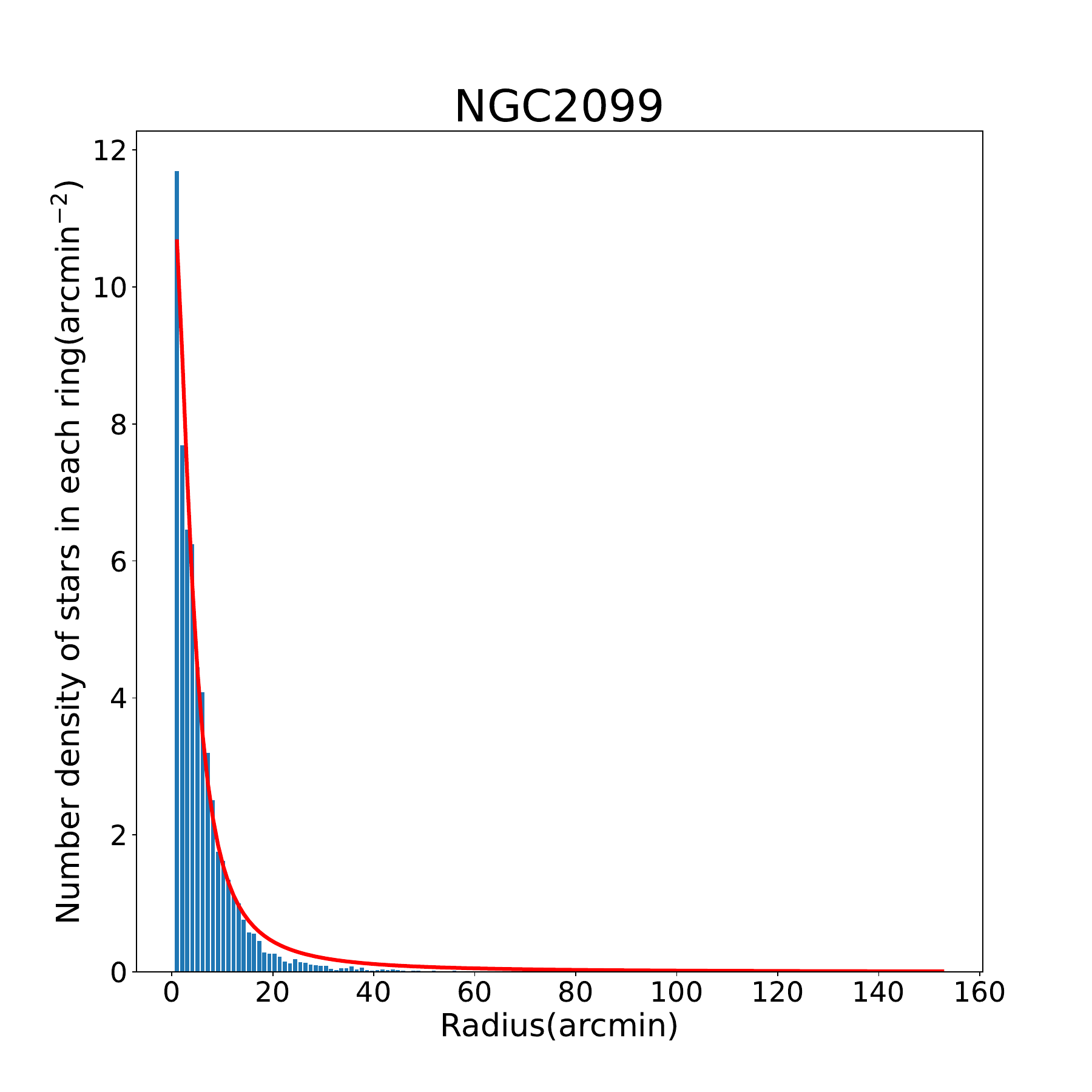}

        \end{subfigure}
        \begin{subfigure}{0.25\textwidth}
        \centering
           \includegraphics[width=\textwidth]{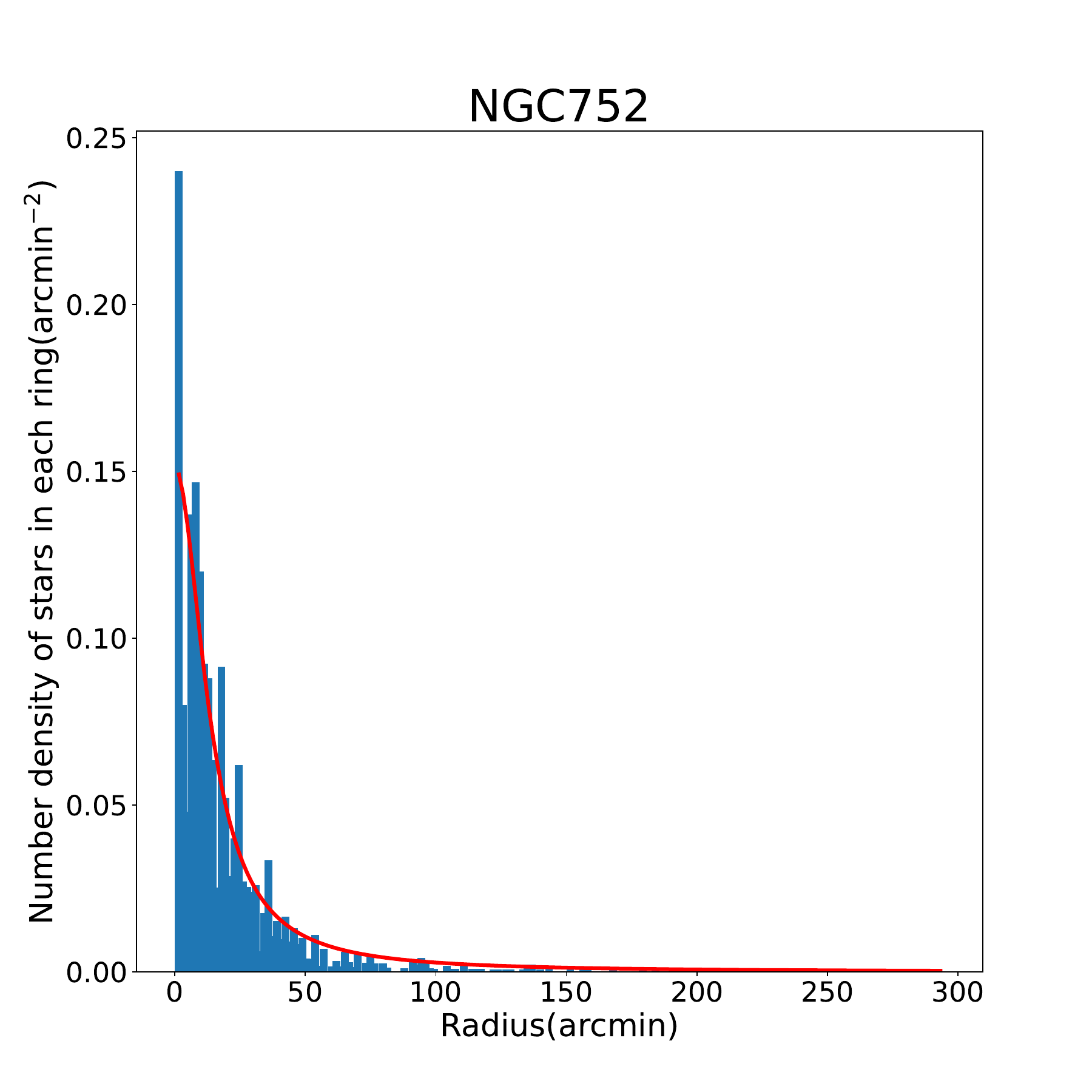}

        \end{subfigure}
        \begin{subfigure}{0.25\textwidth}
        \centering
           \includegraphics[width=\textwidth]{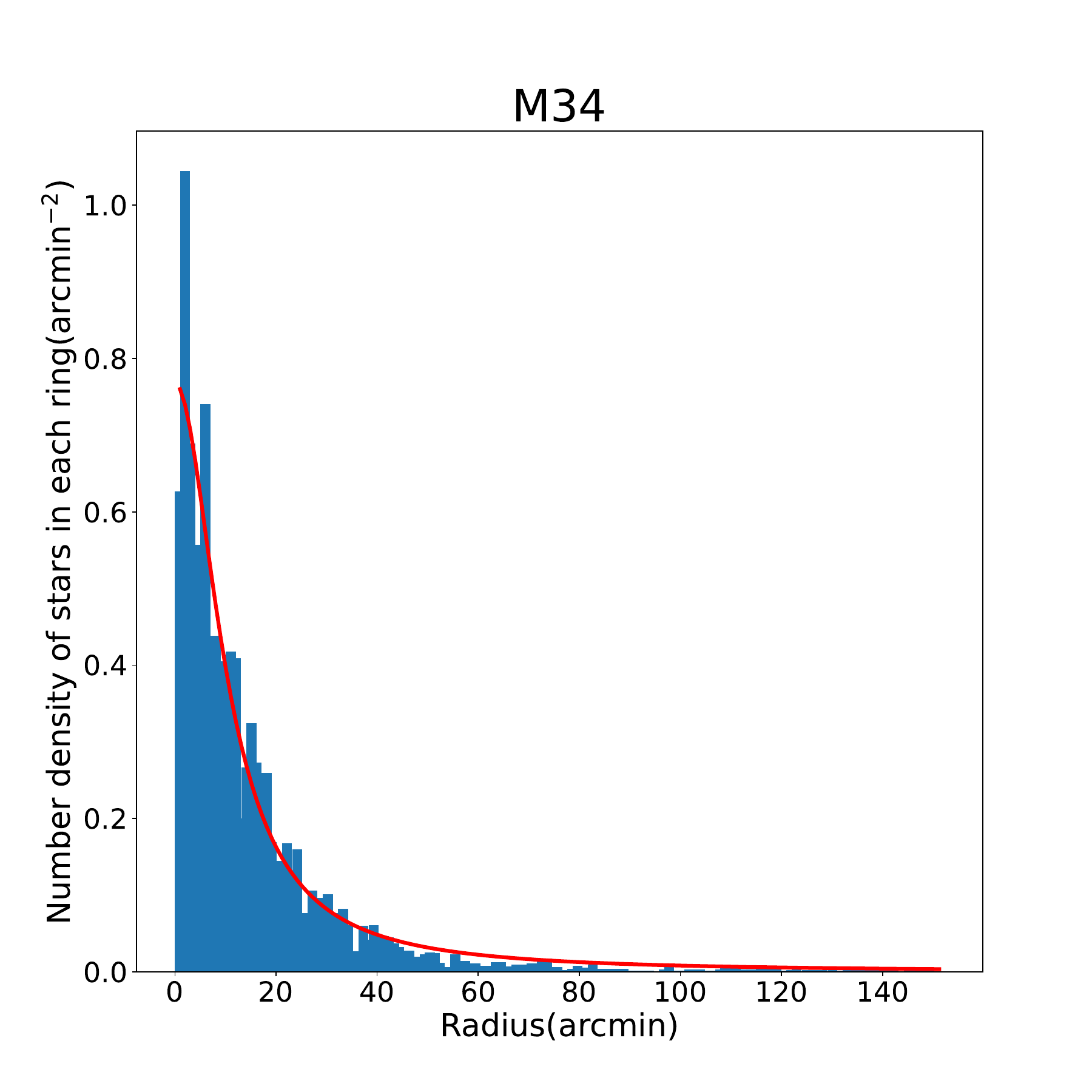}

        \end{subfigure}
        \begin{subfigure}{0.25\textwidth}
        \centering
           \includegraphics[width=\textwidth]{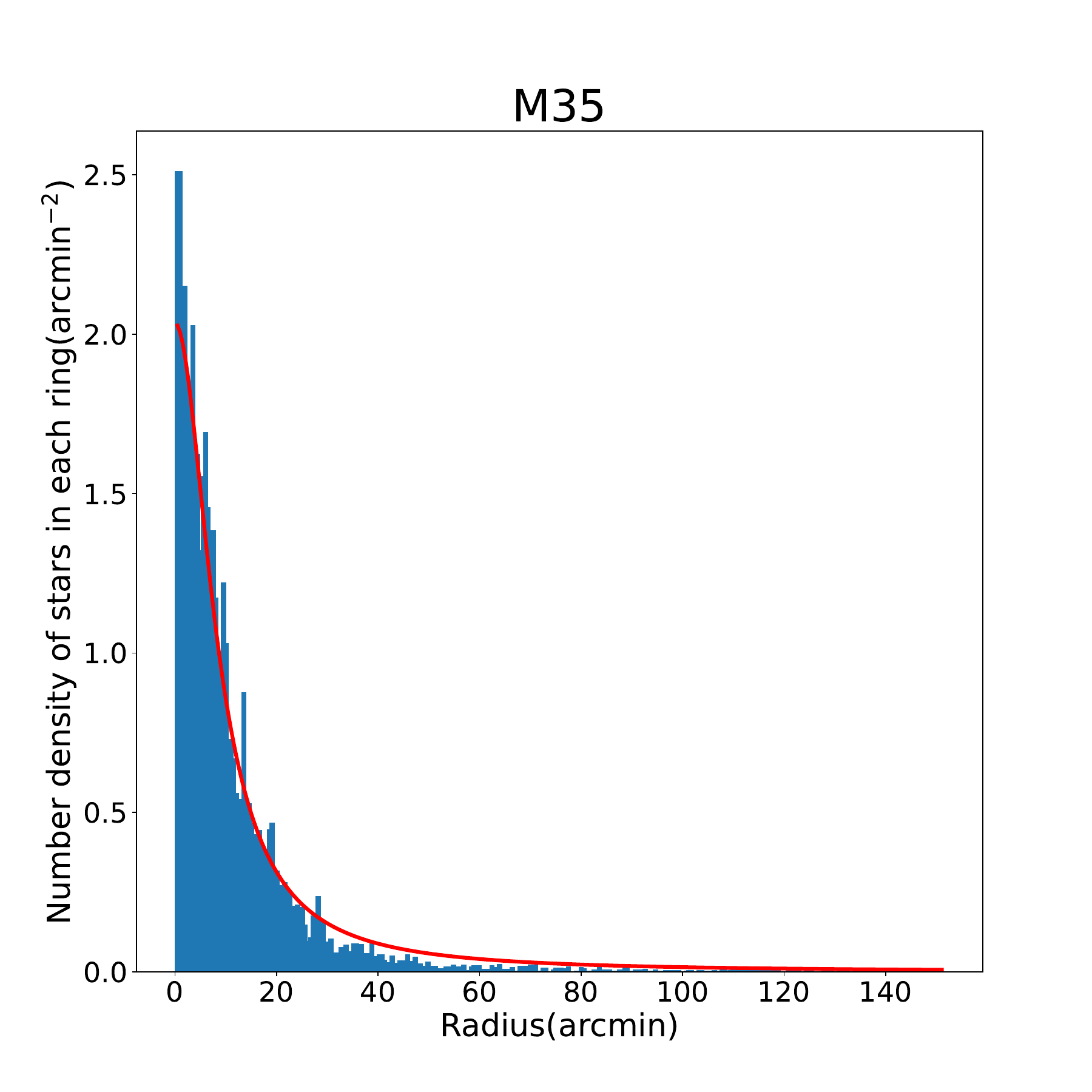}

        \end{subfigure}
        \begin{subfigure}{0.25\textwidth}
        \centering
           \includegraphics[width=\textwidth]{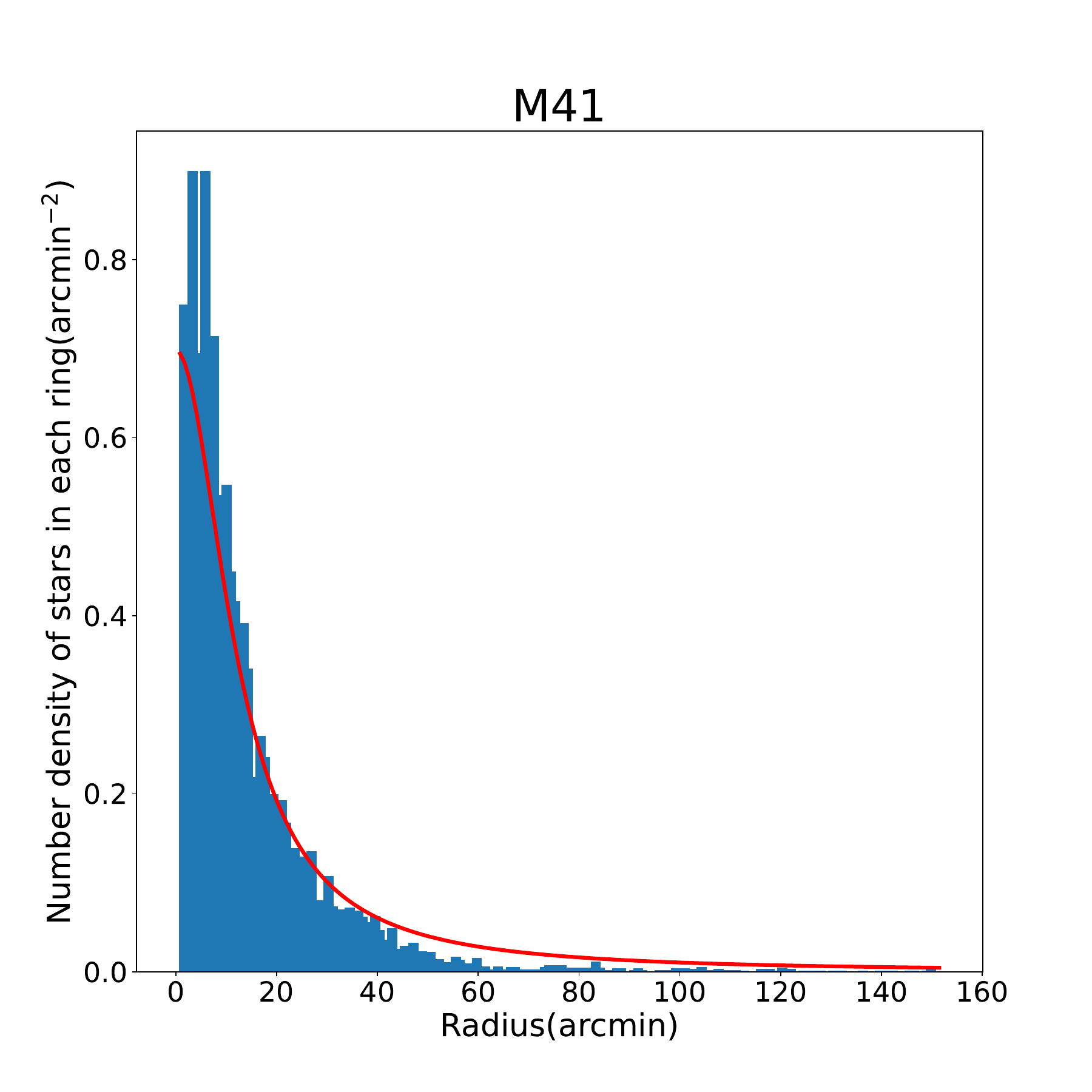}

        \end{subfigure}
        \begin{subfigure}{0.25\textwidth}
        \centering
           \includegraphics[width=\textwidth]{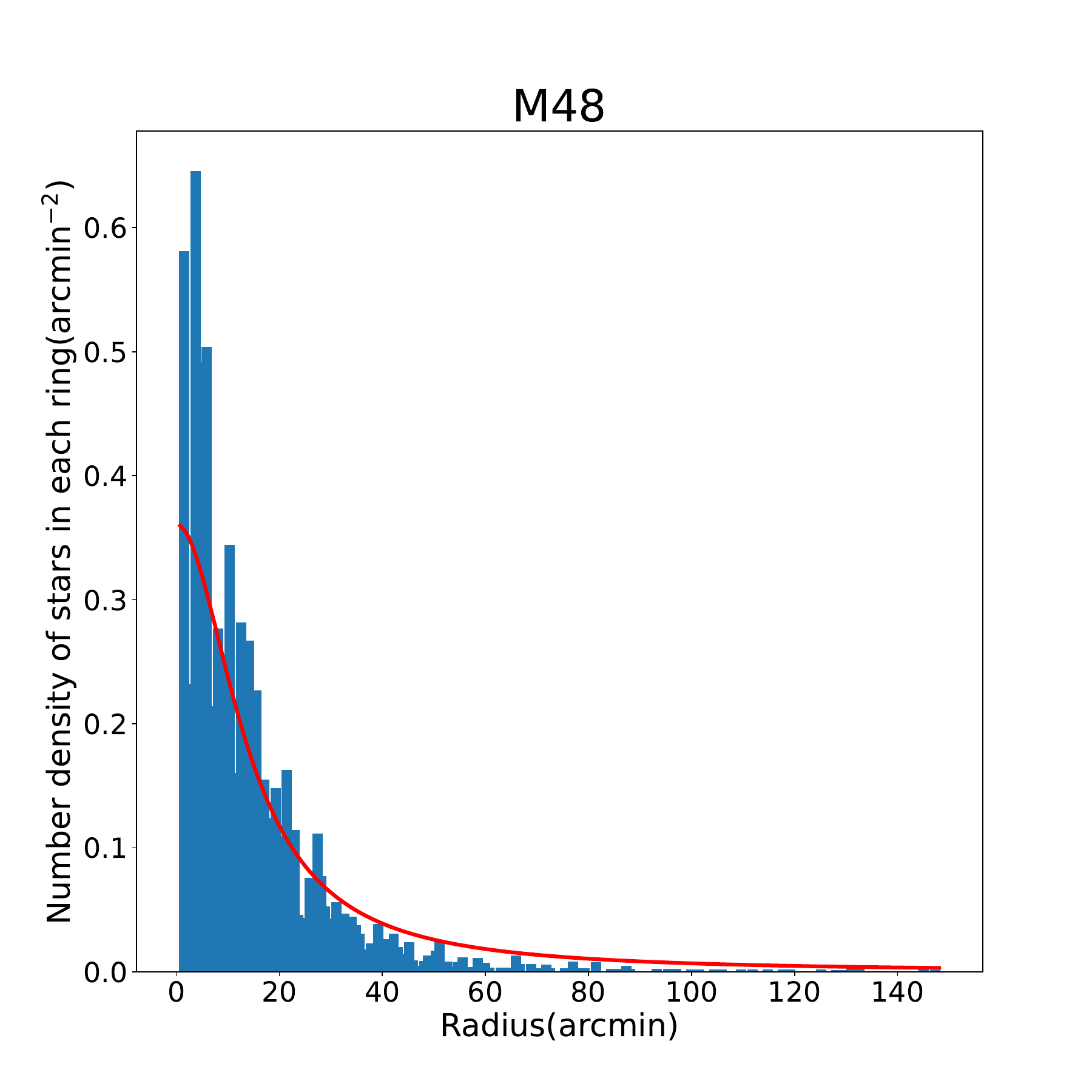}

        \end{subfigure}
        \begin{subfigure}{0.25\textwidth}
        \centering
           \includegraphics[width=\textwidth]{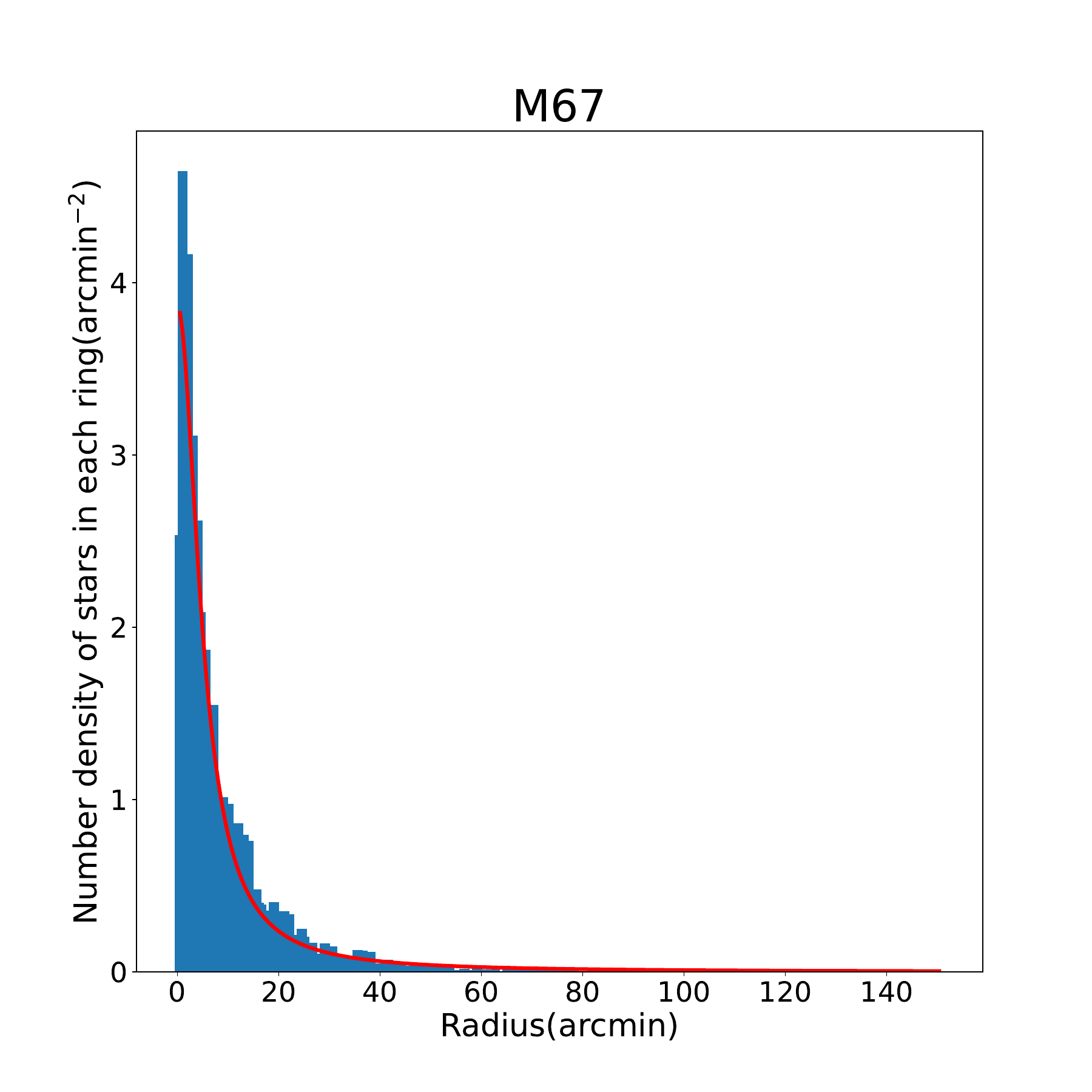}

        \end{subfigure}
        \begin{subfigure}{0.25\textwidth}
        \centering
           \includegraphics[width=\textwidth]{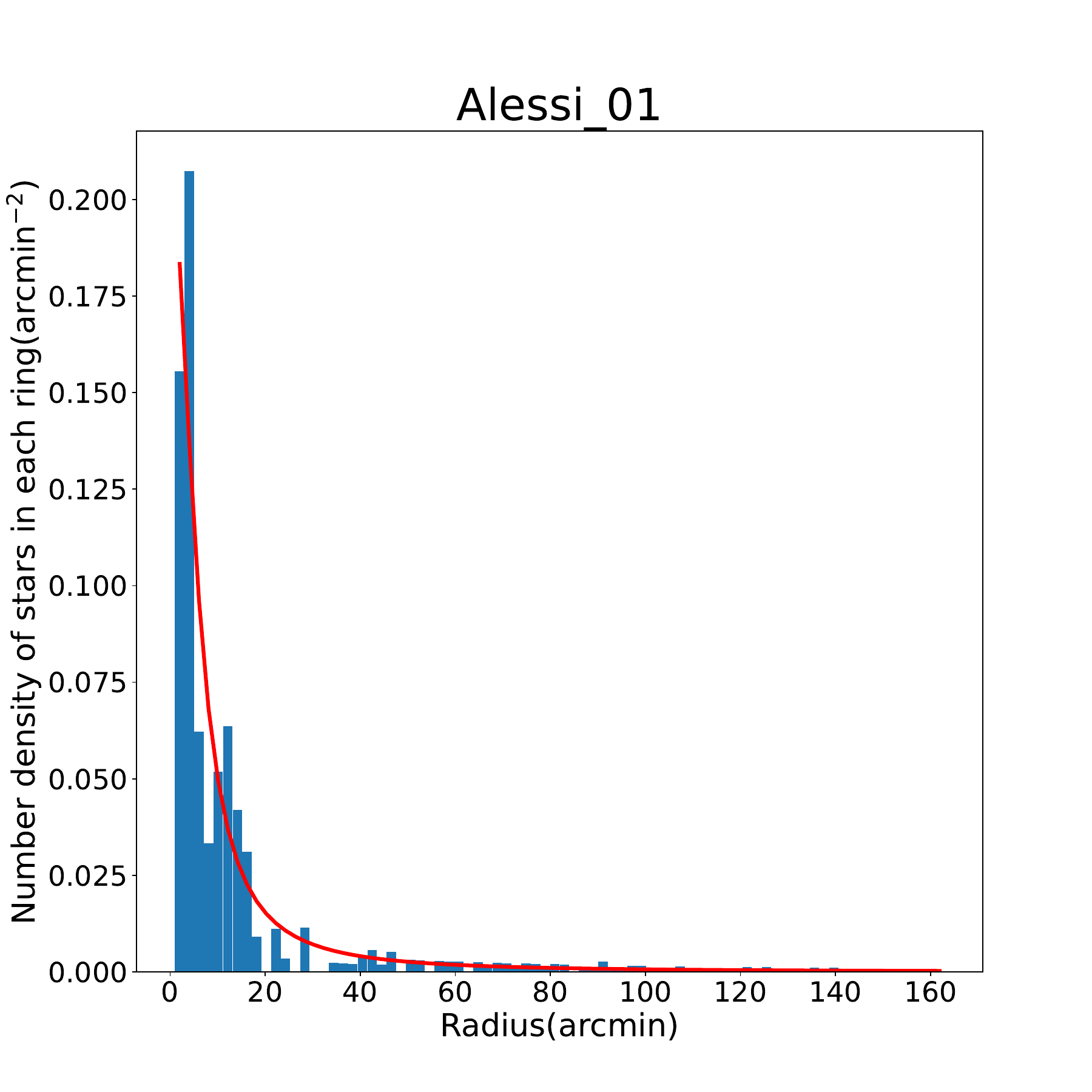}

        \end{subfigure}
  \caption{The fitting of the King profile to cluster density.}
  \label{kingfit.fig}
\end{figure}

\begin{figure}
  \centering
  \captionsetup[subfigure]{labelformat=empty}
        \begin{subfigure}{0.25\textwidth}
        \centering
           \includegraphics[width=\textwidth]{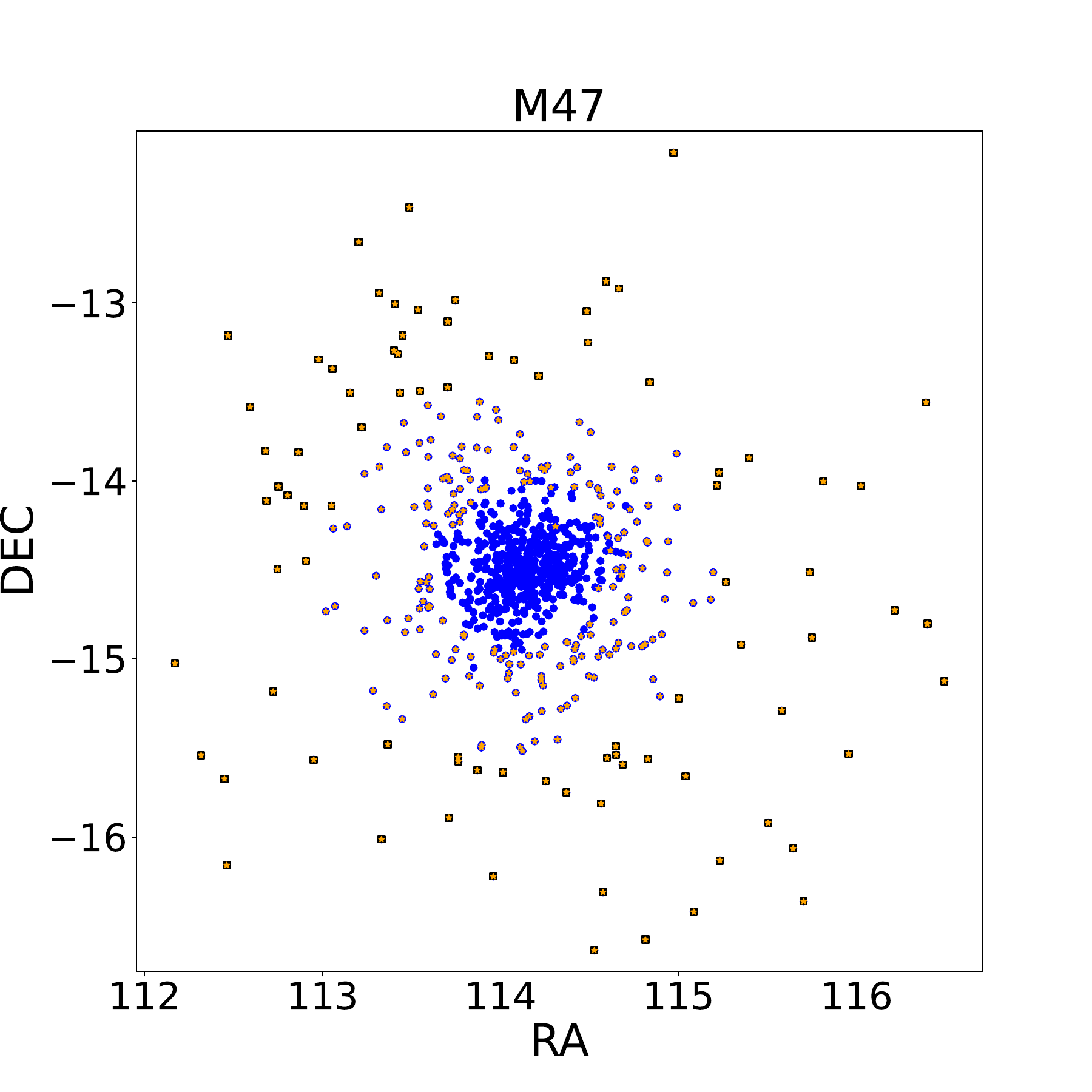}

        \end{subfigure}
        \begin{subfigure}{0.25\textwidth}

                \centering
                \includegraphics[width=\textwidth]{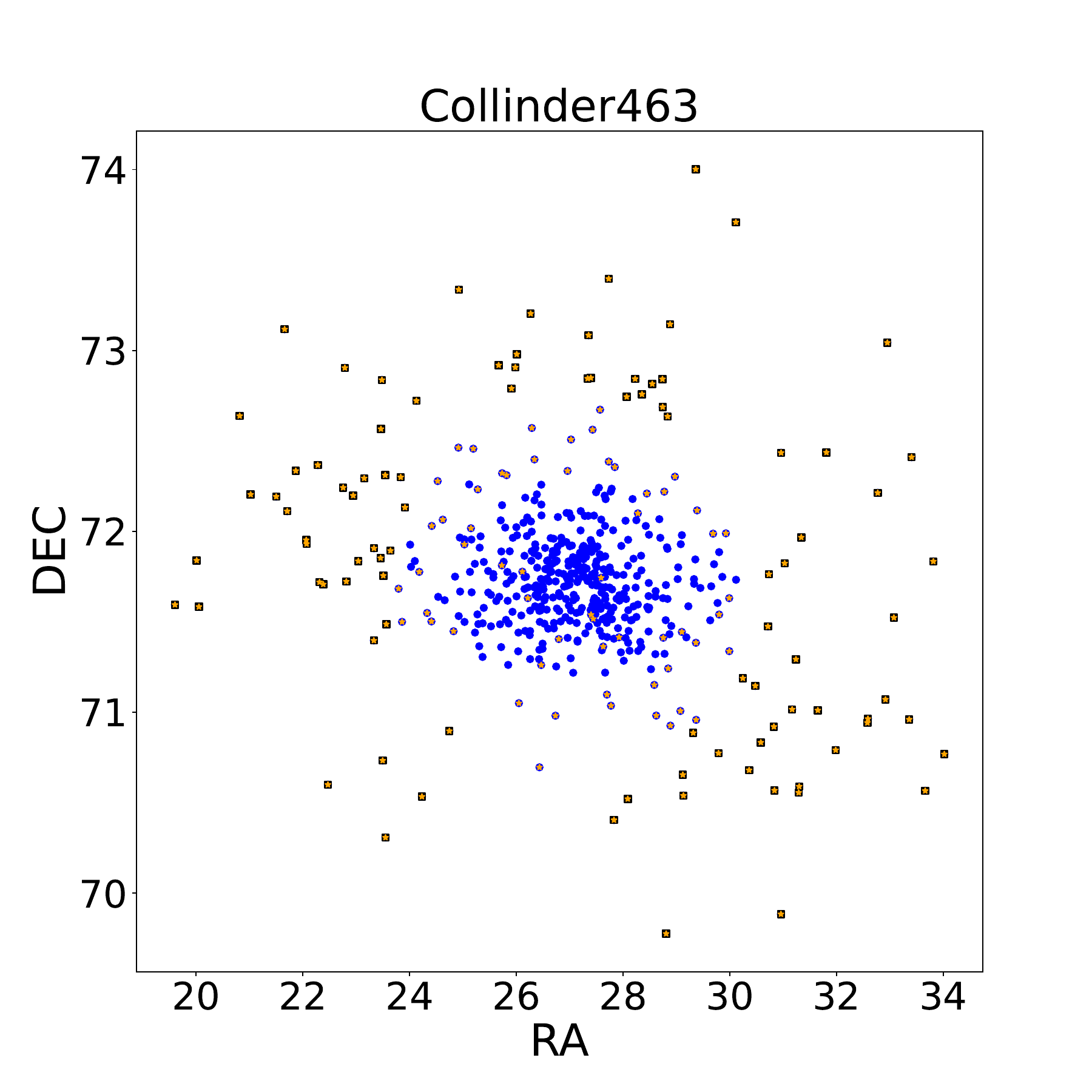}

        \end{subfigure}
        \begin{subfigure}{0.25\textwidth}
                \centering
           \includegraphics[width=\textwidth]{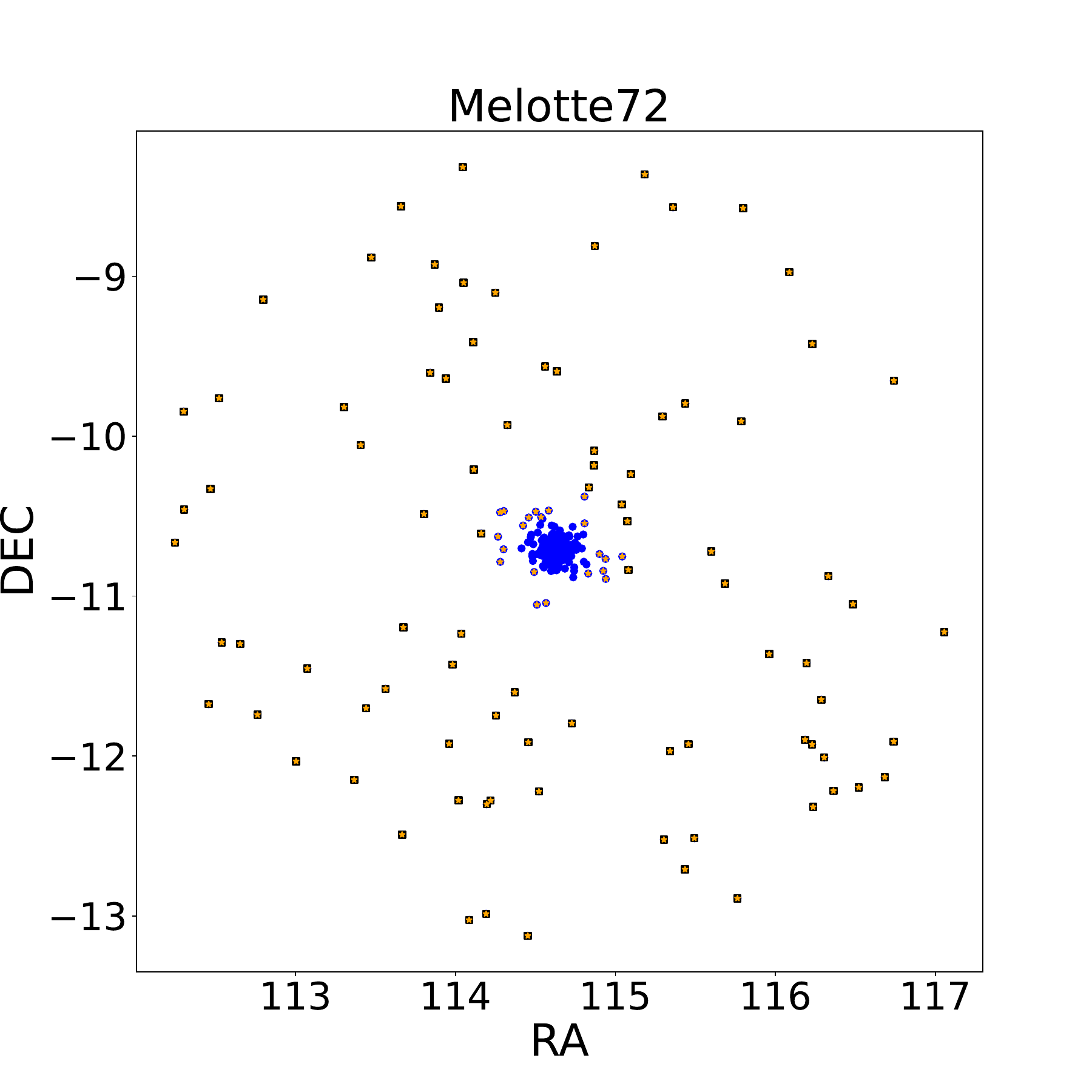}

        \end{subfigure}
        \begin{subfigure}{0.25\textwidth}
                \centering

                \includegraphics[width=\textwidth]{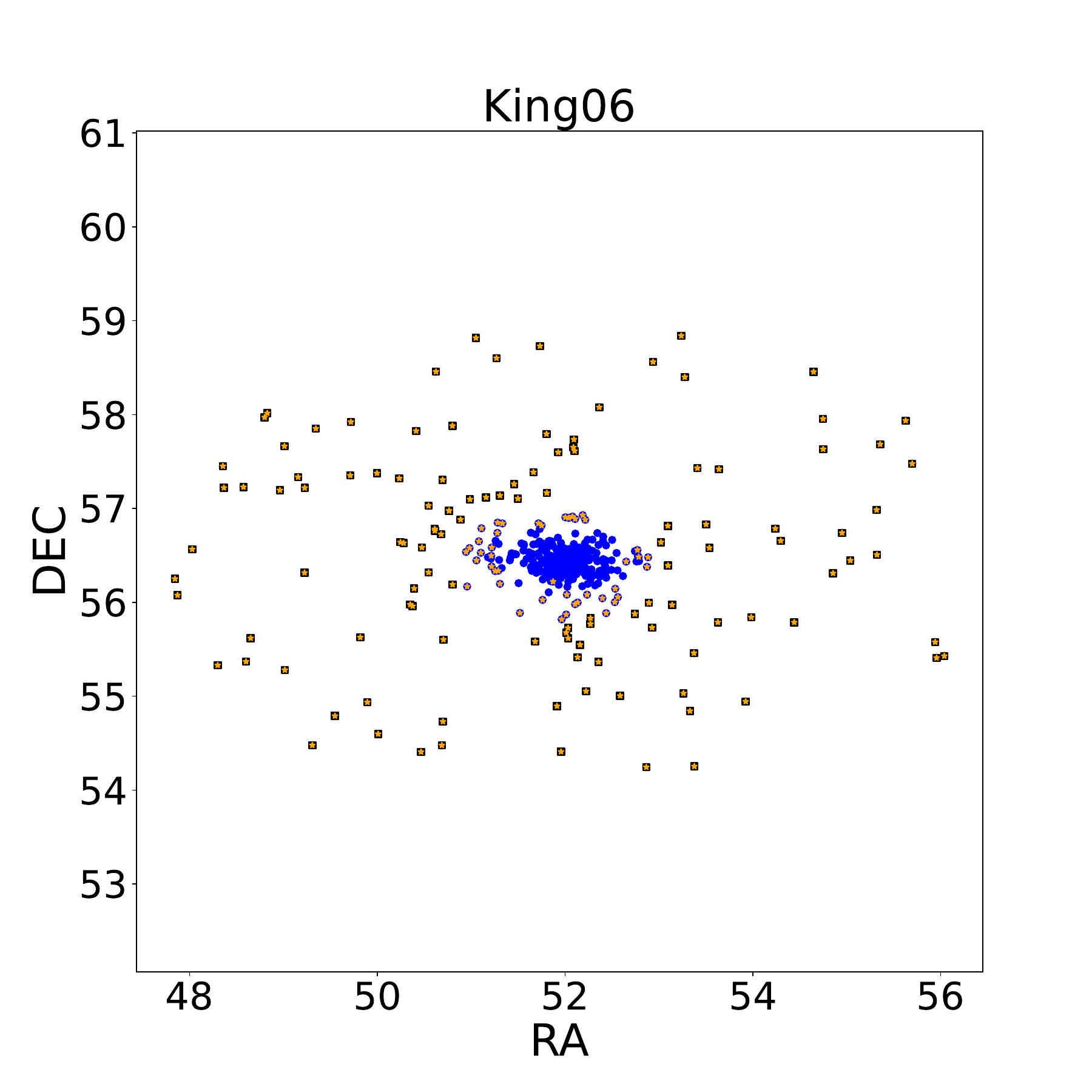}

        \end{subfigure}
        \begin{subfigure}{0.25\textwidth}
                \centering

                \includegraphics[width=\textwidth]{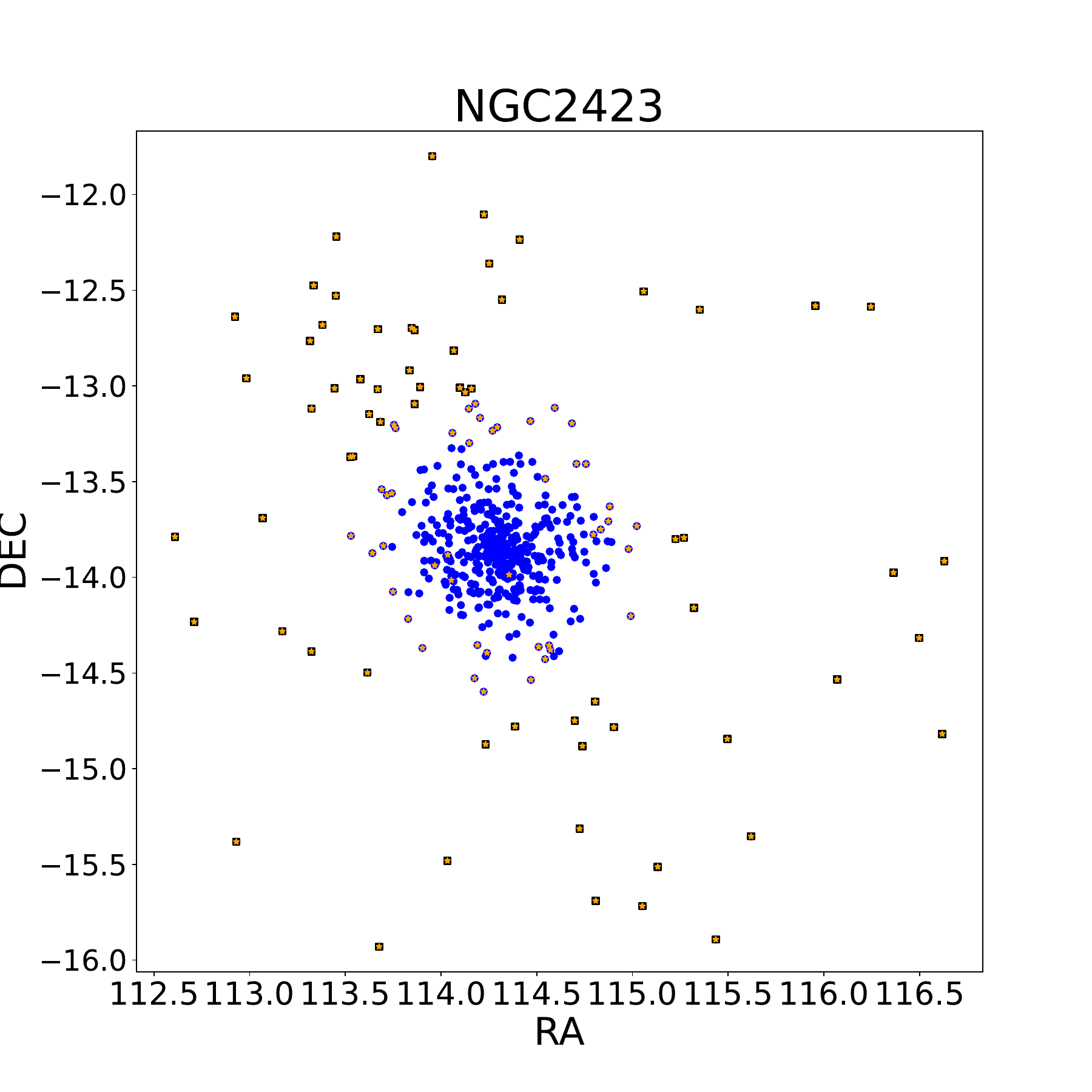}

        \end{subfigure}
        \begin{subfigure}{0.25\textwidth}
                \centering

                \includegraphics[width=\textwidth]{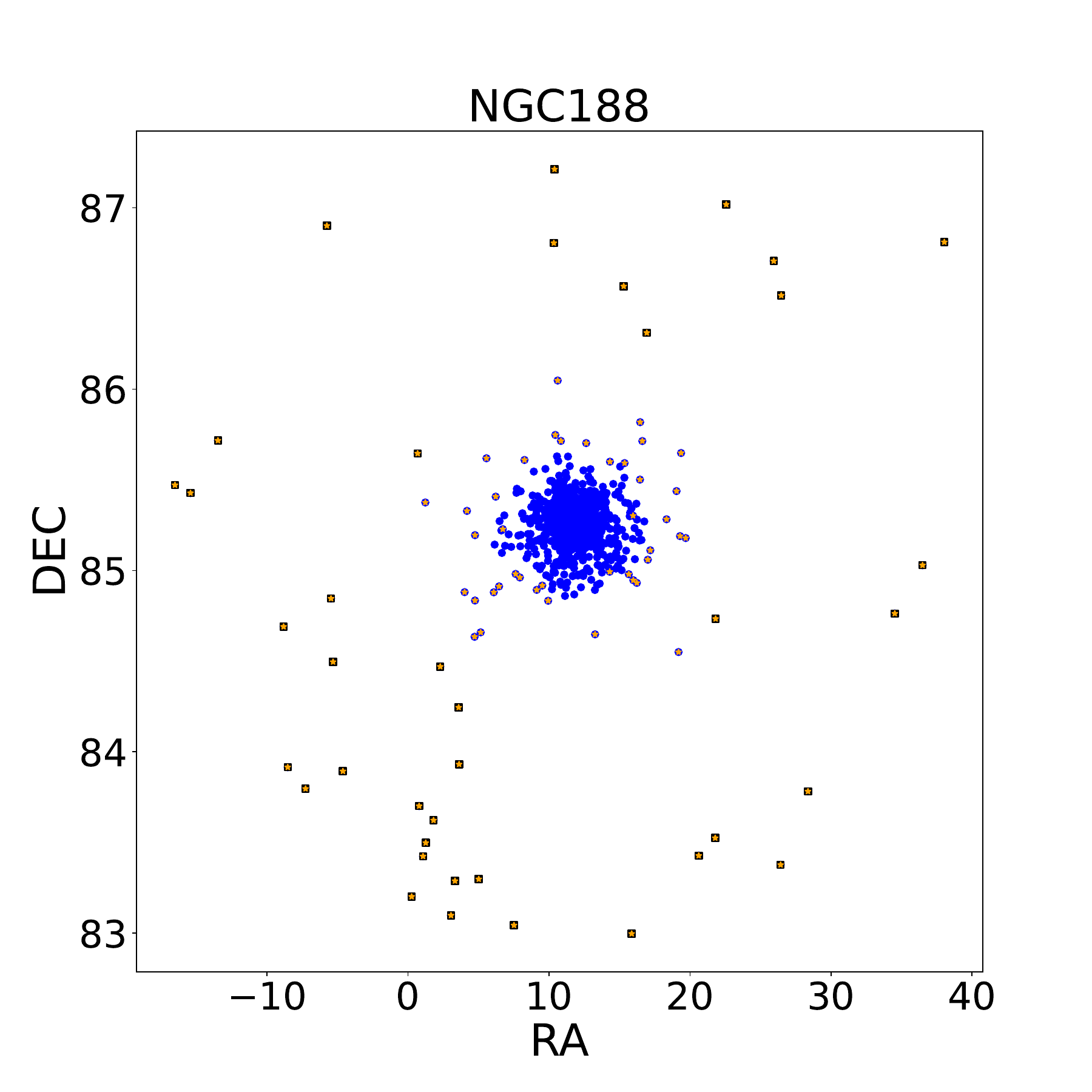}

        \end{subfigure}
        \begin{subfigure}{0.25\textwidth}
        \centering
           \includegraphics[width=\textwidth]{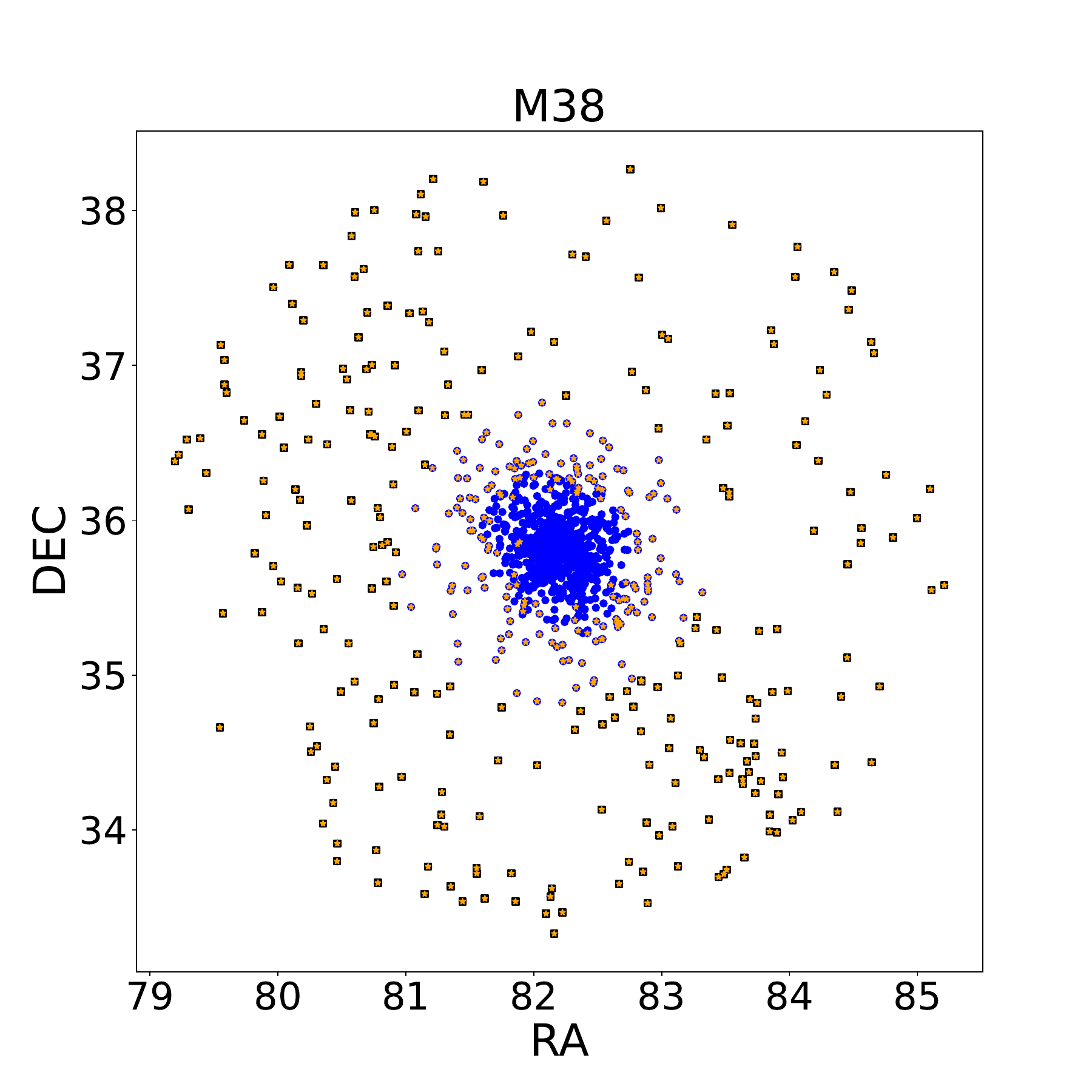}

        \end{subfigure}
        \begin{subfigure}{0.25\textwidth}
        \centering
           \includegraphics[width=\textwidth]{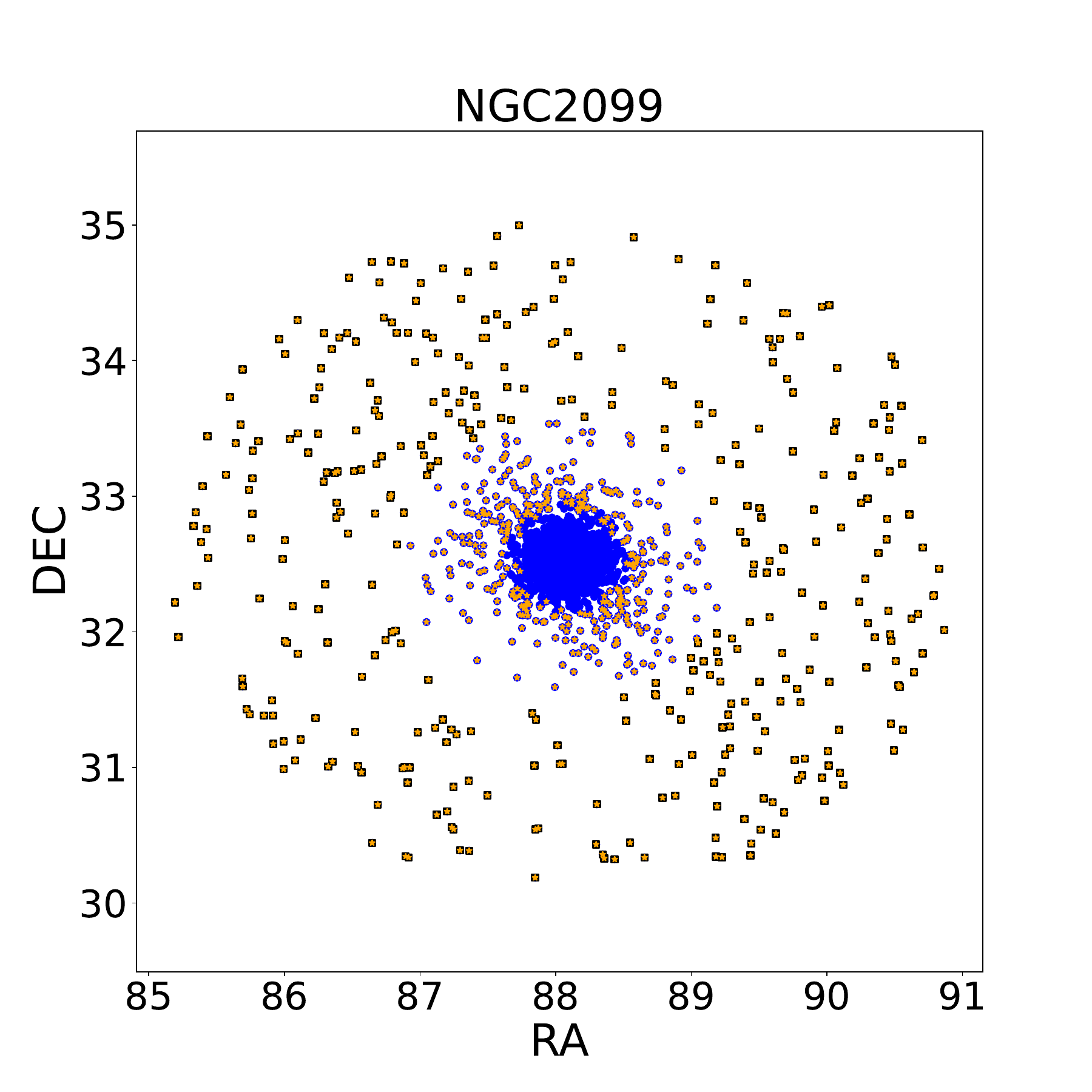}

        \end{subfigure}
        \begin{subfigure}{0.25\textwidth}
        \centering
           \includegraphics[width=\textwidth]{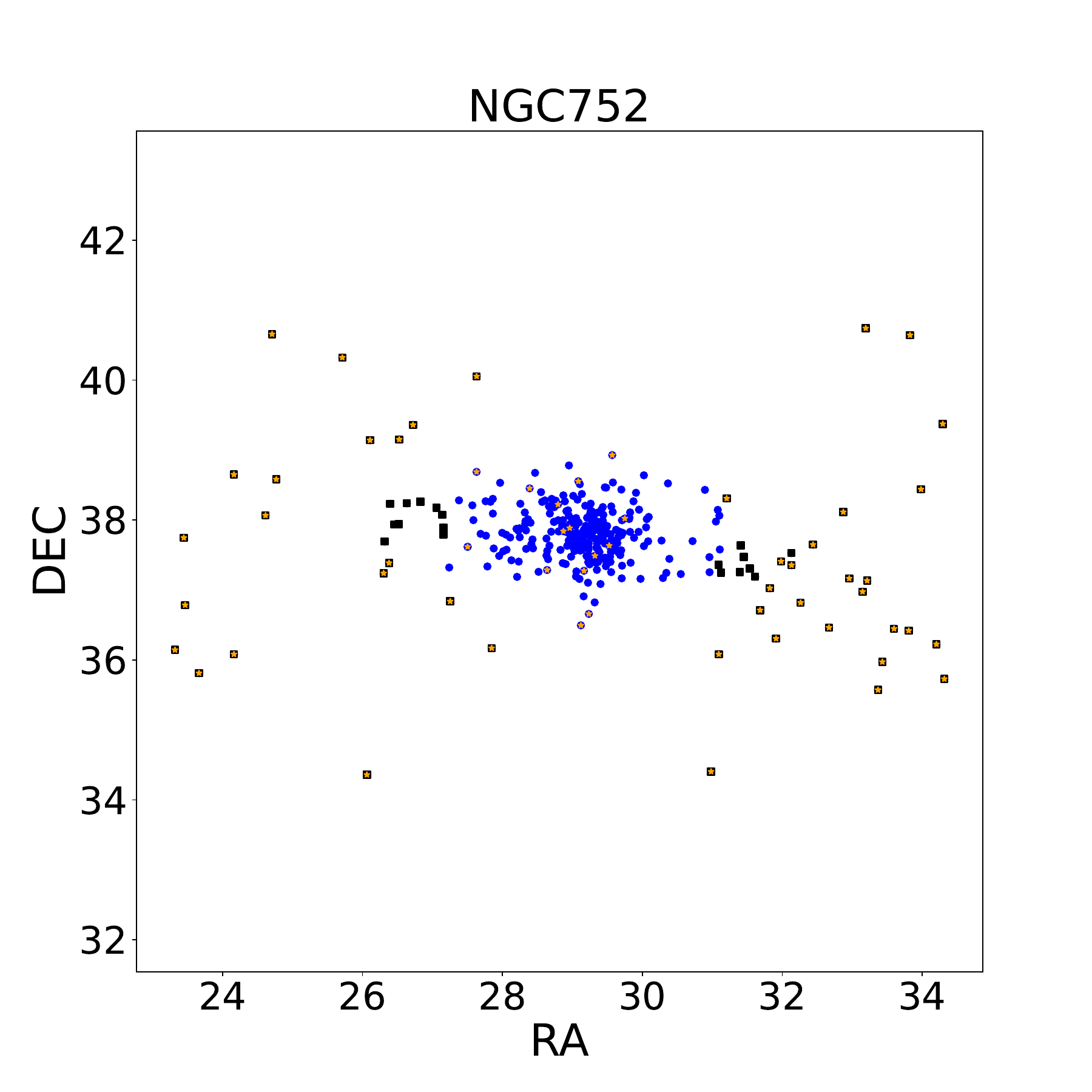}

        \end{subfigure}
        \begin{subfigure}{0.25\textwidth}
        \centering
           \includegraphics[width=\textwidth]{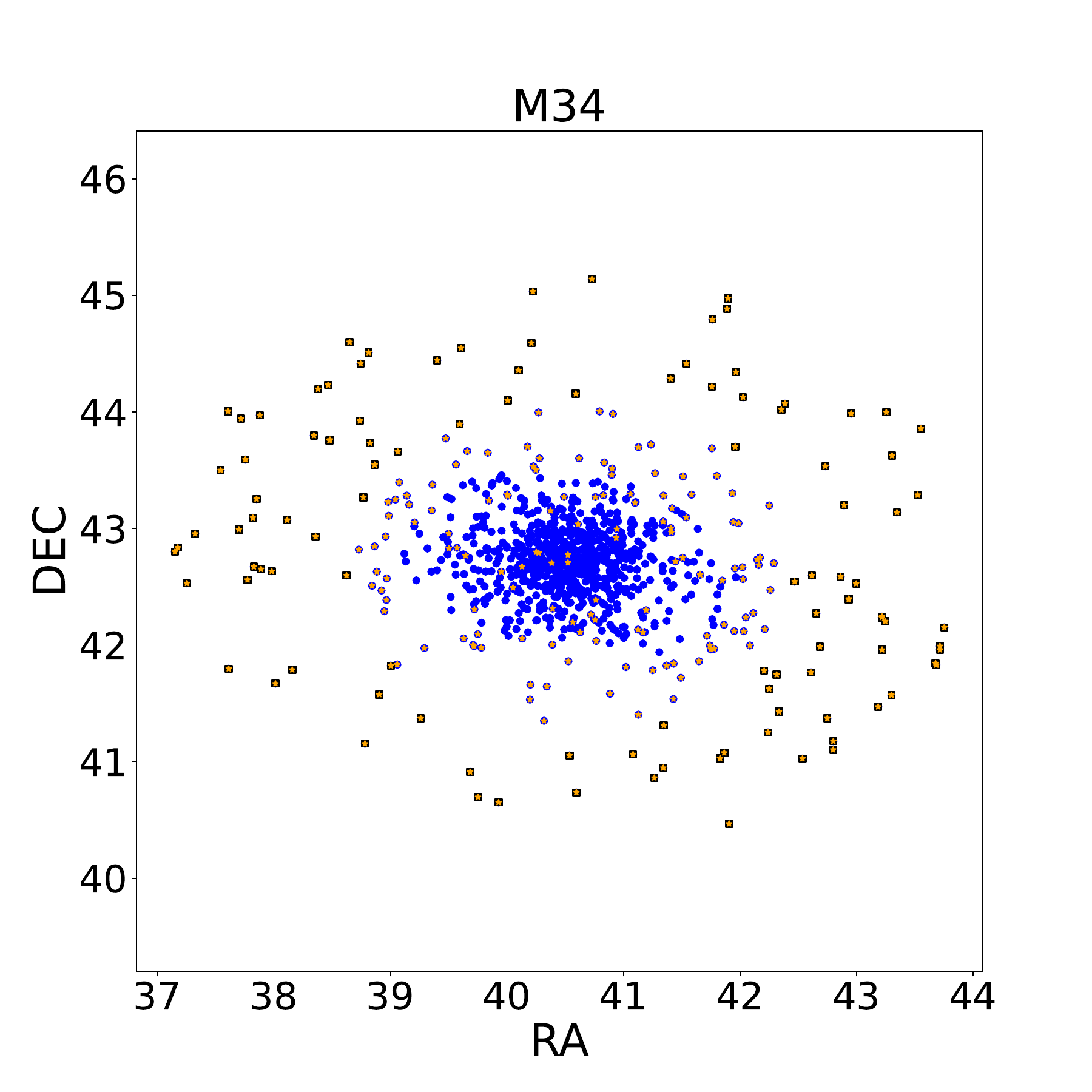}

        \end{subfigure}
        \begin{subfigure}{0.25\textwidth}
        \centering
           \includegraphics[width=\textwidth]{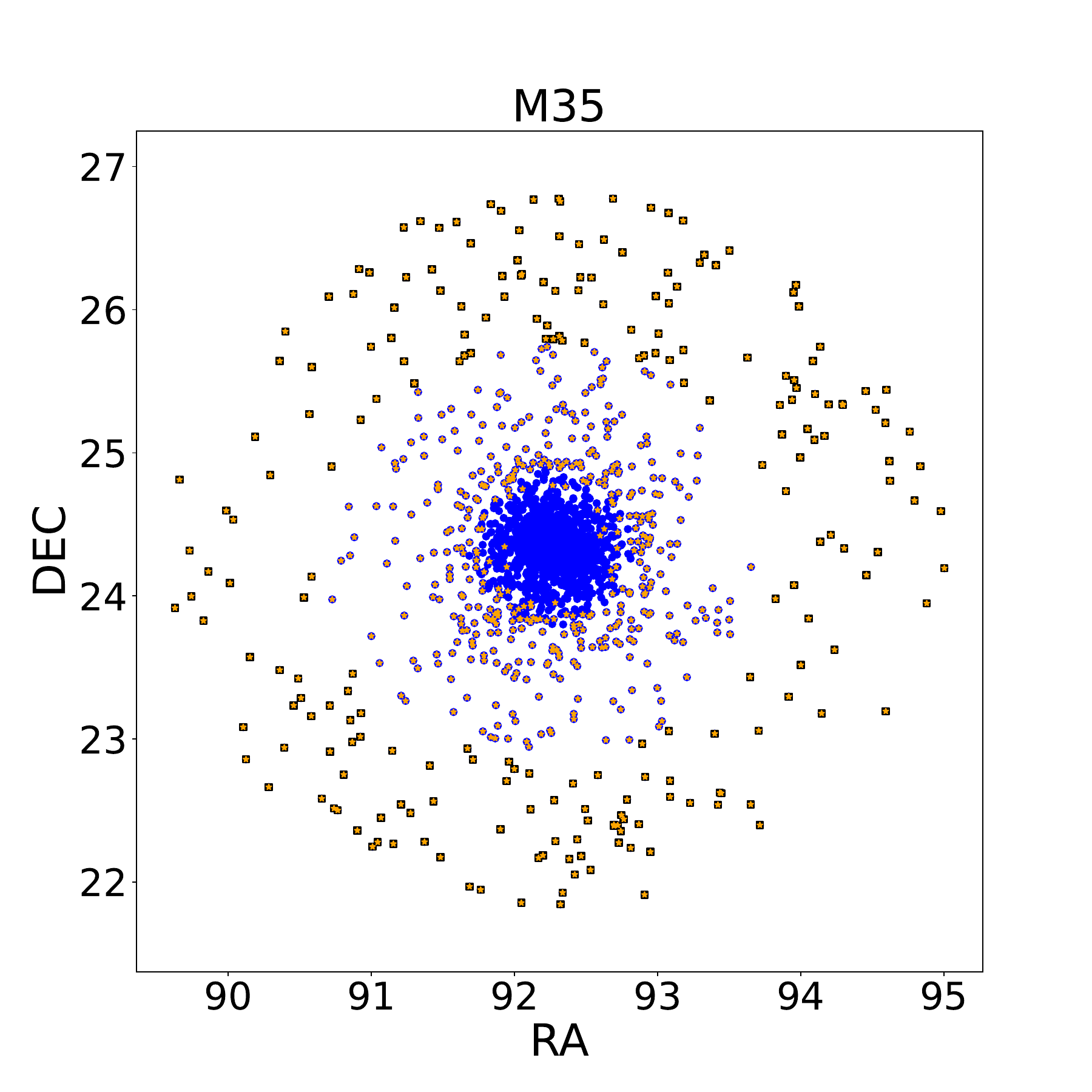}

        \end{subfigure}
        \begin{subfigure}{0.25\textwidth}
        \centering
           \includegraphics[width=\textwidth]{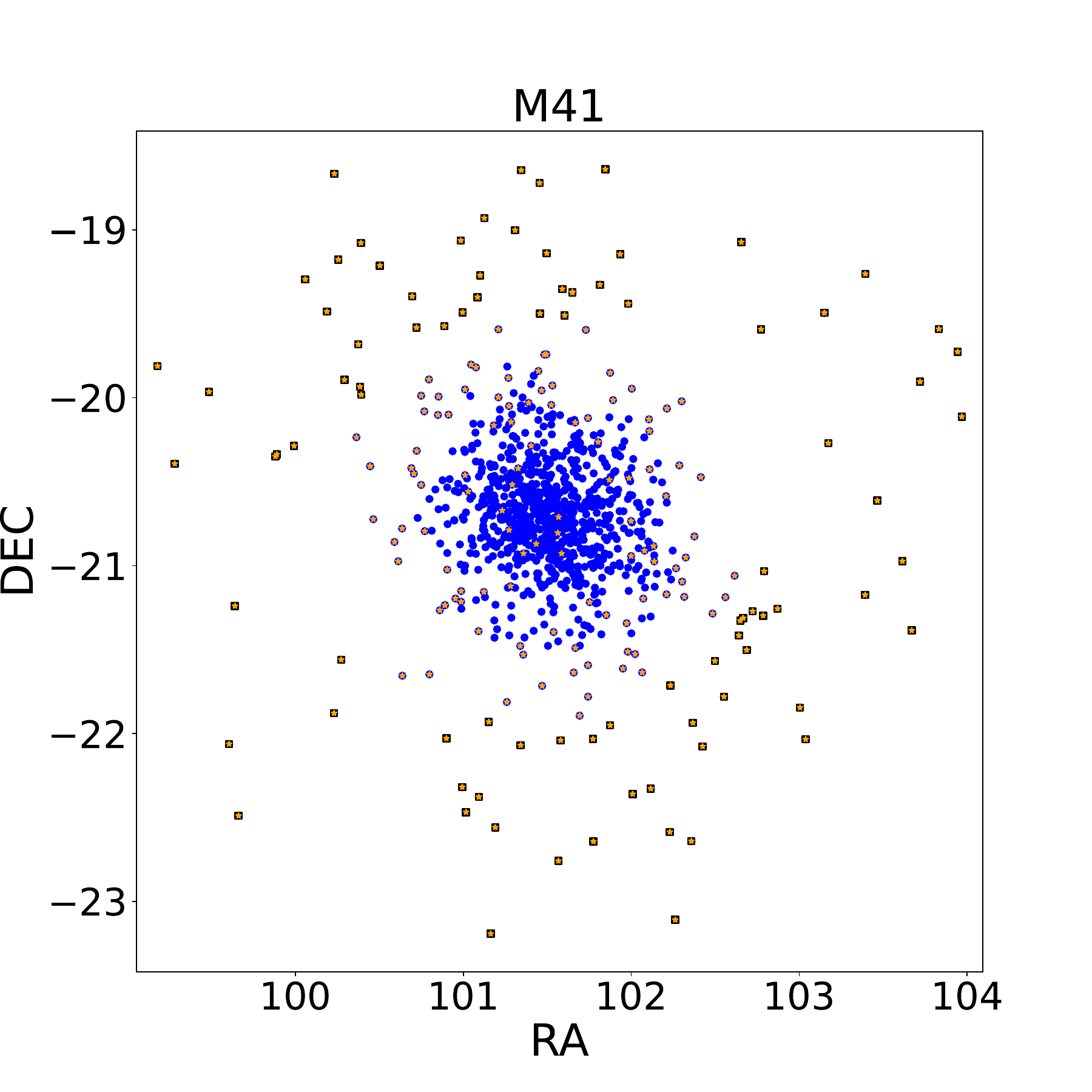}

        \end{subfigure}
        \begin{subfigure}{0.25\textwidth}
        \centering
           \includegraphics[width=\textwidth]{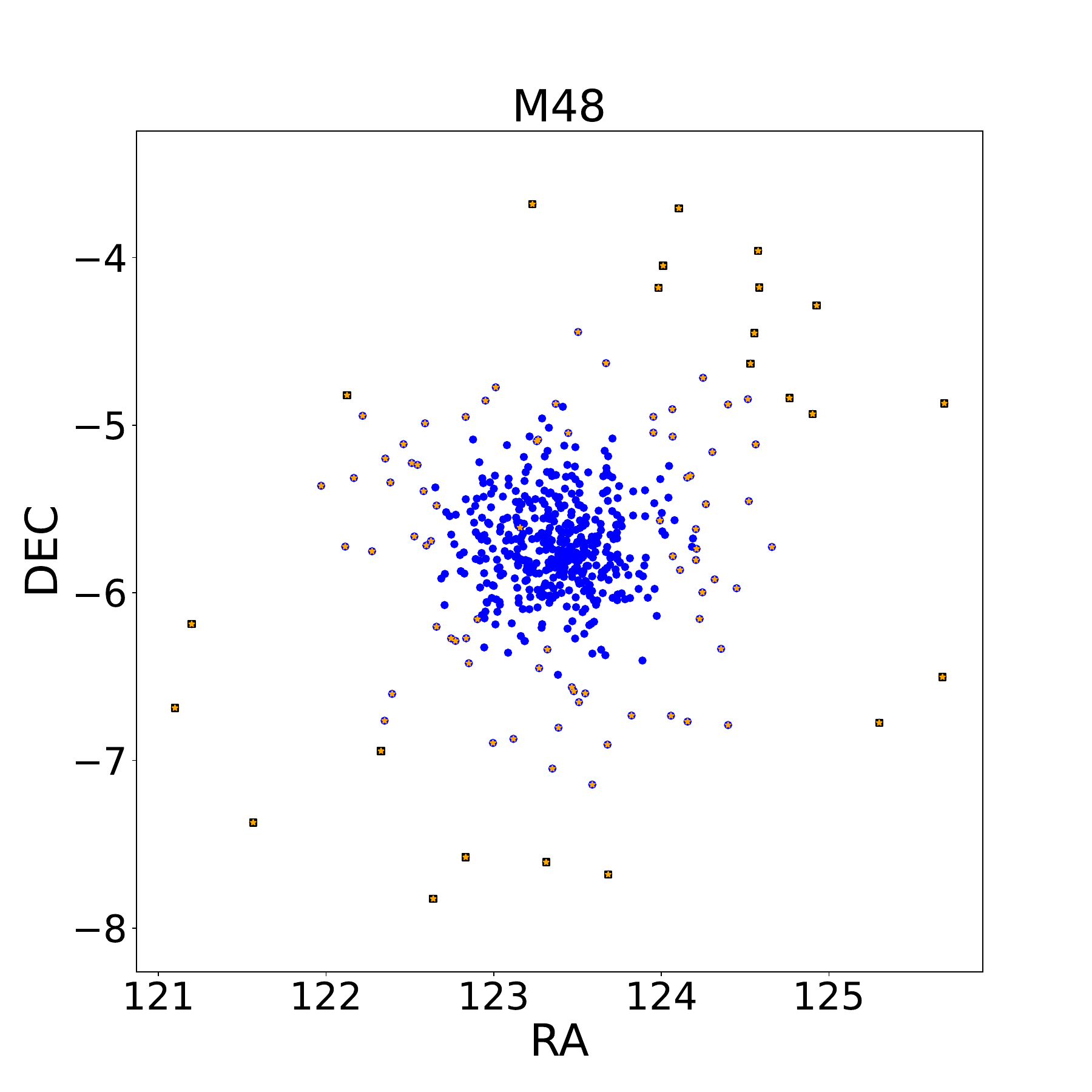}

        \end{subfigure}
        \begin{subfigure}{0.25\textwidth}
        \centering
           \includegraphics[width=\textwidth]{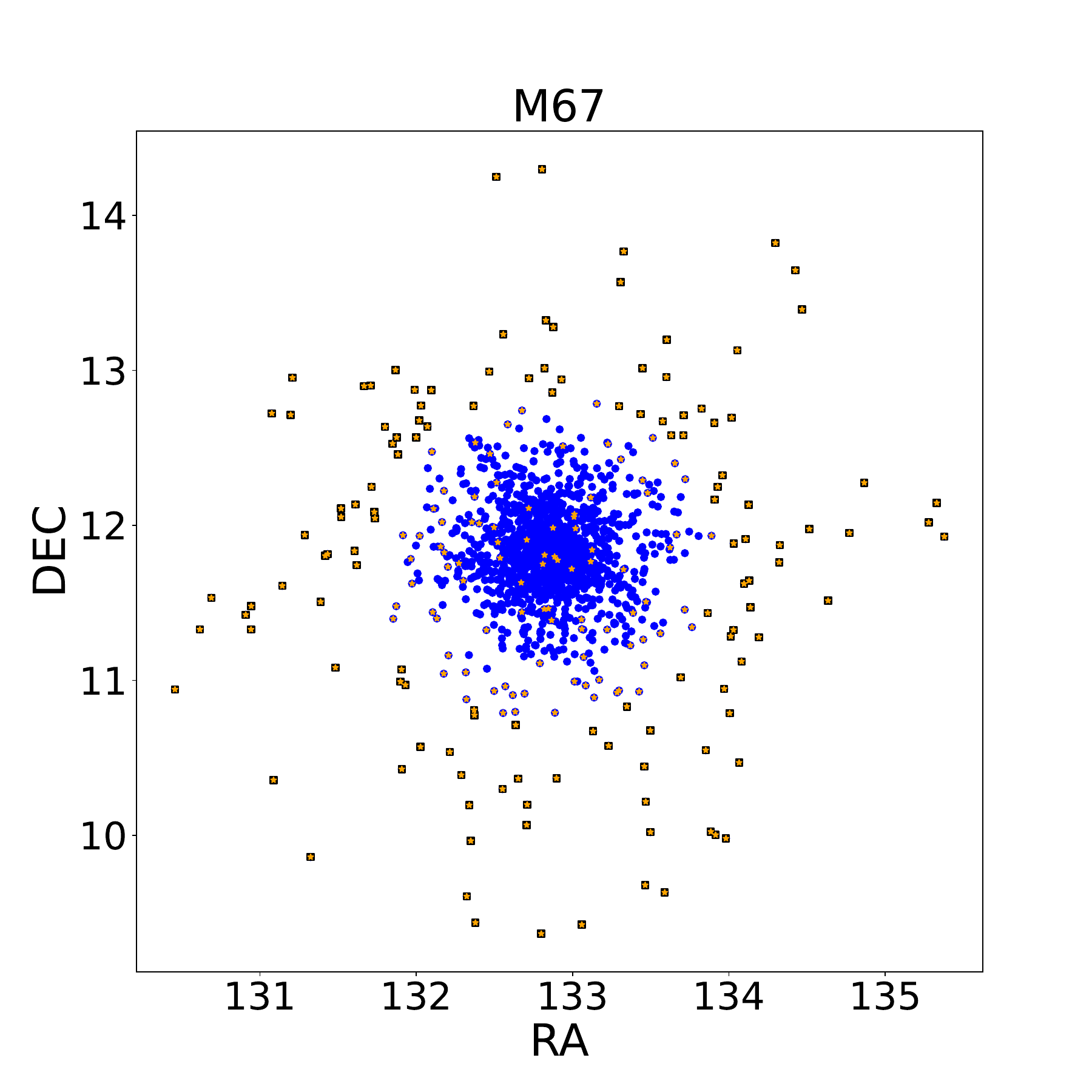}

        \end{subfigure}
        \begin{subfigure}{0.25\textwidth}
        \centering
           \includegraphics[width=\textwidth]{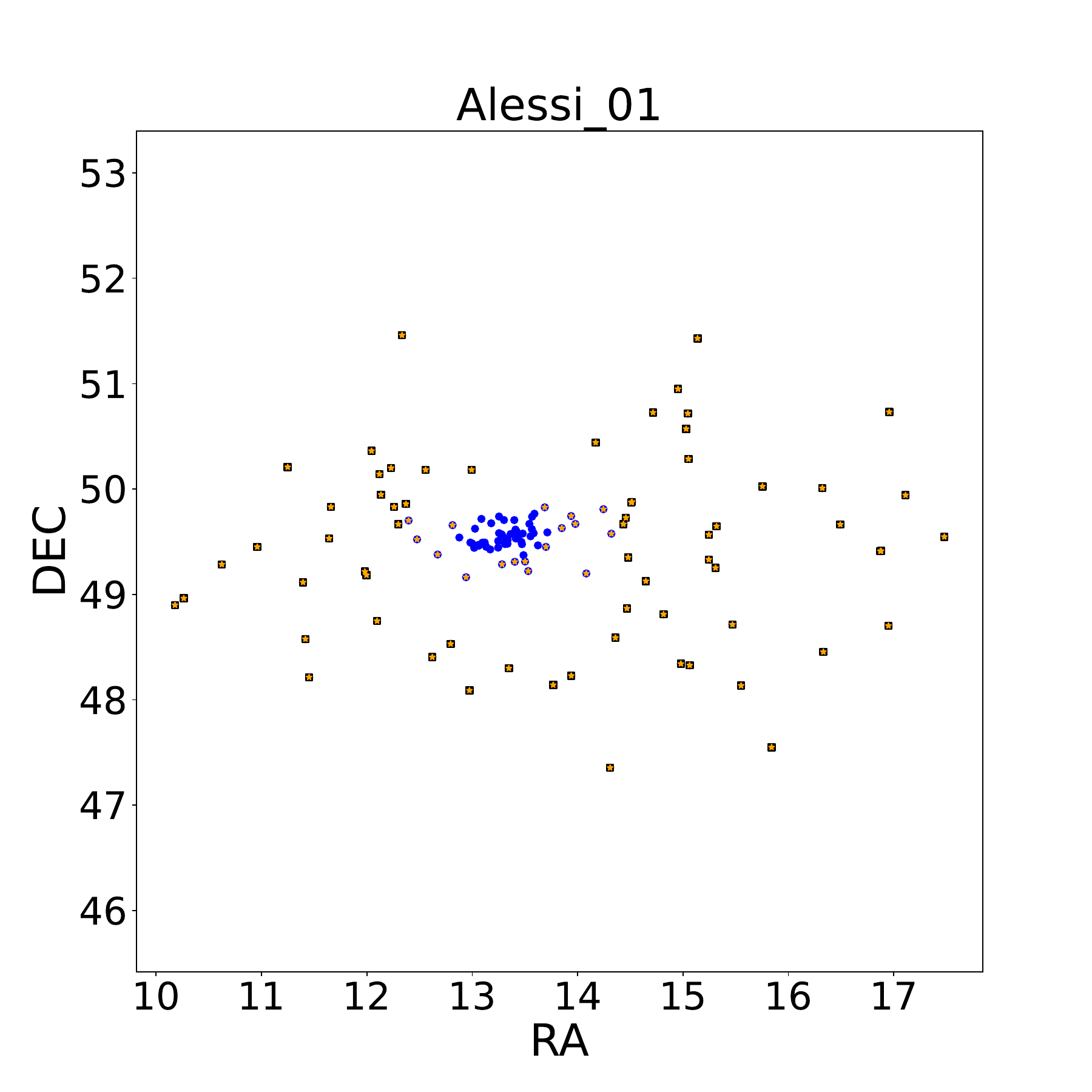}

        \end{subfigure}
  \caption{The position of cluster stars outside and inside the tidal radius. Blue dots represent stars within the tidal radius, while black dots indicate stars distributed outside the tidal radius. Random forest-selected stars are shown as orange stars.}
  \label{position-tidal.fig}
\end{figure}

To determine the distribution of cluster members, we first found the tidal radius by fitting the King profile(\cite{king}). For this, we divided the cluster regions into several concentric rings. Next, we calculated the number density of stars in each ring using Equation~\ref{dens.eq}, where $N_i$ is number of stars in each ring and $r_i$ is the distance from the center of the cluster for each ring.
After that, the King profile was fitted, using Equation~\ref{king.eq} where $f_b$ is surface density background, $f_0$ is peak of density, and $R_C$ is cluster core region. Finally, the tidal radius was calculated by Equation~\ref{tidal.eq} (\cite{tidal_r}), where $\sigma_b$ is surface density background uncertainty.\\
Fig~\ref{kingfit.fig} displays a fitted King profile for the detection members. As seen in Fig~\ref{kingfit.fig}, the number density of stars decreases significantly beyond the tidal radius. The stars within and outside the tidal radius are shown in Fig~\ref{position-tidal.fig}.\\ 
As seen in Fig~\ref{position-tidal.fig}, stars within the tidal radius show dense regions. However, stars beyond the tidal radius exhibit a scattered distribution. Some members that were detected with Random Forest lie inside the tidal radius. The Random Forest detection method has improved tidal radius calculation. Table~\ref{tab_discussion} shows the tidal radius and members within and outside the tidal radius for each cluster.
\noindent
\onecolumn

\begin{figure}
  \centering
  \captionsetup[subfigure]{labelformat=empty}
        \begin{subfigure}{0.25\textwidth}
        \centering
           \includegraphics[width=\textwidth]{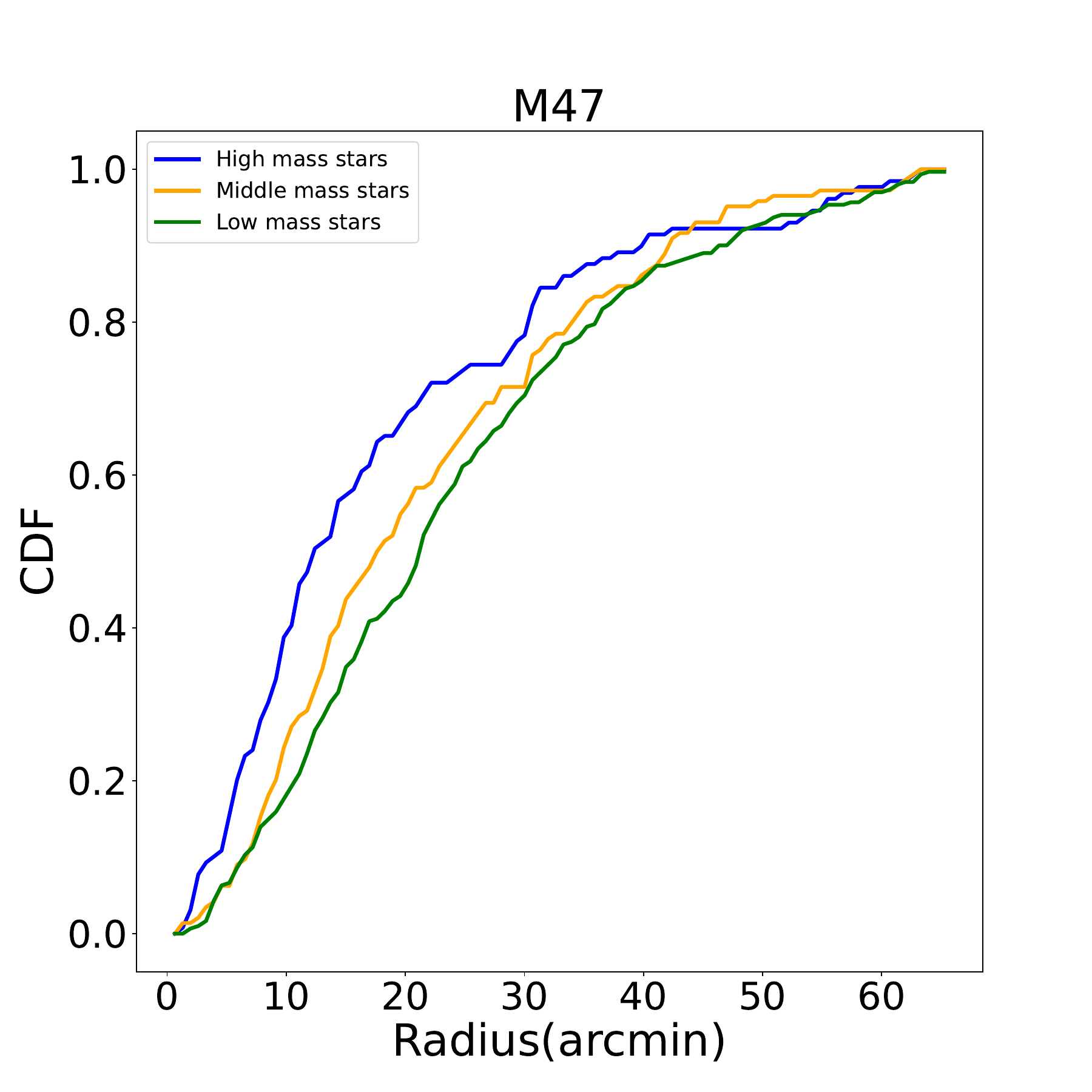}

        \end{subfigure}
        \begin{subfigure}{0.25\textwidth}

                \centering
                \includegraphics[width=\textwidth]{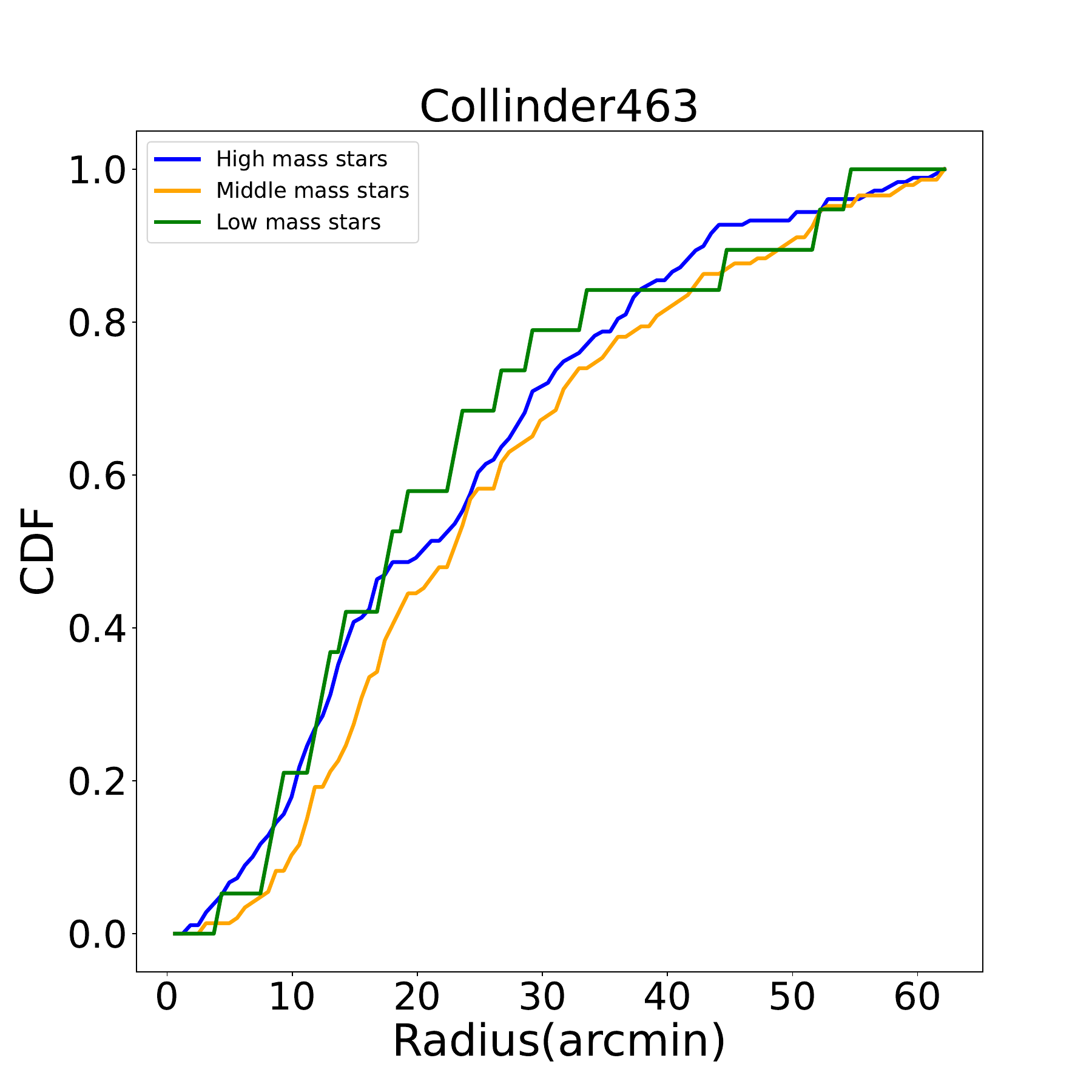}

        \end{subfigure}
        \begin{subfigure}{0.25\textwidth}
                \centering
           \includegraphics[width=\textwidth]{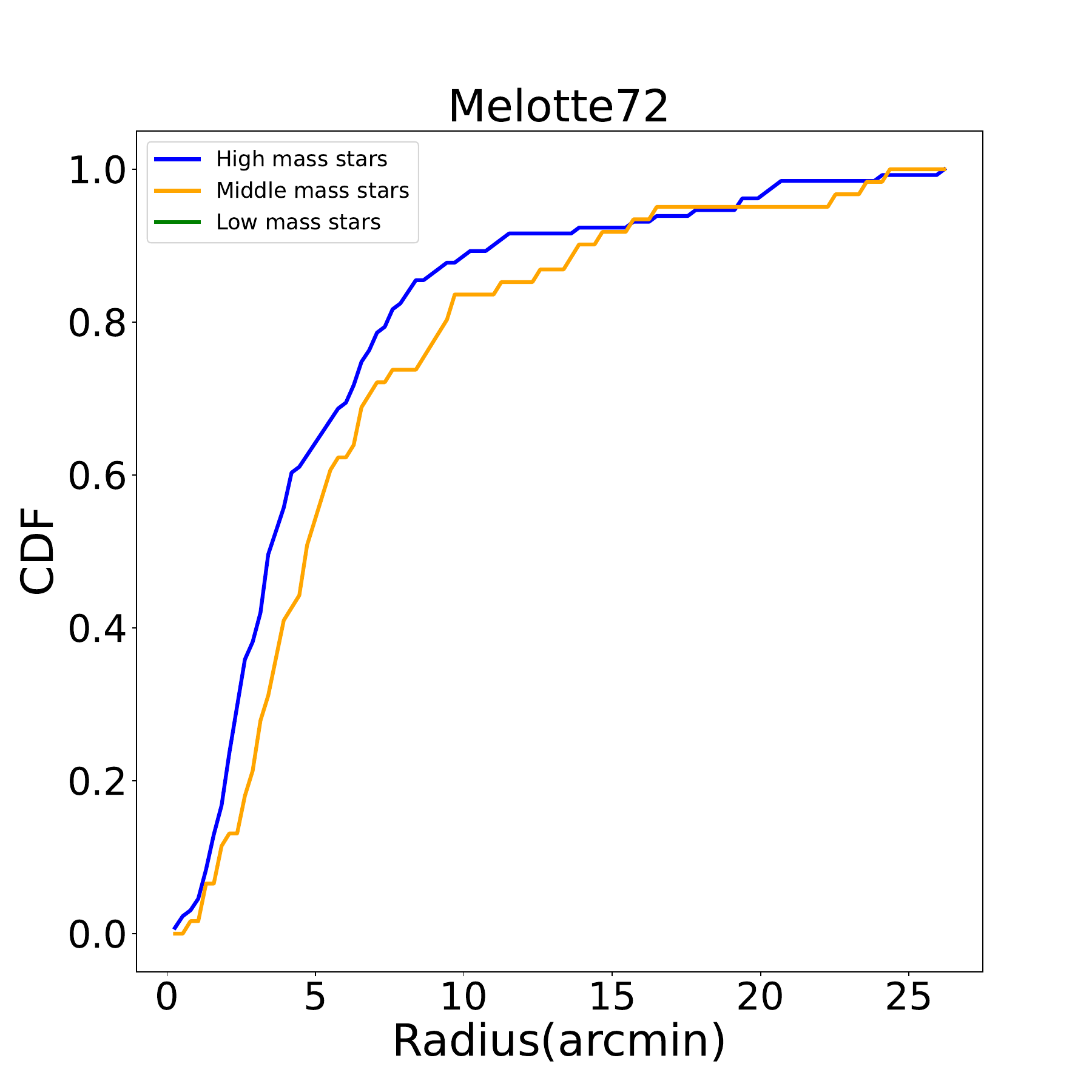}

        \end{subfigure}
        \begin{subfigure}{0.25\textwidth}
                \centering

                \includegraphics[width=\textwidth]{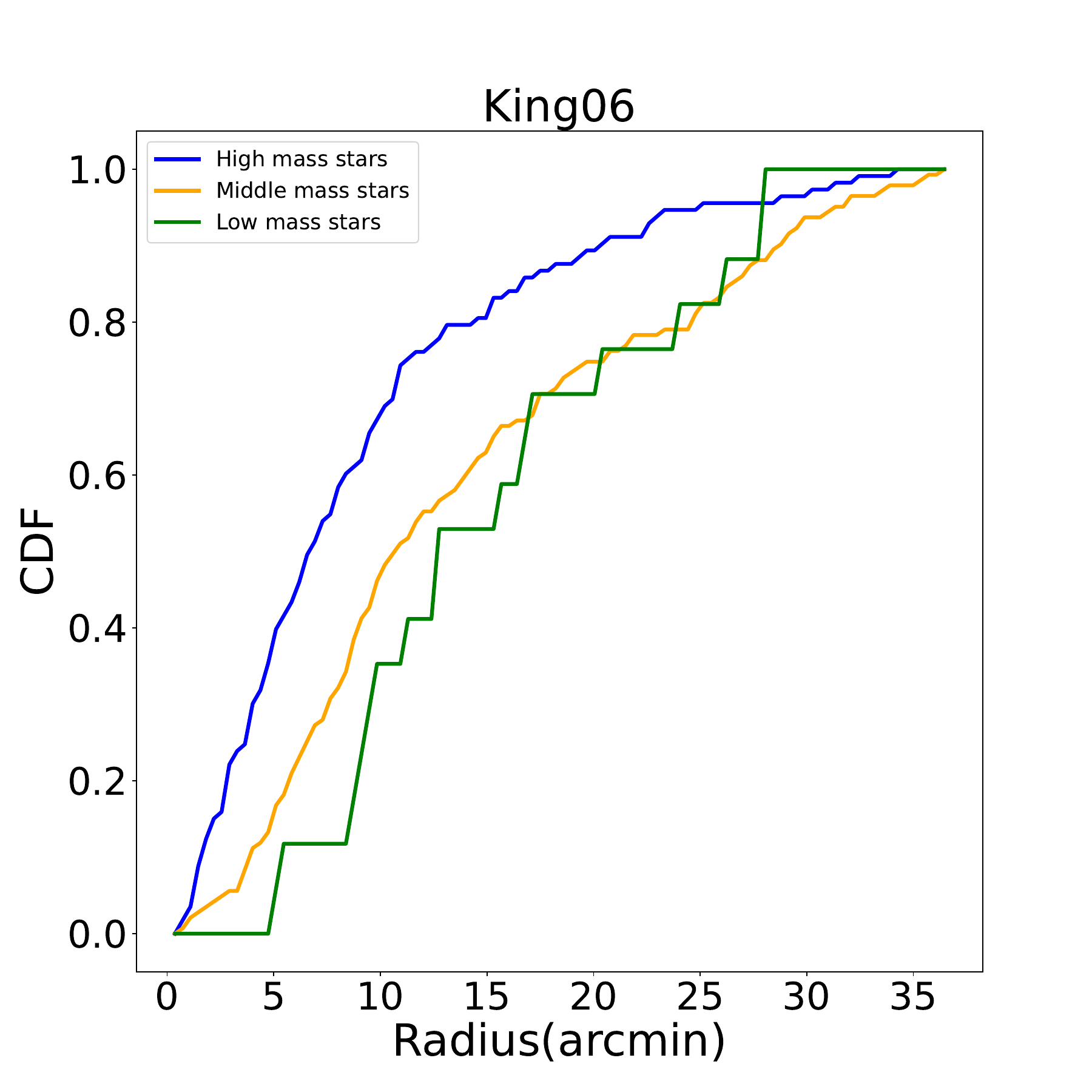}

        \end{subfigure}
        \begin{subfigure}{0.25\textwidth}
                \centering

                \includegraphics[width=\textwidth]{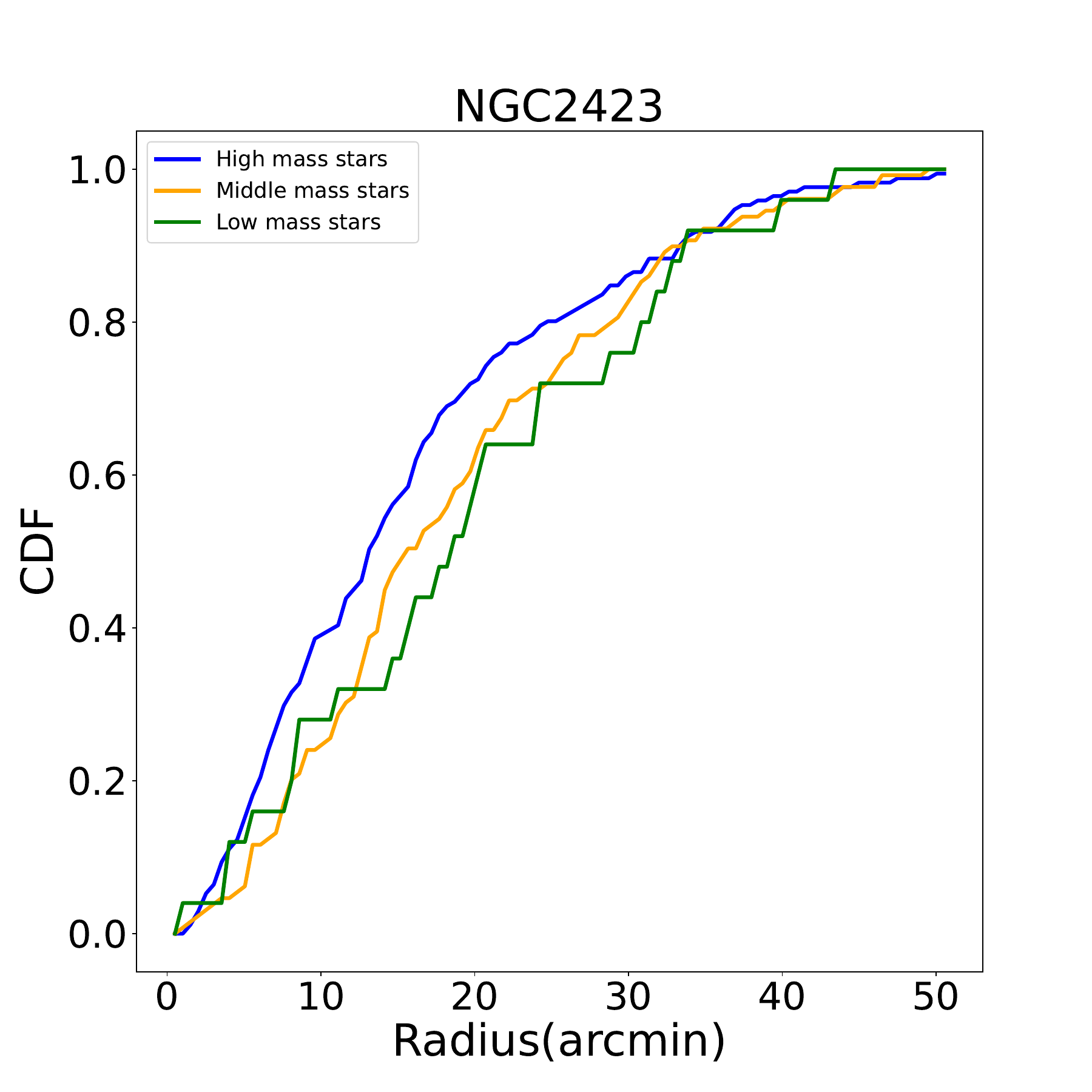}

        \end{subfigure}
        \begin{subfigure}{0.25\textwidth}
                \centering

                \includegraphics[width=\textwidth]{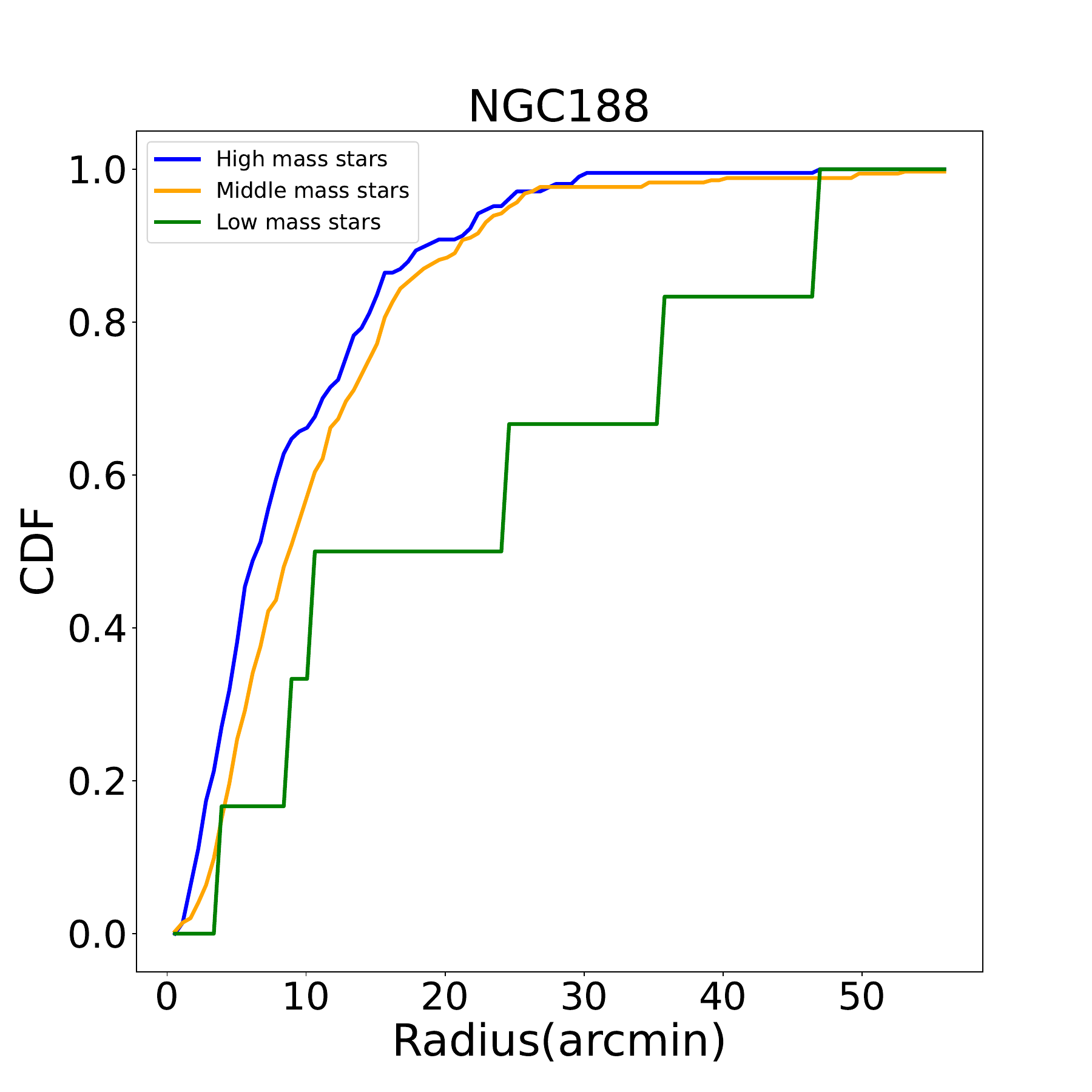}

        \end{subfigure}
        \begin{subfigure}{0.25\textwidth}
        \centering
           \includegraphics[width=\textwidth]{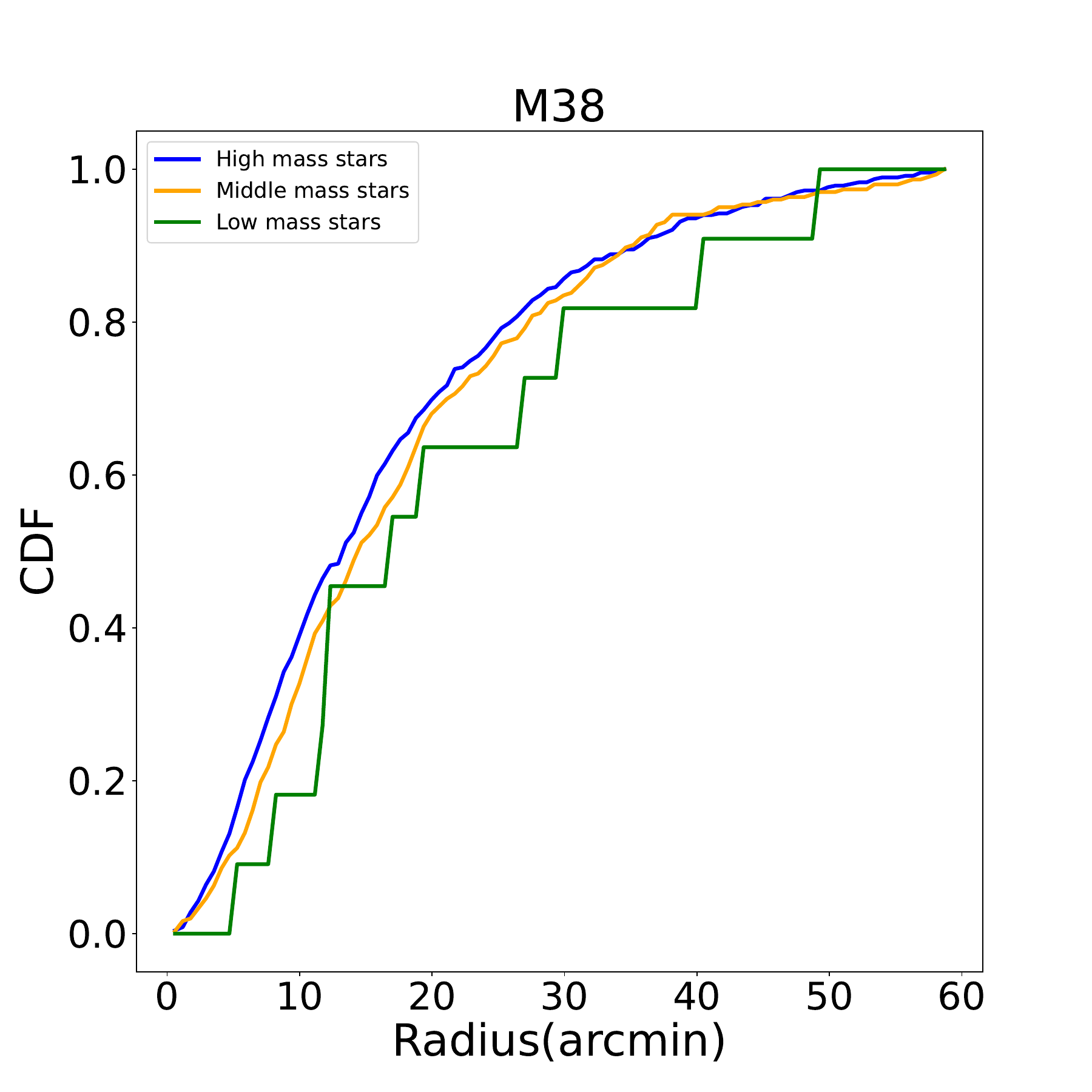}

        \end{subfigure}
        \begin{subfigure}{0.25\textwidth}
        \centering
           \includegraphics[width=\textwidth]{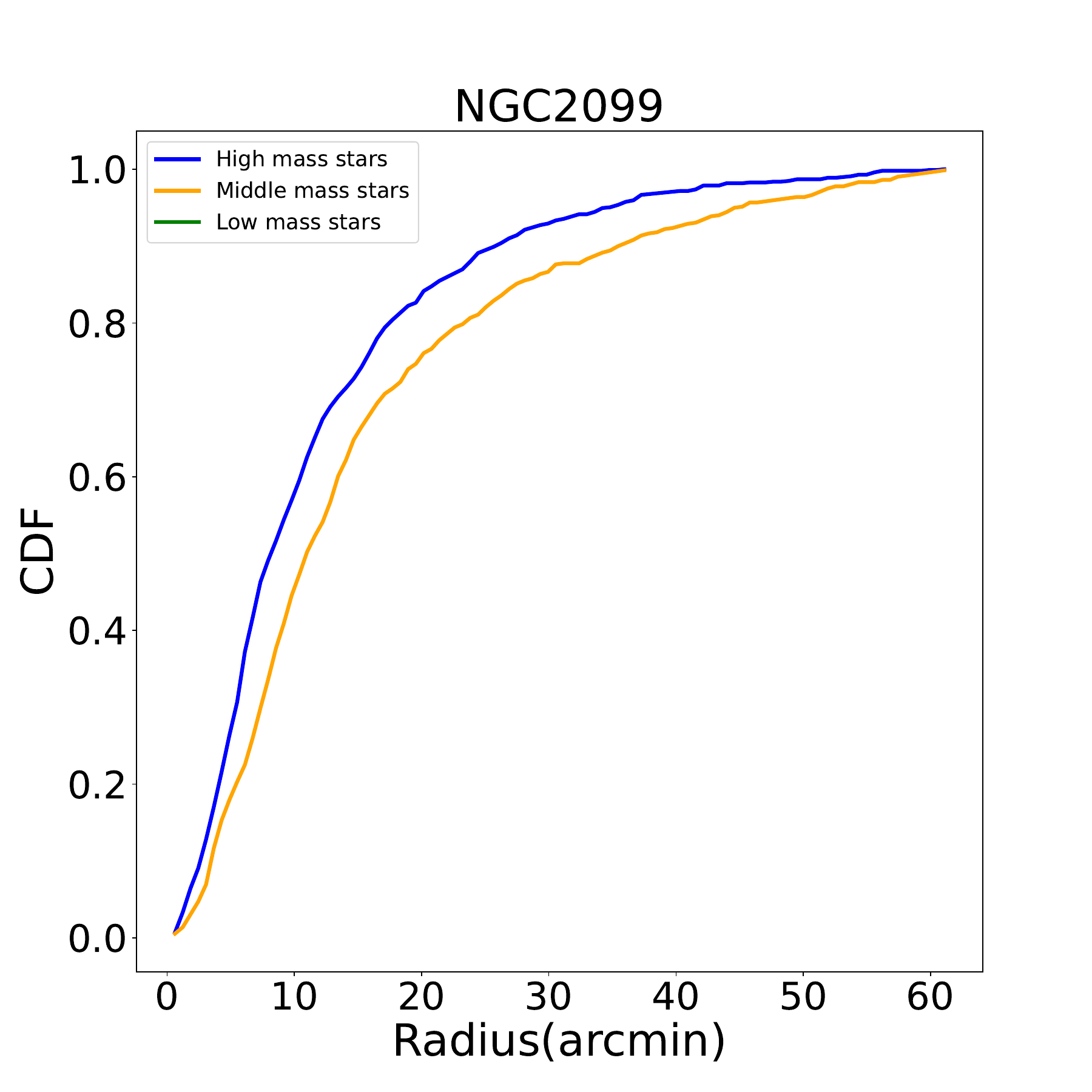}

        \end{subfigure}
        \begin{subfigure}{0.25\textwidth}
        \centering
           \includegraphics[width=\textwidth]{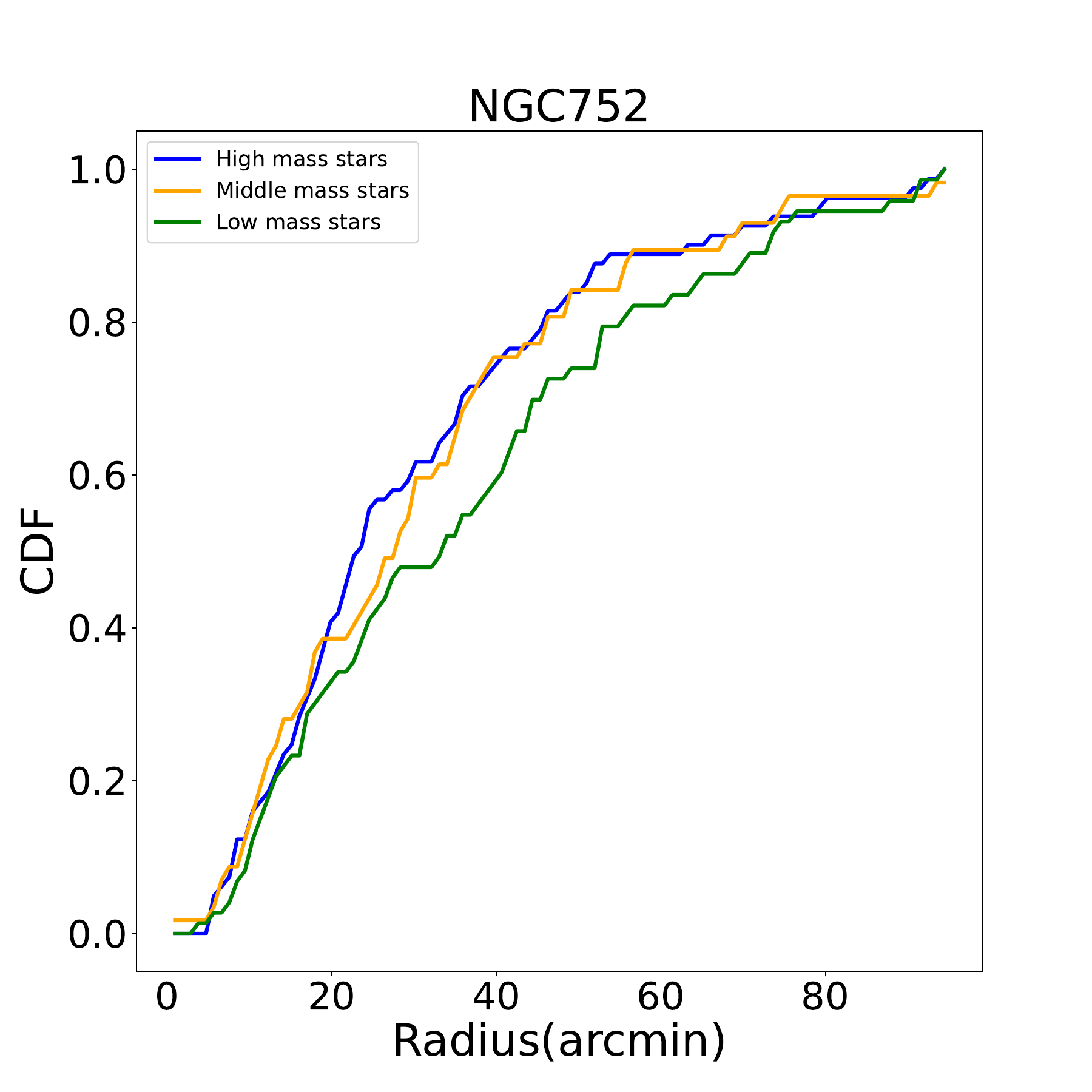}

        \end{subfigure}
        \begin{subfigure}{0.25\textwidth}
        \centering
           \includegraphics[width=\textwidth]{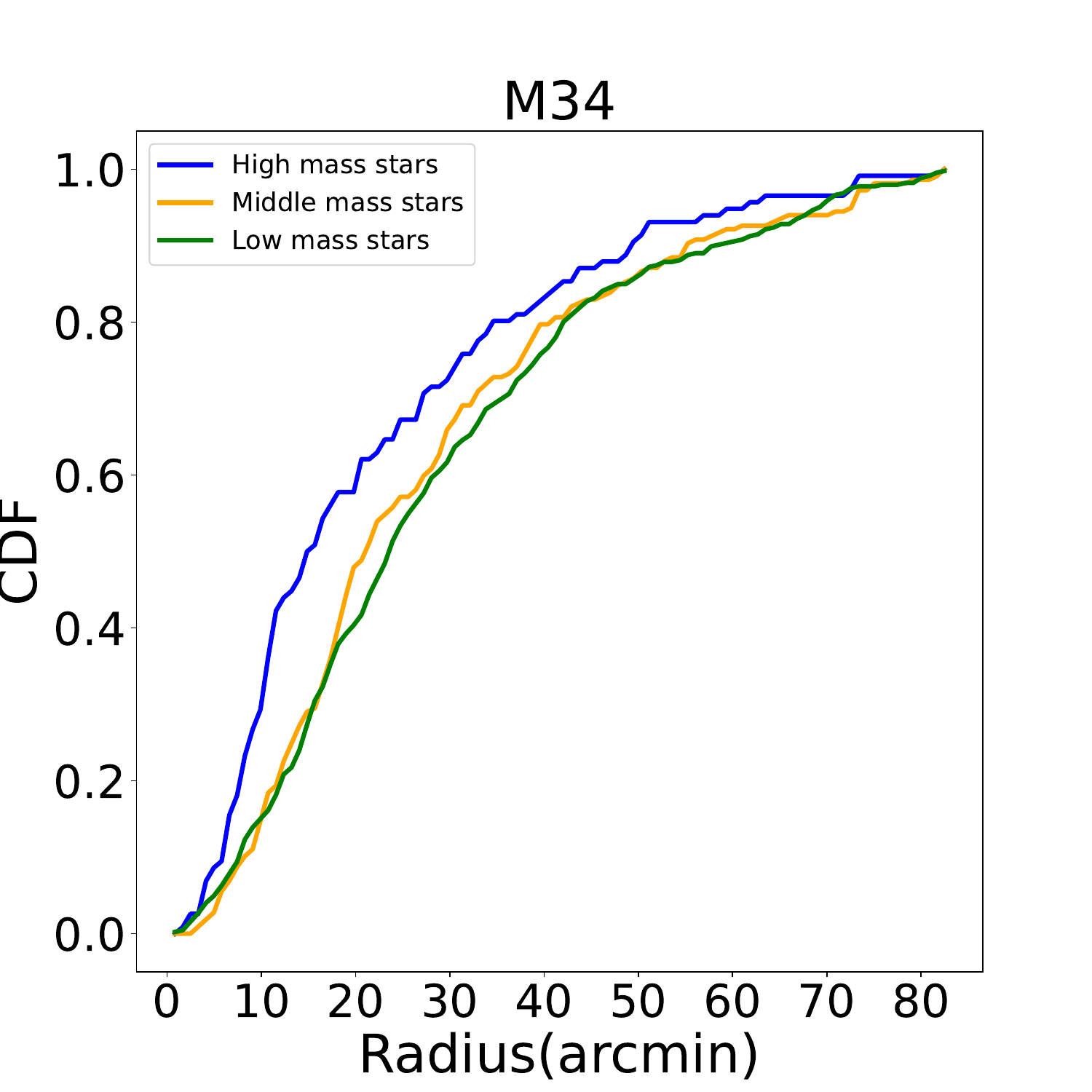}

        \end{subfigure}
        \begin{subfigure}{0.25\textwidth}
        \centering
           \includegraphics[width=\textwidth]{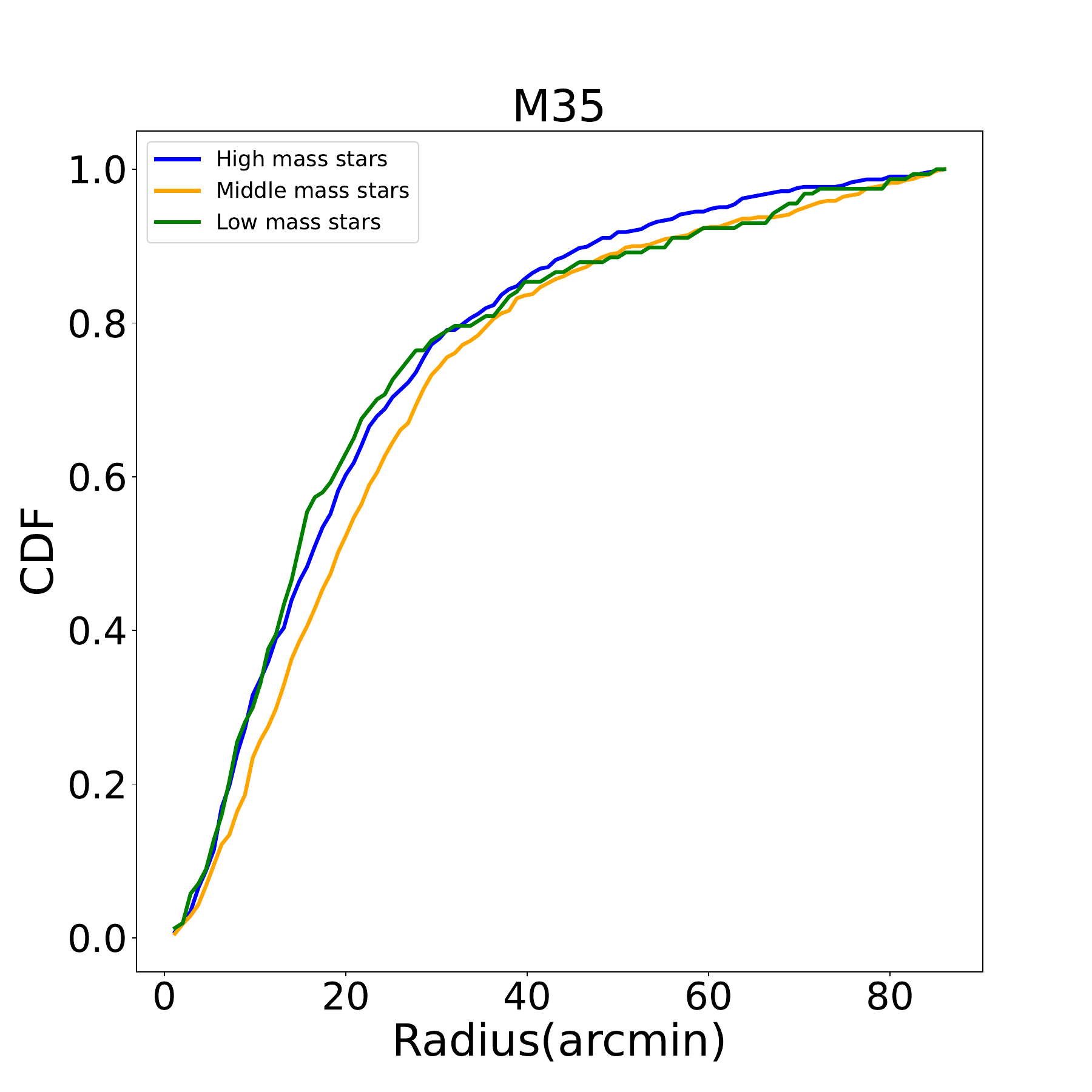}

        \end{subfigure}
        \begin{subfigure}{0.25\textwidth}
        \centering
           \includegraphics[width=\textwidth]{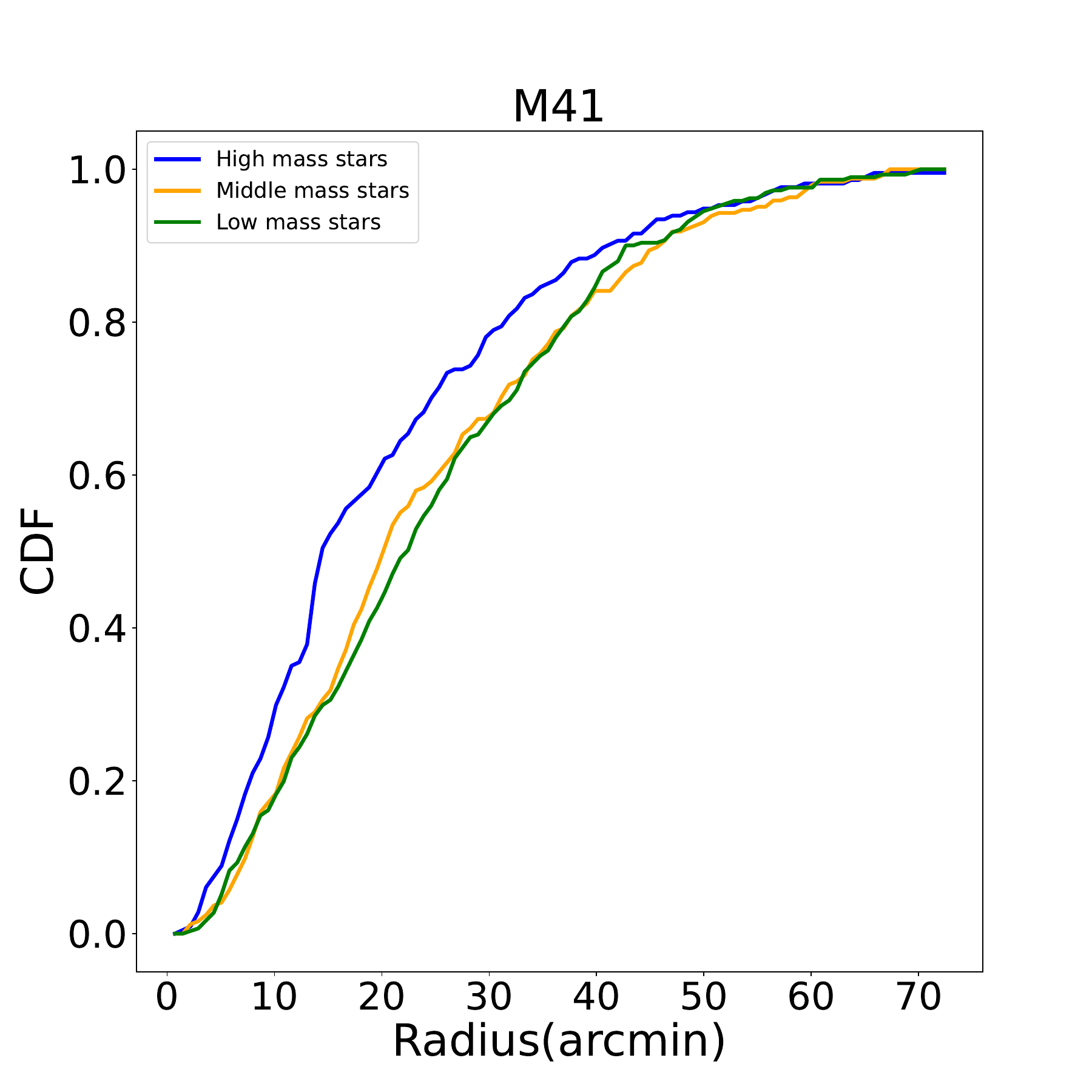}

        \end{subfigure}
        \begin{subfigure}{0.25\textwidth}
        \centering
           \includegraphics[width=\textwidth]{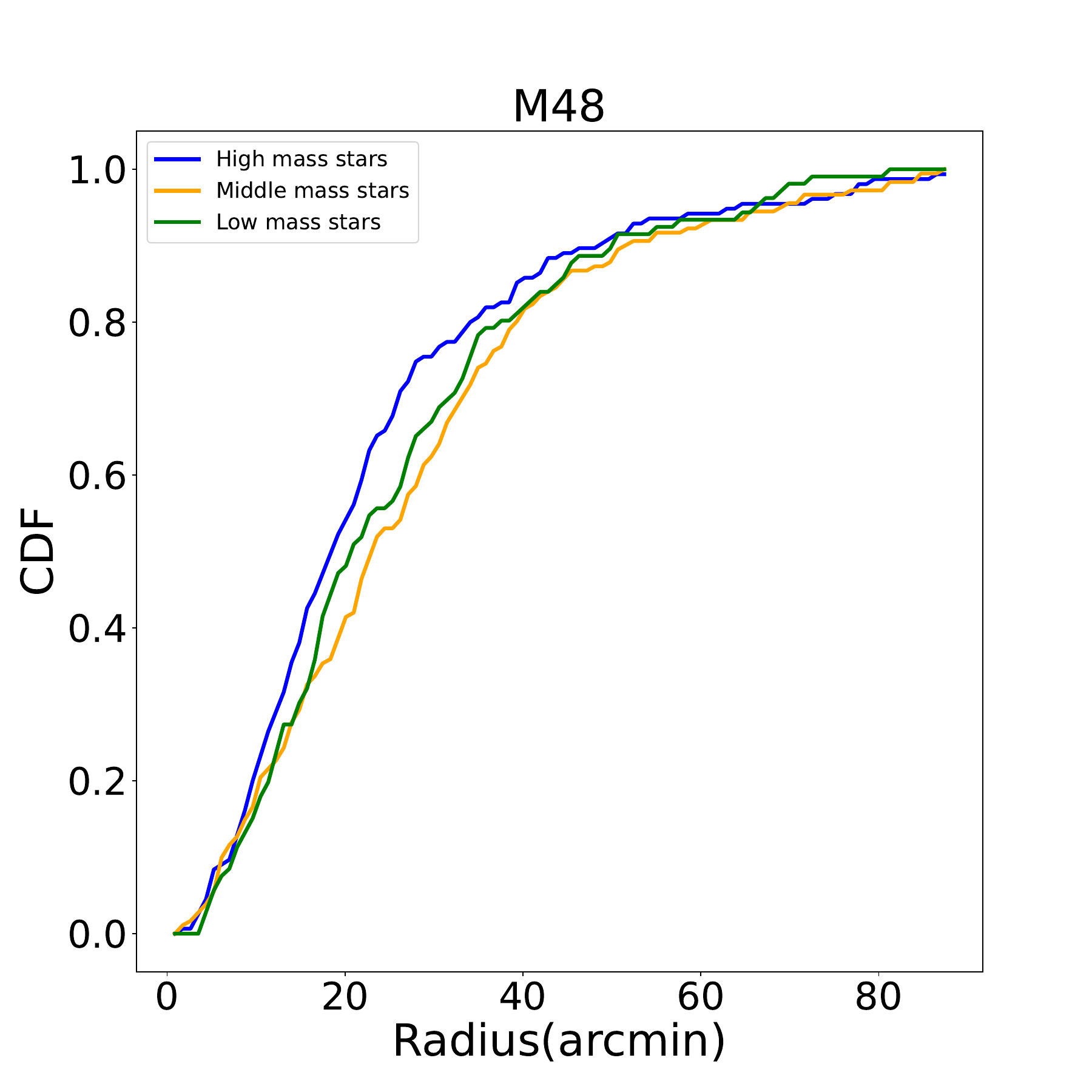}

        \end{subfigure}
        \begin{subfigure}{0.25\textwidth}
        \centering
           \includegraphics[width=\textwidth]{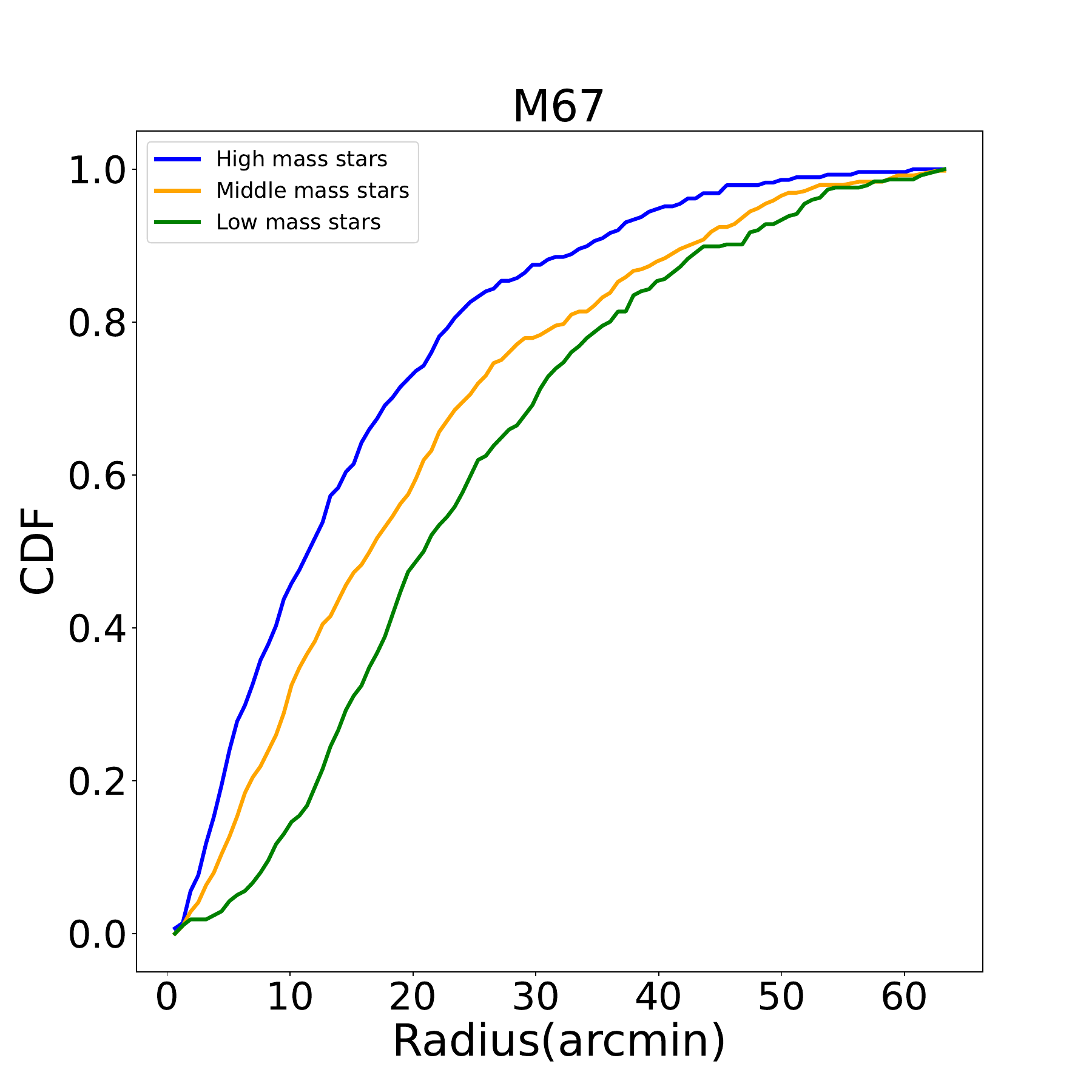}

        \end{subfigure}
  \caption{Cumulative Distribution of Main Sequence Luminosity within the Tidal Radius.}
  \label{mass-tidal.fig}
\end{figure}

\begin{figure}
  \centering
  \captionsetup[subfigure]{labelformat=empty}
        \begin{subfigure}{0.25\textwidth}
        \centering
           \includegraphics[width=\textwidth]{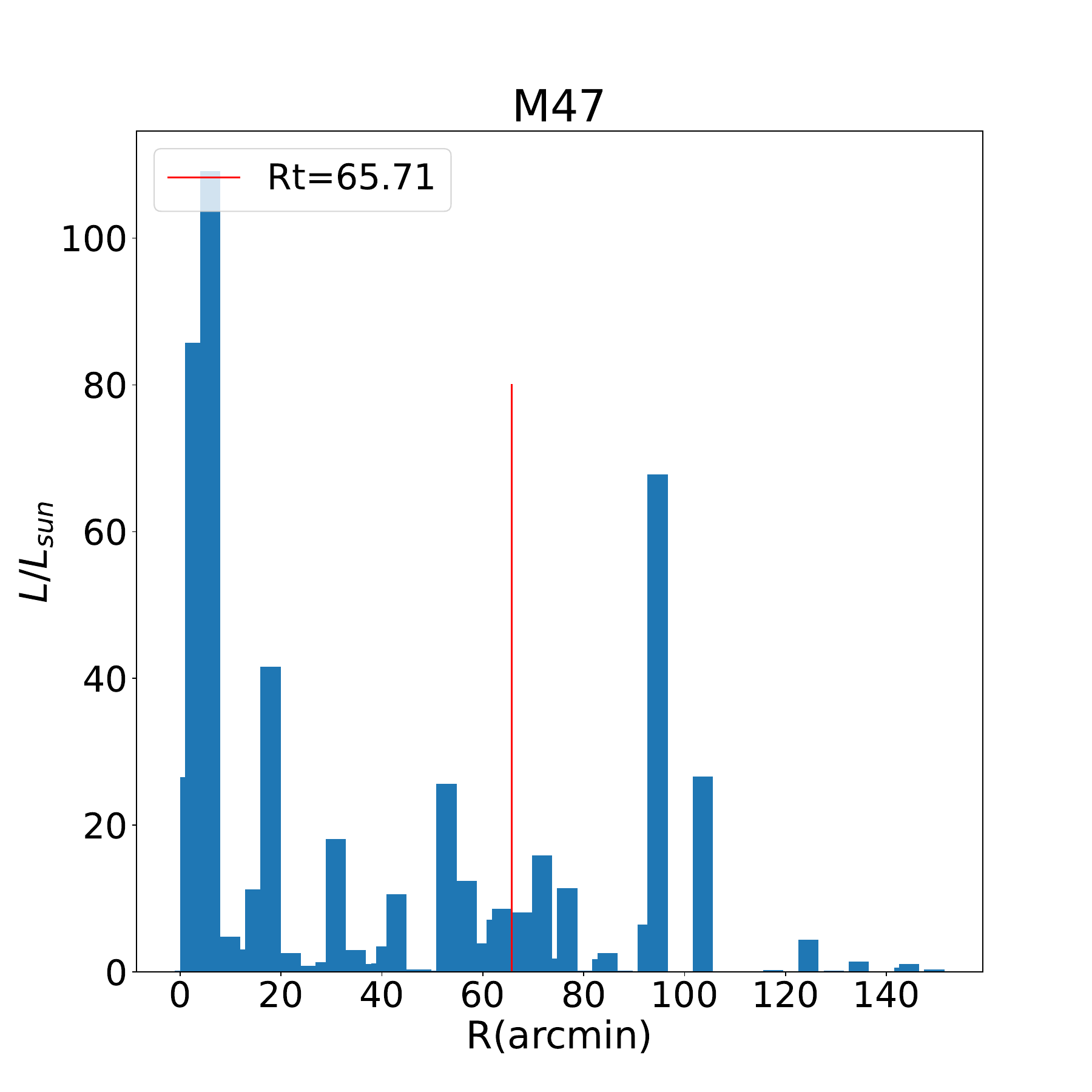}

        \end{subfigure}
        \begin{subfigure}{0.25\textwidth}

                \centering
                \includegraphics[width=\textwidth]{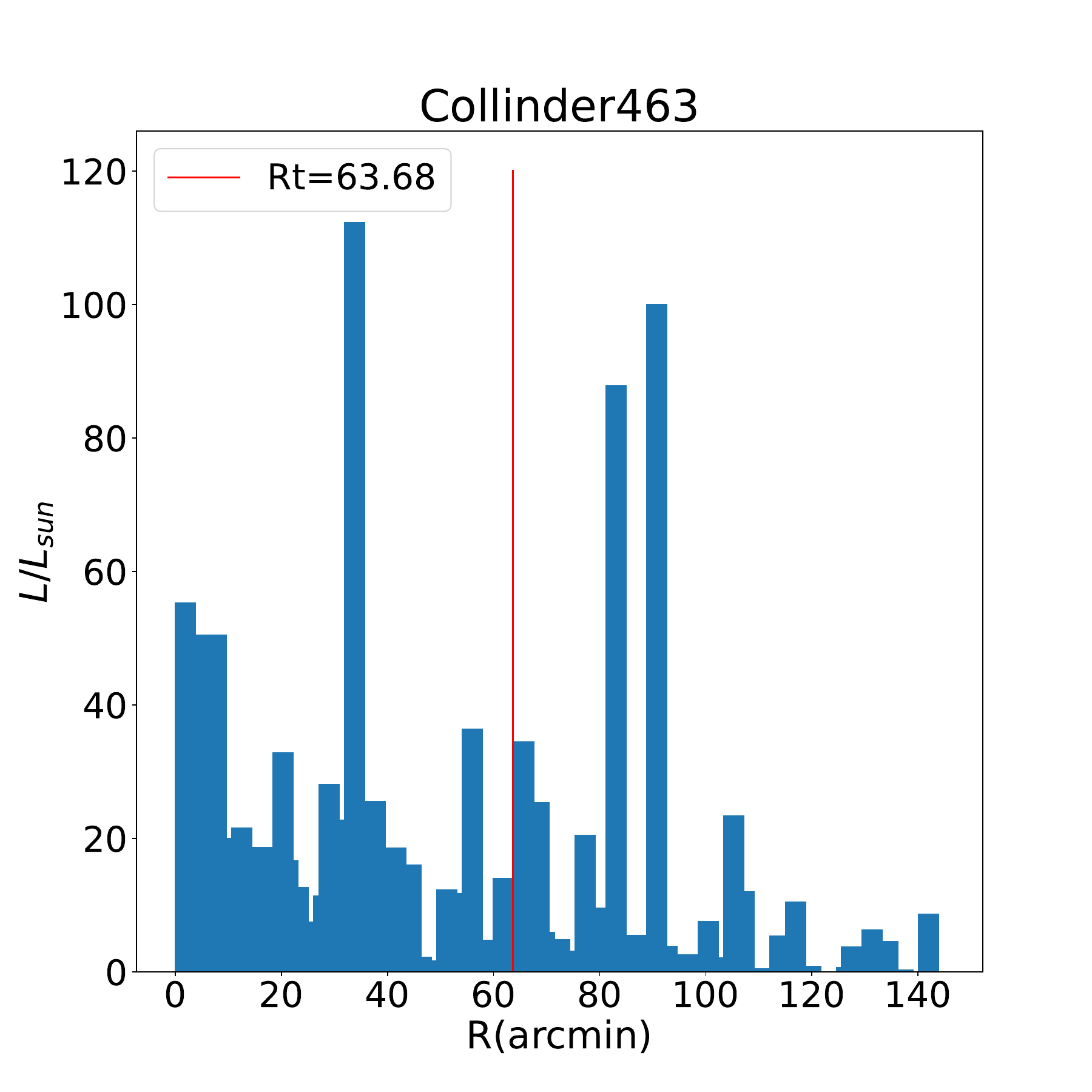}

        \end{subfigure}
        \begin{subfigure}{0.25\textwidth}
                \centering
           \includegraphics[width=\textwidth]{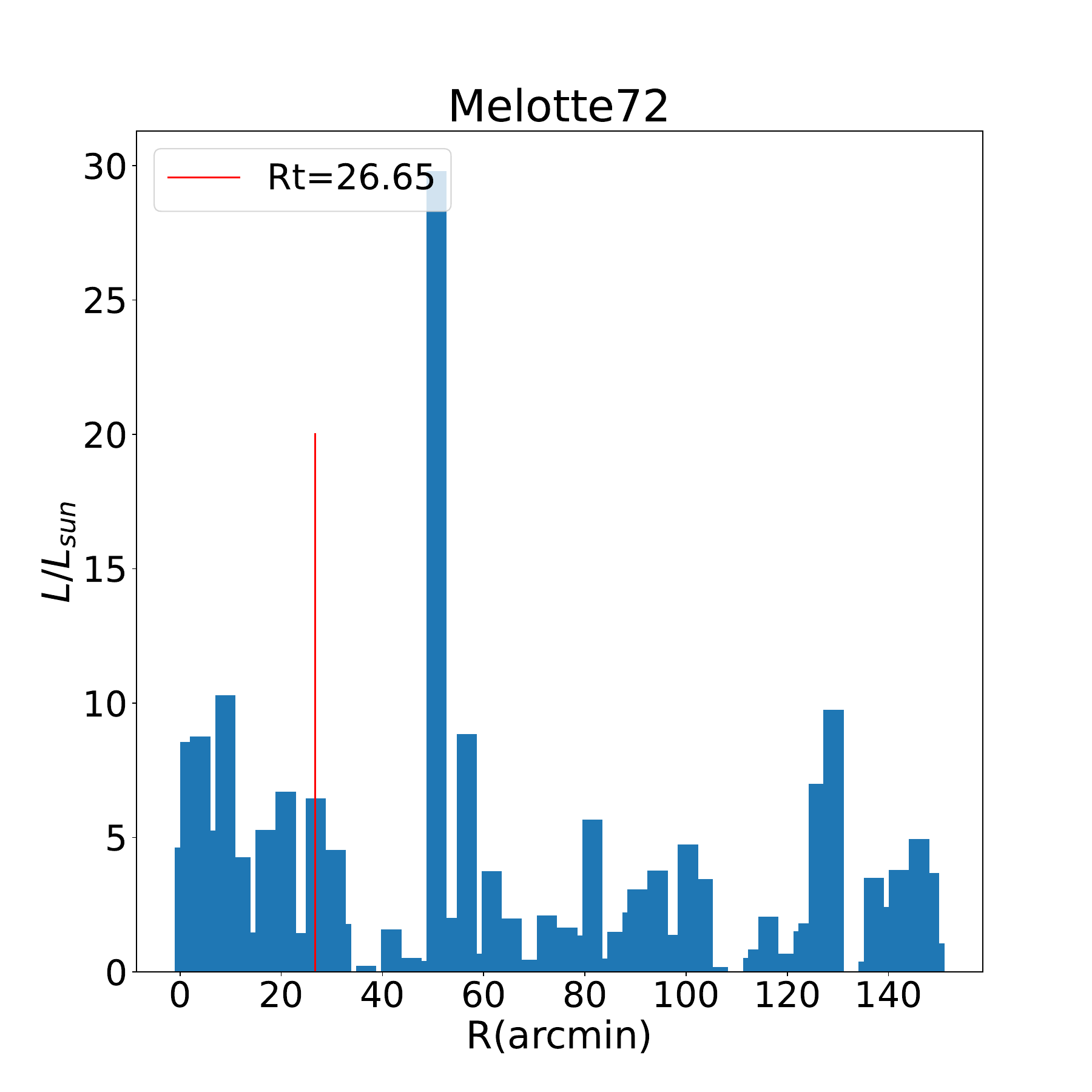}

        \end{subfigure}
        \begin{subfigure}{0.25\textwidth}
                \centering

                \includegraphics[width=\textwidth]{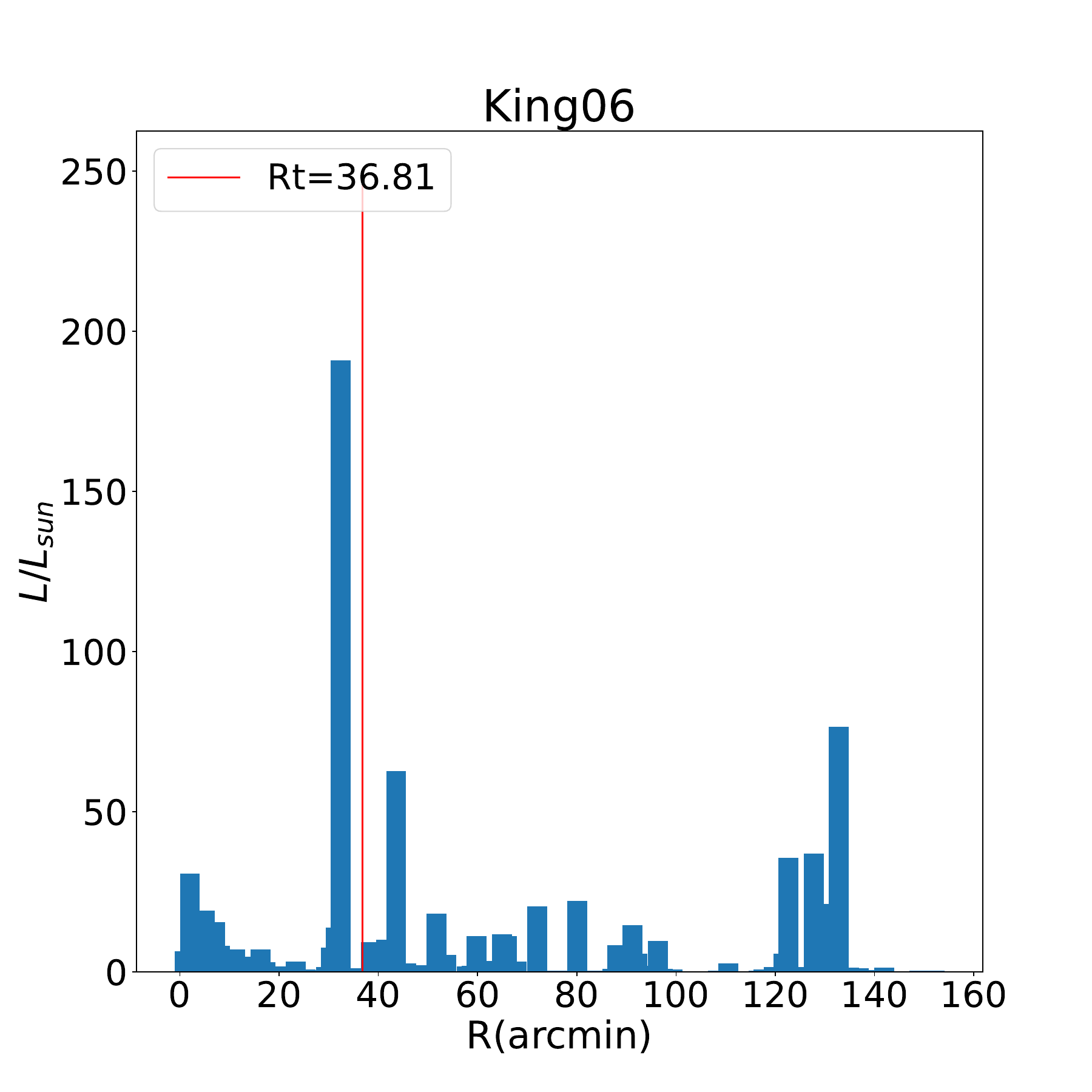}

        \end{subfigure}
        \begin{subfigure}{0.25\textwidth}
                \centering

                \includegraphics[width=\textwidth]{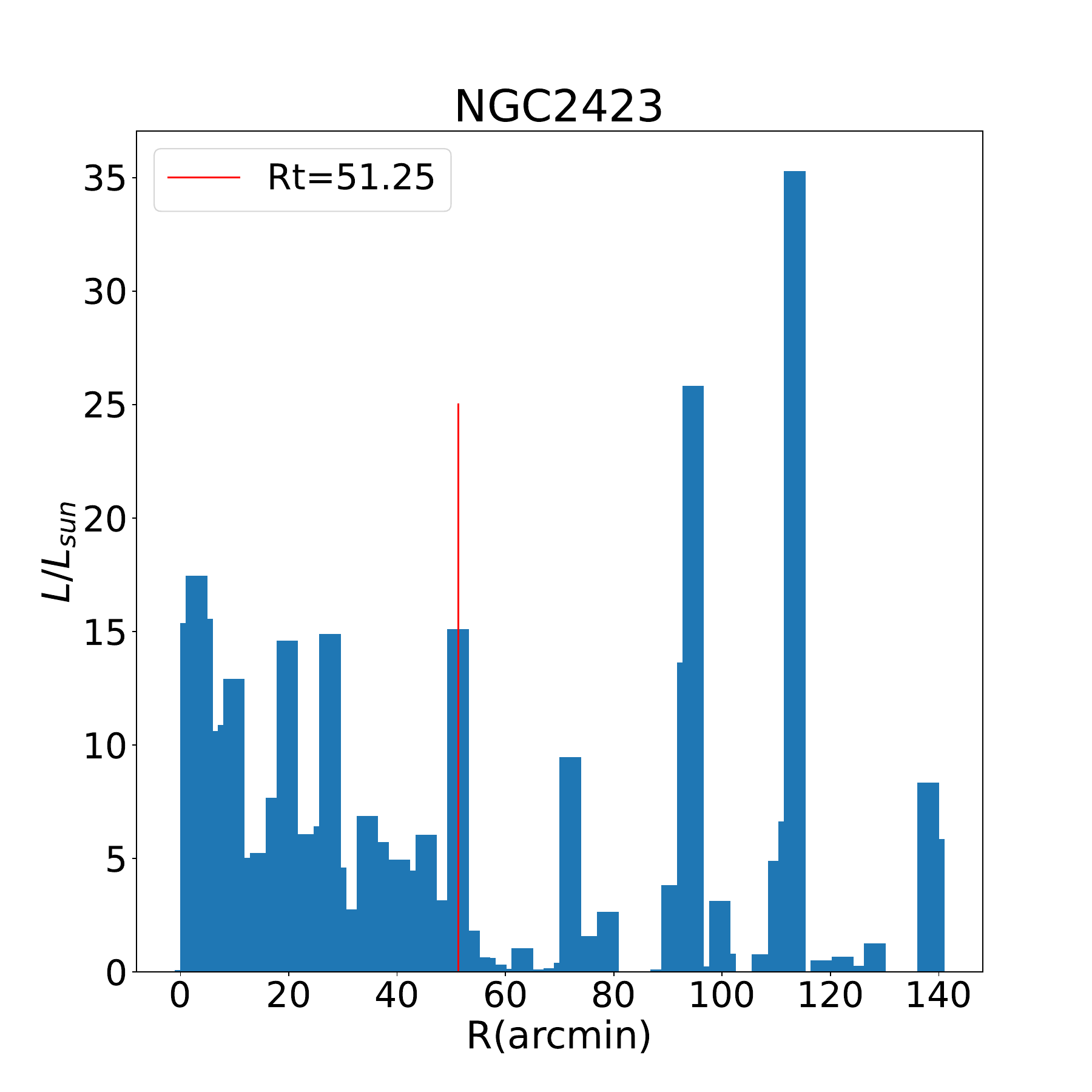}

        \end{subfigure}
        \begin{subfigure}{0.25\textwidth}
                \centering

                \includegraphics[width=\textwidth]{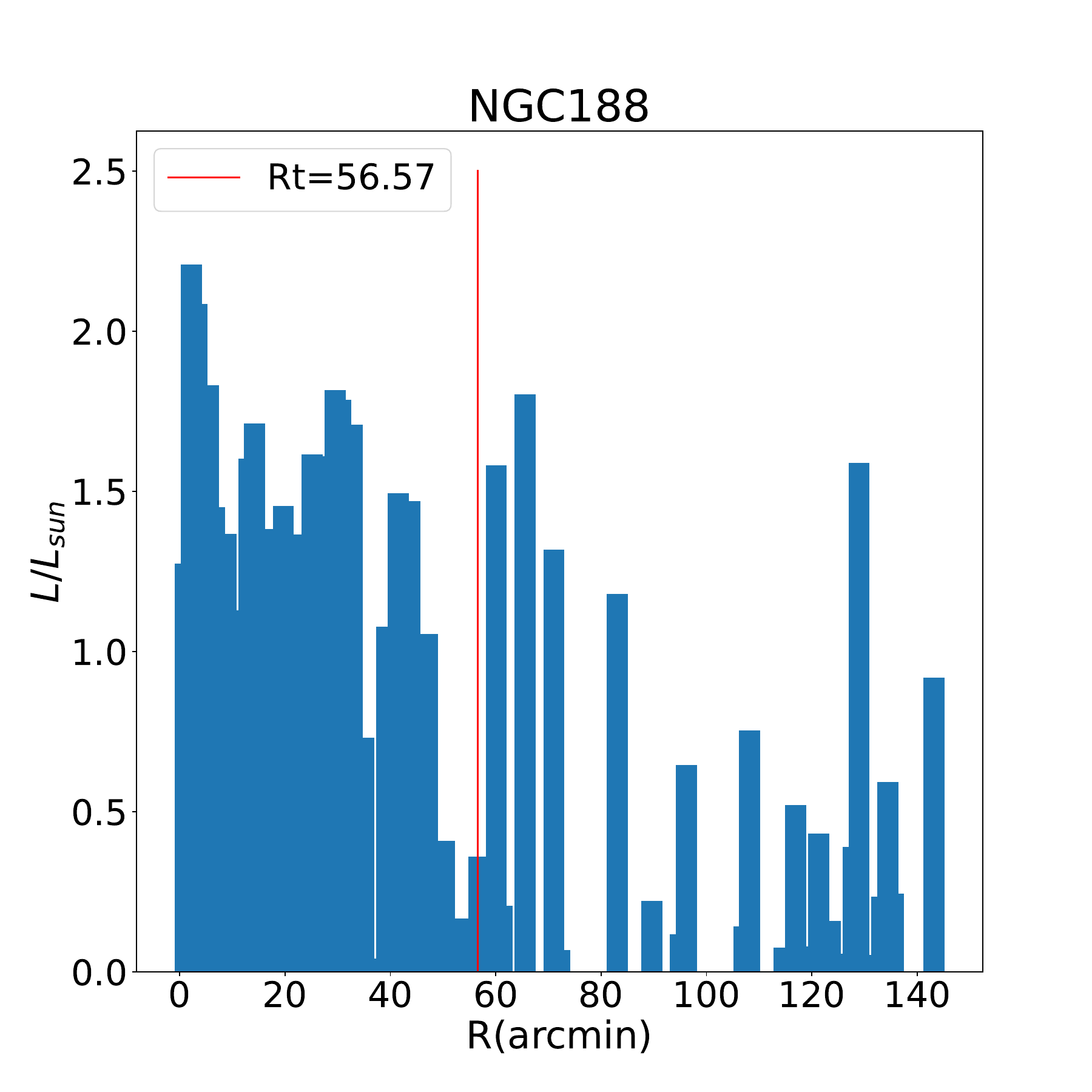}

        \end{subfigure}
        \begin{subfigure}{0.25\textwidth}
        \centering
           \includegraphics[width=\textwidth]{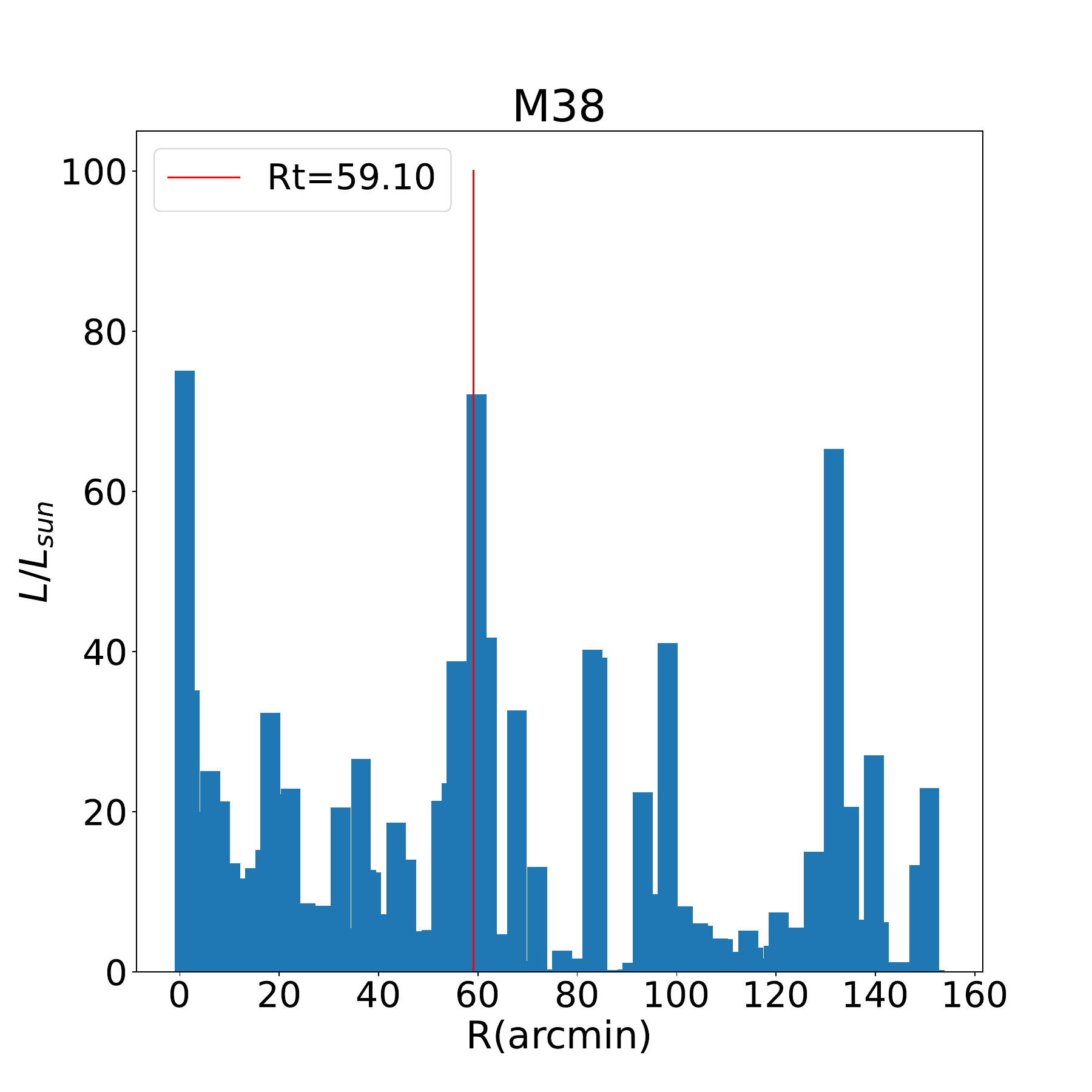}

        \end{subfigure}
        \begin{subfigure}{0.25\textwidth}
        \centering
           \includegraphics[width=\textwidth]{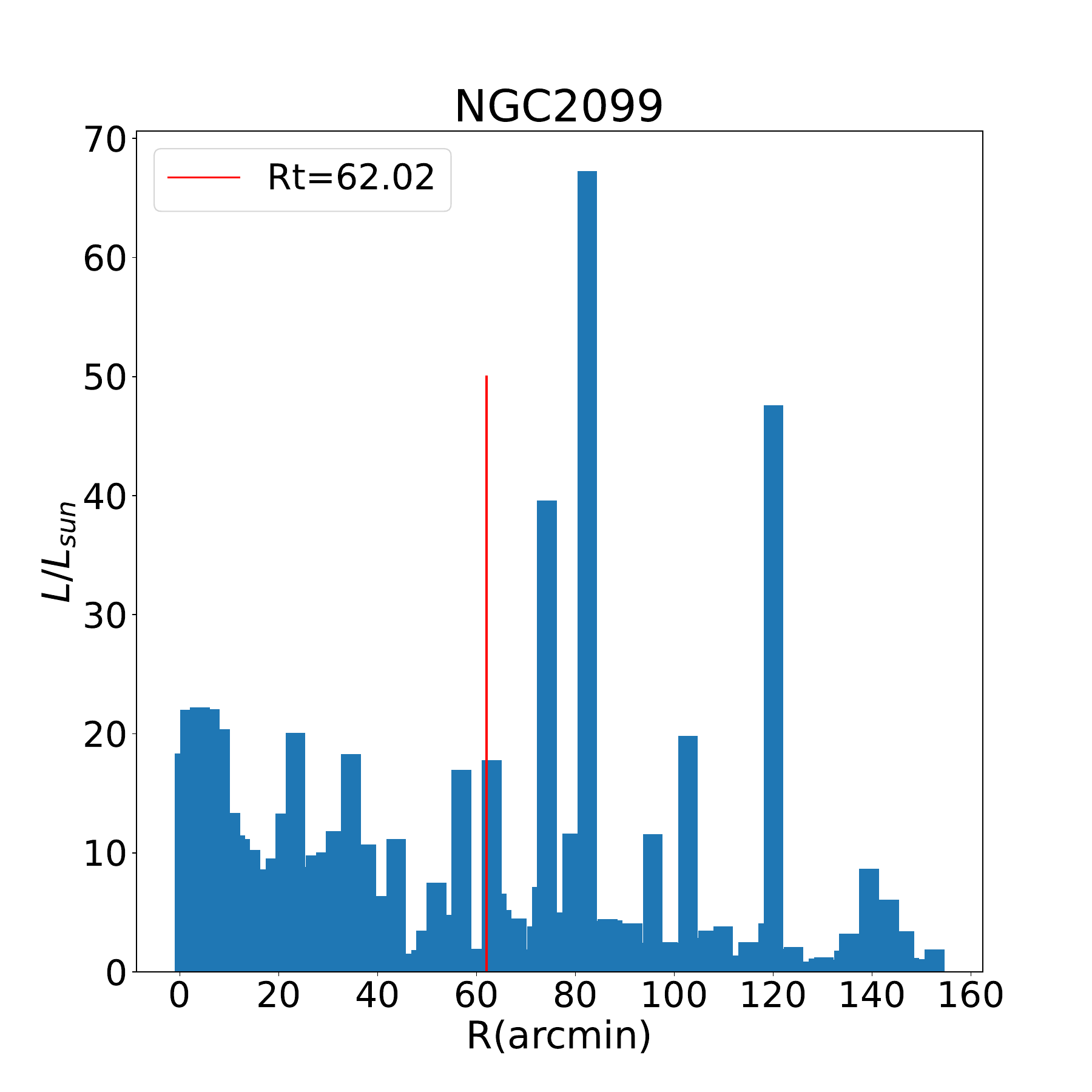}

        \end{subfigure}
        \begin{subfigure}{0.25\textwidth}
        \centering
           \includegraphics[width=\textwidth]{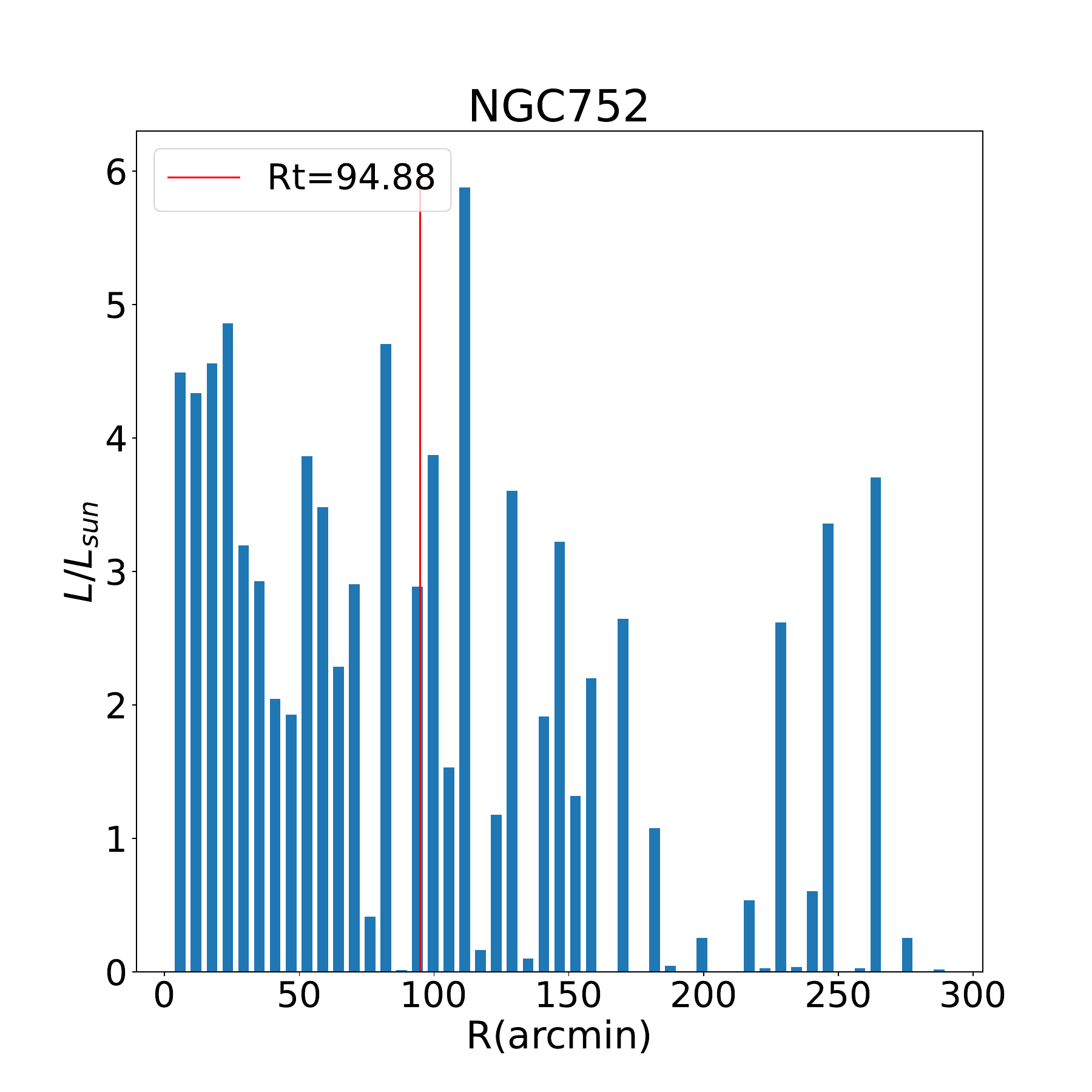}

        \end{subfigure}
        \begin{subfigure}{0.25\textwidth}
        \centering
           \includegraphics[width=\textwidth]{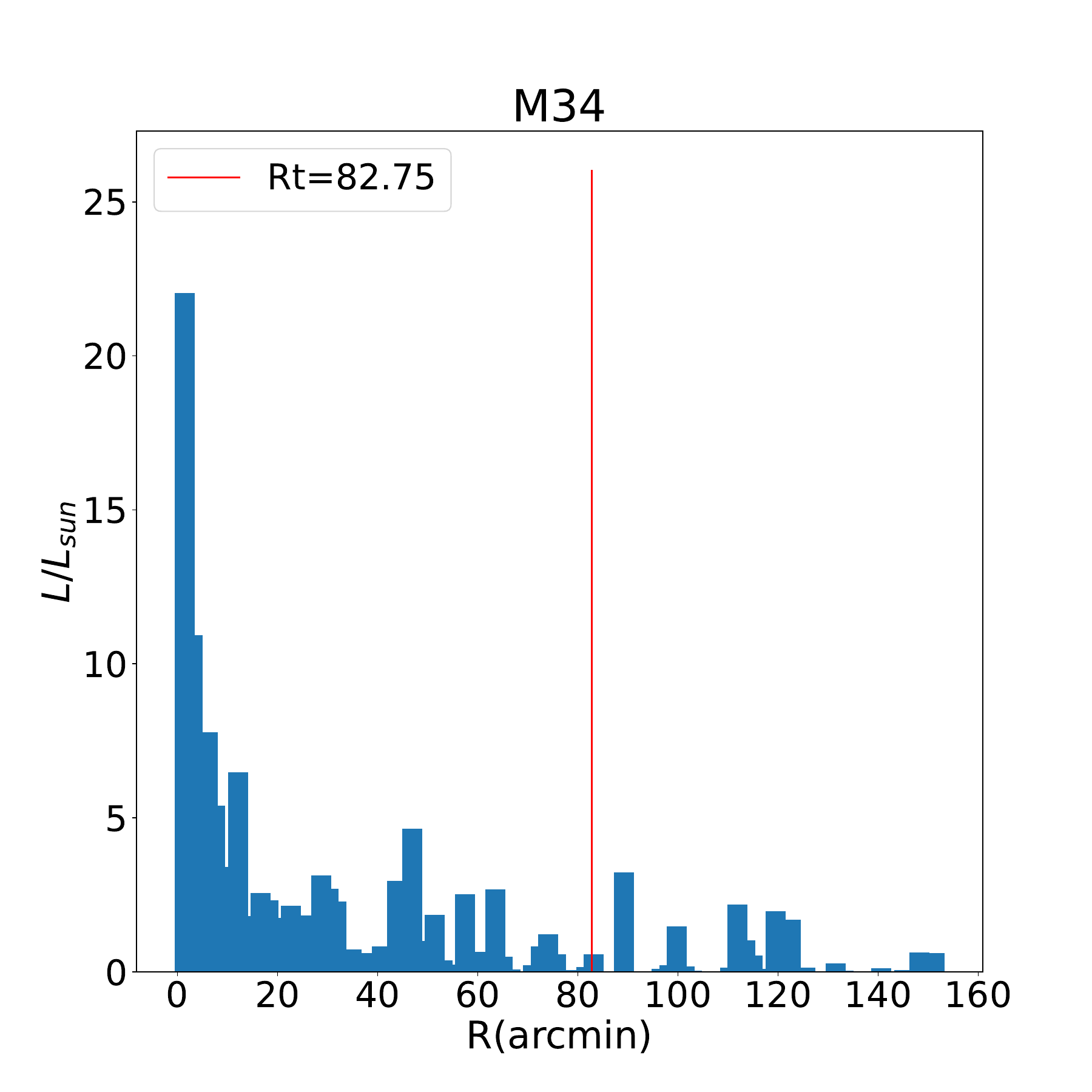}

        \end{subfigure}
        \begin{subfigure}{0.25\textwidth}
        \centering
           \includegraphics[width=\textwidth]{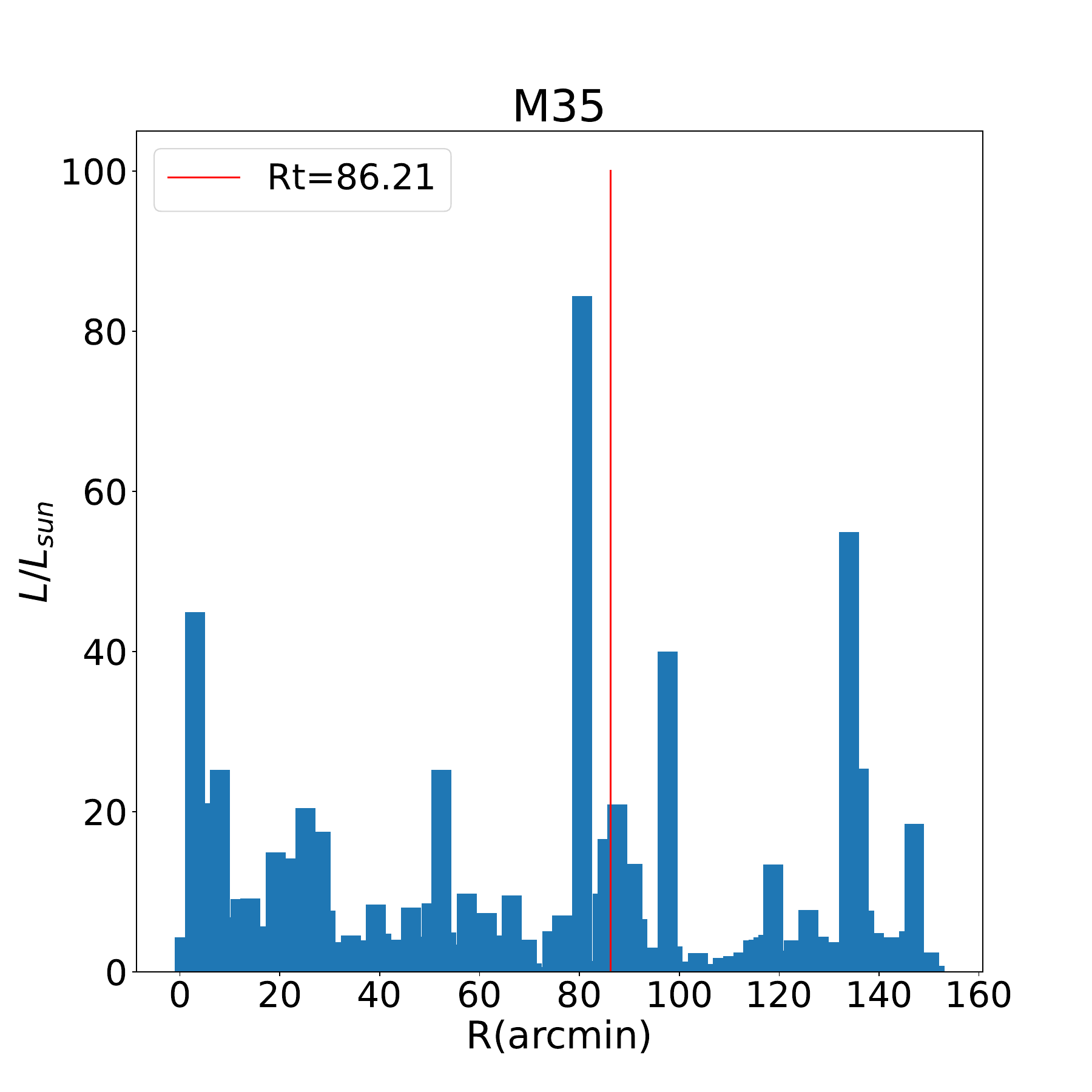}

        \end{subfigure}
        \begin{subfigure}{0.25\textwidth}
        \centering
           \includegraphics[width=\textwidth]{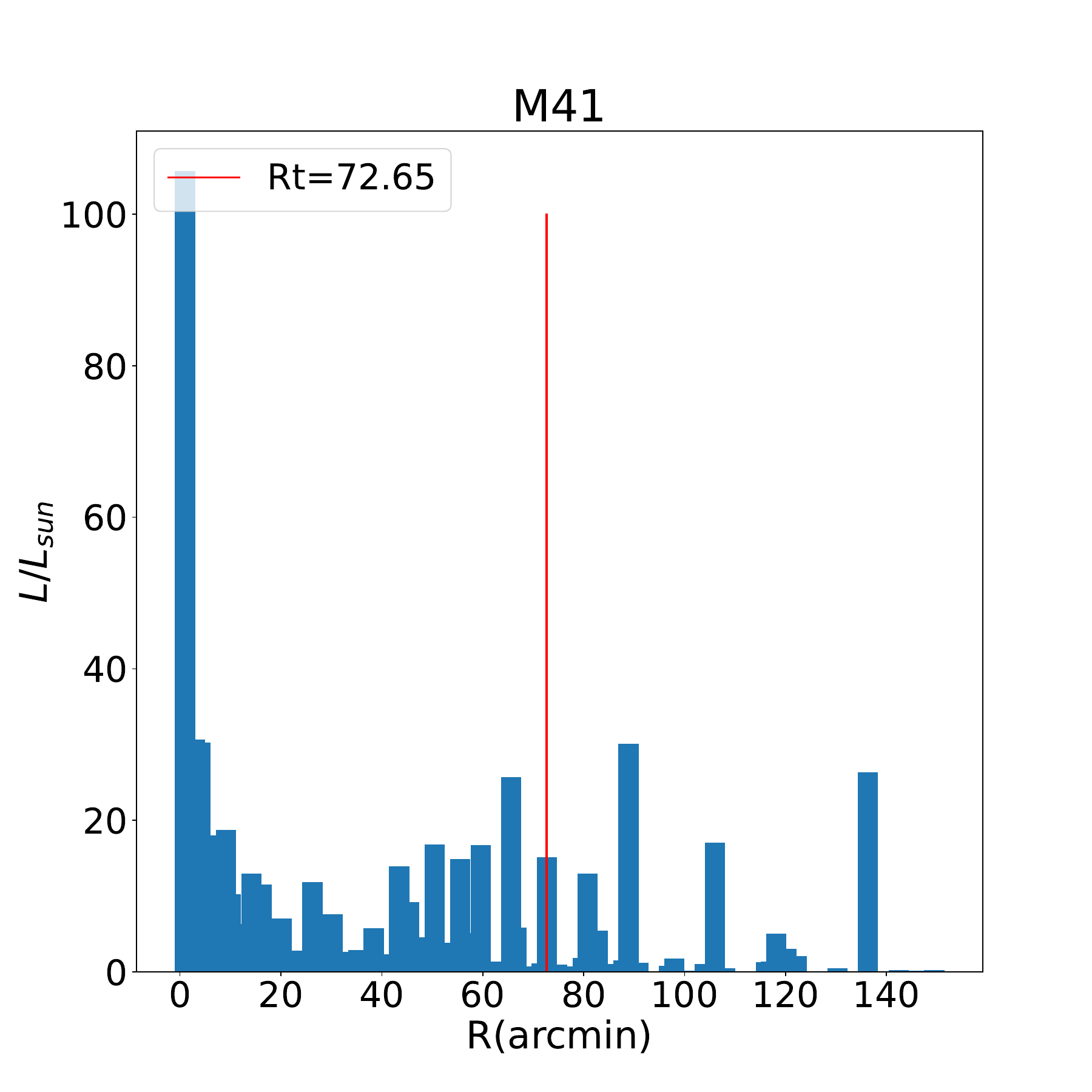}

        \end{subfigure}
        \begin{subfigure}{0.25\textwidth}
        \centering
           \includegraphics[width=\textwidth]{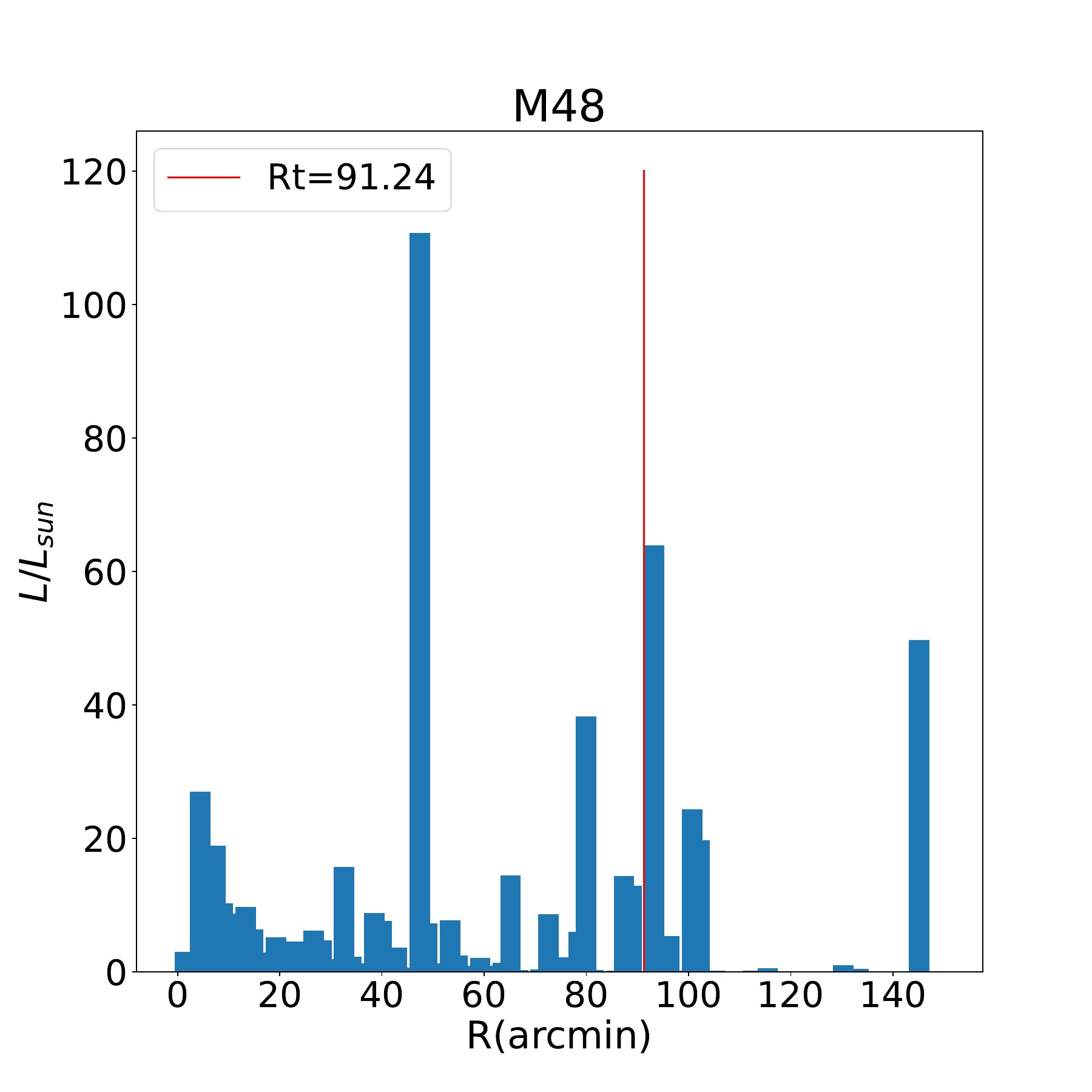}

        \end{subfigure}
        \begin{subfigure}{0.25\textwidth}
        \centering
           \includegraphics[width=\textwidth]{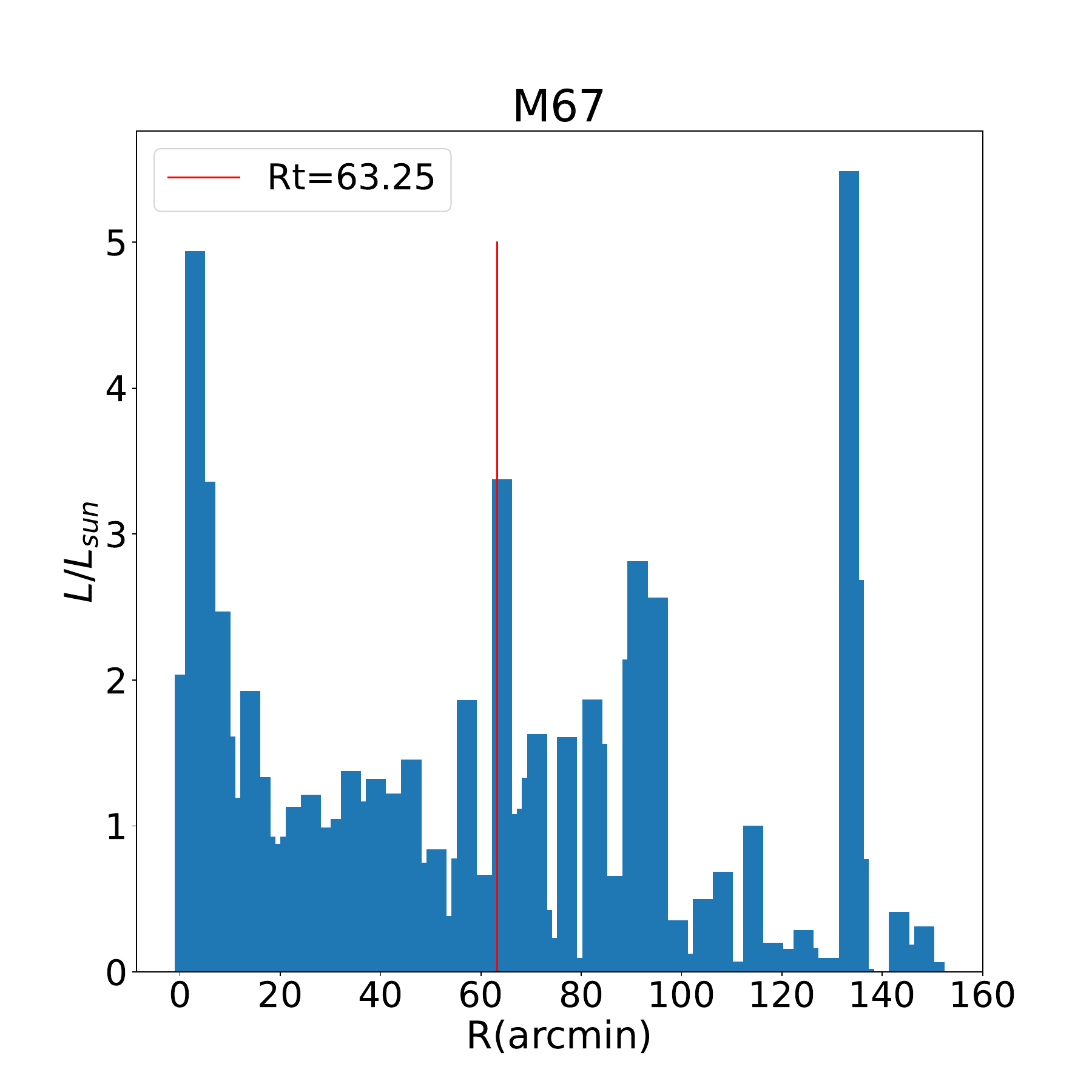}

        \end{subfigure}
  \caption{The average luminosity of main sequence stars across different areas of the cluster is depicted. Additionally, the red line represents the tidal radius.}
  \label{l.fig}
\end{figure}
\twocolumn
\noindent
\begin{equation}\label{dens.eq}
  n(r)=\frac{N_i}{4\pi({r_{i+1}}^2-{r_i}^2)},
\end{equation}
\begin{equation}\label{king.eq}
  f(r)=f_b+\frac{f_0}{1+(\frac{r}{R_C})^{2}},
\end{equation}
\begin{equation}\label{tidal.eq}
  R_t=R_C\sqrt{\frac{f_0}{3\sigma_b}-1},
\end{equation}
As seen in Fig~\ref{CMD dgr.fig}, for stars with a magnitude fainter than 18, the selection of stars by the Random Forest increased, which could be due to mass segregation. Luminosity calculation need information about reddening. We used other works for data about reddening for each cluster. For King\,06, Melotte\,72, M\,38, M\,41, M\,48, M\,67, and NGC\,2423 we used value of $A_v$ from~\cite{mass-extended}. For other clusters, we used value of $E(B-V)$ and approximation $A_v=3.1E(B-V)$(\cite{Rodr3.1}). The references of $A_v$ are shown in~Table~\ref{tab_discussion}. To study mass segregation in clusters, we divided cluster members into three categories: $\frac{L}{Lsun}>2$ (high-mass stars), $0.1<\frac{L}{Lsun}<1$ (middle-mass stars), $\frac{L}{Lsun}<0.05$ (low-mass stars) and after that, the cumulative distribution function (CDF) was calculated. Fig~\ref{mass-tidal.fig} shows the cumulative distribution function (CDF) diagram for each cluster. It should be mentioned that for all clusters, main sequence inside tidal radius was considered. In more distant clusters, low-mass stars have been overlooked.\\
In Collinder\,463, high-mass and middle-mass stars are segregated, but in the case of low-mass stars, it is not observed. In old open clusters, M\,67, mass segregation occurs completely.\\
We calculated cluster mean luminosity in each central ring and showed data in Fig~\ref{l.fig}. As shown in Fig~\ref{l.fig}, the luminosity of clusters has decreased from the center to the outer layer of the cluster. However, one luminosity peak has been observed either inside or outside the cluster's tidal radius, which will be studied in future works.\\
\section{CONCLUSION}\label{con.Data}
For a comprehensive study of star clusters, including aspects such as membership inside and outside of the tidal radius, tidal tail morphology, formation and evolution of stars within clusters, and determination of cluster ages, we require a method capable of identifying reliable members across the wide field of view encompassing these clusters. In our previous work, we successfully identified reliable cluster members by combining two unsupervised machine-learning algorithms: DBSCAN and GMM. Applying our method to 12 distinct open clusters, we demonstrated its effectiveness in identifying reliable members within the tidal radius. However, the method also detected outside members that lay within a range of proper motion, parallax, and color-magnitude diagrams associated with high probability selection members.\\
In the current study, we take a step further from our previous work by incorporating a supervised machine learning algorithm, Random Forest. With this method, we successfully identified outside members of 15 open clusters across the wide field of view, revealing the morphology of clusters at greater distances. Additionally, through fitting the King profile, we calculated the tidal radius and detected members beyond this radius.\\
With a comprehensive view of cluster members, we searched for mass segregation in the understudy cluster and explored cluster luminosity. We found one peak of cluster luminosity far away from the cluster center; in some clusters, the peak is outside the tidal radius. The data obtained using this approach holds significant value for researching cluster's evolution, evaporation processes, interactions between the Galaxy and clusters, and theories related to star formation within these clusters.
\section*{DATA AVAILABILITY}
The data used in this work are Gaia DR3 available at \url{https://gea.esac.esa.int/archive/} and we are ready to send our data to any research request.
\bibliographystyle{mnras}
\bibliography{ref}

\end{document}